\documentclass[a4paper,fleqn,twocolumn]{cas-dc}
\usepackage[numbers]{natbib}
\usepackage{lscape} 
\usepackage{graphicx}

\usepackage{tocbasic}
\DeclareTOCStyleEntry[
  beforeskip=.2em plus 1pt,
  pagenumberformat=\textbf
]{tocline}{chapter}

\usepackage{setspace}

\usepackage{dcolumn}
\usepackage{multirow}
\usepackage{amsmath}
\usepackage{bm}
\usepackage{textcomp}
\usepackage{epsfig}
\usepackage{rotating}
\usepackage{xfrac}
\usepackage{threeparttable}
\usepackage{lipsum} 
\usepackage{longtable}
\usepackage{stfloats}
\usepackage[many]{tcolorbox}
\usepackage{multicol}
\usepackage{widetext}
\newtcolorbox{boxA}{
 sharp corners,
 colback = white,
 colbacklower=white,
 boxrule=0pt,
 colframe = white 
}

\RequirePackage{hyperref}
\hypersetup
{
bookmarks=true,%
unicode=false,%
pdftoolbar=true,
pdfmenubar=true,
pdffitwindow=false,
pdfstartview=FitH,
pdftitle={ADNDT},
pdfauthor={LAM Yi Hua},
pdfsubject={Nuclear Physics},
pdfcreator={LAM Yi Hua},
pdfproducer={LAM Yi Hua},
pdfkeywords={Nuclear Physics},
bookmarksnumbered,%
colorlinks=true,%
linkcolor=blue,%
citecolor=magenta,%
filecolor=green,%
urlcolor=blue,
breaklinks, 
plainpages=false%
}

\usepackage{xcolor}
\definecolor{Mycolor}{HTML}{bfffbf}
\definecolor{blue}{HTML}{4477aa} 
\definecolor{red}{HTML}{ee6677} 
\definecolor{green}{HTML}{228833}
\definecolor{magenta}{HTML}{ee3377}
\definecolor{cyan}{HTML}{66ccee}
\definecolor{yellow}{HTML}{ccbb44}
\definecolor{grey}{HTML}{bbbbbb}

\def\tsc#1{\csdef{#1}{\textsc{\lowercase{#1}}\xspace}}
\tsc{WGM}
\tsc{QE}
\tsc{EP}
\tsc{PMS}
\tsc{BEC}
\tsc{DE}

\begin{document}
\let\WriteBookmarks\relax
\def\floatpagepagefraction{1}
\def\textpagefraction{.001}
\shorttitle{Nuclear ground-state properties probed by the relativistic Hartree-Bogoliubov approach}
\shortauthors{Z. X. Liu, Y. H. Lam, N. Lu, P. Ring} 
\title[mode = title]{\Large Nuclear ground-state properties probed by the relativistic Hartree-Bogoliubov approach$^\dag$}

\author[1,2,3]{Zi Xin Liu}[%
style=,
type=editor,
auid=000,
bioid=1,
prefix=,
suffix=,
role=,
orcid=0000-0001-5652-1516]
\cormark[1]
\ead{liuzixin1908@impcas.ac.cn}
\credit{Conceptualization of this study, Data curation - calculating and listing the data set, Formal analysis, Investigation, Methodology, Resources, Software, Validation, Visualization, Writing - original draft preparation, Writing - review \& editing}
\address[1]{Institute of Modern Physics, Chinese Academy of Sciences, Lanzhou 730000, People's Republic of China}
\address[2]{School of Nuclear Science and Technology, University of Chinese Academy of Sciences, Beijing 100049, People's Republic of China}
\address[3]{School of Physics Science and Technology, Lanzhou University, Lanzhou 730000, People's Republic of China}

\author[1,2]{Yi Hua Lam}[%
style=,
type=editor,
auid=000,
bioid=1,
prefix=,
suffix=,
role=,
orcid=0000-0001-6646-0745]
\cormark[1]
\ead{lamyihua@impcas.ac.cn}
\credit{Conceptualization of this study, Data curation - calculating and listing the data set, Formal analysis, Funding acquisition, Investigation, Methodology, Resources, Software, Supervision, Validation, Visualization, Writing - original draft preparation, Writing - review \& editing, Project administration}

\author[1,4,2]{Ning Lu}[%
style=,
type=editor,
auid=000,
bioid=1,
prefix=,
suffix=,
role=,
orcid=0000-0002-3445-0451]
\credit{Data curation - preparing the experimental data set, Visualization, Writing - Original draft preparation}
\address[4]{School of Nuclear Science and Technology, Lanzhou University, Lanzhou 730000, People's Republic of China}

\author[5]{Peter Ring}[%
style=,
type=editor,
auid=000,
bioid=1,
prefix=,
suffix=,
role=,
orcid=0000-0001-7129-2942]
\credit{Important suggestions, Writing - Original draft preparation}
\address[5]{Fakult{\"a}t f{\"u}r Physik, Technische Universit{\"a}t München, D-85748 Garching, Germany}


\begin{abstract}
Using the relativistic Hartree–Bogoliubov approach with separable pairing force coupled with the latest point-coupling and meson-exchange covariant density functionals, i.e., PC-L3R, PC-X, DD-MEX, and DD-PCX, we systematically explore the ground-state properties of all isotopic chains from oxygen ($Z\!=\!8$) to darmstadtium ($Z\!=\!110$).
These properties consist of the binding energies ($E_\mathrm{b}$), one- and two-neutron separation energies ($S_\mathrm{n}$ and $S_\mathrm{2n}$), root-mean-square radii of matter ($R_\mathrm{m}$), of neutron ($R_\mathrm{n}$), of proton ($R_\mathrm{p}$) and of charge ($R_\mathrm{c}$) distributions, Fermi surfaces ($\lambda$), ground-state spins ($J$) and parities ($\pi$). 
We then use these calculated properties to predict the edges of nuclear landscape and bound nuclei for the isotopic chains of $Z\!=\!8$-$110$. The number of bound nuclei predicted by PC-L3R, PC-X, DD-MEX, and DD-PCX, are $9004$, $9162$, $7112$, and $6799$, respectively. 
These latest covariant density functionals produce a set of rather similar proton drip lines due to the strong repulsive Coulomb force shifting up the single-proton energy of the proton-rich nuclei. PC-L3R and PC-X estimate more extended borders of the neutron-rich region compared with the neutron drip lines estimated by DD-MEX, and DD-PCX. 
Meanwhile, the root-mean-square deviations of one- (two-) neutron separation energies yielded from PC-L3R, PCX, DD-MEX, and DD-PCX are $0.962$~($1.300$)~MeV, $0.920$~($1.483$)~MeV, $1.010$~($1.544$)~MeV, and $0.993$~($1.753$)~MeV, respectively. The deviations of theoretical $S_\mathrm{n}$, $S_\mathrm{2n}$, and charge radii from the available experimental ones increase at the regions further away from the proton magic numbers, indicating the important role of deformation in these regions. 
The root-mean-square deviations of charge radius distributions of comparing the available experimental values with the theoretical counterparts resulted from PC-L3R, PC-X, DD-MEX, and DD-PCX are $0.035$~fm, $0.037$~fm, $0.034$~fm, and $0.035$~fm, respectively. 
We compute and present the root-mean-square radii of neutron distributions for all isotopes of $8\!\leq\! Z\!\leq\!110$ with respect to the empirical $R_{\rm n}$. 
Basically, the systematic trends of the theoretical root-mean-square radii of neutron distributions generated from PC-L3R and PC-X closely follow the empirical $R_{\rm n}$, except for the region of extreme neutron-rich nuclei, whereas DD-PCX produces a trend lower than the empirical $R_{\rm n}$ at the $N\!<\!150$ region. We notice pronounced differences between the empirical and theoretical $R_{\rm n}$ at nuclei near the neutron drip line of the Mg, Ca, and Kr isotopic chains, suggesting the possible existence of the halo or giant halo phenomena. 
\end{abstract}




\received{14 September 2023}
\revised {12 December 2023} 
\accepted{18 December 2023}

\maketitle



{
\small 
\begin{spacing}{0.97}
\tableofcontents
\end{spacing}
}
\clearpage
\newpage


\section{Introduction}
\label{sec:intro}

Nuclear mass and radius are two of the key observables directly manifesting the important characteristics of the nuclear structure. To date, the masses of about 2500 nuclides have been experimentally determined. The proton drip line has been experimentally determined up to protactinium, whereas the neutron-rich boundary is known only up to oxygen~\cite{AME2020}. Knowing the nuclear masses (or binding energies) is of crucial importance not only to understand the nucleon-nucleon interactions, weak interactions, and nuclear structure, but also to investigate the nucleosyntheses at various extreme astrophysical environments. Moreover, nuclear charge radius could be the potential evidence of new magic numbers and the disappearance of traditional magic number with considering the impact of nuclear shell effect on charge radii~\cite{Nat2018Tran}, and the difference between root-mean-square radius of charge and neutron distributions could be used to probe the existence of neutron halo in exotic nuclei at the neutron-rich region.

With entering the era of new generation large-scale radioactive-ion beam facilities, including the Second Generation System On-Line Production of Radioactive Ions (SPIRAL2) at GANIL in France~\cite{NPA2010Gales}, the Radioactive Ion Beam Factory (RIBF) at RIKEN in Japan~\cite{NPA2010Motobayashi}, the Facility for Rare Isotope Beams (FRIB) in the US~\cite{NPA2010Thoennessen}, the Advanced Rare Isotope Laboratory (ARIEL) in Canada, the Facility for Antiproton and Ion Research (FAIR) in Germany~\cite{NPA2010Sturm}, the Rare Isotope Accelerator complex for Online Experiments (RAON) in Korea~\cite{NIMPR2013Tshoo}, and the Heavy-Ion Accelerator Facility (HIAF) in China~\cite{AB2022Zhou}, the advancements of such facilities enable solving those previously challenging experiments, determining more masses and radii of exotic nuclei, revealing the important physics related to the exotic nuclei far from the $\beta$-stability valley, and thus permits us to deepen our understanding of the origins of the chemical elements in the Universe.

Accessing the neutron-rich nuclei far from the valley of stability is still somehow a challenge to the current and foreseeable future experimental capabilities. Therefore, a set of high descriptive power theoretical frameworks could be advantageous for us to predict the unmeasured nuclear properties of exotic nuclei around the neutron-rich edge, to guide future experiments, and to obtain the most reliable nuclear physics inputs for astrophysical interests. Conversely, the capability of a theoretical framework in describing nuclear masses and radii can be probed or even be improved with new and more precise experimental findings \cite{PLB2023Liu}. In addition, obtaining a proven high descriptive power theoretical framework permits us to study the equation of states correlating the saturation properties of the symmetric nuclear matter and astrophysical compact objects.

In recent years, the covariant density functional theory (CDFT) has been introduced to the nuclear physics community \cite{PPNP1996Ring_37_193,ZPA1991Kucharek_339_23,PRL1996Meng_77_3963,PRL1997Poschl_79_3841,JPG2017Lv,PRC2018Liu,ADNDT2018Xia,PLB2018Afanasjev,PRC2021Fan,PRC2021Vale,ADNDT2022Zhang}. Furthermore, CDFT can be an alternative for generating a reliable set of relevant nuclear physics inputs for astrophysical modelings. It receives wide attention as its capability in describing experimentally determined nuclear properties has been proven, especially it obeys basic symmetries of quantum chromodynamics, i.e., preserving the Lorentz invariance, which not only automatically includes the spin-orbit coupling, but also puts stringent restrictions on the number of phenomenological parameters without losing the good agreement with experimental data. Hence, the CDFT has been widely used in studying the pseudospin symmetry in nucleon spectrum~\cite{PRC1998Meng,PRC1998MengLalazissis-58-R45,PRC2010Long,PRC2012Chen}, and spin symmetry in anti-nucleon spectrum~\cite{PRL2003Zhou}.

Over the past three decades, several types of CDFT were proposed, e.g., (i) meson exchange with non-linear meson couplings (NL-ME), (ii) meson exchange with density-dependent meson-nucleon couplings (DD-ME), (iii) non-linear point-coupling model (NL-PC), and (iii) density-dependent point-coupling model (DD-PC). The covariant density functionals commonly used for describing a variety of nuclear phenomena are categorized to the meson-exchange nucleon couplings, i.e., DD-ME1~\cite{PRC2002Niksic} and DD-ME2~\cite{PRC2005Lalazissis_71_024312}, and to the point-coupling interactions, i.e., PC-PK1~\cite{PRC2010Zhao} and DD-PC1~\cite{PRC2008Niksic}. Recently, a number of new point-coupling interactions were proposed, i.e., DD-PCX~\cite{PRC2019Yuksel}, DD-MEX~\cite{PLB2020Taninah}, PC-X~\cite{PLB2020Taninah}, and PC-L3R~\cite{PLB2023Liu}. The DD-PCX interaction was fitted to observables of the selected open-shell nuclei, i.e., the binding energies (34~nuclei), charge radii (26~nuclei), mean pairing gaps (15~nuclei), and to two observables of collective excitations with the constrained isoscalar giant monopole resonance energy and dipole polarizability of $^{208}$Pb~\cite{PRC2019Yuksel}. The fitting protocols of DD-MEX and PC-X interactions are identical to the procedures of the DD-ME2 and PC-PK1 interactions, see Refs.~\cite{PRC2005Lalazissis_71_024312,PRC2010Zhao}. The latest non-linear point-coupling interaction is the PC-L3R interaction, which is stringently constrained by reproducing the binding energies of 91 spherical nuclei, charge radii of 63 nuclei, and 12 sets of mean pairing gaps consisting of a total of 54 nuclei~\cite{PLB2023Liu}. The PC-X, DD-MEX, DD-PCX, and PC-L3R interactions are the covariant density functionals constructed for the relativistic Hartree-Bogoliubov (RHB) approach with a separable pairing force, which is developed to unify the treatment of pairing correlations and mean-field potentials \cite{PR2005Vretenar_409_101}. The RHB approach has been successfully used in the studies of nuclear structure and properties, for instance, the description of nuclear masses~\cite{ADNDT2018Xia} and charge radii~\cite{PRC2021Perera}, the prediction of the neutron drip line~\cite{PRC2021Zhang}, formation of neutron halos~\cite{PRC2002Meng,PRC2010Zhou}, nuclear shape coexistence~\cite{PRC2022Seonghyun}, and ﬁssion barriers of hyperheavy nuclei~\cite{PLB2018Afanasjev,PRC2019Agbemava}.

As the development of radioactive ion beam facilities opens the opportunity of determining the properties of unexplored exotic nuclei, especially the excess neutron region near to the edge of nuclear landscape, predicting such nuclear properties could be useful for constructing new experiments, be applicable to probe the new covariant density functionals, and be beneficial to generate reliable nuclear inputs for astrophysical modelings. 
In this work, we aim to systematically study the ground state properties of atomic nuclei throughout a vast range of nuclear landscape and also to predict the proton and neutron drip lines using the RHB approach based on the latest point-coupling interactions, i.e., PC-X, DD-PCX, and PC-L3R, and the latest meson-exchange interaction, DD-MEX. These properties consist of the nuclear masses (or binding energies), one- and two-nucleon separation energies, charge radii, the root-mean-square radii of neutron and of proton distributions. We adopt the configuration space constructed from the spherical harmonic oscillator wave functions. In Sec.~\ref{sec:theory}, the meson-exchange and point-coupling models are briefly introduced, and the RHB approach are presented as well. Then, we illustrate and discuss the calculated properties of all isotopes from $Z\!=\!8$-$110$ in Sec.~\ref{sec.Res}. The summary is given in Sec.~\ref{sec:summary}. 



\section{Theoretical framework}
\label{sec:theory}

The covariant density functional theory (CDFT) is one of the successful models for studying the nuclear structures of almost the whole nuclear landscape. A covariant density functional is constructed from the nucleon-nucleon interactions based on either the finite-range meson-exchange interaction or the contact interaction between nucleons in the point-coupling representation. 
With the advantages of using covariant density functionals, i.e., (a) the natural inclusion of the nucleon spin degree of freedom, (b) the consistent treatment of isoscalar Lorentz scalar and vector self-energies, which provides a unique parameterization of time-odd components of the nuclear mean field, (c) a natural explanation of the empirical pseudospin symmetry, (d) a new saturation mechanism of nuclear matter with a distinction between scalar and four-vector nucleon self-energies~\cite{PR1999Afanasjev,JPG2015MengJ_42_093101,PR2015Liang}, we can describe and even predict the nuclear masses, charge radii, single- and two-nucleon separation energies, proton and neutron drip lines, and halo nuclei. 

In this section, we give brief descriptions for the meson-exchange and point-coupling models, and the relativistic Hartree-Bogoliubov approach.

\subsection{Meson-exchange models}
\label{sec:ME_model}

A nuclide is described as a system of Dirac fermions coupled to meson exchange through an effective Lagrangian according to the approach of conventional quantum hadrodynamics (QHD)~\cite{AP1974Walecka}. In the finite-range meson-exchange representation, the isoscalar-scalar $\sigma$ meson, isoscalar-vector $\omega$ meson, and isovector-vector $\rho$ meson assemble the minimal set of meson fields for describing the bulk and single-particle nuclear properties.


\subsubsection{The nonlinear meson-exchange model}
\label{sec:NLME}

The Lagrangian density of the nonlinear meson-exchange model generally includes the nucleon field $\psi$, mesons field $\sigma$, $\omega_{\mu}$, and $\bar{\rho}_{mu}$, and the electromagnetic field $A_{\mu}$, and is written as \cite{PPNP1996Ring_37_193,PRL1992Brockmann, NPA1994Sugahara,PRC2004Long}
\begin{eqnarray}
\nonumber \mathcal{L}_{\rm NLME} \!\!&=&\!\! \bar{\psi}[\gamma_{\mu}(i\partial^{\mu}-g_{\omega}\omega^{\mu}-g_{\rho}\vec{\tau}\cdot\vec{\rho}_{\mu})\\
&&\nonumber-(M-g_{\sigma}\sigma)-e\gamma^{\mu}A_{\mu}\frac{1-\tau_{3}}{2}]\psi\\
\nonumber&& + \frac{1}{2}(\partial^{\mu}\sigma\partial_{\mu}\sigma-m_{\sigma}^{2}\sigma^{2})-\frac{1}{3}g_{2}\sigma^{3}-\frac{1}{4}g_{3}\sigma^{4}\\
\nonumber&& - \frac{1}{4}\Omega^{\mu\nu}\Omega_{\mu\nu}+\frac{1}{2}m^{2}_{\omega}\omega^{\mu}\omega_{\mu}+\frac{1}{4}c_{3}(\omega^{\mu}\omega_{\mu})^2\\
&&\nonumber -\frac{1}{4}\vec{R}^{\mu\nu}\cdot\vec{R}_{\mu\nu}+\frac{1}{2}m^{2}_{\rho}\vec{\rho}^{\mu}\cdot\vec{\rho}_{\mu}  
  +\frac{1}{4}d_{3}[\vec{\rho}^{\mu}\cdot\vec{\rho}_{\mu}]^{2}\\
  &&-\frac{1}{4}F^{\mu\nu}F_{\mu\nu} \label{L:NLME}
\end{eqnarray}
where $M$ is the mass of the nucleon, and $m_{\sigma}$, $m_{\omega}$, $m_{\rho}$ (and $g_{\sigma}$, $g_{\omega}$, $g_{\rho}$) are the masses (the coupling constants) of the $\sigma$, $\omega$, and $\rho$ mesons, respectively. $\bar{\tau}$ is the isospin vector with the third component, $\tau_{3}$. The non-linear parameters $g_{2}$ and $g_{3}$ are adjusted to the surface properties of finite nuclei~\cite{PPNP1996Ring_37_193} and $c_{3}$ is fitted to reproduce the vector and scalar potentials of the density-dependent calculations of the Dirac-Brueckner approach~\cite{PRL1992Brockmann, NPA1994Sugahara}. To improve the density dependence of the isospin-dependent part of the potentials, the nonlinear self-coupling ($d_{3}$) of the isovector meson ($\rho$) is required~\cite{PRC2004Long}.
The anti-symmetric field tensors of the vector mesons are $\Omega^{\mu\nu}$ and $\vec{R}^{\mu\nu}$, of which $\Omega^{\mu\nu}\!=\!\partial^{\mu}\omega^{\nu}-\partial^{\nu}\omega^{\mu}$ and $\vec{R}^{\mu\nu}\!=\!\partial^{\mu}\vec{\rho}^{\nu}-\partial^{\nu}\vec{\rho}^{\mu}$. The electromagnetic field is $F^{\mu\nu}\!=\!\partial^{\mu}A^{\nu}-\partial^{\nu}A^{\mu}$.

The Hamiltonian density can be obtained from the Lagrangian density in Eq.~(\ref{L:NLME}) using the general Legendre transformation~\cite{PPNP1996Ring_37_193}. 
With the no-sea approximation and the mean-field approximation, the energy functional of the nuclear system, i.e., by integrating the Hamiltonian density over the coordinate space, ${\bm r}$, is obtained as \cite{PRC2004Long} 
\begin{eqnarray}
 \nonumber  && E_{\rm {RMF}}[\hat{\rho},\sigma,\omega^{\nu},\vec{\rho}^{\nu},A^{\nu}] \\ \nonumber &=& \!\!\int\!\! d^{3}r \mathbf{Tr}[\beta (\bm{r} \cdot \bm {p} +M + g_{\sigma}\sigma+g_{\omega}\omega^{\mu}\gamma_{\mu}+ g_{\rho}\vec{\tau}\cdot\vec{\rho}_{\mu}\gamma_{\mu} \\
  &&  \nonumber+ \frac{1-\tau_{3}}{2}eA_{\mu}\gamma^{\mu})\hat{\rho}]\\
  \nonumber  &~& +\!\!\int\!\! d^{3}r [\frac{1}{2}\partial^{0}\sigma\partial_{0}\sigma-\frac{1}{2}\partial^{i}\sigma\partial_{i}\sigma+ \frac{1}{2}m_{\sigma}^{2}\sigma^{2}+\frac{1}{3}g_{2}\sigma^{3}+\frac{1}{4}g_{3}\sigma^{4}]\\
  \nonumber &~&- \!\!\int\!\! d^{3}r [\frac{1}{4}\Omega^{0\nu}\Omega_{0\nu}- \frac{1}{4}\Omega^{i\nu}\Omega_{i\nu}+ \frac{1}{2}m^{2}_{\omega}\omega_{\nu}\omega^{\nu}+\frac{1}{4}c_{3}(\omega^{\nu}\omega_{\nu})^2] \\
  \nonumber &~& -  \!\!\int\!\! d^{3}r  [ \frac{1}{4}\vec{R}^{0\nu}\cdot\vec{R}_{0\nu}-\frac{1}{4}\vec{R}^{i\nu}\cdot\vec{R}_{i\nu}+\frac{1}{2}m^{2}_{\rho}\vec{\rho}^{\nu}\cdot\vec{\rho}_{\nu}
  \\
   &~&+ \frac{1}{4}d_{3}(\vec{\rho}^{\nu}\cdot\vec{\rho}_{\nu})^{2}] - \!\!\int\!\! d^{3}r [\frac{1}{4}F^{0\nu}F_{0\nu}-\frac{1}{4}F^{i\nu}F_{i\nu}]
\end{eqnarray}
where the relativistic single-nucleon density matrix $\hat{\rho}$ is defined as: 
\begin{equation}\label{SN:rho}
  \hat{\rho}(\bm{r, r'}) = \sum_{i}^{A}|\psi_{i}(\bm{r})\rangle\langle\psi_{i}(\bm{r'})|,
\end{equation}      
and the trace operation involves a sum over the Dirac indices and an integral in the $r$-space.

\subsubsection{The density-dependent meson-exchange model}
\label{sec:DDME}

Instead of introducing terms involving self-interactions of the meson field as proposed in the nonlinear meson-exchange model, the coupling constants can be defined as density-dependent factors. For the density-dependent meson-exchange (DD-ME) model, these coupling constants are self-consistently regulated by the nuclear wave functions. 
The Lagrangian density of this model is written as \cite{PRC2002Niksic,PRC2005Lalazissis_71_024312}
\begin{eqnarray}
\label{L:DDME}
\nonumber  \mathcal{L}_{\rm DDME} \!\!&=&\!\! \bar{\psi}[\gamma_{\mu}(i\partial^{\mu}-\Gamma_{\omega}\omega^{\mu}-\Gamma_{\rho}\vec{\tau}\cdot\vec{\rho}_{\mu})-(M-\Gamma_{\sigma}\sigma)\\
&&\nonumber -e\gamma^{\mu}A_{\mu}\frac{1-\tau_{3}}{2}]\psi+ \frac{1}{2}(\partial^{\mu}\sigma\partial_{\mu}\sigma-m_{\sigma}^{2}\sigma^{2}) \\
\nonumber   &~& -\frac{1}{4}\Omega^{\mu\nu}\Omega_{\mu\nu}+\frac{1}{2}m^{2}_{\omega}\omega^{\mu}\omega_{\mu}
 -\frac{1}{4}\vec{R}^{\mu\nu}\cdot\vec{R}_{\mu\nu} \\
  &~&+\frac{1}{2}m^{2}_{\rho}\vec{\rho}^{\mu}\cdot\vec{\rho}_{\mu} -\frac{1}{4}F^{\mu\nu}F_{\mu\nu}
\end{eqnarray}

The symbols in Eq.~(\ref{L:DDME}) inherit the same meanings as in Eq.~(\ref{L:NLME}), whereas the coupling constants, $\Gamma_{\sigma}$, $\Gamma_{\omega}$, and $\Gamma_{\rho}$ depend on the (nucleon) density and read  
\begin{equation}
\Gamma_{i}(\rho) = \Gamma_{i}(\rho_{\rm sat})f_{i}(x)\, ,
\end{equation}
of~which $i\!=\!\sigma$, $\omega$, and $\rho$ mesons. 
$f(x)$ is a function of $x\!=\!\rho_\mathrm{b}/\rho_{\rm sat}$, where $\rho_\mathrm{b}$ is the baryon density and $\rho_{\rm sat}$ is the baryon density at saturation in the symmetric nuclear matter. 
For $\sigma$ or $\omega$ meson, the respective $f(x)$ is 
\begin{equation}
  f_j(x)=a_j\frac{1+b_j(x+d_j)^2}{1+c_j(x+d_j)^2} \, ,~\mathrm{of which}~j = \sigma, \omega \, ,
\end{equation}
and for $\rho$ meson, the respective $f(x)$ is
\begin{equation}
f_{\rho}(x) = e^{[-a_{\rho}(x-1)]} \, .
\end{equation}

The DD-ME1~\cite{PRC2002Niksic} and DD-ME2~\cite{PRC2005Lalazissis_71_024312} interactions are the commonly used covariant density functionals of the DD-ME model. Recently, Taninah \emph{et al}.~\cite{PLB2020Taninah} proposed the DD-MEX interaction, which are fitted by the simulated annealing method.

\subsection{Point-coupling models}
\label{sec:PC_model}

In order to avoid the heavy-meson exchange associated in the short-distance dynamics that cannot be resolved at low energies in the meson-exchange models, two types of point-coupling model are proposed for QHD \cite{AP1974Walecka}. In the point-coupling models, nucleon-nucleon interactions are represented by the effective point-like interactions, without exchanging mesons. The effective Lagrangian of point-coupling model consists of isoscalar–scalar, isoscalar–vector, isovector-scalar, and isovector-vector four-fermion interactions.


\subsubsection{The nonlinear point-coupling model}
\label{sec:NLPC}

In the nonlinear point-coupling model, the effective Lagrangian density is expressed as~\cite{PRC1992Nikolaus,PRC2002Burvenich} 
\begin{eqnarray}
\label{L:NLPC}
\nonumber  \mathcal{L}_{\rm NLPC}\!\!\! &=&\!\!\! \bar{\psi}(i\gamma_{\mu}\partial^{\mu}-M)\psi -\frac{1}{2}\alpha_{S}(\bar{\psi}\psi)(\bar{\psi}\psi) \\
\nonumber &-&\!\!\!\frac{1}{2}\alpha_{V}(\bar{\psi}\gamma_{\mu}\psi)(\bar{\psi}\gamma^{\mu}\psi)-\frac{1}{2}\alpha_{TV}(\bar{\psi}\vec{\tau}\gamma_{\mu}\psi)\!\cdot\!(\bar{\psi}\vec{\tau}\gamma^{\mu}\psi) \\
\nonumber&-&\!\!\!\frac{1}{3}\beta_{S}(\bar{\psi}\psi)^{3}-\frac{1}{4}\gamma_{V}[(\bar{\psi}\gamma_{\mu}\psi)
  (\bar{\psi}\gamma^{\mu}\psi)]^2-\frac{1}{4}\gamma_{S}(\bar{\psi}\psi)^{4} \\
\nonumber &-&\!\!\! \frac{1}{2}\delta_{S}\partial_{\nu}(\bar{\psi}\psi)\partial^{\nu}
  (\bar{\psi}\psi)-\frac{1}{2}\delta_{V}\partial_{\nu}(\bar{\psi}\gamma_{\mu}\psi)\partial^{\nu}
  (\bar{\psi}\gamma^{\mu}\psi)\\
 \nonumber &-&\!\!\! \frac{1}{2}\delta_{TV}\partial_{\nu}(\bar{\psi}\vec{\tau}\gamma_{\mu}\psi)\partial^{\nu}(\bar{\psi}\vec{\tau}\gamma_{\mu}\psi)\\
   &-&\!\!\! \frac{1}{4}F^{\mu\nu}F_{\mu\nu} -e\frac{(1-\tau_{3})}{2} \bar{\psi}\gamma^{\mu}\psi A_{\mu} \, .
\end{eqnarray}
The subscripts $S$, $V$, and $TV$ denote the scalar, vector, and isovector nucleon fields, respectively. The $\alpha_{S}$, $\alpha_{V}$, $\alpha_{TV}$, $\beta_{S}$, $\gamma_{S}$, $\gamma_{V}$, $\delta_{S}$, $\delta_{V}$, and $\delta_{TV}$ are the respective coupling constants, of which the symbols $\alpha_{i}$, $\beta_{i}$, $\gamma_{i}$, and $\delta_{i}$ are four-fermion (or second-order) terms, third-order terms, fourth-order terms, and derivative couplings, respectively. The rotational symmetry of each coupling constant is indicated by the subscript. $A_{\mu}$ and $F_{\mu\nu}$ are the four-vector potential and field strength tensor of the electromagnetic field, respectively. 

The commonly used point-coupling parameterized interactions are PC-PK1~\cite{PRC2010Zhao}, PC-LA~\cite{PRC1992Nikolaus}, and PC-F1~\cite{PRC2002Burvenich}. Recently, the PC-L3R~\cite{PLB2023Liu} and the PCX~\cite{PLB2020Taninah} interactions were proposed. The construction of PC-L3R has considered  more observables in the fit and further optimized the separable pairing force. 

\subsubsection{The density-dependent point-coupling model}

The Lagrangian density of the density-dependent point-coupling (DD-PC) model is written as
\begin{eqnarray}
\label{L:DDPC}
\nonumber  \mathcal{L}_{\rm DDPC} \!\!\!\!\!&=&\!\!\!\!\! \bar{\psi}(i\gamma_{\mu}\partial^{\mu}-M)\psi  \\
\nonumber &-&\!\!\!\!\!\frac{1}{2}G_{S}(\hat{\rho})(\bar{\psi}\psi)(\bar{\psi}\psi)-\frac{1}{2}G_{V}(\hat{\rho})(\bar{\psi}\gamma_{\mu}\psi)(\bar{\psi}\gamma^{\mu}\psi)\\
\nonumber&-&\!\!\!\!\!\frac{1}{2}G_{TV}(\hat{\rho})(\bar{\psi}\vec{\tau}\gamma_{\mu}\psi)\cdot(\bar{\psi}\vec{\tau}\gamma^{\mu}\psi)\\
&-&\!\!\!\!\! \frac{1}{2}D_{S}\partial_{\nu}(\bar{\psi}\psi)\partial^{\nu}
(\bar{\psi}\psi)\!-\!e\frac{(1-\tau_{3})}{2}\bar{\psi}\gamma^{\mu}\psi \!A_{\mu} \, . 
\end{eqnarray}
The density-dependent point-coupling model includes only the second-order interaction terms with density-dependent couplings $G_{i}(\hat{\rho})$. The microscopic density-dependent scalar and vector self-energies are quantified using the following practical ansatz functional form of couplings
\begin{equation}
  G_{i}(\hat{\rho}) = a_{i}+(b_{i}+c_{i}x)e^{-d_{i}x}~~~~~~ {\rm for}~~~~i=S,~V,~TV \, .
\end{equation}
The derivative term, $D_{S}\partial_{\nu}(\bar{\psi}\psi)\partial^{\nu}(\bar{\psi}\psi)$, in Eq.~(\ref{L:DDPC}) accounts for the leading effects of finite-range interactions that are crucial for the quantitative description of nuclear density distribution. The coupling parameter, $D_{S}$, has been estimated, for instance, the in-medium chiral perturbation calculation of inhomogeneous nuclear matter estimates $D_{S}\!=\!$ $-0.85$ to $-0.7$~fm$^4$ \cite{NPA2005Fritsch}. 

One of the most commonly used covariant density functionals for the DD-PC model is the DD-PC1 interaction \cite{PRC2008Niksic}. Recently, the DD-PCX interaction was developed by Y\"uksel \emph{et al}. \cite{PRC2019Yuksel}, to closely reproduce the properties of collective excitations, i.e., giant monopole resonance energies and dipole polarizability of $^{208}$Pb.

\begin{figure*}
\centering
\begin{minipage}{0.505\linewidth}
\centering
\includegraphics[width=\linewidth, angle=0]{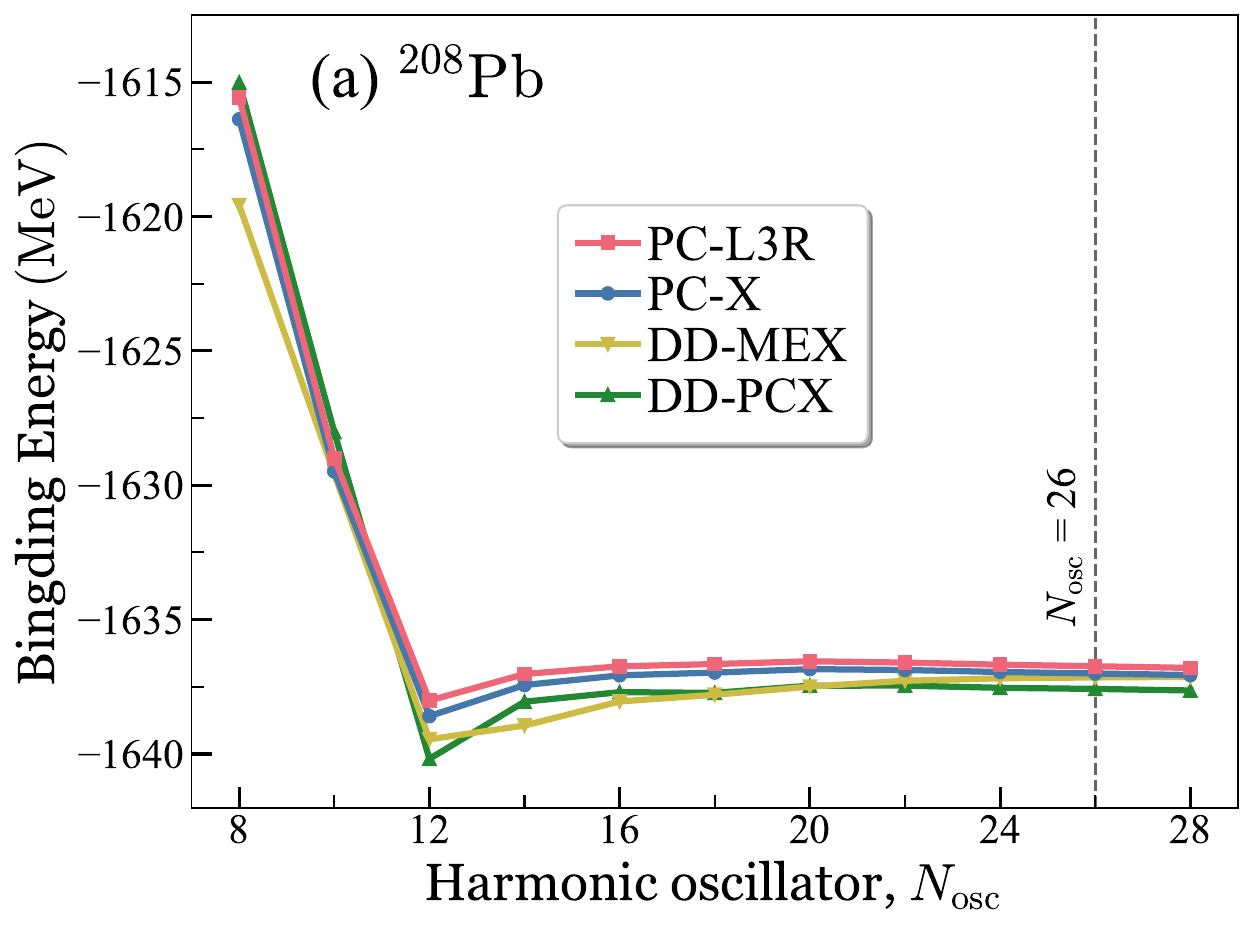}
\end{minipage}
\begin{minipage}{0.495\linewidth}
\includegraphics[width=\linewidth, angle=0]{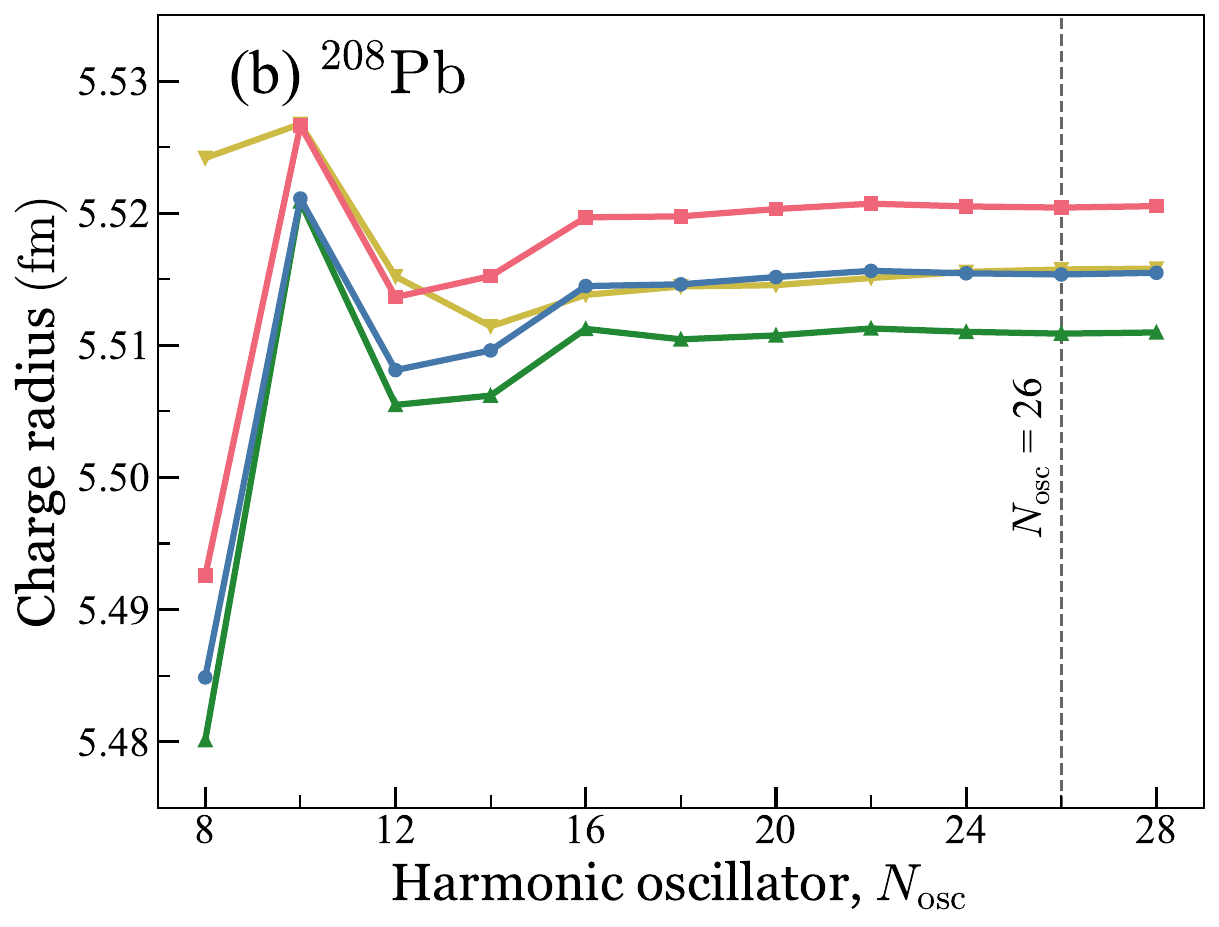}
\end{minipage}
\vspace{-2mm}
\begin{minipage}{0.5\linewidth}
\centering
\includegraphics[width=\linewidth, angle=0]{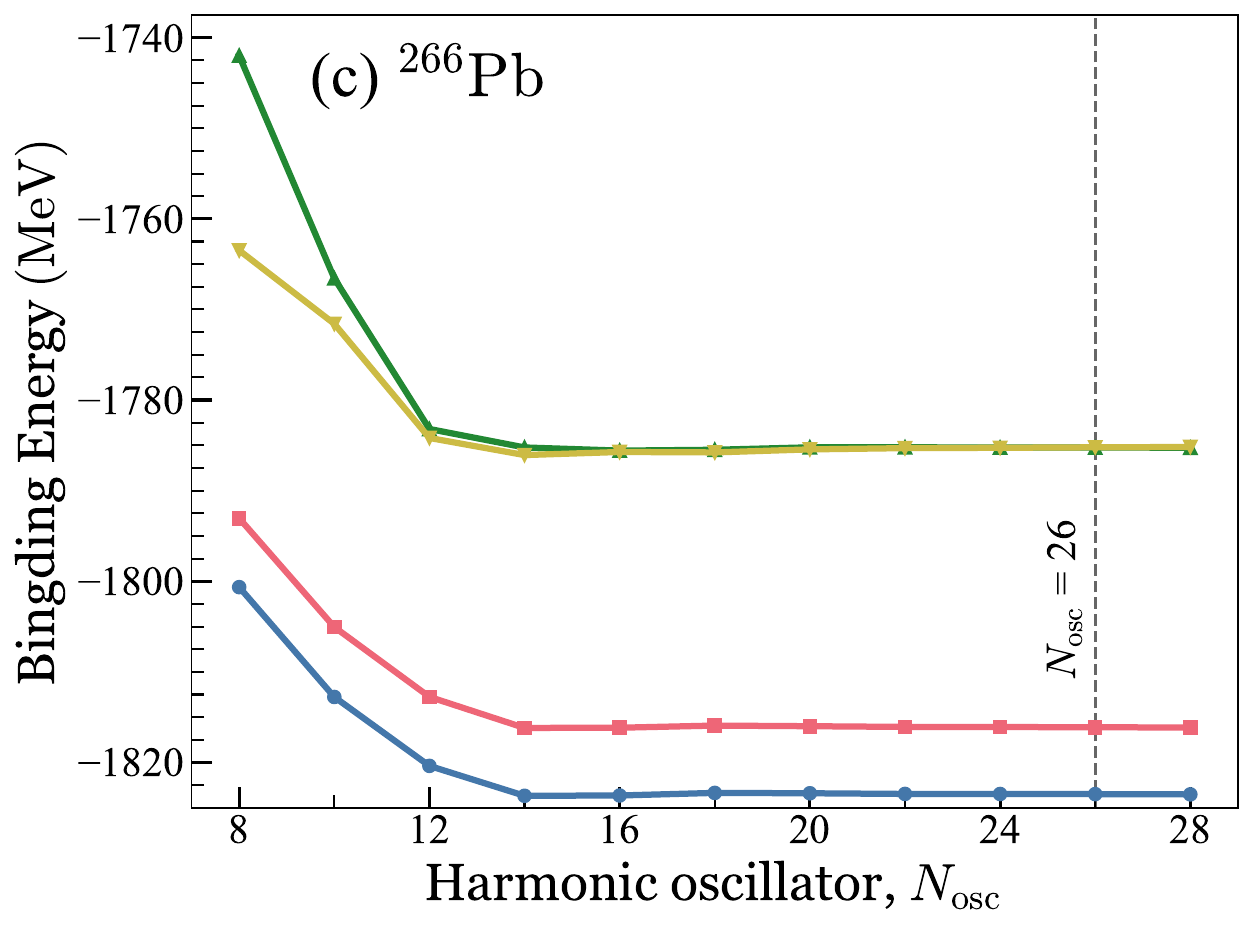}
\end{minipage}
\begin{minipage}{0.5\linewidth}
\includegraphics[width=\linewidth, angle=0]{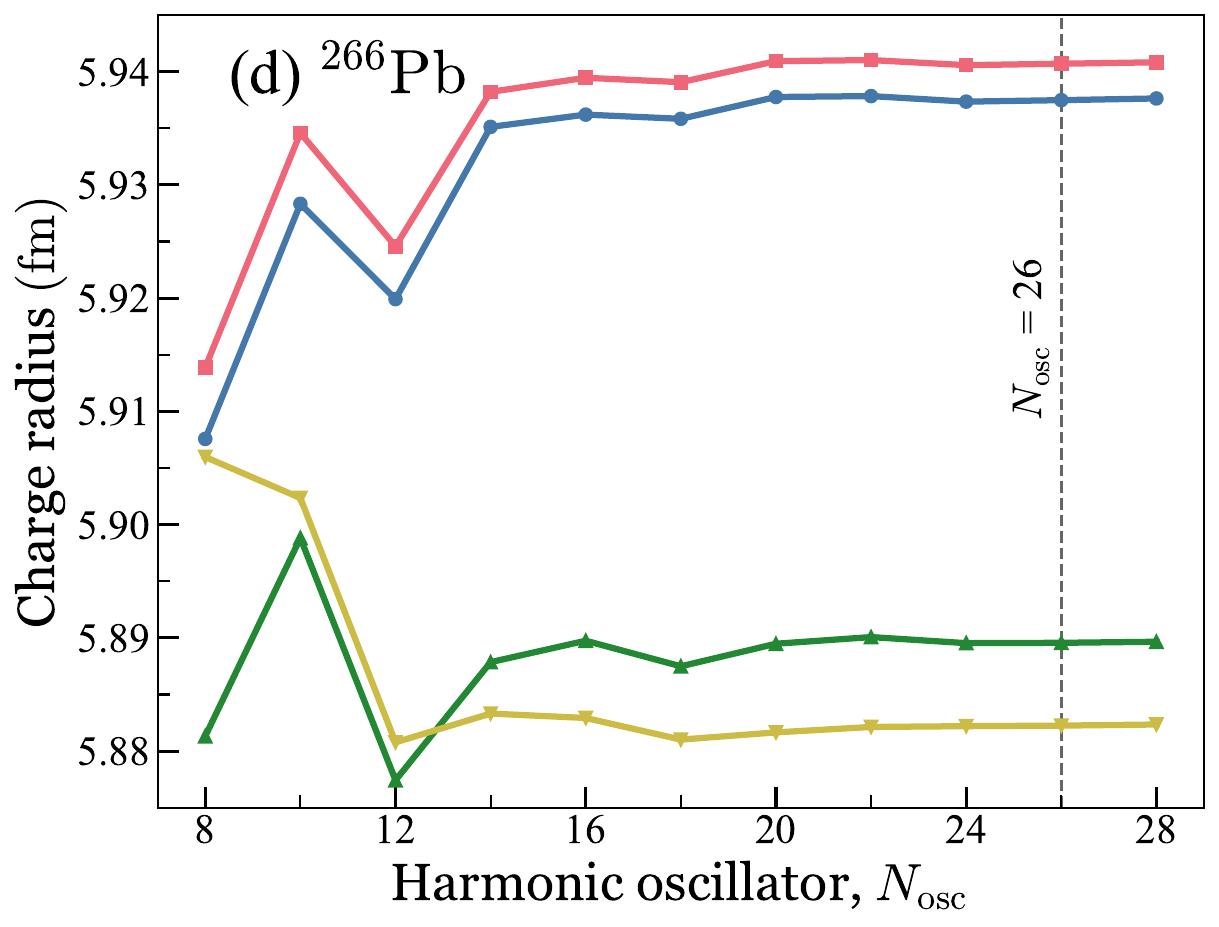}
\end{minipage}
\caption{Assessment of binding-energy and charge-radii convergences. The calculated binding energies (left column) and charge radii (right column) of $^{208}$Pb and $^{266}$Pb as the functions of harmonic oscillator shells $N_\mathrm{osc}$. The results are based on the RHB approach with the PC-L3R (red squares), PC-X (blue dots), DD-MEX (yellow down triangles), and DD-PCX (green up triangles) interactions.}
\label{fig:N_osc}
\end{figure*}%

\subsection{Relativistic Hartree–Bogoliubov approach}
\label{RHB}

Recently, the relativistic Hartree-Bogoliubov (RHB)  approach has received wide attentions for its successful description of many nuclear phenomena~\cite{PLB2023Liu,ADNDT2022Zhang,PRC1998Lalazissis,PPNP2006MengJ_57_470,PRC2009Tian_064301,PRC2016Agbemava}. 
The RHB model can be derived within the framework of covariant density functional theory, which unifies the self-consistent mean field and a pairing field. RHB had been developed to describe the properties of open-shell nuclei.
With implementing the effective Lagrangian density described by either Eqs.~(\ref{L:NLME}), or (\ref{L:DDME}), or (\ref{L:NLPC}), or (\ref{L:DDPC}), the RHB equation for nucleons is derived by the variational procedure,
\begin{equation}\label{eq:RHB}
  \int d^3 \bm{r}' \left(
   \begin{array}{cc}
    h_\mathrm{D}-\lambda_{\tau} &  \Delta \\
     -\Delta^{\ast} &  -h^{\ast}_\mathrm{D}+\lambda_{\tau} \\
  \end{array}
\right)\left(
   \begin{array}{c}
    U_{k}\\
    V_{k}\\
  \end{array}
\right)=E_{k}\left(
   \begin{array}{c}
    U_{k}\\
    V_{k}\\
  \end{array}
\right),
\end{equation}
where $\lambda_{\tau}$ ($\tau\!=\!\mathrm{n}, \mathrm{p}$) is the chemical potential, $U_{k}$ and $V_{k}$ are quasiparticle wave functions, $h_\mathrm{D}$ is the Dirac Hamiltonian and $E_{k}$ is the quasiparticle energy,
\begin{equation}\label{Dirac-H}
  h_\mathrm{D}(\bm r)=\bm{\alpha}\cdot\bm{p}+V(\bm r)+ \beta(M+S(\bm r)) \, ,
\end{equation}
where $\bm{\alpha}$ and $\beta$ are the Dirac matrices, $p$ is the momentum operator, and $S(\bm r)$ and $V(\bm r)$ are the scalar and vector potentials, respectively. The local densities are written as
\begin{eqnarray}
  \rho_{S}(\bm r) &=& \sum_{k>0}\bar{V}_{k}(\bm r)V_{k}(\bm r) \,, \nonumber\\
  \rho_{V} (\bm r)&=&  \sum_{k>0} V^{\dag}_{k}(\bm r)V_{k}(\bm r) \,, {\rm and} \\
  \rho_{TV}(\bm r) &=& \sum_{k>0} V^{\dag}_{k}(\bm r)\tau_{3}V_{k}(\bm r) \,. \nonumber
\end{eqnarray}
The pairing field $\Delta$ in Eq.~(\ref{eq:RHB}) reads
\begin{equation}
\label{eq:Delta}
\Delta_{n_{1}n'_{1}}=\frac{1}{2}\sum_{n_{2}n'_{2}}\langle n_{1}n'_{1}|V^{pp}|n_{2}n'_{2} \rangle \kappa_{n_{2}n'_{2}} \, .
\end{equation}
The separable form of pairing force is used in Eq.~(\ref{eq:Delta}), which is applied for the study of pairing properties in nuclei close to and far from stability \cite{PLB2009Tian_676_44}. Compared to using the finite range Gogny force, using the separable pairing force can reduce the computational costs significantly. 
The separable form of the pairing force is defined as 
\begin{equation}
\label{eq:Vpp}
V^{pp}(\bm{r}_{1},\bm{r}_{2},\bm{r}'_{1},\bm{r}'_{2})=-G\delta(\bm{R}-\bm{R}')
P(\bm{r})P(\bm{r}')\frac{1}{2}(1-P^{\sigma}) \, ,
\end{equation}
where the center of mass, $\bm{R}=(\bm{r_{1}}+\bm{r_{2}})/2$, the relative coordinates, $\bm{r} = \bm{r_{1}}-\bm{r_{2}}$, and the form factor $P(\bm r)$ is of Gaussian shape
\begin{equation}
\label{eq:Gaussian}
P(\bm r)=\frac{1}{(4\pi a^{2})^{3/2}}e^{-r^{2}/4a^{2}} \, .
\end{equation}

For calculating the odd-nucleon systems (odd-A or odd-odd nuclei), we consider the blocking effects of the unpaired nucleon(s), which is included with the equal filling approximation~\cite{ManybodyProb1980,PRC2008Martin,CPL2012Li,PRC2019Sun}. The same procedure was applied in the recent works related to evaluating the PC-L3R interaction~\cite{PLB2023Liu} and to constraining the proton threshold of $^{66}$Se for type-I X-ray burst calculations \cite{APJ2022Lam}.

\begin{figure*}
\centering
\begin{minipage}{0.7\textwidth}
\centering
\includegraphics[width=\linewidth, angle=0]{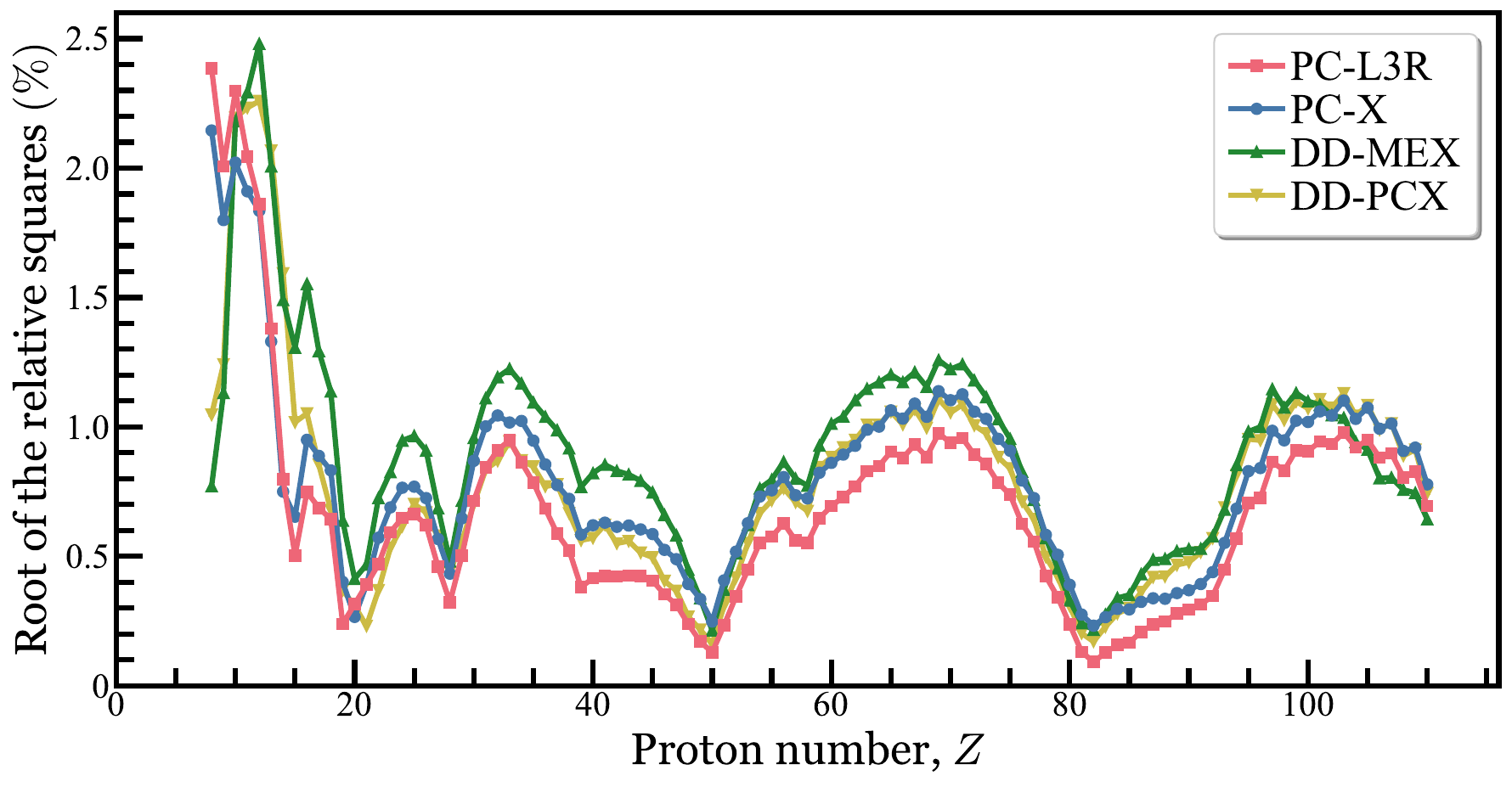}
\end{minipage}
\caption{The root of the relative squares (rrs) of comparing the available experimental and theoretical binding energies of 2471 nuclei calculated from the RHB approach with the PC-L3R~\cite{PLB2023Liu}, PC-X~\cite{PLB2020Taninah}, DD-MEX~\cite{PLB2020Taninah}, and DD-PCX~\cite{PRC2019Yuksel} interactions.}
\label{fig:rrs_BE}
\end{figure*}%

\begin{figure*}
\centering
\begin{minipage}{0.5\textwidth}
\centering
\frame{\includegraphics[width=\linewidth, angle=0]{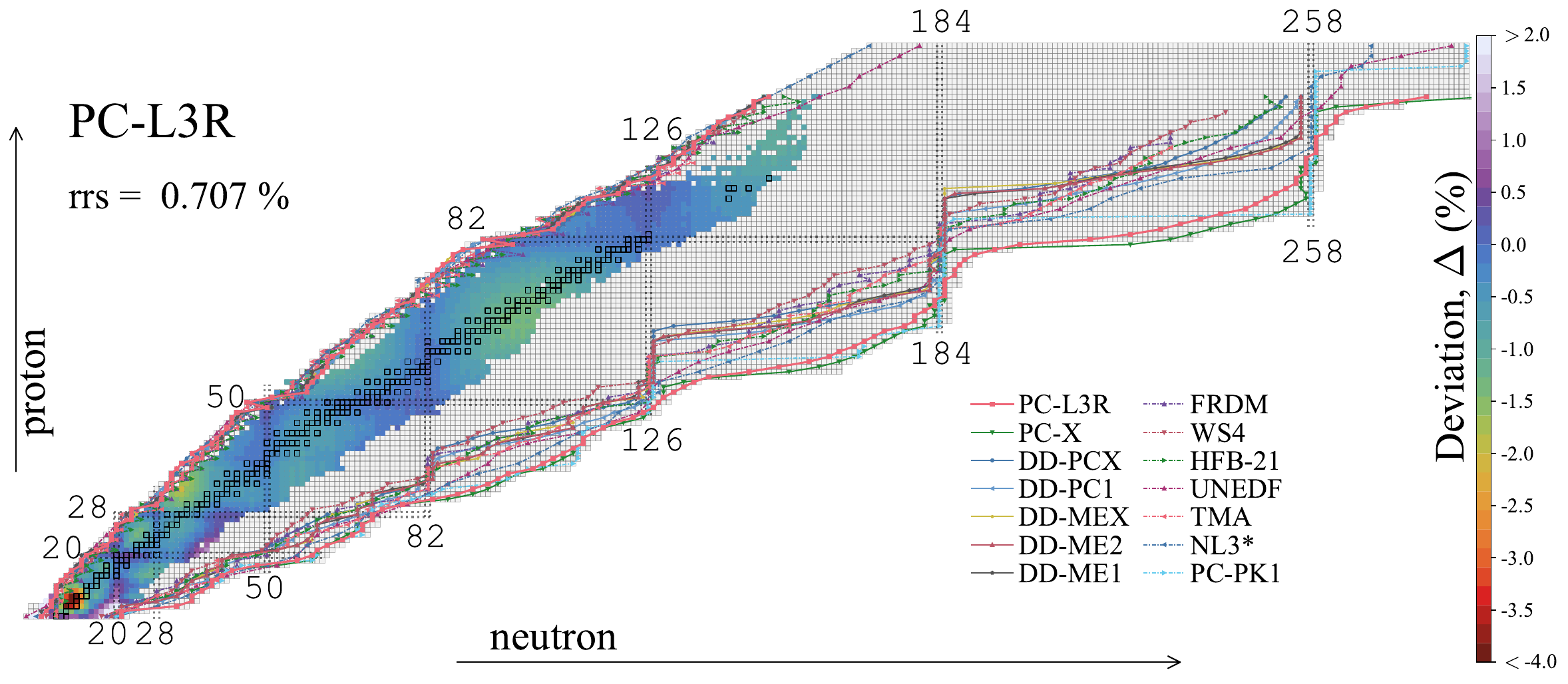}}
\label{fig:PCL3R_AME2020}
\end{minipage}
\begin{minipage}{0.5\textwidth}
\centering
\frame{\includegraphics[width=\linewidth, angle=0]{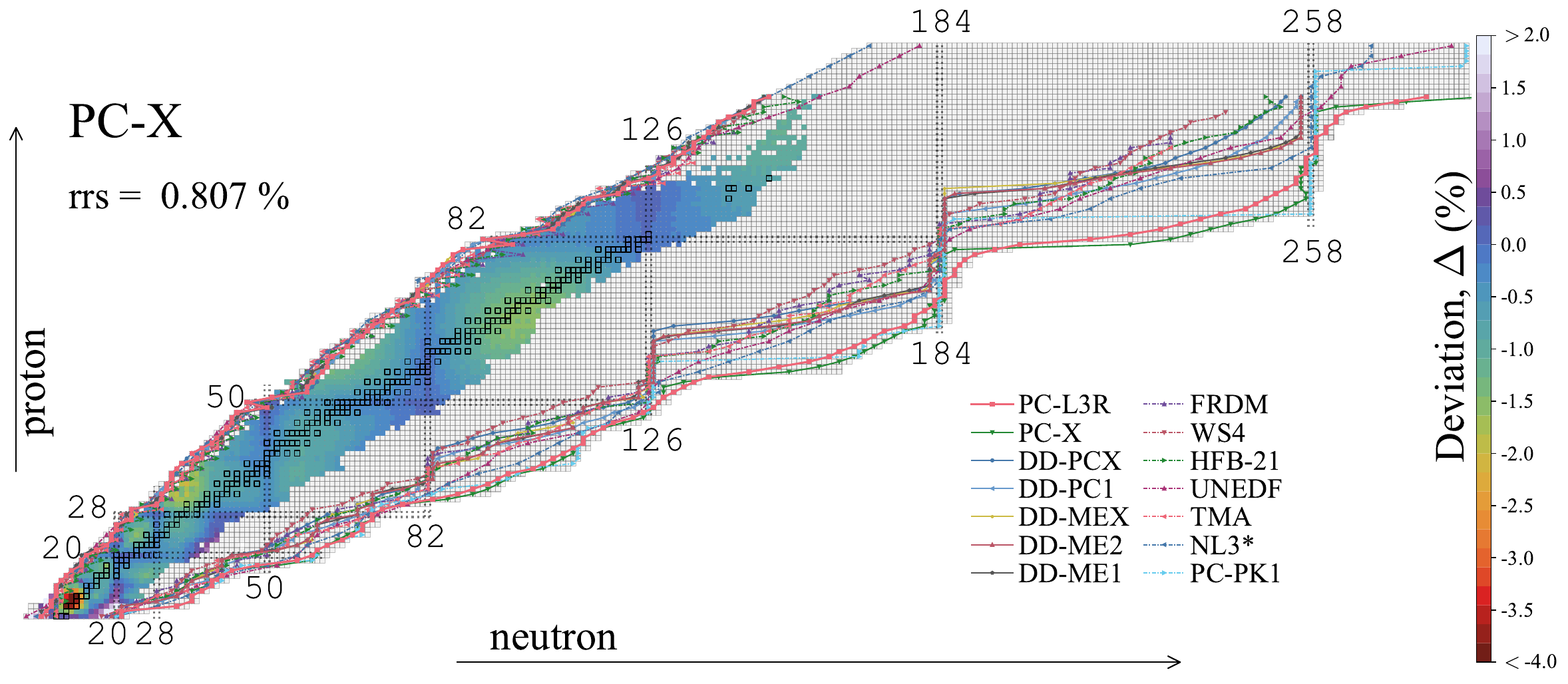}}
\label{fig:PCX_AME2020}
\end{minipage}
\centering
\begin{minipage}{0.5\textwidth}
\centering
\frame{\includegraphics[width=\linewidth, angle=0]{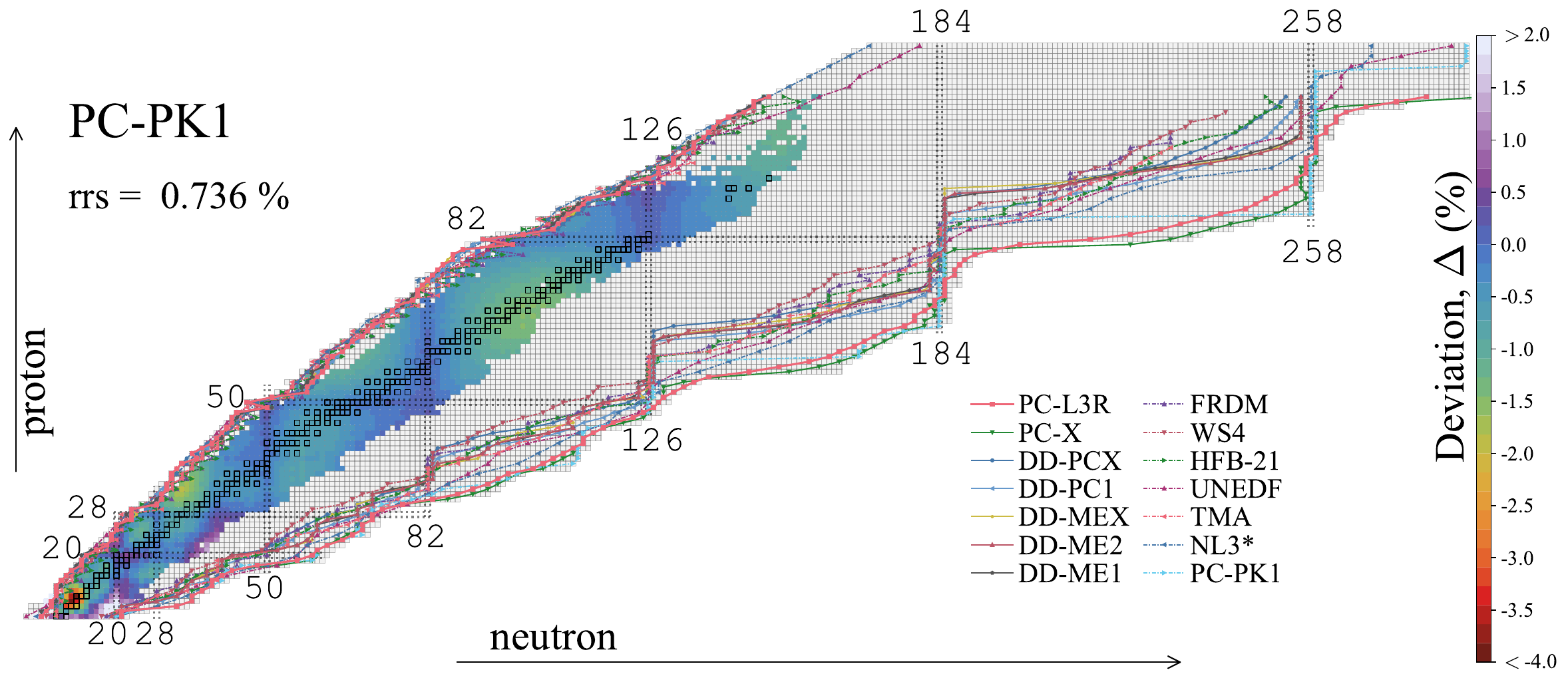}}
\label{fig:PCPK1_AME2020}
\end{minipage}
\begin{minipage}{0.5\textwidth}
\centering
\frame{\includegraphics[width=\linewidth, angle=0]{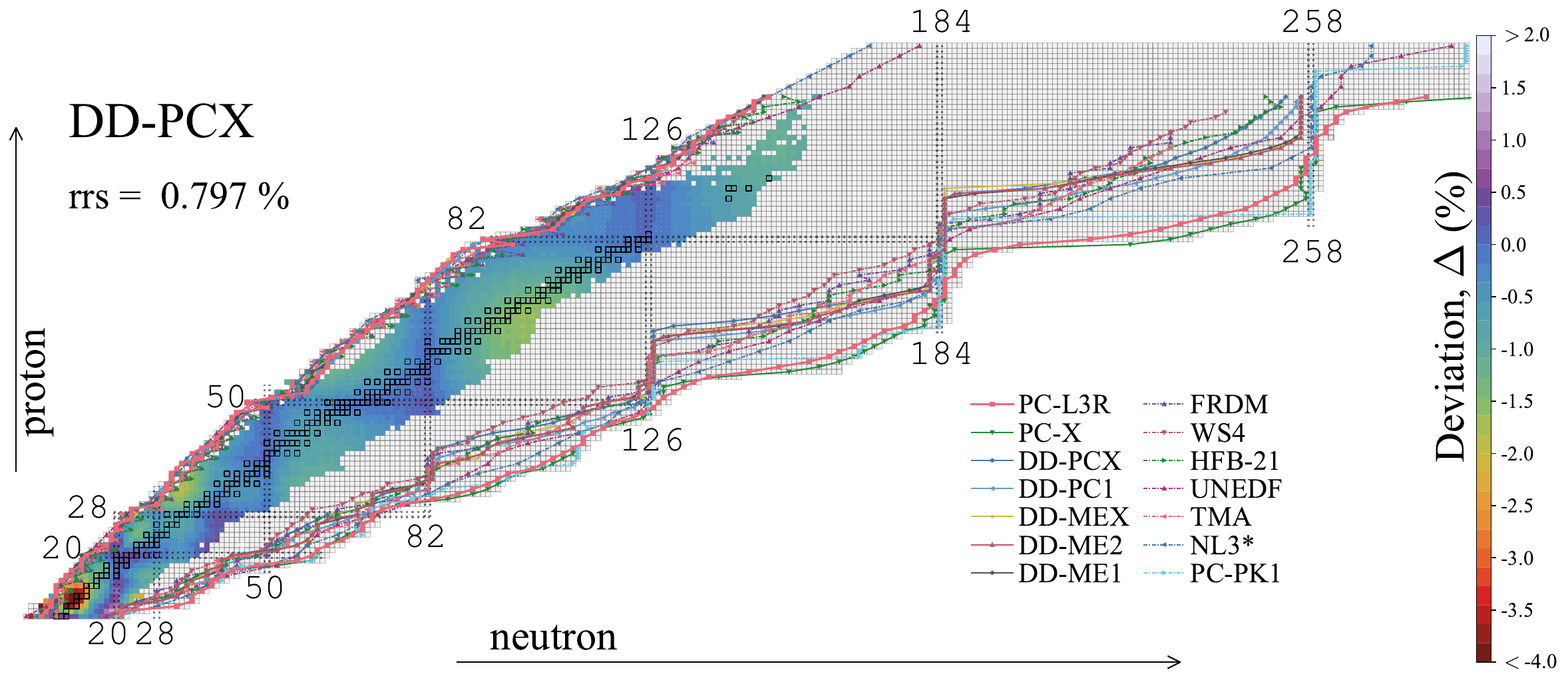}}
\label{fig:DDPCX_AME2020}
\end{minipage}
\begin{minipage}{0.5\textwidth}
\centering
\frame{\includegraphics[width=\linewidth, angle=0]{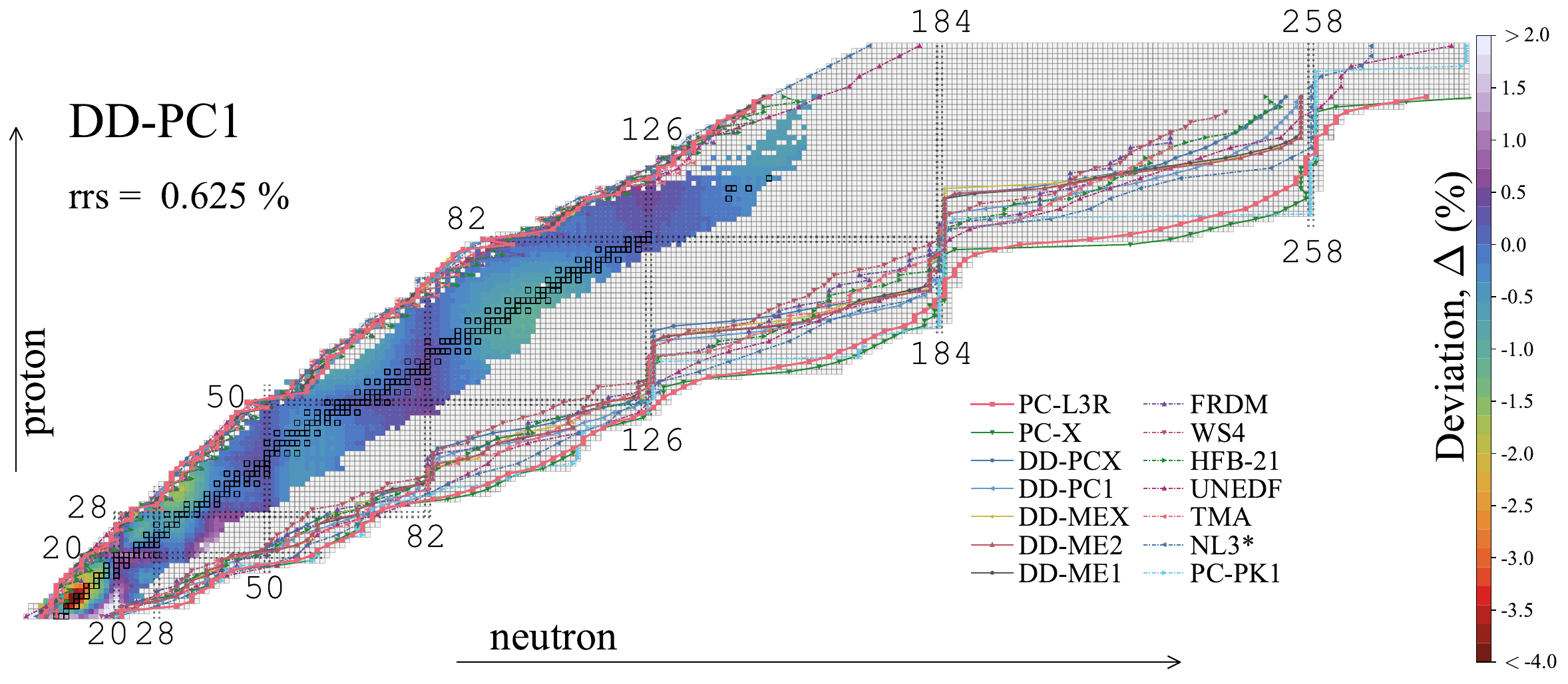}}
\label{fig:DDPC1_AME2020}
\end{minipage}
\begin{minipage}{0.5\textwidth}
\centering
\frame{\includegraphics[width=\linewidth, angle=0]{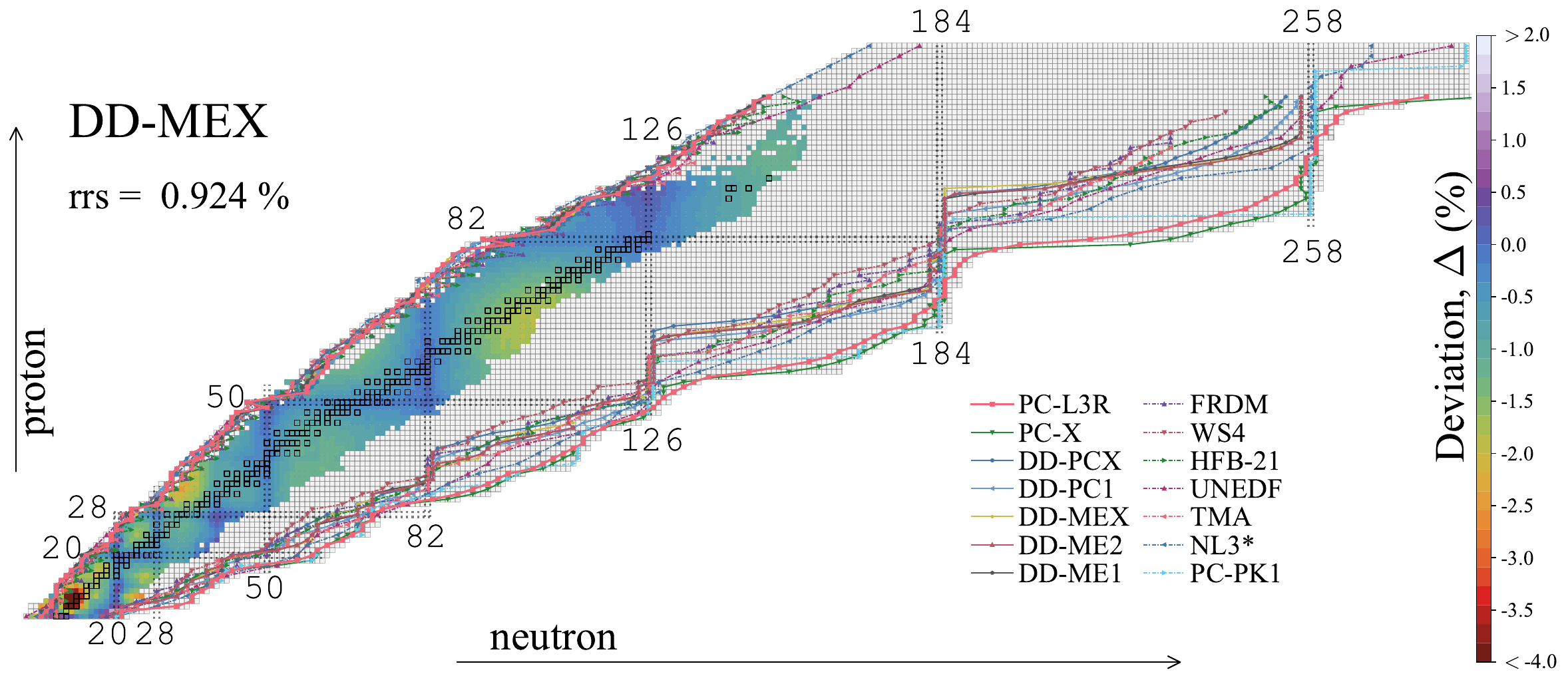}}
\label{fig:DDMEX_AME2020}
\end{minipage}
\begin{minipage}{0.5\textwidth}
\centering
\frame{\includegraphics[width=\linewidth, angle=0]{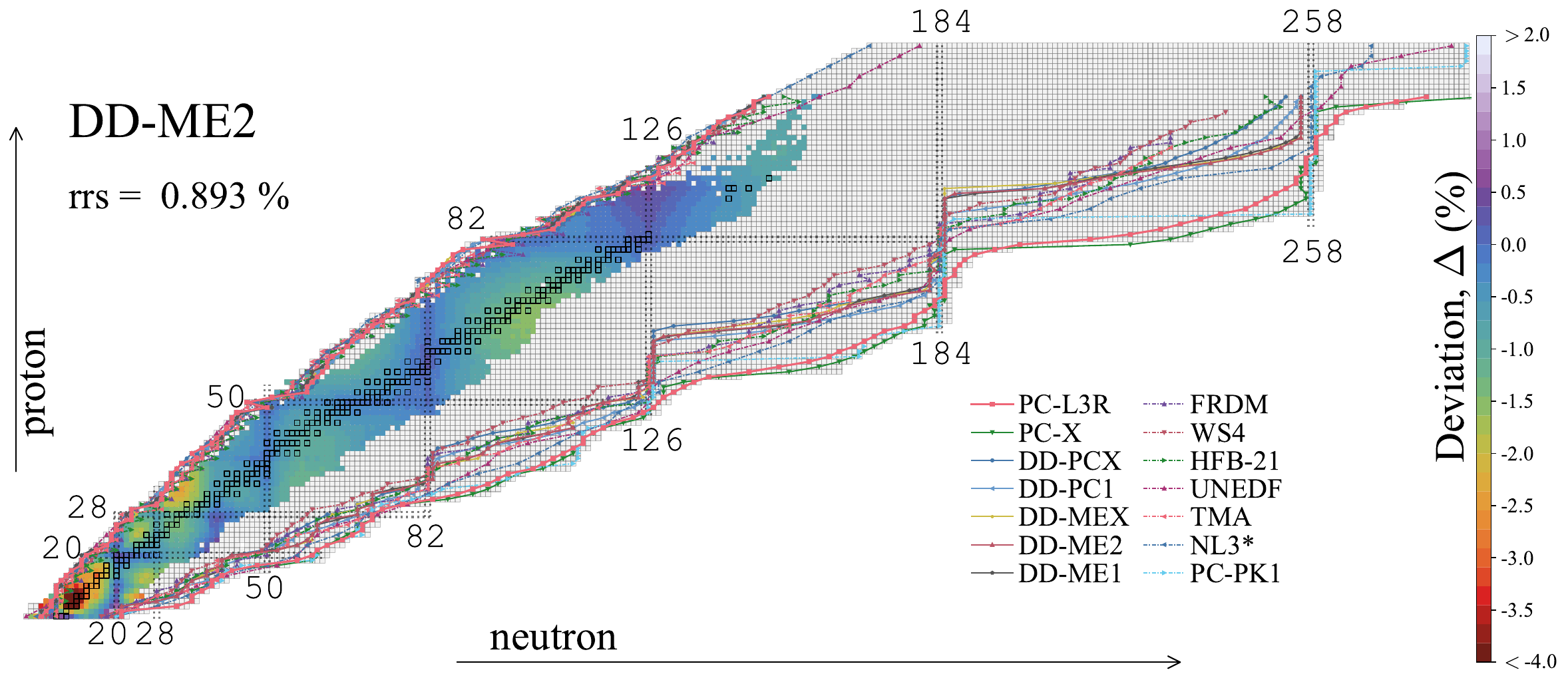}}
\label{fig:DDME2_AME2020}
\end{minipage}
\begin{minipage}{0.5\textwidth}
\centering
\frame{\includegraphics[width=\linewidth, angle=0]{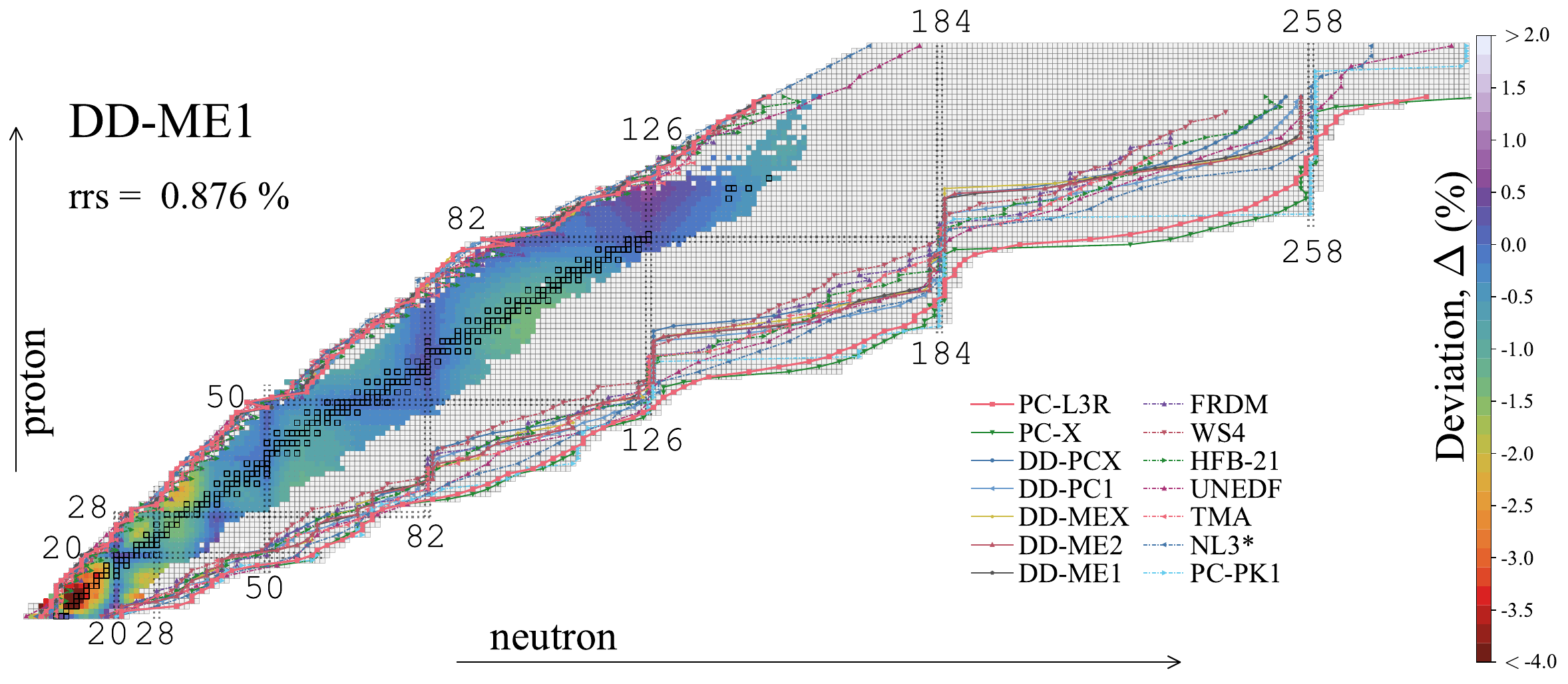}}
\label{fig:DDME1_AME2020}
\end{minipage}
\centering
\begin{minipage}{\textwidth}
\caption{\footnotesize The relative deviations of the available experimental binding energies of 2471 nuclei with the ones calculated from the RHB approach with the PC-L3R~\cite{PLB2023Liu}, or PC-X~\cite{PLB2020Taninah}, or DD-PCX~\cite{PRC2019Yuksel}, or DD-PC1~\cite{PRC2008Niksic}, or DD-MEX~\cite{PLB2020Taninah}, or DD-ME2~\cite{PRC2005Lalazissis_71_024312}, or DD-ME1~\cite{PRC2002Niksic} interactions, and from the RCHB approach with the PC-PK1 interaction~\cite{ADNDT2018Xia}. The available theoretical PC-PK1 binding energies of 2447 nuclei are quoted from the work of Xia \emph{et al}.~\cite{ADNDT2018Xia} and are compared with experiment (second row of left column). 
The relative deviation between theoretical and experimental value, $\Delta\!=\!(E^{\rm{Theo.}}_{\rm b}$-$E^{\rm{Exp.}}_{\rm b}$)/$E^{\rm{Exp.}}_{\rm b}\!\times\!100\%$ for each nucleus is represented by the color tone referring to the right color scale in the nuclear chart of each panel. 
Meanwhile, the value of root of the relative squares (rrs; Eq.~(\ref{eq:rrs})) of comparing the available experimental binding energies and theoretical ones yielded from the RHB with each interaction is shown in the respective panel.  
The blank squared nuclei are excluded in the present comparison due to either unknown or ambiguous experimental data. 
The proton and neutron drip lines (dashed color lines) predicted from FRDM~\cite{MOLLER1995}, WS4~\cite{PLB2014Wang}, HFB-21~\cite{PRC2010Goriely}, UNEDF~\cite{Nature2012Erler}, TMA~\cite{PTP2005Geng}, NL3*~\cite{PLB2013Afanasjev}, and PC-PK1~\cite{ADNDT2018Xia} are plotted for comparison with the drip lines (solid color lines) calculated from the RHB approach with PC-L3R, PC-X, DD-PCX, DD-PC1, DD-MEX, DD-ME2, and DD-ME1. These solid and dashed color lines are merely used to guide the eyes, and do not refer to the color scale.}
\label{fig:nuclear_chart}
\end{minipage}
\end{figure*}%
\bigskip
\bigskip

\section{Results and discussions}
\label{sec.Res}

We perform systematic calculations for all nuclei from $Z\!=\!8$~to~$110$, and from the proton drip line to the neutron drip line using the RHB approach with the PC-L3R~\cite{PLB2023Liu}, PC-X~\cite{PLB2020Taninah}, DD-MEX~\cite{PLB2020Taninah}, and DD-PCX~\cite{PRC2019Yuksel} interactions and the respective separable pairing force. The separable pairing force D1S is adopted for the PC-X and DD-MEX interactions, of which the pairing strength parameters are $G\!=\!728$~MeV$\cdot$fm$^{3}$ and $a\!=\!0.644$~fm \cite{PLB2009Tian_676_44}. The PC-L3R and DD-PCX interactions implements the further optimized separable pairing force, which takes into account the different proton and neutron pairing strengths.

Meanwhile, we assess the convergence of binding energies and charge radii with respect to the number of harmonic oscillator shell, $N_\mathrm{osc}$. Figure~\ref{fig:N_osc} presents a typical convergence of binding energies and charge radii for the RHB approach with the PC-L3R, PC-X, DD-MEX, and DD-PCX interactions, based on a series of $N_\mathrm{osc}\!=\!8$, $\dots$, $28$, for instance, $^{208}$Pb and $^{266}$Pb. For $N_\mathrm{osc}\!=\!26$, the RHB solution for $^{208}$Pb produces a relative accuracy of less than $0.0035$~\% (or $0.0025$~\%) in the calculated binding energy (or charge radii) compared to the ones calculated from $N_\mathrm{osc}\!=\!28$ (Figs.~\ref{fig:N_osc}~(a) and \ref{fig:N_osc}~(b)). A set of similar result are obtained for $^{266}$Pb (Figs.~\ref{fig:N_osc}~(c) and \ref{fig:N_osc}~(d)). The RHB solution for $^{266}$Pb leads to a relative accuracy of less than $0.002$~\% (or $0.0025$~\%) in the calculated binding energy (or charge radii) compared to the ones calculated from $N_\mathrm{osc}\!=\!28$. Henceforth, we adopt $N_\mathrm{osc}\!=\!26$ for the present work.

In Tables~\ref{tab:BindingEnergy},~\ref{tab:radii}, and~\ref{tab:Fermi_surface}, we systematically tabulate the ground state properties of these nuclei generated from the RHB approach with the PC-L3R~\cite{PLB2023Liu}, PC-X~\cite{PLB2020Taninah}, DD-MEX~\cite{PLB2020Taninah}, and DD-PCX~\cite{PRC2019Yuksel} interactions. (Hereinafter, the name of an interaction labels the RHB approach associated with this interaction, wherever appropriate).

Table~\ref{tab:BindingEnergy} lists the name of elements, proton numbers $Z$, mass numbers $A$, neutron numbers $N$, binding energies $E_{b}$ from the available experimentally determined $2471$ nuclear masses~\cite{AME2020} and from theoretical calculations, single-neutron separation energies $S_\mathrm{n}$, and two-neutron separation energies $S_\mathrm{2n}$. The extrapolated nuclear masses of AME2020 are not included in the present study. Meanwhile, we also present the root-mean-square deviation values (rms) of comparing the theoretical and experimental binding energies, single-neutron separation energies, and two-neutron separation energies for each isotopic chain. In Table~\ref{tab:radii}, the calculated root-mean-square radii of matter ($R_\mathrm{m}$), of neutron ($R_\mathrm{n}$), of proton ($R_\mathrm{p}$), and of charge ($R_\mathrm{c}$) distributions are presented. The available experimental $R_\mathrm{c}$ (932 data points) \cite{ADNDT2013Angeli} are given. The rms of charge radii of comparing the theoretical and available experimental radii for each element are also displayed. Table~\ref{tab:Fermi_surface} shows the neutron Fermi surfaces $\lambda_\mathrm{n}$, proton Fermi surfaces $\lambda_\mathrm{p}$, ground-state spins and parities of neutron $J^{\pi}(N)$ and proton $J^{\pi}(P)$. 

\subsection{Binding energies (nuclear masses)}
\label{sub:Mass}

In this section, we study the influence of different covariant density functionals on the binding energies and borders of the nuclear landscape based on the same RHB approach. Enhancing the capability of the RHB approach and covariant density functionals on reproducing the experimental binding energies in high accuracy is beyond the scope of the present work. 

We define that a nucleus is considered to be bound if the two-neutron and proton separation energies of this nucleus are positive and the proton and neutron Fermi surfaces, $\lambda_{i}\!<\!0, (i\!=\!\mathrm{n},\mathrm{p})$. In Table~\ref{tab:BindingEnergy}, for the isotopic chains from oxygen (O, $Z\!=\!8$) to darmstadtium (Ds, $Z\!=\!110$), PC-L3R~\cite{PLB2023Liu}, or PC-X~\cite{PLB2020Taninah}, or DD-MEX~\cite{PLB2020Taninah}, or DD-PCX~\cite{PRC2019Yuksel} estimate a total of $9004$, $9162$, $7112$, and $6799$ bound nuclei, respectively. 



Compared to the density-dependent interactions DD-MEX and DD-PCX, the point-coupling interactions (PC-L3R and PC-X) predict more bound nuclei, especially at the neutron-rich region. For the proton-rich region, both density-dependent and point-coupling interactions yield similar predictions. A rather similar estimation is obtained by the PC-PK1 interaction \cite{ADNDT2018Xia} for the proton-rich region. Compared with the available experimentally determined nuclear masses of $2471$ nuclei, the PC-L3R, PC-X, DD-MEX, and DD-PCX reproduces the experimental binding energies with the root-mean-square deviations (root of the relative square) of $7.081$~MeV ($0.707\%$), $8.618$~MeV ($0.807\%$), $9.762$~MeV ($0.924\%$), and $8.620$~MeV ($0.797\%$), respectively. The root-mean-square deviation values (rms) and root of the relative squares (rrs) for each isotope chains are listed in Table~\ref{tab:BindingEnergy}. The rms and rrs are defined as,
\begin{equation}
\label{eq:rms}
\hspace{-6mm}
\mathrm{rms} = \left[\left(\sum_{i}^{N}(O_{i}^{\rm Theo.}-O_{i}^{\rm Exp.})^{2}\right)/N\right]^{1/2}\, ,~\mathrm{and} 
\end{equation}
\begin{equation}
\label{eq:rrs}
\hspace{-5mm}
\mathrm{rrs} = \left[\left(\sum_{i}^{N}(O_{i}^{\rm Theo.}-O_{i}^{\rm Exp.})^{2}/(O_{i}^{\rm Exp.})^{2}\right)/N\right]^{1/2}\, ,
\end{equation} 
respectively, of which $O_{i}$ are the physical observables. Each type of observables with the label $i$ is associated with a set of relevant $N$ data points. 
Meanwhile, the rms of PC-PK1~\cite{ADNDT2018Xia}, DD-PC1~\cite{PRC2008Niksic}, DD-ME2~\cite{PRC2005Lalazissis_71_024312}, and DD-ME1~\cite{PRC2002Niksic} are calculated to be $8.038$~MeV, $4.958$~MeV, $7.894$~MeV, and $6.728$~MeV, respectively. DD-PC1 is capable of generating the lowest rms among the rms values yielded from these covariant density functionals as the construction of DD-PC1 is assessed for reproducing the properties of deformed nuclei at the medium-heavy ($A\!\approx\!150$-$180$) and heavy nuclear regions ($A\!\approx\!230$-$250$) \cite{PRC2008Niksic}.


Compared with the region close to magic number, the rrs of the region of neutron (proton) number further away from the magic number increases significantly up to 2.480~\% (Fig.~\ref{fig:rrs_BE}). Such significant increment is due to the consideration of spherical symmetry for the present calculations. Therefore, the comparisons of rms for the nuclei with either neutron or proton magic number could be useful to evaluate the capability of covariant density functionals on reproducing the binding energies of (almost) spherical nuclei. For the isotopes with $Z\!=\!8$, $20$, $28$, $50$, $82$ and isotones with $N\!=\!8$, $20$, $28$, $50$, $82$, $126$ (a total of $220$ nuclei), the corresponding rms are reduced to $1.628$~MeV, $2.238$~MeV, $2.155$~MeV, and $1.786$~MeV for PC-L3R, PC-X, DD-MEX, and DD-PCX, respectively. 


The investigation of the borders of the nuclear landscape has attracted worldwide interest in both experimental and theoretical nuclear physics communities because the proton and neutron drip lines serve as the natural laboratories probing our fundamental understanding of nuclear forces. Figure~\ref{fig:nuclear_chart} illustrates the nuclear landscapes and borders from O ($Z\!=\!8$) to Ds ($Z\!=\!110$) explored by the RHB approaches with the PC-L3R~\cite{PLB2023Liu}, or PC-X~\cite{PLB2020Taninah}, or DD-PCX~\cite{PRC2019Yuksel}, or DD-PC1~\cite{PRC2008Niksic}, or DD-MEX~\cite{PLB2020Taninah}, or DD-ME2~\cite{PRC2005Lalazissis_71_024312}, or DD-ME1~\cite{PRC2002Niksic} interactions. The relative binding energy differences, $(E^{\rm Theo.}_{\rm b}-E^{\rm Exp.}_{\rm b})/E^{\rm Exp.}_{\rm b}$, for the measured $2471$ nuclei yielded from PC-L3R, or PC-X, or DD-PCX, or DD-PC1, or DD-MEX, or DD-ME2, or DD-ME1 are plotted according to the attached color scale. Meanwhile, the comparison of the borders of nuclear landscape predicted by the mass tables FRDM~\cite{MOLLER1995}, WS4~\cite{PLB2014Wang}, HFB-21~\cite{PRC2010Goriely}, UNEDF~\cite{Nature2012Erler}, TMA~\cite{PTP2005Geng}, NL3*~\cite{PLB2013Afanasjev}, and PC-PK1~\cite{ADNDT2018Xia} are given as well. The proton drip lines of these mass models are rather similarly distributed not far away from the valley of stability and close to the experimentally determined proton drip line. This is mainly due to the strong repulsive Coulomb force effectively shifting up the single-proton energy of the proton-rich nuclei. For the neutron drip line, the RHB approaches with the nonlinear point-coupling interactions, PC-L3R, PC-X, or PC-PK1, estimate more extended borders. We give the respective discussion in the following sections. 

\begin{figure*}
\centering
\begin{minipage}{0.7\textwidth}
\centering
\includegraphics[width=\linewidth, angle=0]{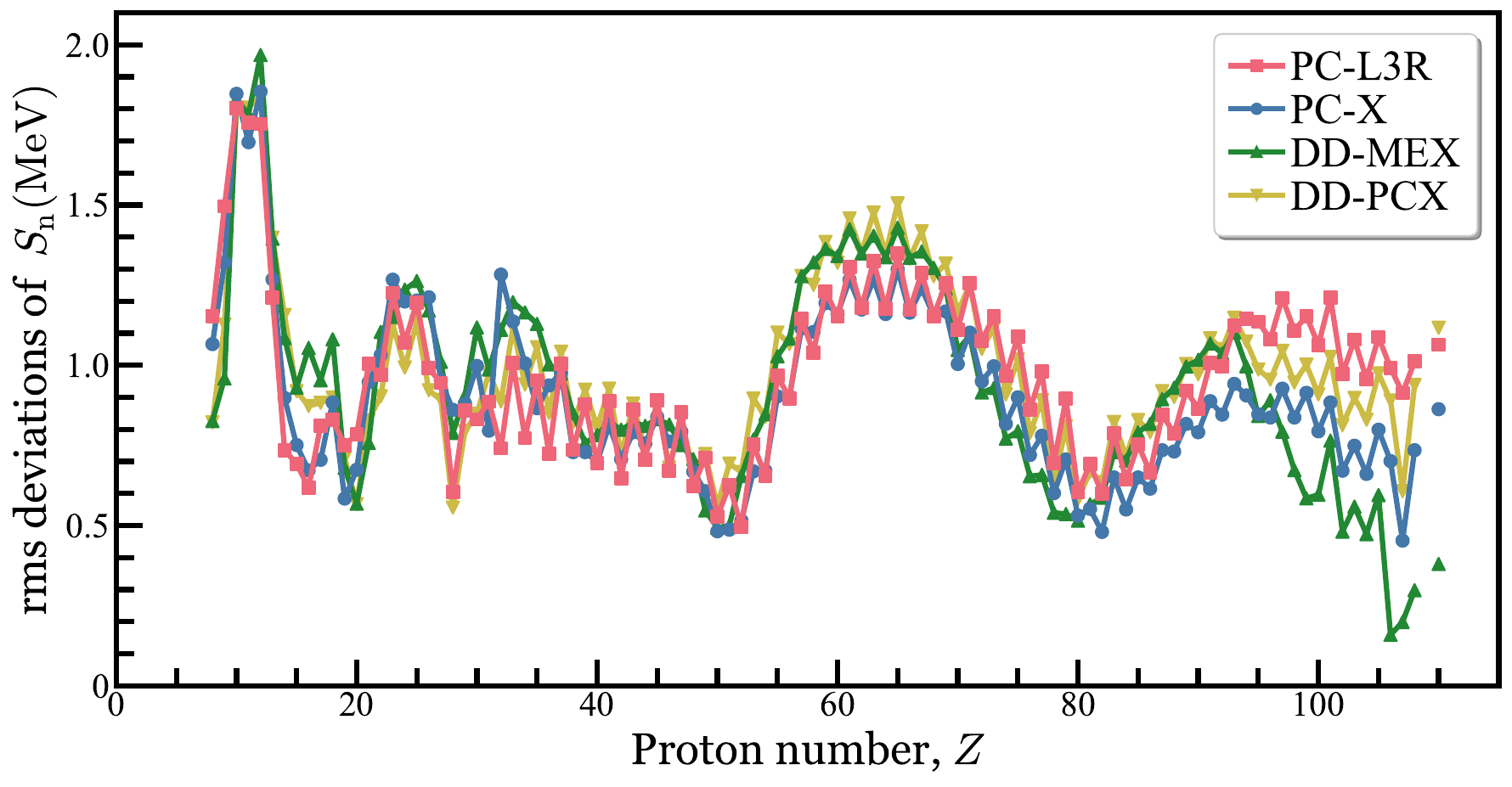}
\end{minipage}
\begin{minipage}{0.7\textwidth}
\centering
\includegraphics[width=\linewidth, angle=0]{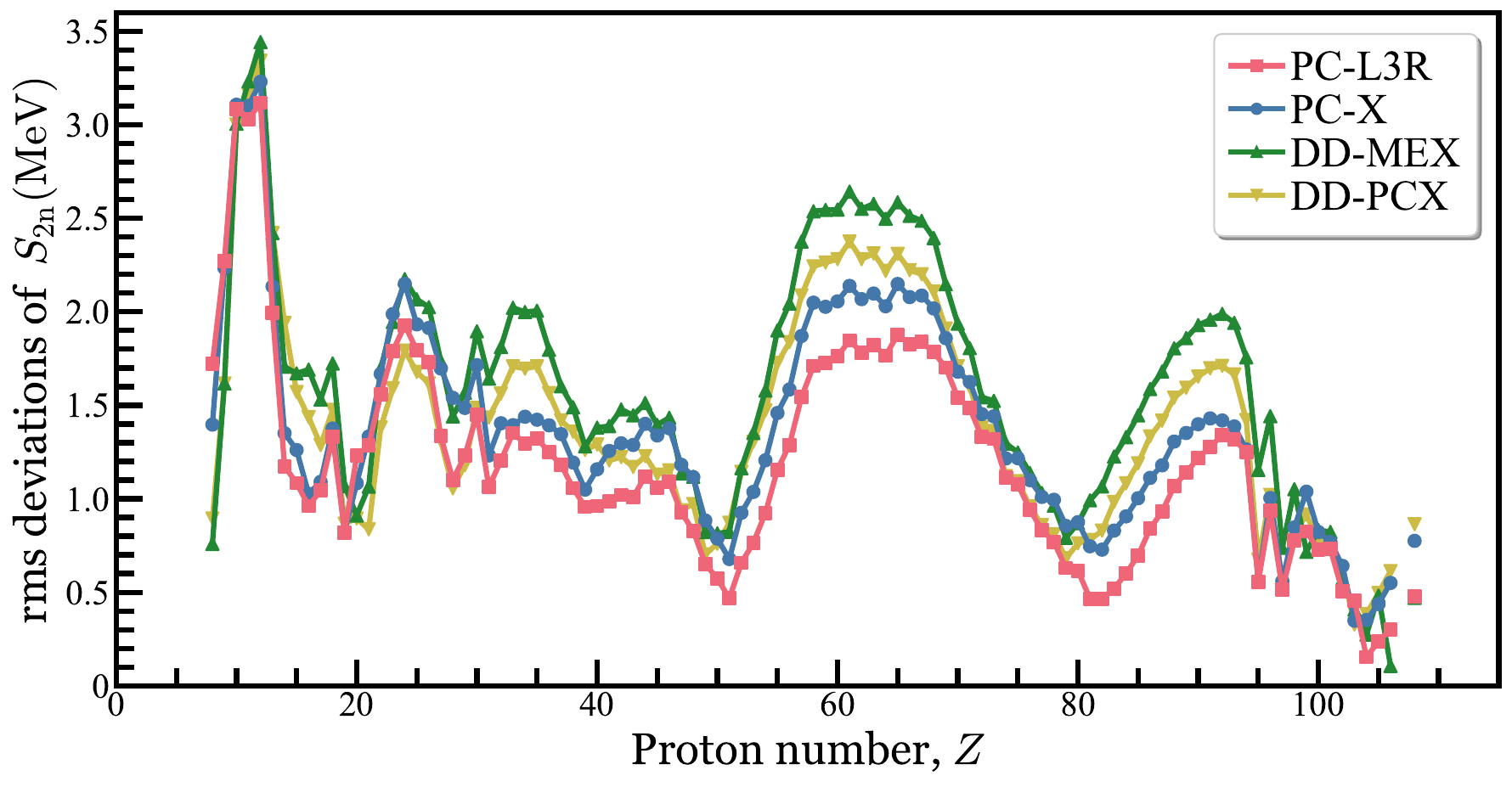}
\end{minipage}
\caption{The root-mean-square deviation values (rms) of comparing the available experimental one-neutron (top panel) and two-neutron (bottom panel) separation energies of nuclei with the ones calculated from the RHB approach with the PC-L3R~\cite{PLB2023Liu}, PC-X~\cite{PLB2020Taninah}, DD-MEX~\cite{PLB2020Taninah}, and DD-PCX~\cite{PRC2019Yuksel} interactions. The total numbers of experimental one- and two-neutron separation energies are 2323 and 2237, respectively.}
\label{fig:S_rms}
\end{figure*}
 
\begin{table}[pos=t!,width=\textwidth]
\centering
\caption{\label{tab:proton_dripline}\footnotesize The neutron number of the isotopes along the proton drip line \\
predicted by the RHB approach with the PC-L3R~\cite{PLB2023Liu}, \\
or PC-X~\cite{PLB2020Taninah}, or DD-MEX~\cite{PLB2020Taninah}, or DD-PCX~\cite{PRC2019Yuksel} interactions.}
\label{tab:Neu-dp}
\tiny
\begin{tabular}{@{}l@{\hspace{1mm}}c@{\hspace{1mm}}c@{\hspace{1mm}}c@{\hspace{1mm}}c@{\hspace{1mm}}c@{\hspace{1mm}}c@{\hspace{1mm}}c@{\hspace{1mm}}c@{\hspace{1mm}}c@{\hspace{1mm}}c@{}}
\toprule
\midrule
$\rm {Element}$ & PC-L3R &  PC-X & DD-MEX & DD-PCX & ~~~~ & $\rm {Element}$& PC-L3R & PC-X & DD-MEX & DD-PCX  \\

\midrule
$_{8  }$O    &      $5   $    &     $ 5  $   &  $ 5  $ & $ 5     $ &   & $_{9  }$F    &  $ 6   $  &  $ 6   $ &   $ 6   $ &  $ 6    $          \\
$_{10 }$Ne   &      $7   $    &     $ 7  $   &  $ 7  $ & $ 7     $ &   & $_{11 }$Na   &  $ 7   $  &  $ 7   $ &   $ 8   $ &  $ 8    $          \\
$_{12 }$Mg   &      $8   $    &     $ 8  $   &  $ 8  $ & $ 8     $ &   & $_{13 }$Al   &  $ 8   $  &  $ 8   $ &   $ 8   $ &  $ 8    $          \\
$_{14 }$Si   &      $9   $    &     $ 9  $   &  $ 9  $ & $ 9     $ &   & $_{15 }$P    &  $ 10  $  &  $ 10  $ &   $ 10  $ &  $ 11   $          \\
$_{16 }$S    &      $11  $    &     $ 11 $   &  $ 11 $ & $ 12    $ &   & $_{17 }$Cl   &  $ 12  $  &  $ 12  $ &   $ 12  $ &  $ 12   $          \\
$_{18 }$Ar   &      $13  $    &     $ 13 $   &  $ 13 $ & $ 13    $ &   & $_{19 }$K    &  $ 15  $  &  $ 15  $ &   $ 15  $ &  $ 16   $          \\
$_{20 }$Ca   &      $16  $    &     $ 16 $   &  $ 16 $ & $ 16    $ &   & $_{21 }$Sc   &  $ 16  $  &  $ 16  $ &   $ 16  $ &  $ 16   $          \\
$_{22 }$Ti   &      $18  $    &     $ 18 $   &  $ 19 $ & $ 19    $ &   & $_{23 }$V    &  $ 19  $  &  $ 19  $ &   $ 19  $ &  $ 19   $          \\
$_{24 }$Cr   &      $20  $    &     $ 20 $   &  $ 20 $ & $ 20    $ &   & $_{25 }$Mn   &  $ 20  $  &  $ 20  $ &   $ 20  $ &  $ 20   $          \\
$_{26 }$Fe   &      $21  $    &     $ 21 $   &  $ 21 $ & $ 21    $ &   & $_{27 }$Co   &  $ 23  $  &  $ 23  $ &   $ 23  $ &  $ 23   $          \\
$_{28 }$Ni   &      $23  $    &     $ 23 $   &  $ 24 $ & $ 24    $ &   & $_{29 }$Cu   &  $ 24  $  &  $ 24  $ &   $ 24  $ &  $ 24   $          \\
$_{30 }$Zn   &      $28  $    &     $ 28 $   &  $ 28 $ & $ 28    $ &   & $_{31 }$Ga   &  $ 29  $  &  $ 29  $ &   $ 29  $ &  $ 29   $          \\
$_{32 }$Ge   &      $29  $    &     $ 29 $   &  $ 30 $ & $ 29    $ &   & $_{33 }$As   &  $ 30  $  &  $ 30  $ &   $ 30  $ &  $ 30   $          \\
$_{34 }$Se   &      $31  $    &     $ 31 $   &  $ 31 $ & $ 31    $ &   & $_{35 }$Br   &  $ 31  $  &  $ 34  $ &   $ 31  $ &  $ 32   $          \\
$_{36 }$Kr   &      $32  $    &     $ 32 $   &  $ 32 $ & $ 32    $ &   & $_{37 }$Rb   &  $ 33  $  &  $ 33  $ &   $ 33  $ &  $ 33   $          \\
$_{38 }$Sr   &      $34  $    &     $ 34 $   &  $ 34 $ & $ 34    $ &   & $_{39 }$Y    &  $ 36  $  &  $ 36  $ &   $ 36  $ &  $ 36   $          \\
$_{40 }$Zr   &      $37  $    &     $ 37 $   &  $ 37 $ & $ 37    $ &   & $_{41 }$Nb   &  $ 37  $  &  $ 38  $ &   $ 38  $ &  $ 38   $          \\
$_{42 }$Mo   &      $40  $    &     $ 40 $   &  $ 40 $ & $ 40    $ &   & $_{43 }$Tc   &  $ 41  $  &  $ 41  $ &   $ 41  $ &  $ 41   $          \\
$_{44 }$Ru   &      $42  $    &     $ 42 $   &  $ 42 $ & $ 42    $ &   & $_{45 }$Rh   &  $ 43  $  &  $ 43  $ &   $ 43  $ &  $ 43   $          \\
$_{46 }$Pd   &      $43  $    &     $ 43 $   &  $ 44 $ & $ 44    $ &   & $_{47 }$Ag   &  $ 44  $  &  $ 44  $ &   $ 44  $ &  $ 45   $          \\
$_{48 }$Cd   &      $45  $    &     $ 45 $   &  $ 45 $ & $ 46    $ &   & $_{49 }$In   &  $ 50  $  &  $ 51  $ &   $ 51  $ &  $ 51   $          \\
$_{50 }$Sn   &      $51  $    &     $ 51 $   &  $ 51 $ & $ 52    $ &   & $_{51 }$Sb   &  $ 52  $  &  $ 52  $ &   $ 52  $ &  $ 53   $          \\
$_{52 }$Te   &      $58  $    &     $ 58 $   &  $ 58 $ & $ 59    $ &   & $_{53 }$I    &  $ 59  $  &  $ 59  $ &   $ 59  $ &  $ 59   $          \\
$_{54 }$Xe   &      $60  $    &     $ 60 $   &  $ 60 $ & $ 60    $ &   & $_{55 }$Cs   &  $ 61  $  &  $ 61  $ &   $ 61  $ &  $ 61   $          \\
$_{56 }$Ba   &      $61  $    &     $ 61 $   &  $ 62 $ & $ 62    $ &   & $_{57 }$La   &  $ 63  $  &  $ 63  $ &   $ 63  $ &  $ 63   $          \\
$_{58 }$Ce   &      $63  $    &     $ 64 $   &  $ 64 $ & $ 64    $ &   & $_{59 }$Pr   &  $ 64  $  &  $ 65  $ &   $ 65  $ &  $ 65   $          \\
$_{60 }$Nd   &      $66  $    &     $ 66 $   &  $ 66 $ & $ 66    $ &   & $_{61 }$Pm   &  $ 67  $  &  $ 67  $ &   $ 67  $ &  $ 68   $          \\
$_{62 }$Sm   &      $68  $    &     $ 68 $   &  $ 69 $ & $ 69    $ &   & $_{63 }$Eu   &  $ 70  $  &  $ 70  $ &   $ 70  $ &  $ 71   $          \\
$_{64 }$Gd   &      $71  $    &     $ 71 $   &  $ 71 $ & $ 72    $ &   & $_{65 }$Tb   &  $ 72  $  &  $ 72  $ &   $ 72  $ &  $ 72   $          \\
$_{66 }$Dy   &      $74  $    &     $ 74 $   &  $ 74 $ & $ 74    $ &   & $_{67 }$Ho   &  $ 75  $  &  $ 75  $ &   $ 75  $ &  $ 75   $          \\
$_{68 }$Er   &      $76  $    &     $ 76 $   &  $ 76 $ & $ 77    $ &   & $_{69 }$Tm   &  $ 77  $  &  $ 77  $ &   $ 77  $ &  $ 78   $          \\
$_{70 }$Yb   &      $79  $    &     $ 79 $   &  $ 79 $ & $ 79    $ &   & $_{71 }$Lu   &  $ 80  $  &  $ 80  $ &   $ 80  $ &  $ 80   $          \\
$_{72 }$Hf   &      $81  $    &     $ 81 $   &  $ 81 $ & $ 81    $ &   & $_{73 }$Ta   &  $ 82  $  &  $ 82  $ &   $ 82  $ &  $ 82   $          \\
$_{74 }$W    &      $83  $    &     $ 83 $   &  $ 83 $ & $ 84    $ &   & $_{75 }$Re   &  $ 84  $  &  $ 84  $ &   $ 84  $ &  $ 85   $          \\
$_{76 }$Os   &      $85  $    &     $ 86 $   &  $ 85 $ & $ 86    $ &   & $_{77 }$Ir   &  $ 86  $  &  $ 87  $ &   $ 86  $ &  $ 87   $          \\
$_{78 }$Pt   &      $88  $    &     $ 88 $   &  $ 88 $ & $ 89    $ &   & $_{79 }$Au   &  $ 89  $  &  $ 89  $ &   $ 88  $ &  $ 90   $          \\
$_{80 }$Hg   &      $91  $    &     $ 91 $   &  $ 91 $ & $ 92    $ &   & $_{81 }$Tl   &  $ 99  $  &  $ 99  $ &   $ 99  $ &  $ 99   $          \\
$_{82 }$Pb   &      $100 $    &     $ 101 $  &  $ 101$ & $ 101   $ &   & $_{83 }$Bi   &  $ 101 $  &  $ 102 $ &   $ 102 $ &  $ 103  $          \\
$_{84 }$Po   &      $108 $    &     $ 108 $  &  $ 109$ & $ 109   $ &   & $_{85 }$At   &  $ 109 $  &  $ 109 $ &   $ 110 $ &  $ 110  $          \\
$_{86 }$Rn   &      $110 $    &     $ 111 $  &  $ 111$ & $ 112   $ &   & $_{87 }$Fr   &  $ 112 $  &  $ 112 $ &   $ 112 $ &  $ 113  $          \\
$_{88 }$Ra   &      $113 $    &     $ 113 $  &  $ 113$ & $ 114   $ &   & $_{89 }$Ac   &  $ 114 $  &  $ 115 $ &   $ 114 $ &  $ 116  $          \\
$_{90 }$Th   &      $116 $    &     $ 116 $  &  $ 115$ & $ 117   $ &   & $_{91 }$Pa   &  $ 120 $  &  $ 120 $ &   $ 119 $ &  $ 121  $          \\
$_{92 }$U    &      $121 $    &     $ 121 $  &  $ 120$ & $ 122   $ &   & $_{93 }$Np   &  $ 122 $  &  $ 122 $ &   $ 121 $ &  $ 123  $          \\
$_{94 }$Pu   &      $126 $    &     $ 126 $  &  $ 126$ & $ 126   $ &   & $_{95 }$Am   &  $ 128 $  &  $ 128 $ &   $ 128 $ &  $ 127  $          \\
$_{96 }$Cm   &      $129 $    &     $ 129 $  &  $ 129$ & $ 129   $ &   & $_{97 }$Bk   &  $ 131 $  &  $ 131 $ &   $ 130 $ &  $ 130  $          \\
$_{98 }$Cf   &      $132 $    &     $ 132 $  &  $ 132$ & $ 132   $ &   & $_{99 }$Es   &  $ 133 $  &  $ 134 $ &   $ 133 $ &  $ 133  $          \\
$_{100}$Fm   &      $135 $    &     $ 135 $  &  $ 134$ & $ 135   $ &   & $_{101}$Md   &  $ 136 $  &  $ 136 $ &   $ 135 $ &  $ 136  $          \\
$_{102}$No   &      $137 $    &     $ 138 $  &  $ 137$ & $ 138   $ &   & $_{103}$Lr   &  $ 139 $  &  $ 139 $ &   $ 138 $ &  $ 139  $          \\
$_{104}$Rf   &      $140 $    &     $ 141 $  &  $ 139$ & $ 141   $ &   & $_{105}$Db   &  $ 142 $  &  $ 142 $ &   $ 141 $ &  $ 142  $          \\
$_{106}$Sg   &      $144 $    &     $ 144 $  &  $ 143$ & $ 144   $ &   & $_{107}$Bh   &  $ 145 $  &  $ 145 $ &   $ 144 $ &  $ 145  $          \\
$_{108}$Hs   &      $147 $    &     $ 147 $  &  $ 147$ & $ 147   $ &   & $_{109}$Mt   &  $ 148 $  &  $ 149 $ &   $ 149 $ &  $ 149  $          \\
$_{110}$Ds   &      $150 $    &     $ 151 $  &  $ 151$ & $ 151   $ &   &    &    &   &    &            \\
\bottomrule
\end{tabular}
\end{table}

\begin{table}[pos=h!,width=\textwidth]
\centering
\caption{\label{tab:neutron_dripline}\footnotesize 
The neutron number of the isotopes along the neutron drip \\
line predicted by the RHB approach with the PC-L3R~\cite{PLB2023Liu}, \\
or PC-X~\cite{PLB2020Taninah}, or DD-MEX~\cite{PLB2020Taninah}, or DD-PCX~\cite{PRC2019Yuksel} interactions.}
\tiny
\begin{tabular}{@{}l@{\hspace{1mm}}c@{\hspace{1mm}}c@{\hspace{1mm}}c@{\hspace{1mm}}c@{\hspace{1mm}}c@{\hspace{1mm}}c@{\hspace{1mm}}c@{\hspace{1mm}}c@{\hspace{1mm}}c@{\hspace{1mm}}c@{}}
\toprule
\midrule
$\rm {Element}$ & PC-L3R &  PC-X & DD-MEX & DD-PCX &~~~~  & $\rm {Element}$& PC-L3R & PC-X & DD-MEX & DD-PCX  \\
\midrule
$_{8  }$O    &      $21  $    &     $ 21 $   &  $ 20 $ & $ 18    $ &   & $_{9  }$F    &  $ 21  $  &  $ 21  $ &   $ 21  $ &  $ 21   $          \\
$_{10 }$Ne   &      $32  $    &     $ 32 $   &  $ 21 $ & $ 21    $ &   & $_{11 }$Na   &  $ 35  $  &  $ 35  $ &   $ 28  $ &  $ 25   $          \\
$_{12 }$Mg   &      $35  $    &     $ 35 $   &  $ 32 $ & $ 31    $ &   & $_{13 }$Al   &  $ 37  $  &  $ 35  $ &   $ 32  $ &  $ 32   $          \\
$_{14 }$Si   &      $39  $    &     $ 38 $   &  $ 34 $ & $ 32    $ &   & $_{15 }$P    &  $ 39  $  &  $ 38  $ &   $ 35  $ &  $ 34   $          \\
$_{16 }$S    &      $41  $    &     $ 41 $   &  $ 37 $ & $ 37    $ &   & $_{17 }$Cl   &  $ 43  $  &  $ 41  $ &   $ 39  $ &  $ 38   $          \\
$_{18 }$Ar   &      $48  $    &     $ 52 $   &  $ 41 $ & $ 41    $ &   & $_{19 }$K    &  $ 56  $  &  $ 59  $ &   $ 41  $ &  $ 41   $          \\
$_{20 }$Ca   &      $59  $    &     $ 60 $   &  $ 43 $ & $ 43    $ &   & $_{21 }$Sc   &  $ 60  $  &  $ 60  $ &   $ 52  $ &  $ 49   $          \\
$_{22 }$Ti   &      $62  $    &     $ 62 $   &  $ 56 $ & $ 51    $ &   & $_{23 }$V    &  $ 65  $  &  $ 64  $ &   $ 59  $ &  $ 53   $          \\
$_{24 }$Cr   &      $67  $    &     $ 66 $   &  $ 59 $ & $ 54    $ &   & $_{25 }$Mn   &  $ 69  $  &  $ 68  $ &   $ 60  $ &  $ 56   $          \\
$_{26 }$Fe   &      $69  $    &     $ 68 $   &  $ 60 $ & $ 58    $ &   & $_{27 }$Co   &  $ 71  $  &  $ 71  $ &   $ 62  $ &  $ 60   $          \\
$_{28 }$Ni   &      $71  $    &     $ 71 $   &  $ 63 $ & $ 62    $ &   & $_{29 }$Cu   &  $ 73  $  &  $ 73  $ &   $ 68  $ &  $ 68   $          \\
$_{30 }$Zn   &      $77  $    &     $ 77 $   &  $ 68 $ & $ 68    $ &   & $_{31 }$Ga   &  $ 83  $  &  $ 86  $ &   $ 71  $ &  $ 71   $          \\
$_{32 }$Ge   &      $88  $    &     $ 91 $   &  $ 71 $ & $ 73    $ &   & $_{33 }$As   &  $ 91  $  &  $ 94  $ &   $ 76  $ &  $ 78   $          \\
$_{34 }$Se   &      $94  $    &     $ 96 $   &  $ 80 $ & $ 80    $ &   & $_{35 }$Br   &  $ 96  $  &  $ 98  $ &   $ 83  $ &  $ 82   $          \\
$_{36 }$Kr   &      $98  $    &     $ 100$   &  $ 83 $ & $ 83    $ &   & $_{37 }$Rb   &  $ 103 $  &  $ 105 $ &   $ 83  $ &  $ 83   $          \\
$_{38 }$Sr   &      $105 $    &     $ 107$   &  $ 86 $ & $ 83    $ &   & $_{39 }$Y    &  $ 107 $  &  $ 109 $ &   $ 92  $ &  $ 83   $          \\
$_{40 }$Zr   &      $109 $    &     $ 111$   &  $ 93 $ & $ 83    $ &   & $_{41 }$Nb   &  $ 111 $  &  $ 111 $ &   $ 94  $ &  $ 84   $          \\
$_{42 }$Mo   &      $113 $    &     $ 113$   &  $ 96 $ & $ 88    $ &   & $_{43 }$Tc   &  $ 113 $  &  $ 113 $ &   $ 98  $ &  $ 91   $          \\
$_{44 }$Ru   &      $115 $    &     $ 115$   &  $ 100$ & $ 95    $ &   & $_{45 }$Rh   &  $ 117 $  &  $ 117 $ &   $ 106 $ &  $ 102  $          \\
$_{46 }$Pd   &      $119 $    &     $ 121$   &  $ 111$ & $ 107   $ &   & $_{47 }$Ag   &  $ 123 $  &  $ 125 $ &   $ 110 $ &  $ 110  $          \\
$_{48 }$Cd   &      $125 $    &     $ 126$   &  $ 113$ & $ 112   $ &   & $_{49 }$In   &  $ 127 $  &  $ 127 $ &   $ 113 $ &  $ 117  $          \\
$_{50 }$Sn   &      $128 $    &     $ 128$   &  $ 117$ & $ 122   $ &   & $_{51 }$Sb   &  $ 129 $  &  $ 129 $ &   $ 125 $ &  $ 124  $          \\
$_{52 }$Te   &      $131 $    &     $ 129$   &  $ 124$ & $ 124   $ &   & $_{53 }$I    &  $ 133 $  &  $ 129 $ &   $ 127 $ &  $ 124  $          \\
$_{54 }$Xe   &      $135 $    &     $ 133$   &  $ 127$ & $ 126   $ &   & $_{55 }$Cs   &  $ 138 $  &  $ 140 $ &   $ 127 $ &  $ 127  $          \\
$_{56 }$Ba   &      $146 $    &     $ 154$   &  $ 127$ & $ 127   $ &   & $_{57 }$La   &  $ 152 $  &  $ 160 $ &   $ 127 $ &  $ 127  $          \\
$_{58 }$Ce   &      $158 $    &     $ 164$   &  $ 127$ & $ 127   $ &   & $_{59 }$Pr   &  $ 162 $  &  $ 167 $ &   $ 127 $ &  $ 127  $          \\
$_{60 }$Nd   &      $164 $    &     $ 169$   &  $ 127$ & $ 127   $ &   & $_{61 }$Pm   &  $ 167 $  &  $ 171 $ &   $ 127 $ &  $ 127  $          \\
$_{62 }$Sm   &      $169 $    &     $ 173$   &  $ 127$ & $ 127   $ &   & $_{63 }$Eu   &  $ 171 $  &  $ 175 $ &   $ 129 $ &  $ 127  $          \\
$_{64 }$Gd   &      $173 $    &     $ 177$   &  $ 137$ & $ 127   $ &   & $_{65 }$Tb   &  $ 177 $  &  $ 181 $ &   $ 143 $ &  $ 131  $          \\
$_{66 }$Dy   &      $179 $    &     $ 181$   &  $ 152$ & $ 143   $ &   & $_{67 }$Ho   &  $ 179 $  &  $ 183 $ &   $ 158 $ &  $ 151  $          \\
$_{68 }$Er   &      $181 $    &     $ 183$   &  $ 162$ & $ 156   $ &   & $_{69 }$Tm   &  $ 183 $  &  $ 182 $ &   $ 164 $ &  $ 162  $          \\
$_{70 }$Yb   &      $183 $    &     $ 185$   &  $ 169$ & $ 169   $ &   & $_{71 }$Lu   &  $ 185 $  &  $ 185 $ &   $ 173 $ &  $ 175  $          \\
$_{72 }$Hf   &      $185 $    &     $ 185$   &  $ 181$ & $ 179   $ &   & $_{73 }$Ta   &  $ 185 $  &  $ 185 $ &   $ 182 $ &  $ 182  $          \\
$_{74 }$W    &      $185 $    &     $ 185$   &  $ 182$ & $ 182   $ &   & $_{75 }$Re   &  $ 187 $  &  $ 185 $ &   $ 182 $ &  $ 182  $          \\
$_{76 }$Os   &      $187 $    &     $ 185$   &  $ 183$ & $ 182   $ &   & $_{77 }$Ir   &  $ 188 $  &  $ 185 $ &   $ 182 $ &  $ 182  $          \\
$_{78 }$Pt   &      $189 $    &     $ 185$   &  $ 183$ & $ 184   $ &   & $_{79 }$Au   &  $ 191 $  &  $ 185 $ &   $ 185 $ &  $ 182  $          \\
$_{80 }$Hg   &      $195 $    &     $ 187$   &  $ 185$ & $ 184   $ &   & $_{81 }$Tl   &  $ 200 $  &  $ 222 $ &   $ 185 $ &  $ 184  $          \\
$_{82 }$Pb   &      $214 $    &     $ 230 $  &  $ 185$ & $ 184   $ &   & $_{83 }$Bi   &  $ 222 $  &  $ 234 $ &   $ 185 $ &  $ 185  $          \\
$_{84 }$Po   &      $228 $    &     $ 238 $  &  $ 185$ & $ 185   $ &   & $_{85 }$At   &  $ 232 $  &  $ 242 $ &   $ 185 $ &  $ 185  $          \\
$_{86 }$Rn   &      $236 $    &     $ 247 $  &  $ 185$ & $ 185   $ &   & $_{87 }$Fr   &  $ 240 $  &  $ 251 $ &   $ 185 $ &  $ 185  $          \\
$_{88 }$Ra   &      $245 $    &     $ 253 $  &  $ 185$ & $ 185   $ &   & $_{89 }$Ac   &  $ 247 $  &  $ 255 $ &   $ 185 $ &  $ 185  $          \\
$_{90 }$Th   &      $249 $    &     $ 257 $  &  $ 185$ & $ 185   $ &   & $_{91 }$Pa   &  $ 251 $  &  $ 257 $ &   $ 185 $ &  $ 188  $          \\
$_{92 }$U    &      $253 $    &     $ 256 $  &  $ 185$ & $ 201   $ &   & $_{93 }$Np   &  $ 255 $  &  $ 256 $ &   $ 207 $ &  $ 207  $          \\
$_{94 }$Pu   &      $255 $    &     $ 256 $  &  $ 214$ & $ 213   $ &   & $_{95 }$Am   &  $ 257 $  &  $ 256 $ &   $ 220 $ &  $ 219  $          \\
$_{96 }$Cm   &      $257 $    &     $ 256 $  &  $ 228$ & $ 225   $ &   & $_{97 }$Bk   &  $ 257 $  &  $ 259 $ &   $ 237 $ &  $ 229  $          \\
$_{98 }$Cf   &      $257 $    &     $ 259 $  &  $ 243$ & $ 231   $ &   & $_{99 }$Es   &  $ 259 $  &  $ 259 $ &   $ 248 $ &  $ 234  $          \\
$_{100}$Fm   &      $259 $    &     $ 259 $  &  $ 251$ & $ 237   $ &   & $_{101}$Md   &  $ 259 $  &  $ 259 $ &   $ 253 $ &  $ 239  $          \\
$_{102}$No   &      $259 $    &     $ 259 $  &  $ 256$ & $ 241   $ &   & $_{103}$Lr   &  $ 261 $  &  $ 259 $ &   $ 256 $ &  $ 243  $          \\
$_{104}$Rf   &      $261 $    &     $ 259 $  &  $ 256$ & $ 245   $ &   & $_{105}$Db   &  $ 263 $  &  $ 259 $ &   $ 256 $ &  $ 246  $          \\
$_{106}$Sg   &      $263 $    &     $ 259 $  &  $ 256$ & $ 248   $ &   & $_{107}$Bh   &  $ 265 $  &  $ 259 $ &   $ 256 $ &  $ 249  $          \\
$_{108}$Hs   &      $269 $    &     $ 263 $  &  $ 256$ & $ 251   $ &   & $_{109}$Mt   &  $ 275 $  &  $ 277 $ &   $ 256 $ &  $ 252  $          \\
$_{110}$Ds   &      $281 $    &     $ 293 $  &  $ 256$ & $ 253   $ &   &               &        &     &    &            \\
\bottomrule
\end{tabular}
\end{table}

\subsection{Single- and two- nucleon separation energies}
\label{sub:nucleon_separation_energies}
The one- and two-neutron separation energies $S_{\rm n}$ and $S_{\rm 2n}$ are defined as
\begin{eqnarray}
S_{\rm n}(Z,N)  &=& E_{\rm b}(Z,N)- E_{\rm b}(Z,N-1)\,\,\mathrm{and}\\
S_{\rm 2n}(Z,N) &=& E_{\rm b}(Z,N)- E_{\rm b}(Z,N-2)\, ,
\end{eqnarray}
respectively.
The one- and two-proton separation energies $S_{\rm p}$ and $S_{\rm 2p}$ are defined as
\begin{eqnarray}
S_{\rm p}(Z,N)  &=& E_{\rm b}(Z,N)- E_{\rm b}(Z-1,N)\,\,\mathrm{and}\\
S_{\rm 2p}(Z,N) &=& E_{\rm b}(Z,N)- E_{\rm b}(Z-2,N)\, ,
\end{eqnarray}
respectively.
$E_{\rm b}(Z, N)$ is the binding energy of the nucleus with the proton number $Z$ and the neutron number $N$. 
These quantities provide the information of whether a nucleus is stable against one or two nucleon emissions, and thus define the nucleon drip lines. In this work, we consider the aforementioned criteria of deciding whether a nucleus is bound, based on the following two points: (a) both one- and two-nucleon separation energies of this nucleus must be positive, and (b) the proton and neutron Fermi surfaces, $\lambda_{i}\!<\!0, (i\!=\!\mathrm{n},\mathrm{p})$. For each isotopic chain, the last bound nucleus in the proton-rich (neutron-rich) side is the proton- (neutron-) drip-line nucleus.

The one- and two-neutron separation energies, $S_{\rm n}$ and $S_{\rm 2n}$, of the available experimental data and theoretical calculations based on PC-L3R, PC-X, DD-MEX and DD-PCX are listed in Table~\ref{tab:BindingEnergy}. 
The one- (two-) neutron separation energies of 2323 (2237) nuclei have been experimentally determined and compiled in AME2020~\cite{AME2020}. 
We find that the rms of $S_{\rm n}$ ($S_{\rm 2n}$) of comparing experimental data and theoretical results generated from PC-L3R, PCX, DD-MEX and DD-PCX are $0.962$~MeV ($1.300$~MeV), $0.920$~MeV ($1.483$~MeV), $0.993$~MeV ($1.753$~MeV) and $1.010$~MeV ($1.544$~MeV), respectively (Table~\ref{tab:BindingEnergy}). Overall, the rms of $S_\mathrm{n}$ are within the range of $0.920$-$1.010$~MeV, of which the differences of rms values among these RHB approaches are less than 0.1~MeV. 
Nevertheless, the $S_{\rm 2n}$ is more sensitive to the covariant density functional used in the RHB approach. For instance, among all covariant density functionals, PC-L3R yields the lowest rms value of $S_{\rm 2n}$, i.e., $1.304$~MeV, which is $0.449$~MeV lower than the $S_{\rm 2n}$ rms of DD-MEX.

The rms of one-neutron (top panel) and two-neutron (bottom panel) separation energies of isotopic chains generated from PC-L3R, PC-X, DD-MEX and DD-PCX are plotted in Fig.~\ref{fig:S_rms}. For the $8\!\le\!Z\!<\!14$ region, all theoretical models produce a set of similar high rms values of $S_{\rm n}$ and $S_{\rm 2n}$, for instance, DD-MEX produces $1.968$~MeV (rms of $S_{\rm n}$) and $3.442$~MeV (rms of $S_{\rm 2n}$) for isotopes of magnesium, $Z\!=\!12$. The deviations of theoretical $S_{\rm n}$ and $S_{\rm 2n}$ values are larger than experimental ones for the region further away from the proton magic numbers 8, 20, 28, 50, and 82 (Fig.~\ref{fig:S_rms}). PC-L3R produces a lower $S_{\rm 2n}$ rms value, $\sim\!\!0.5$~MeV, at the $Z\!>\!20$ region, especially at the regions close to $Z\!=\!20$, $28$, $50$, and $82$. These deviations are expected to be reduced when the deformation effects are taken into account.


\subsubsection{Proton and neutron drip lines}
\label{sub:proton_neutron_driplines}

Recently, Yang \emph{et al}.~\cite{PRC2021Yang} investigated the proton and neutron drip lines of even-even nuclei of $8\!\!\!\le\!\!\!Z\!\!\!\le\!\!\!104$ using the triaxial relativistic Hartree-Bogoliubov approach. The beyond-mean-field dynamical correlation energies based on the microscopically mapped five-dimensional collective Hamiltonian (5DCH) without additional free parameters were taken into account. With the consideration of triaxial deformation for these even-even nuclei, the calculated binding energies are close to experimental data. They found that almost no triaxially deformed even-even nuclei near the neutron drip line. The odd-nucleon system cannot be assessed by the triaxial RHB approach due to the theoretical limitation.

The work of Zhang \emph{et al}.~\cite{ADNDT2022Zhang} explores the ground-state properties of even-even nuclei of $8\!\!\le\!\!Z\!\!\le\!\!110$ by using the axial-symmetry relativistic Hartree-Bogoliubov theory in continuum (RHBc). Their work predicts $2583$ even-even bound nuclei and successfully provides a low root-mean-square deviation value of $1.518$~MeV by comparing the theoretical and experimental nuclear masses of 637 nuclei~\cite{AME2020}. By comparing the neutron drip line of Zhang \emph{et al}.~(2022) with the neutron drip line predicted by the spherical-symmetry relativistic continuum Hartree-Bogoliubov (RCHB) approach from Xia \emph{et al}.~\cite{ADNDT2018Xia}, overall, it indicates that the inclusion of the deformation effects does not necessarily extend the neutron drip line. A similar finding was also found by In \emph{et al}.~\cite{IJMPE2021In} for the even-even nuclei of $8\!\!\le\!\!Z\!\!\le\!\!20$. This finding is also valid for the heavier isotopic chains like Sm, Gd, and Dy~\cite{PRC2021Pan}. The works of Zhang \emph{et al}.~\cite{ADNDT2022Zhang}, In \emph{et al}.~\cite{IJMPE2021In}, and Pan \emph{et al}.~\cite{PRC2021Pan} indicate that the expansion or reduction of the neutron drip line depends on the evolution of the degree of deformation along an isotopic chain towards the drip line. Hence, the predicted neutron drip line could be more extended toward the more neutron-rich side if the deformation increases towards the drip line, and vice versa. The exploration of the nuclear landscape of these works are only limited to even-even nuclei, however.

Continuum effects are considered in some previous works, for instance, the recent works of Xia \emph{et al}.~\cite{ADNDT2018Xia}, Zhang \emph{et al}.~\cite{ADNDT2022Zhang}, and Ravli{\'c} \emph{et al}.~\cite{Ravlic2023}. The work of Xia \emph{et al}.~\cite{ADNDT2018Xia} is based on the spherical-symmetry relativistic continuum Hartree-Bogoliubov (RCHB) approach, which is developed by using the Greens function technique and by implementing the appropriate asymptotic behavior of nuclear densities at large coordinate space $r$~\cite{NPA1998Meng}. Compared to the RHB approaches without the consideration of such continuum effects, the RCHB approach predicts more neutron-rich bound nuclei, expanding the neutron drip line. Nonetheless, the proton drip lines obtained by the RCHB and RHB approaches are rather similar, due to the Coulomb repulsive interaction among protons and the Coulomb barrier~\cite{ADNDT2018Xia}. The recent finding of Ravli{\'c} \emph{et al}.~\cite{Ravlic2023} indicates that the neutron drip line is more reduced for the RHB approach with the consideration of finite-temperature (FT) continuum effect at temperature, $T\!=\!1$-$2$~MeV. The finite-temperature continuum effect becomes more pronounced with increasing temperature, particularly for $T\!\gtrapprox\!1$~MeV. Therefore, for an accurate description of the neutron drip line at $T\!\approx\!1$~MeV and above, the finite-temperature continuum effect is non-negligible (see next section for more detailed description).

Other than the study based on the triaxial RHB approach with 5DCH~\cite{PRC2021Yang} discussed above, some previous theoretical works indicate that the nuclear binding energies of transitional and shape-coexistent even-even nuclei at the region between the $N\!=\!126$ and $N\!=\!184$ shell closures are uncertain due to the existence of several nuclear binding energy minima in the solutions of axial-symmetry RHB approach~\cite{PRC2014Agbemava}. These even-even nuclei have flat potential energy surfaces in the axial deformations, which lead to several energy minima and thus the binding energies are not well defined. 
To specifically calculate these transitional and shape-coexistent even-even nuclei, the methods going beyond mean field have to be considered/implemented to improve the description of single-particle energies that affect the potential energy surfaces \cite{PPNP2011Niksic,PRC2013Fu,PRC2013Yao}.

Overall, the previous studies of nuclear landscape based on the triaxial RHB with 5DCH \cite{PRC2021Yang}, axially deformed RHBc \cite{ADNDT2022Zhang}, and FT-RHB in axial symmetry \cite{Ravlic2023}, are constrained in even-even nuclei due to the theoretical limitation. The works based on the spherical-symmetry RCHB \cite{ADNDT2018Xia} and axially deformed RHBc \cite{ADNDT2022Zhang} approaches only assessed the PC-PK1 interaction without comparing other covariant density functionals in equal footing. The present work studies both even-nucleon and odd-nucleon systems and compares the nuclear landscapes and properties yielded from covariant density functionals of density-dependent meson-exchange models, non-linear point-coupling models, and density-dependent point-coupling models. The present work provides a valuable initial estimate of the nuclear landscape of odd-nucleon system that can be further explored in future by considering axial-symmetry and/or triaxial-symmetry calculations and by implementing the FT-RHB approach \cite{Ravlic2023}, continuum effects, and beyond mean-field approach for super heavy nuclei.

Tables~\ref{tab:proton_dripline} and \ref{tab:neutron_dripline} show the nuclides of proton and neutron drip lines predicted by the RHB approach with the PC-L3R~\cite{PLB2023Liu}, PC-X~\cite{PLB2020Taninah}, DD-MEX~\cite{PLB2020Taninah}, or DD-PCX~\cite{PRC2019Yuksel} interactions. The prediction of proton drip lines from these interactions is close to one another and only up to a difference of 3 neutron numbers. 
Compared with PC-L3R and PC-X, the density-dependent models (i.e., DD-MEX and DD-PCX) predict fewer bound nuclei for the neutron-rich side. This systematic trends become even more pronounced with increasing of neutron number, causing a less extensive neutron-rich region. For instance, for the region heavier than $Z\!=\!82$, the neutron drip lines of PC-L3R and PC-X are more extensive than the neutron drip lines of density-dependent models. The drip line of PC-X is the most extensive at around $Z\!=\!82$-$93$, of which more than 45 bound neutron-rich nuclei are predicted, compared with the density-dependent model. Such prediction of more bound neutron-rich nuclei is partly due to the stiffness of the symmetry energy, $E_\mathrm{sym}$, of these interactions: 
$E_\mathrm{sym}^\mathrm{PC-L3R}$~$\!=\!35.8$~MeV, 
$E_\mathrm{sym}^\mathrm{PC-X}$~$\!=\!35.2$~MeV, 
$E_\mathrm{sym}^\mathrm{DD-MEX}$~$\!=\!32.3$~MeV, and $E_\mathrm{sym}^\mathrm{DD-PCX}$~$\!=\!31.1$~MeV
. Interactions with a stiffer symmetry energy tend to predict more bound nuclei, resulting in the systematic shift of the neutron drip line towards more neutron-rich nuclei~\cite{PRC2010Kazuhiro,PRC2023Ravli}. 


\begin{figure}[pos=t] 
\centering
\includegraphics[width=\linewidth, angle=0]{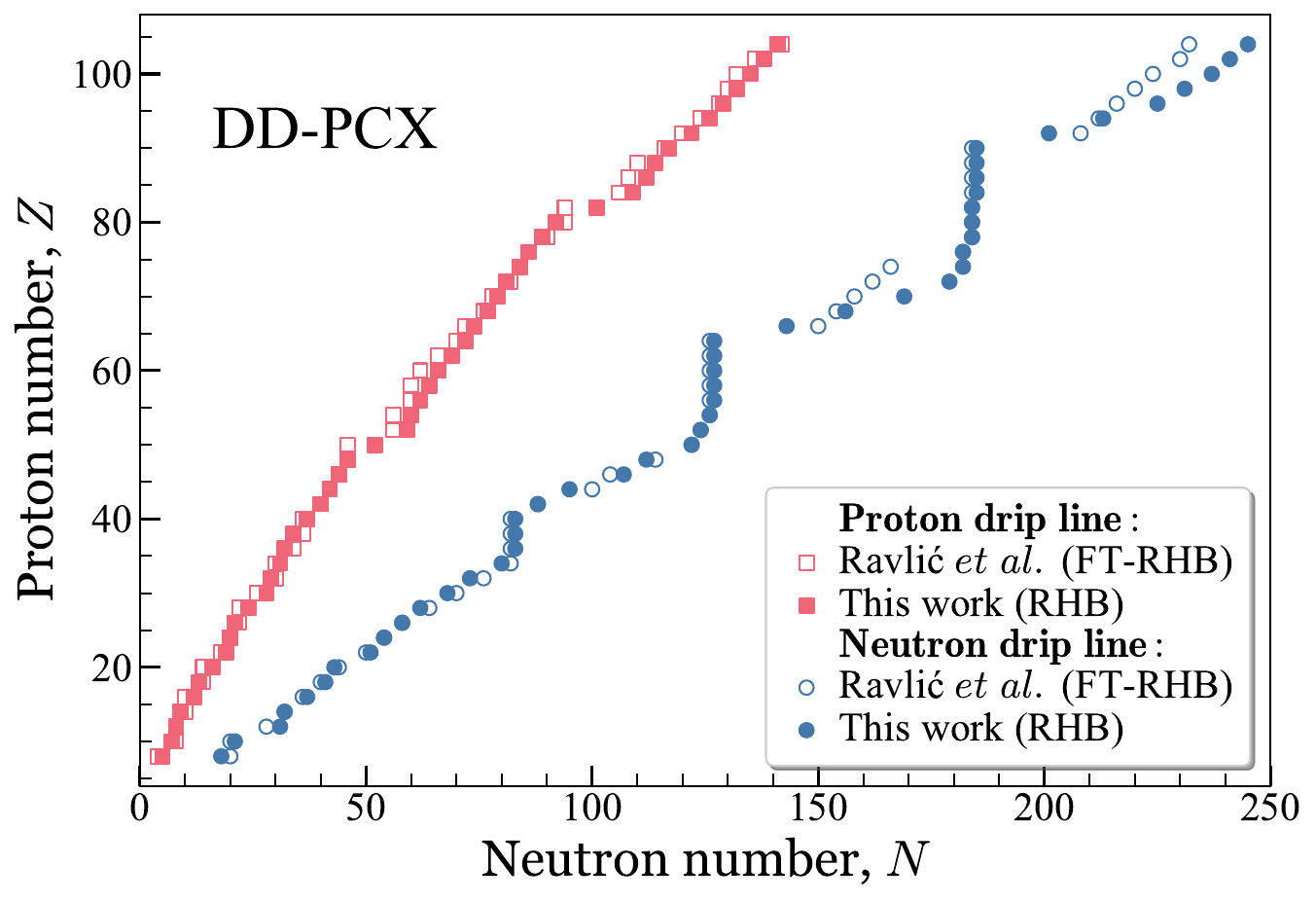}
\caption{The location of the two-proton (red squares) and two-neutron (blue circles) drip lines between oxygen ($Z\!=\!8$) and rutherfordium ($Z\!=\!104$) produced from the RHB approach with the DD-PCX interaction (filled symbols) and yielded from the RHB approach with the DD-PCX interaction in the axially-deformed harmonic oscillator basis with the consideration of finite temperature of $T\!=\!0$~MeV and continuum states (empty symbols) \citep{Ravlic2023}.}
\label{fig:Dripline_DDPCX}
\end{figure}%

\begin{figure*}
\centering
\begin{minipage}{0.8\textwidth}
\centering
\includegraphics[width=\linewidth, angle=0]{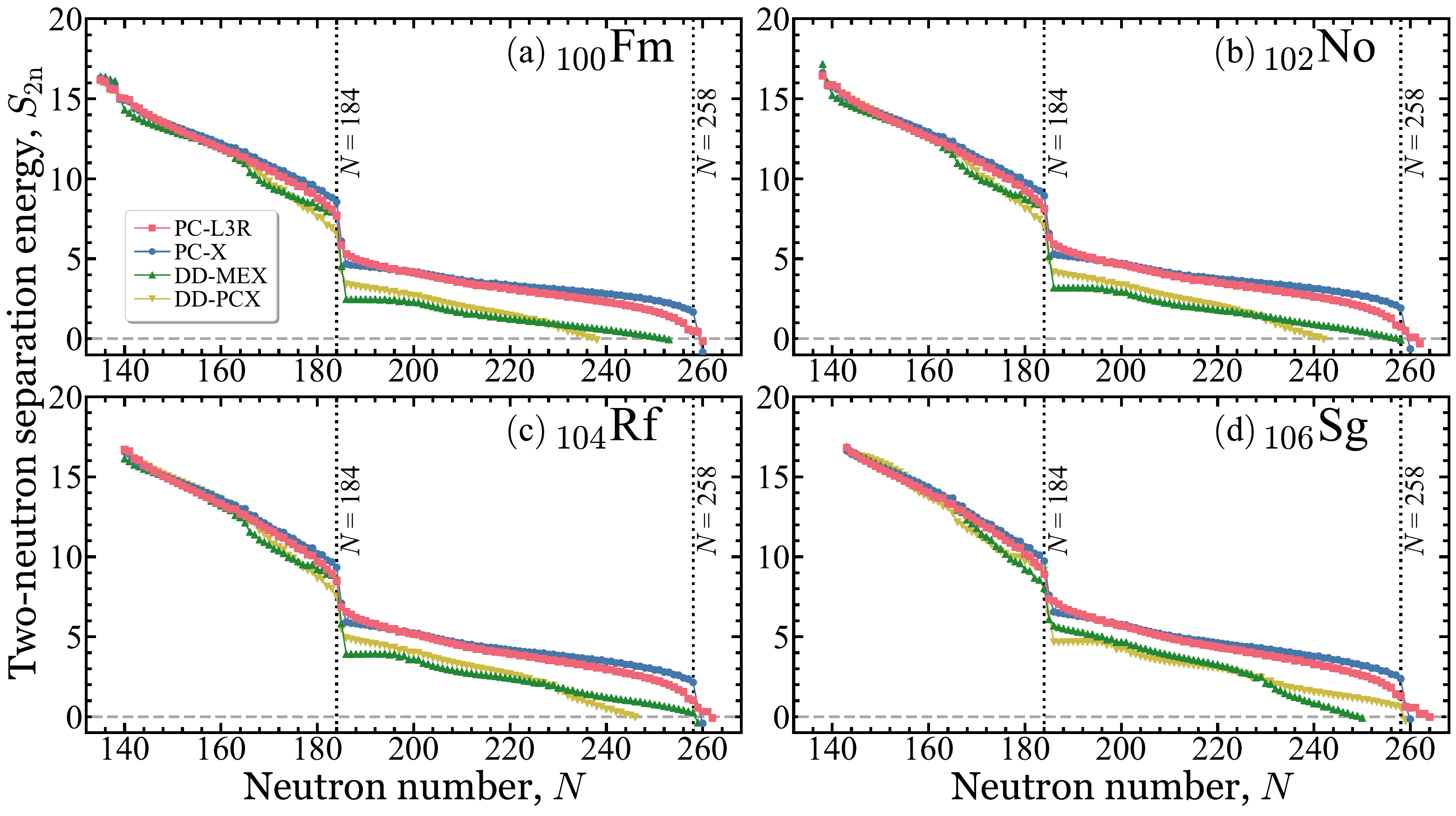}
\end{minipage}
\caption{Two-neutron separation energies, $S_\mathrm{2n}$, of the fermium (a), nobelium (b), rutherfordium (c), and seaborgium (d) isotopic chains as the functions of neutron number, obtained from the RHB approach with the PC-L3R~\cite{PLB2023Liu}, PC-X~\cite{PLB2020Taninah}, DD-MEX~\cite{PLB2020Taninah}, or DD-PCX~\cite{PRC2019Yuksel} interactions.}
\label{fig:new_magic_no}
\end{figure*}%


\subsubsection{Comparison of proton and neutron drip lines predicted by the RHB approaches in the spherical harmonic oscillator basis and in the axially-deformed harmonic oscillator basis}
\label{sub:comparison_FTRHB}

The recent work of {Ravli\'c} \emph{et al}.~\cite{Ravlic2023} estimating the proton and neutron drip lines implements the RHB approach with the DD-PCX interaction in the axially-deformed harmonic oscillator basis with the consideration of finite temperature (FT). The properties of even-even nuclei are obtained from solving the two coupled FT-RHB equations of the nucleus and vapor system and the vapor system only. To describe nuclei at finite temperature, the Bonche-Levit-Vautherin continuum subtraction procedure \cite{Bonche1984,Bonche1985} is used to separate the continuum states from the bound states.

We compare the proton and neutron drip lines predicted from the present work based on DD-PCX with the ones from {Ravli\'c} \emph{et al}.~\cite{Ravlic2023} (Fig.~\ref{fig:Dripline_DDPCX}), which are based on $N_\mathrm{osc}\!=\!20$ and temperature of $T\!=\!0$~MeV. We find that both proton drip lines are rather similar (red filled and empty squares in Fig.~~\ref{fig:Dripline_DDPCX}), whereas the neutron drip line of this work is more extensive than the one of FT-RHB at the regions of $Z\!=\!44$-$48$ and $N\!=\!100$-$106$, at $Z\!=\!66$-$74$ and $N\!=\!150$-$182$, and at $Z\!=\!92$-$105$ and $N\!=\!200$-$250$ (blue filled and empty circles in Fig.~~\ref{fig:Dripline_DDPCX}). The present comparison agrees with two previous sets of results, see Refs.~\cite{ADNDT2018Xia, ADNDT2022Zhang}. With comparing the drip lines of temperature $T\!=\!0$~MeV, the less extensive neutron drip line predicted by FT-RHB could be due to the deformation effect. This comparison renders the impact of the considered deformation effect that can be taken into account for calculating the neutron drip line of heavy nuclei. 



\subsubsection{New neutron magic numbers for super heavy nuclei}
\label{sub:new_magic_numbers}

In the region of super heavy nuclei, the neutron magic-number shell at $N\!=\!184$ is evidently shown by the $S_\mathrm{2n}$ of the fermium, nobelium, rutherfordium, and seaborgium isotopic chains (Fig.~\ref{fig:new_magic_no}), obtained from PC-L3R, PC-X, DD-MEX, and DD-PCX. With comparing the $S_\mathrm{2n}$ difference of neighboring isotopes, $\Delta S_\mathrm{2n} \!=\! S_\mathrm{2n}(Z,N+2) \!-\! S_\mathrm{2n}(Z,N)$, along these isotopic chains, the density-dependent interactions (i.e., DD-MEX and DD-PCX) indicate a more significant $N\!=\!184$ shell structure than PC-L3R and PC-X, of which the $\Delta S_\mathrm{2n}$ values of DD-MEX and especially DD-PCX are obviously larger than the $\Delta S_\mathrm{2n}$ values of PC-L3R and PC-X at $N\!=\!184$ (Table~\ref{tab:BindingEnergy}).


PC-X and DD-MEX predict a shell structure at $N\!=\!258$, similar to the structure at $N\!=\!184$. For PC-L3R, the shell structures may appear at $N\!=\!256$ or at $N\!=\!258$. The different locations of the shell structure could be due to the choice of the pairing strengths and deformation effect. The last bound neutron-rich isotopes of Fm, No, Rf, and Sg predicted by DD-PCX are lighter than $N\!=\!256$, and thus there is no theoretical estimate from DD-PCX for the neutron magic-number shell at $N\!=\!256$ or $258$. In addition, the super heavy nuclei region exist some bound nuclei beyond the primary neutron drip line in several regions, forming peninsulas of stability adjacent to the nuclear mainland. The deformation plays a decisive role in the formation of these stability peninsulas~\cite{PLB2018Afanasjev,PRC2019Agbemava,PRC2014Agbemava}.

\subsection{Nucleon distributions}
\label{sub:nucleon_distributions}

We study the nucleon distributions calculated from the RHB approach with covariant density functionals, PC-L3R~\cite{PLB2023Liu}, PC-X~\cite{PLB2020Taninah}, DD-MEX~\cite{PLB2020Taninah}, and DD-PCX~\cite{PRC2019Yuksel} interactions, by comparing the root-mean-square (rms) radii of matter ($R_\mathrm{m}$), of neutron ($R_\mathrm{n}$), of proton ($R_\mathrm{p}$), and of charge ($R_\mathrm{c}$) distributions (Table~\ref{tab:radii}). 
The rms radii of neutron, proton, and charge distributions, in unit of fm, are expressed as
\begin{eqnarray}
R_\mathrm{n}\!\!\!\! &=& \!\!\!\!\langle r_{n}^{2}\rangle^{1/2} = \Big\{ \int d^{3}\bm{r} [r^{2}\rho_{\mathrm{n},V}(\bm{r})]  \Big\}^{1/2},  \\
R_\mathrm{p} \!\!\!\!&=&\!\!\!\! \langle r_{p}^{2}\rangle^{1/2} = \Big\{ \int d^{3}\bm{r} [r^{2}\rho_{\mathrm{p},V}(\bm{r})]  \Big\}^{1/2},~\mathrm{and}\\
R_\mathrm{c}\!\!\!\! &=&\!\!\!\! \sqrt{\langle r_{p}^{2}\rangle+0.64} \, ,~\mathrm{respectively}.
\end{eqnarray}
The available experimental data \cite{ADNDT2013Angeli} of charge radius $R_\mathrm{c}$ are also shown.

\begin{figure*}
\centering
\begin{minipage}{0.7\textwidth}
\centering
\includegraphics[width=\linewidth, angle=0]{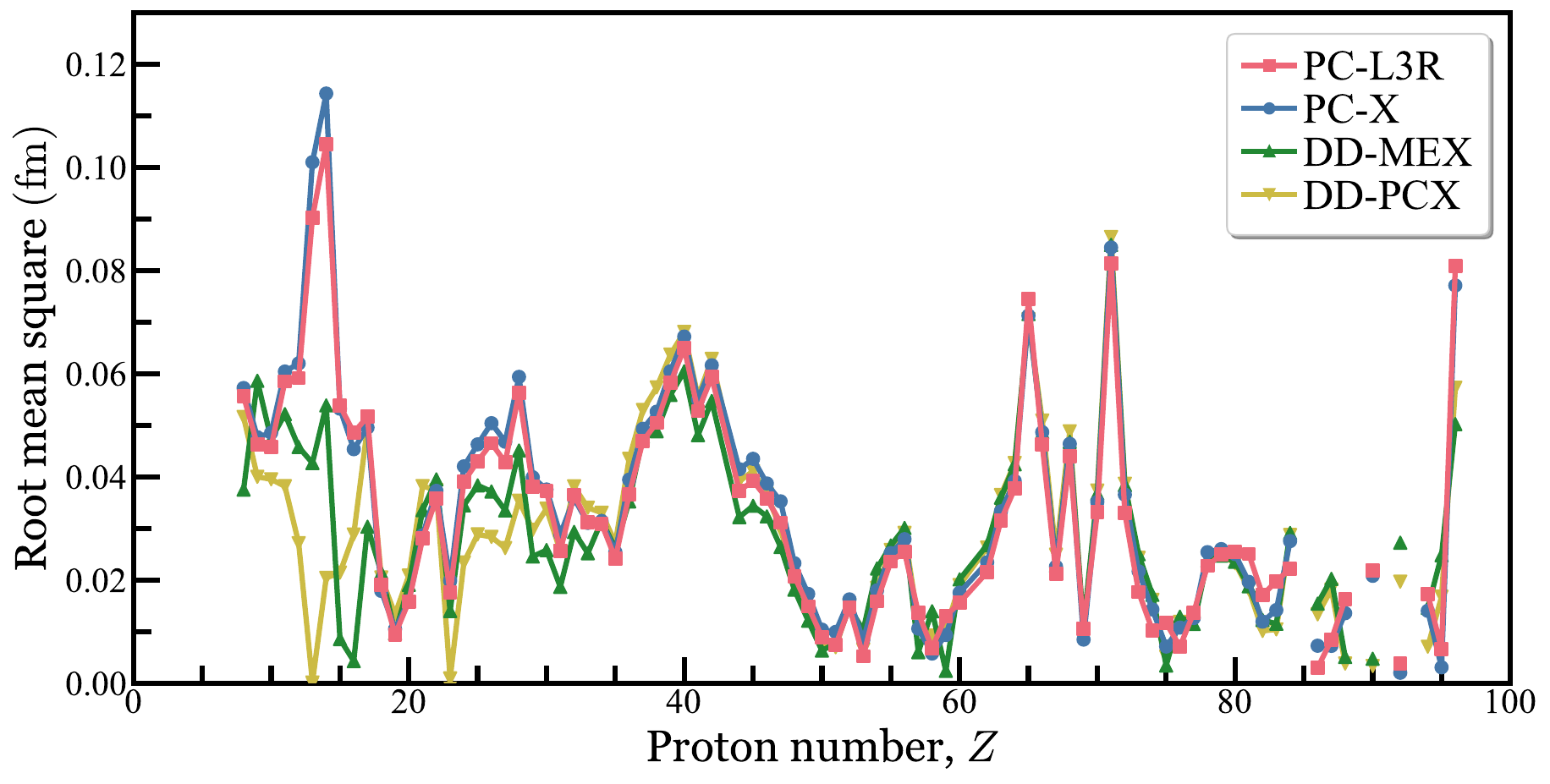}
\end{minipage}
\caption{The root-mean-square deviation values (rms) of comparing the available experimental charge radii of $932$ nuclei \cite{ADNDT2013Angeli} with the theoretical counterparts from the RHB approach with the PC-L3R~\cite{PLB2023Liu}, PC-X~\cite{PLB2020Taninah}, DD-MEX~\cite{PLB2020Taninah}, and DD-PCX~\cite{PRC2019Yuksel} interactions. For each $Z\!=\!13$ and $23$, there is only one available experimental data. There is no available experimental data for the $Z\!=\!85$, $89$, $91$, and $93$ nuclei.}
\label{fig:rms_Rc}
\end{figure*}%

\subsubsection{Charge radii}
\label{sub:charge_radii}

The nuclear charge radius is one of the most obvious and important nuclear properties that indicates the influence of effective interactions on nuclear structure. The charge radii for each isotope resulted from the RHB approach with the PC-L3R, or PC-X, or DD-MEX, or DD-PCX interactions and the available experimental counterpart are presented in Table~\ref{tab:radii}. 
The deviations of theoretical charge radii from the experimental counterparts are in the range of about $\pm0.05$~fm. 
The rms of comparing the available experimental charge radii ($932$ nuclei) with the theoretical charge radii calculated from PC-L3R, PC-X, DD-MEX, and DD-PCX, are $0.035$~fm, $0.037$~fm, $0.034$~fm, and $0.035$~fm, respectively.
The rrs of charge radii for PC-L3R, PC-X, DD-MEX, and DD-PCX are in the range of about $0.765$-$0.864$~\%. 
The trend of the deviations is similar to the ones of $S_{\rm n}$ and $S_{\rm 2n}$ plotted in Fig.~\ref{fig:S_rms}, in such a way that large deviations happen at the regions away from the neutron magic number $20$, $28$, $50$ , $82$, and $126$ (Fig.~\ref{fig:rms_Rc}). These deviations are expected to be reduced once the deformation effects are included in the RHB calculations. The investigations and results with the consideration of deformation effects in the RHB approach will be published elsewhere.


\subsubsection{Neutron radii}
\label{sub:neutron_radii}

The calculated root-mean-square radii of neutron distributions $R_\mathrm{n}$ for all nuclei of $8\!\leq\!Z\!\leq\!110$ are shown as a function of neutron number $N$ in Figs.~\ref{fig:Rn_PCL3R_all},~\ref{fig:Rn_PCX_all},~\ref{fig:Rn_DDPCX_all}, and~\ref{fig:Rn_DDMEX_all}  with respect to the empirical $R_{\rm n}\!=\! r_{0} N^{1/3}$, of which several $r_{0}$ are fitted in the RHB calculations based on the PC-L3R, or PC-X, or DD-MEX, or DD-PCX interactions (Table~\ref{tab:r0}). 
The empirical $R_{\rm n}$ curves of $r_{0}$ ($^{40}$Ca) are steeper than the curves of $r_{0}$ ($^{132}$Sn) and of $r_{0}$ ($^{208}$Pb) at the $N\!<\!50$ region. 
The systematic trends of the theoretical root-mean-square radii of neutron distributions produced from PC-L3R and PC-X closely follow the empirical $R_{\rm n}$, except for the region of extreme neutron-rich nuclei. 
At the $N\!<\!50$ region, the $R_{n}$ of PC-L3R and of PC-X are closer to the empirical $R_{\rm n}$ of $r_{0}$ ($^{40}$Ca), but at the $N\!>\!50$ region, these theoretical $R_\mathrm{n}$ are closer to the empirical $R_{\rm n}$ of $r_{0}$ ($^{208}$Pb). 
The DD-MEX and DD-PCX produce a similar trend of $R_\mathrm{n}$ compared with the $R_\mathrm{n}$ of PC-L3R and of PC-X but their $R_\mathrm{n}$ are lower than the empirical $R_{\rm n}$ of $r_{0}$ ($^{208}$Pb) at the $N\!>\!150$ region, especially the $R_\mathrm{n}$ of DD-PCX. This is mainly due to the properties of DD-PCX of having a lower slope of symmetry energy than the ones of PC-L3R, PC-X, and DD-MEX. See Refs.~\cite{PLB2023Liu,Yuksel2021,Huang2020} for the plots of symmetric nuclear matter of PC-L3R, PC-X, DD-MEX, and DD-PCX.

\begin{figure*}
\centering
\begin{minipage}{0.33\textwidth}
\centering
\includegraphics[width=\linewidth, angle=0]{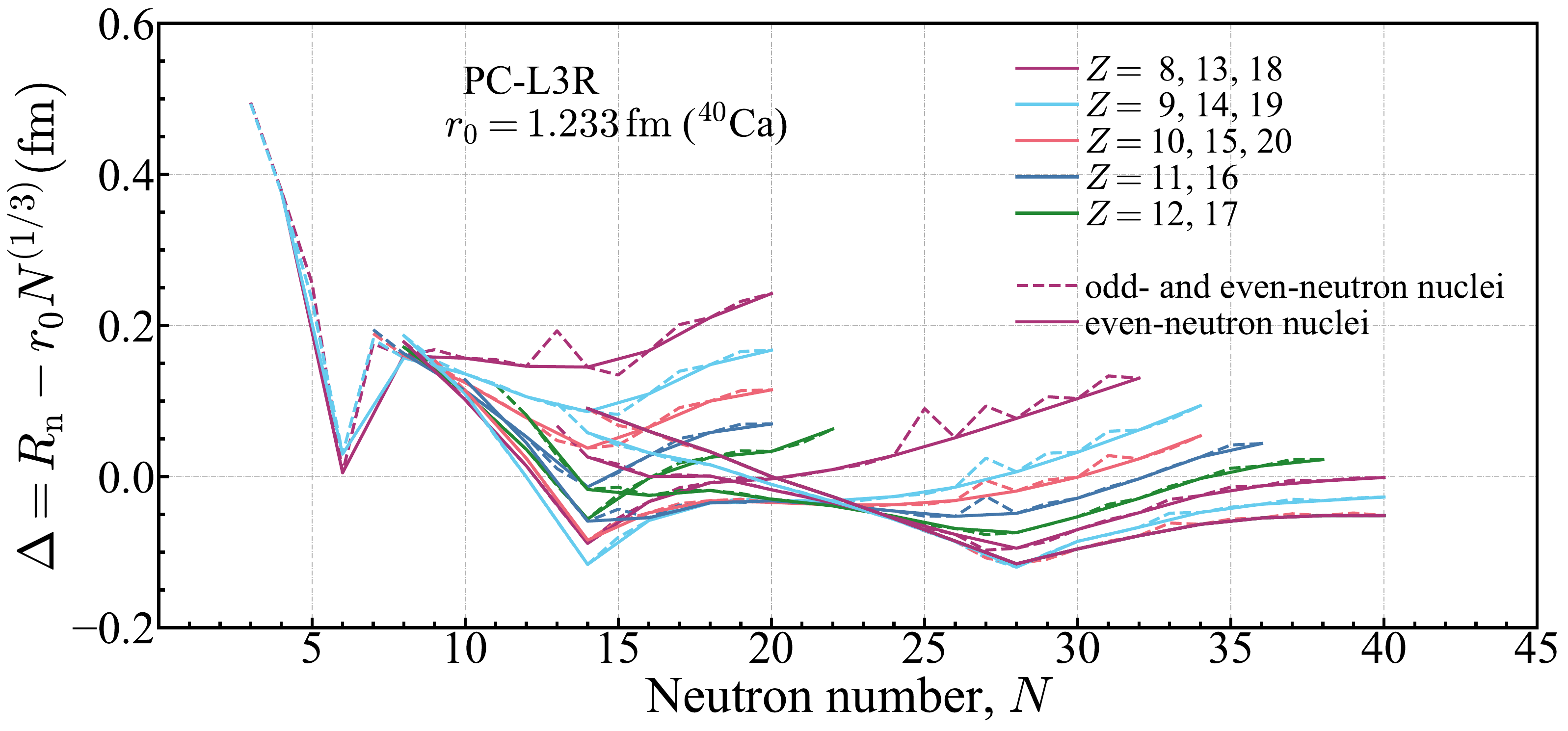}
\label{fig:RN-PCL3R-20-8-20}
\end{minipage}
\begin{minipage}{0.33\textwidth}
\centering
\includegraphics[width=\linewidth, angle=0]{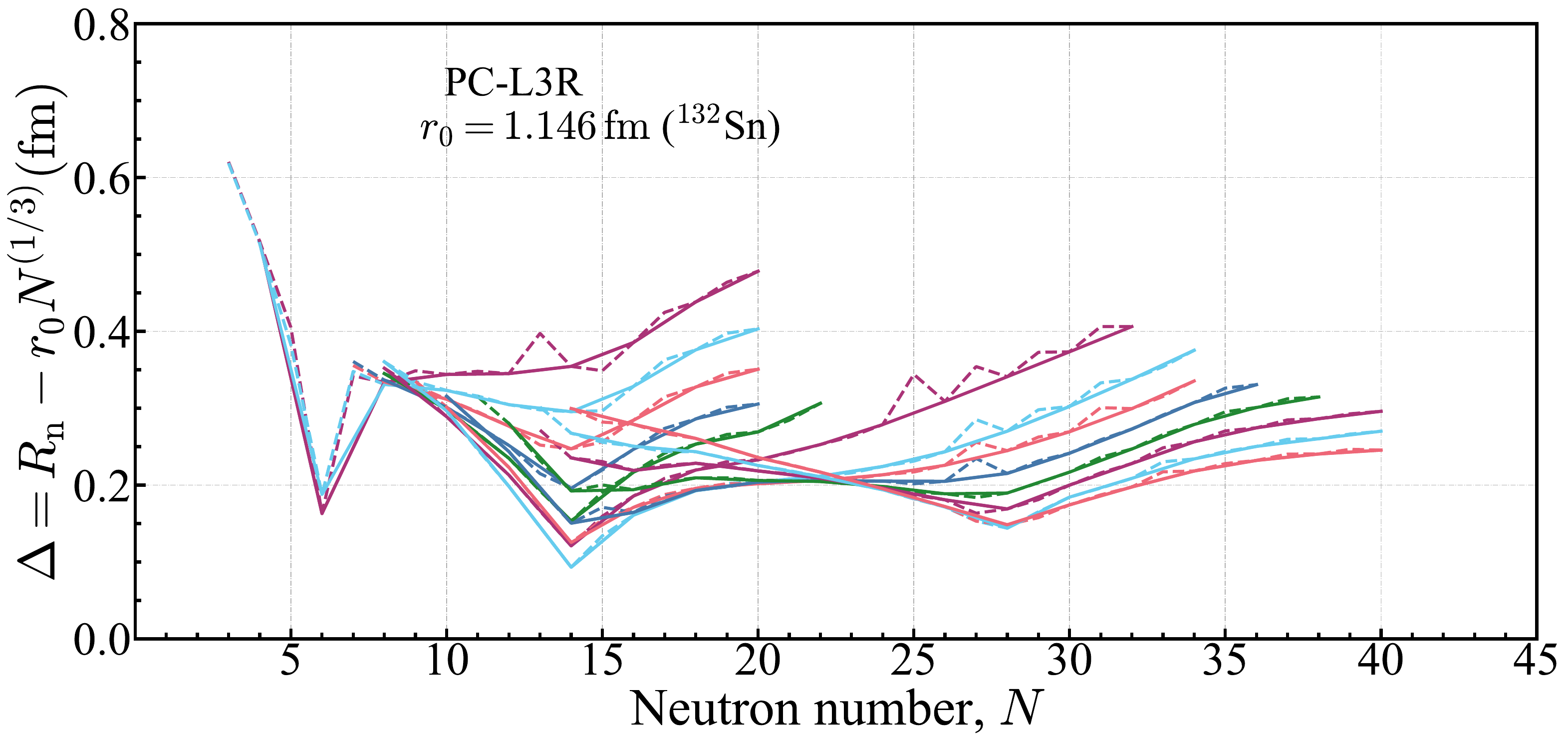}
\label{fig:RN-PCL3R-82-8-20}
\end{minipage}
\begin{minipage}{0.33\textwidth}
\centering
\includegraphics[width=\linewidth, angle=0]{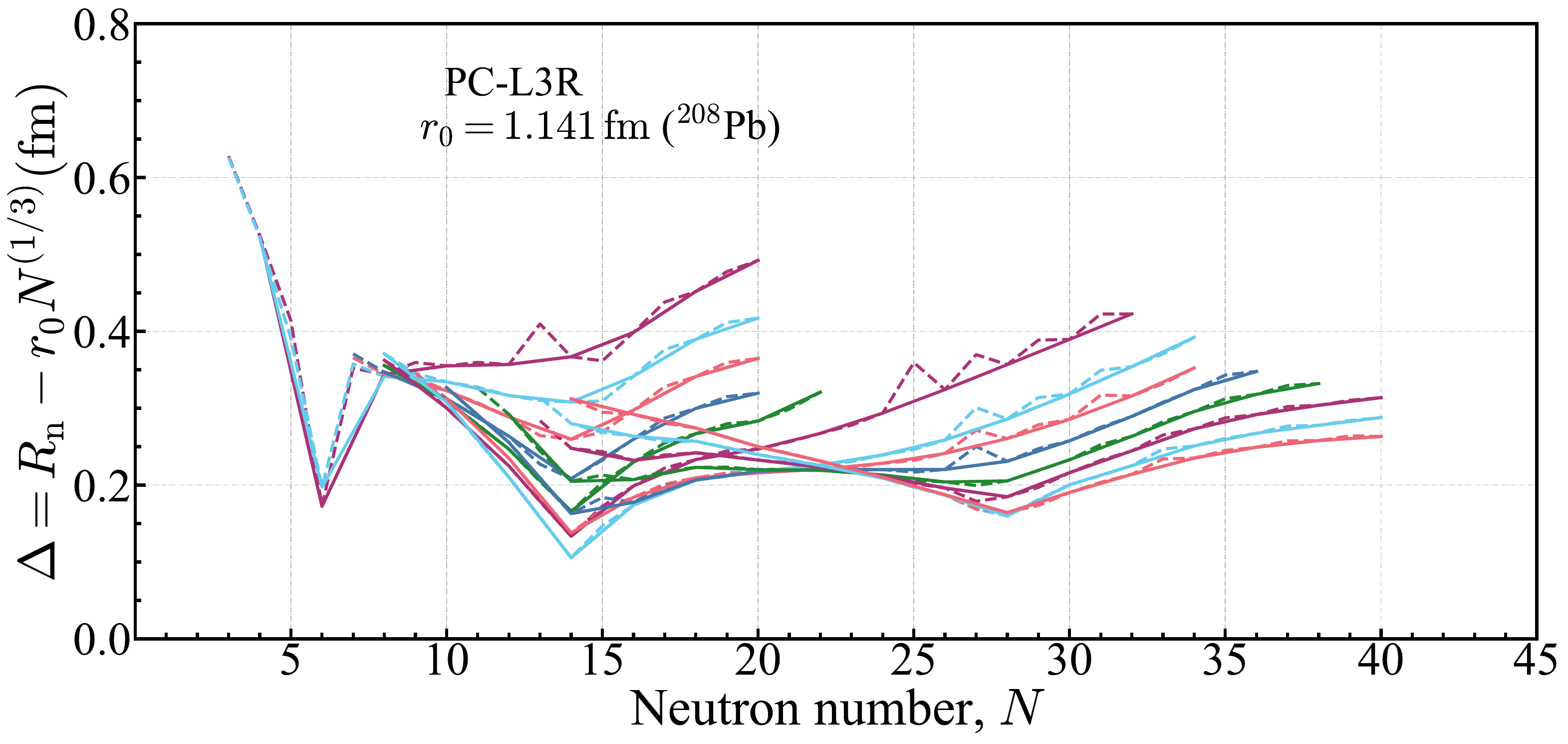}
\label{fig:RN-PCL3R-126-8-20}
\end{minipage}
\vspace{-2mm}
\begin{minipage}{0.33\textwidth}
\centering
\includegraphics[width=\linewidth, angle=0]{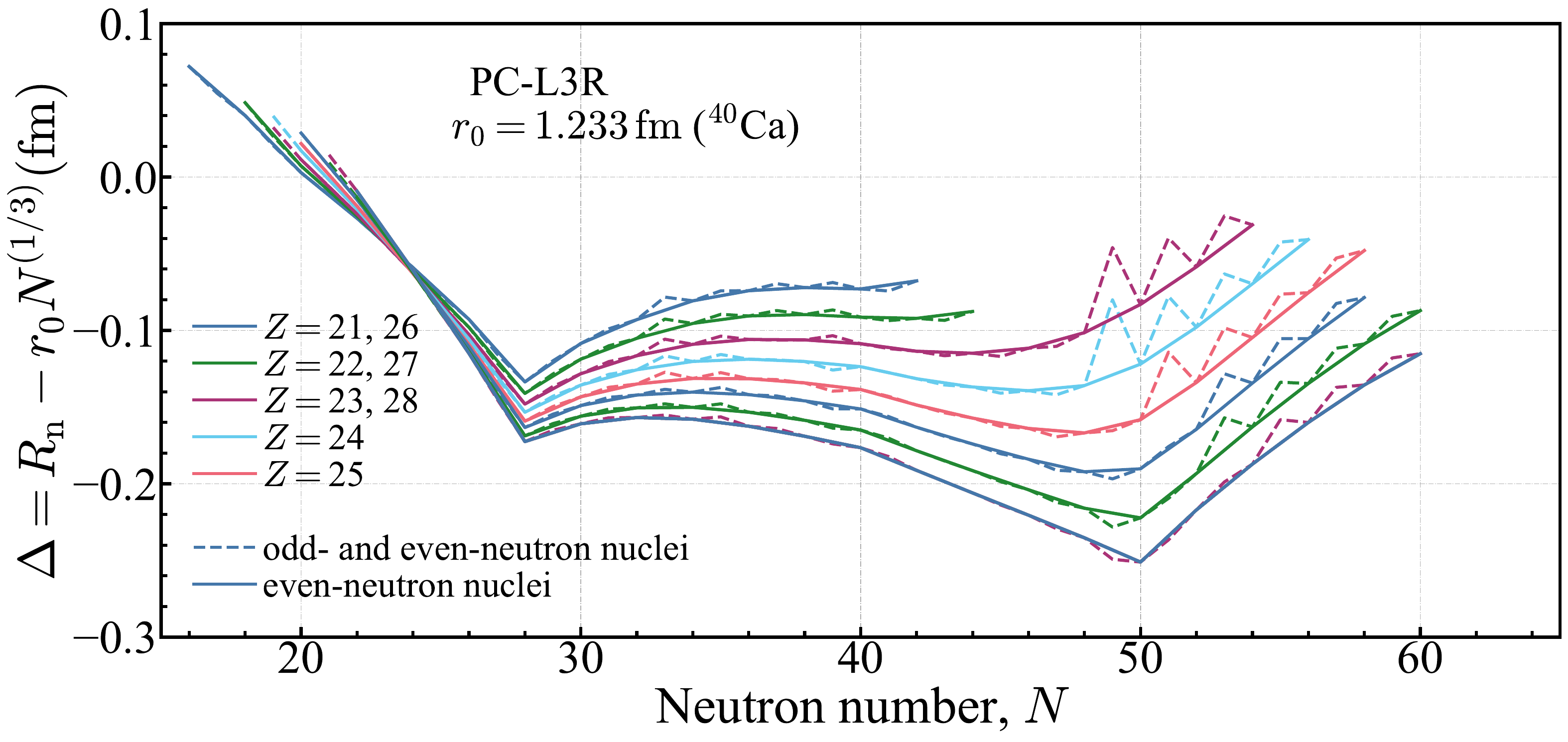}
\label{fig:RN-PCL3R-20-20-28}
\end{minipage}
\begin{minipage}{0.33\textwidth}
\centering
\includegraphics[width=\linewidth, angle=0]{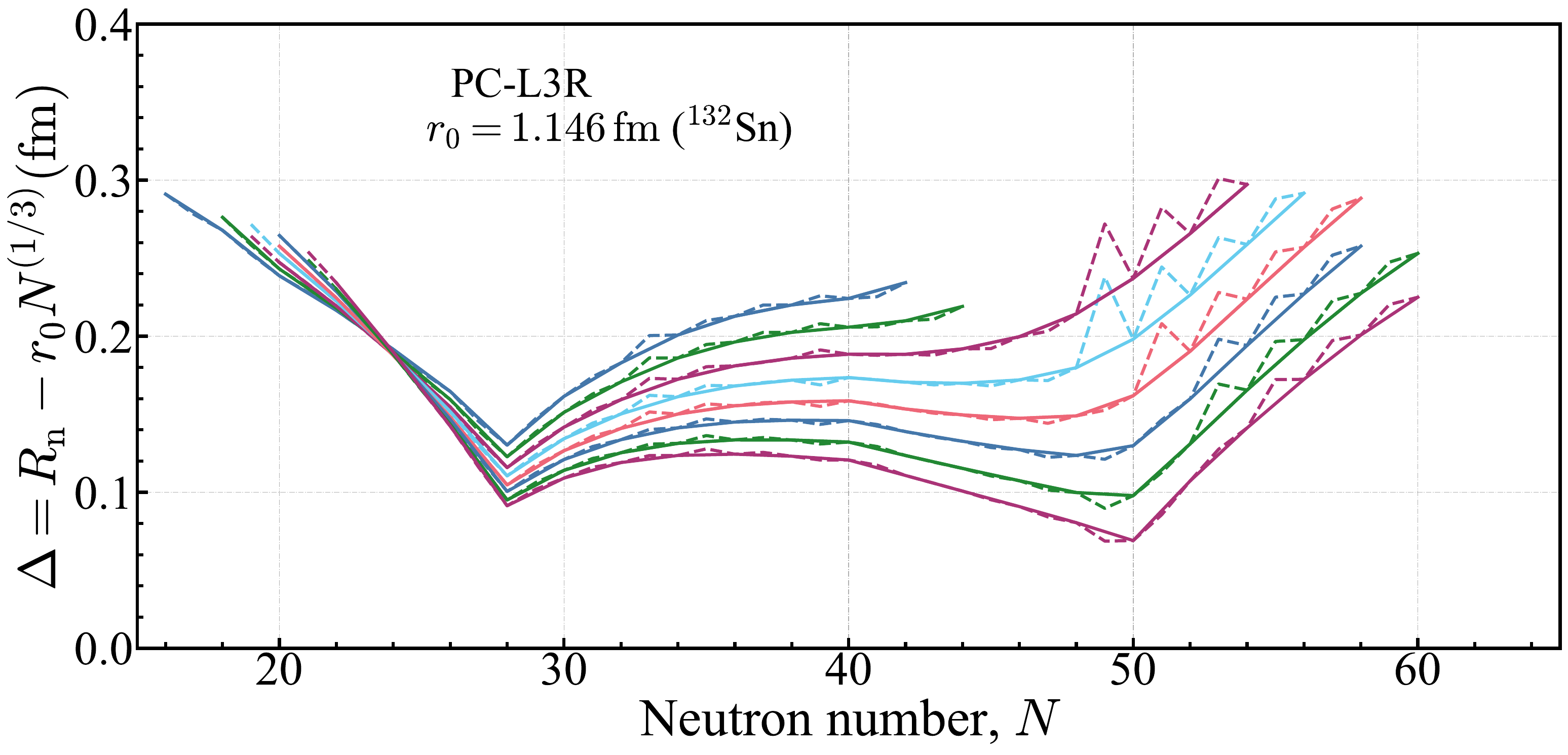}
\label{fig:RN-PCL3R-82-20-28}
\end{minipage}
\begin{minipage}{0.33\textwidth}
\centering
\includegraphics[width=\linewidth, angle=0]{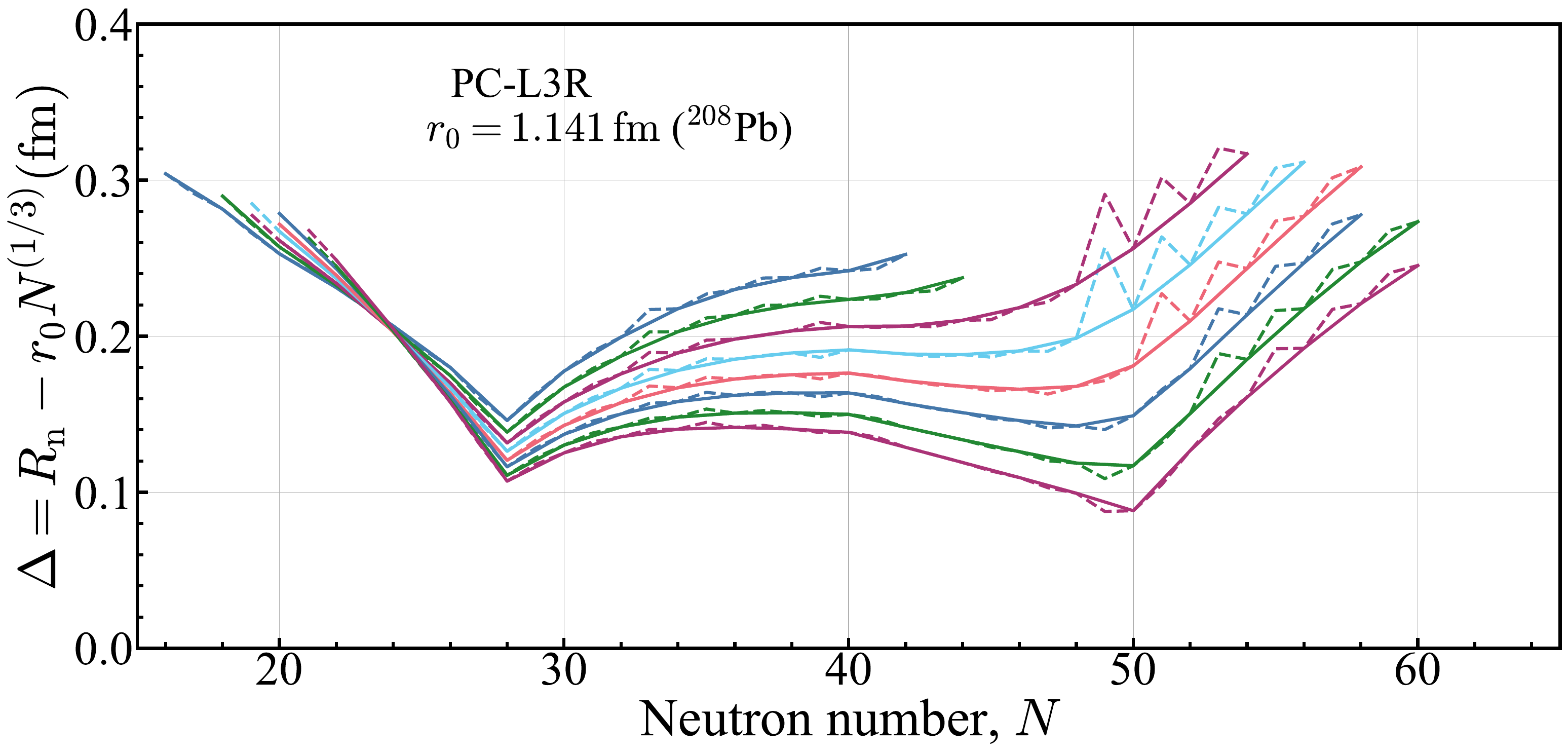}
\label{fig:RN-PCL3R-126-20-28}
\end{minipage}
\vspace{-2mm}
\begin{minipage}{0.33\textwidth}
\centering
\includegraphics[width=\linewidth, angle=0]{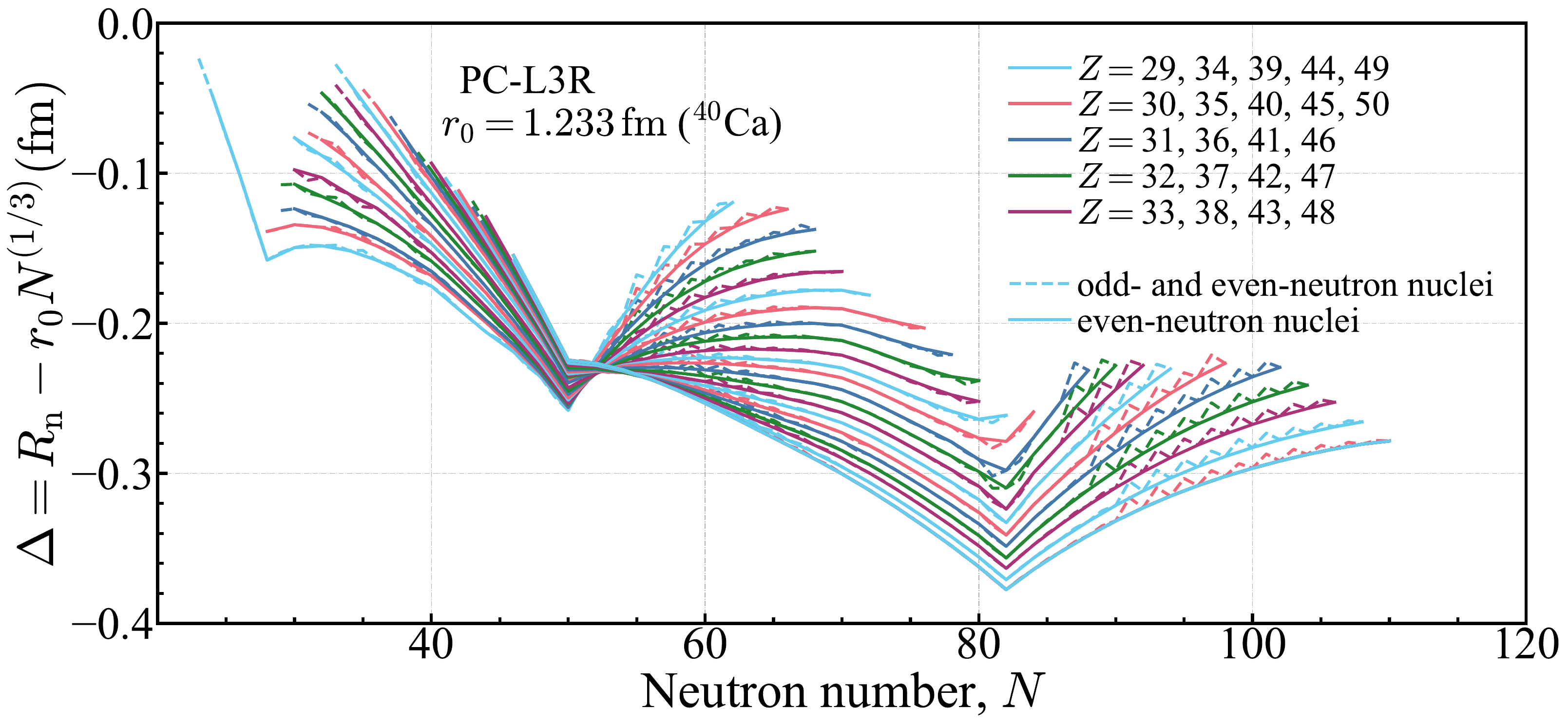}
\label{fig:RN-PCL3R-20-28-50}
\end{minipage}
\begin{minipage}{0.33\textwidth}
\centering
\includegraphics[width=\linewidth, angle=0]{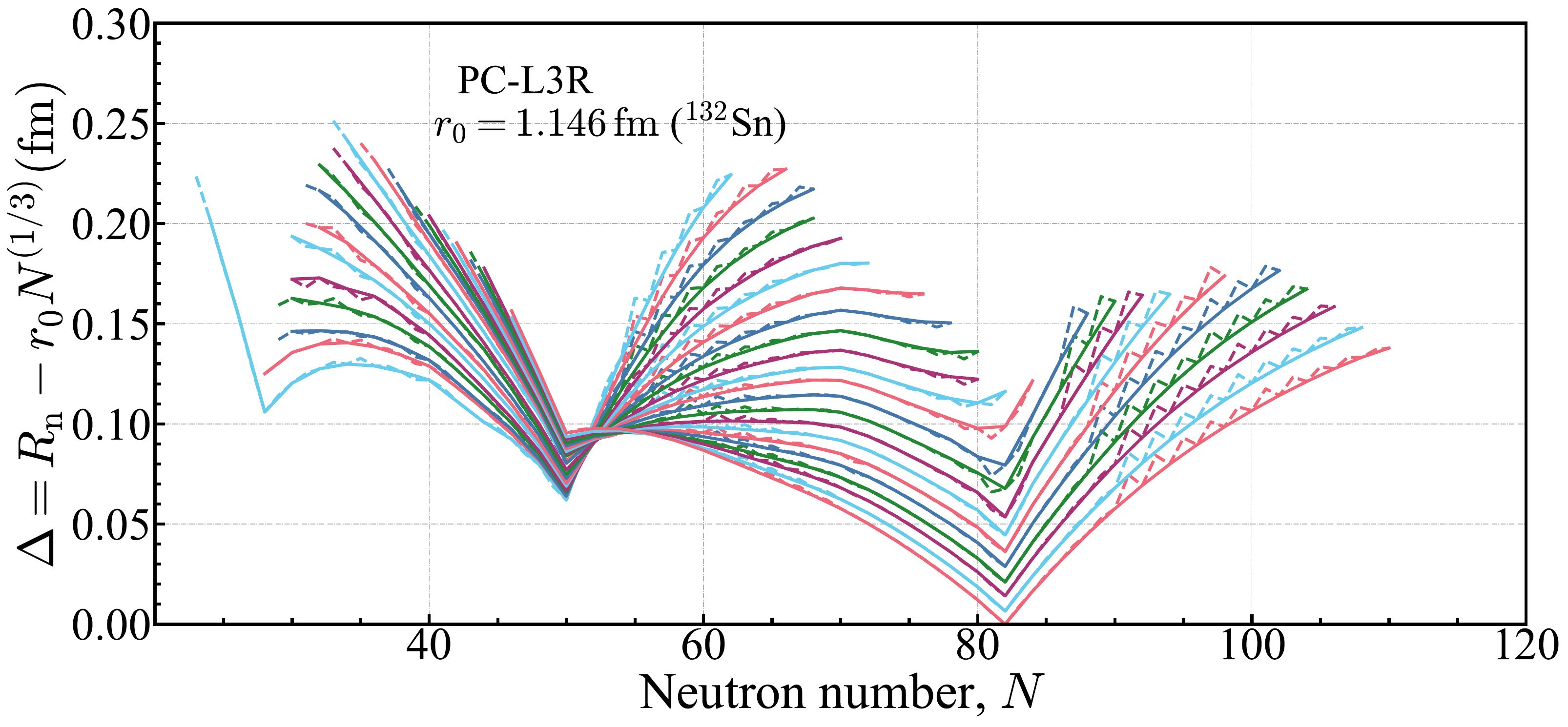}
\label{fig:RN-PCL3R-82-28-50}
\end{minipage}
\begin{minipage}{0.33\textwidth}
\centering
\includegraphics[width=\linewidth, angle=0]{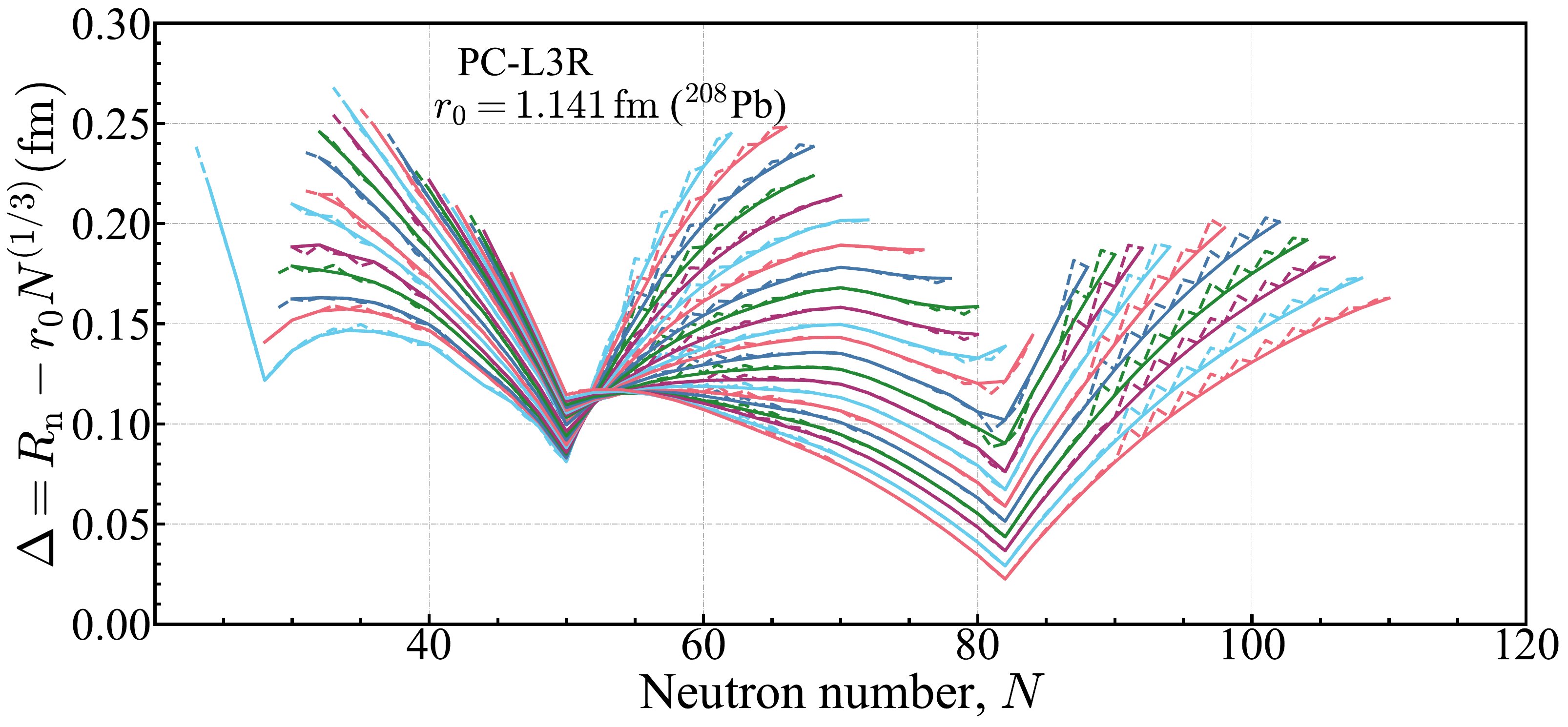}
\label{fig:RN-PCL3R-126-28-50}
\end{minipage}
\vspace{-2mm}
\begin{minipage}{0.33\textwidth}
\centering
\includegraphics[width=\linewidth, angle=0]{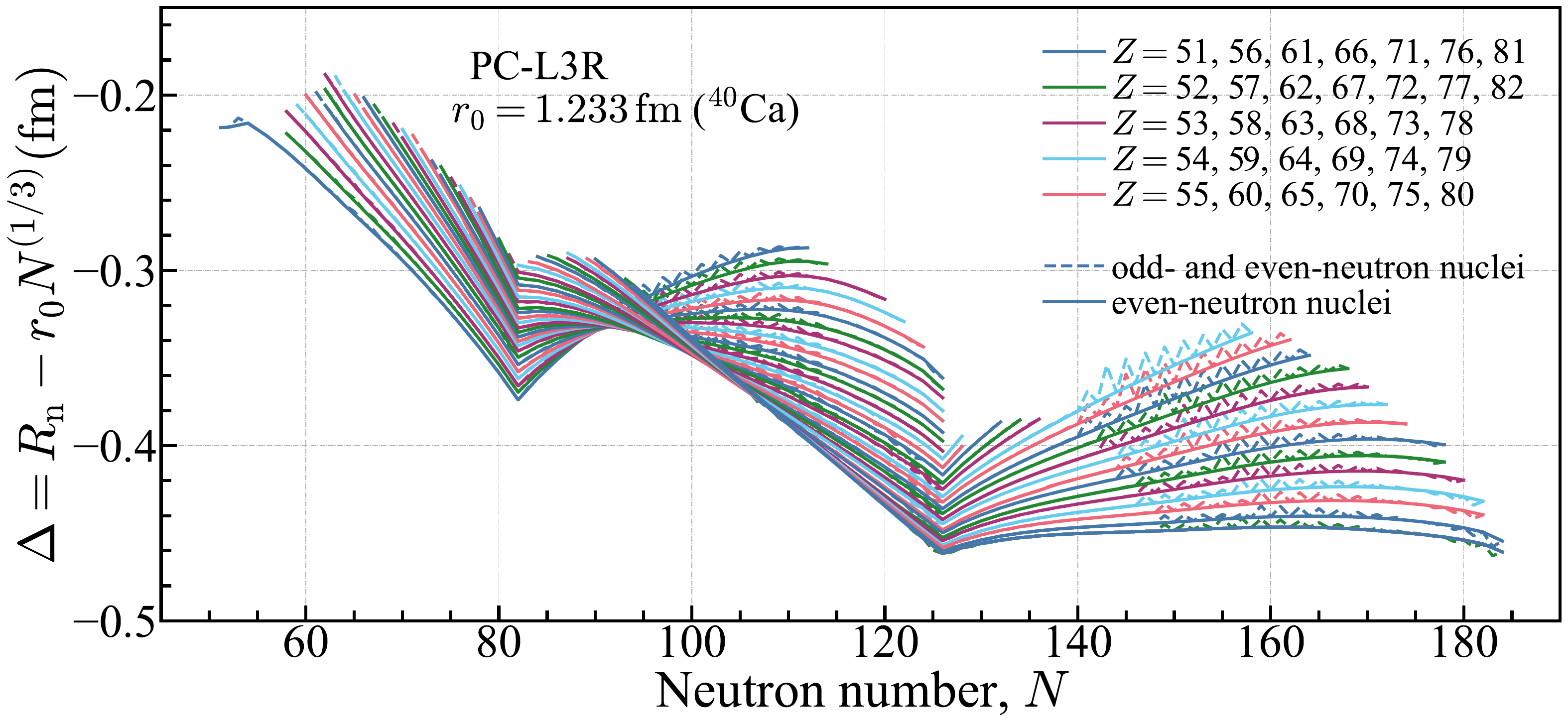}
\label{fig:RN-PCL3R-20-50-82}
\end{minipage}
\begin{minipage}{0.33\textwidth}
\centering
\includegraphics[width=\linewidth, angle=0]{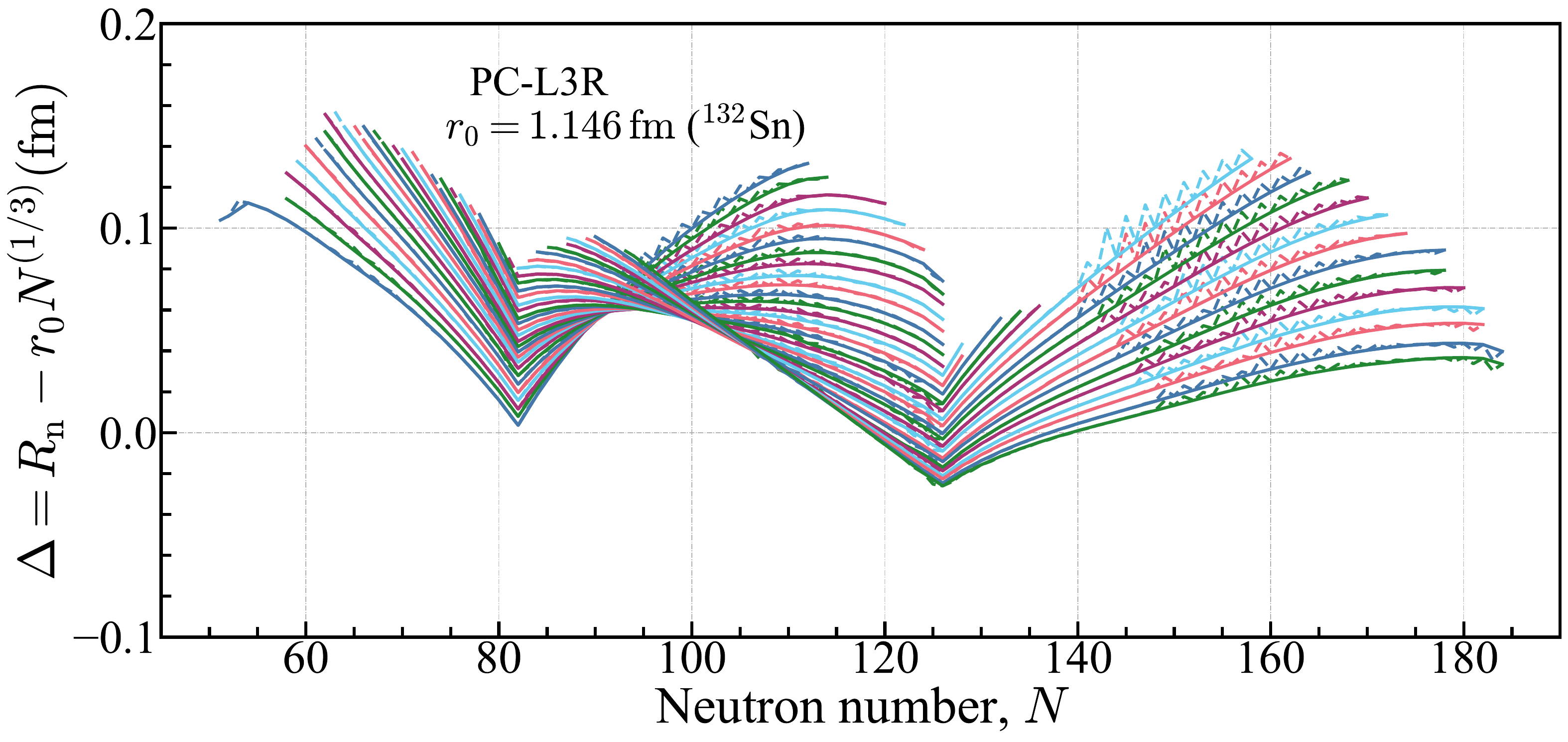}
\label{fig:RN-PCL3R-82-50-82}
\end{minipage}
\begin{minipage}{0.33\textwidth}
\centering
\includegraphics[width=\linewidth, angle=0]{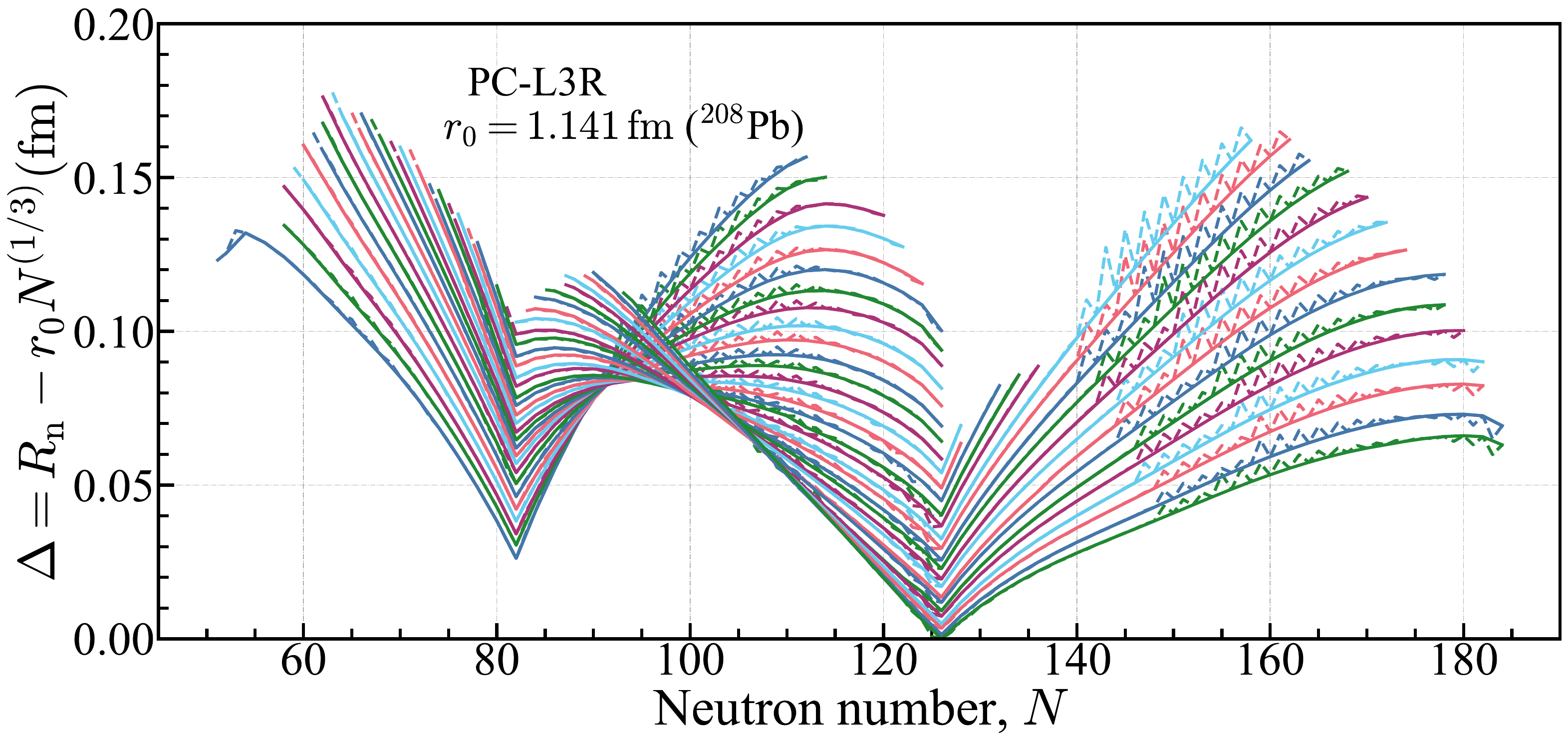}
\label{fig:RN-PCL3R-126-50-82}
\end{minipage}
\vspace{-2mm}
\begin{minipage}{0.33\textwidth}
\centering
\includegraphics[width=\linewidth, angle=0]{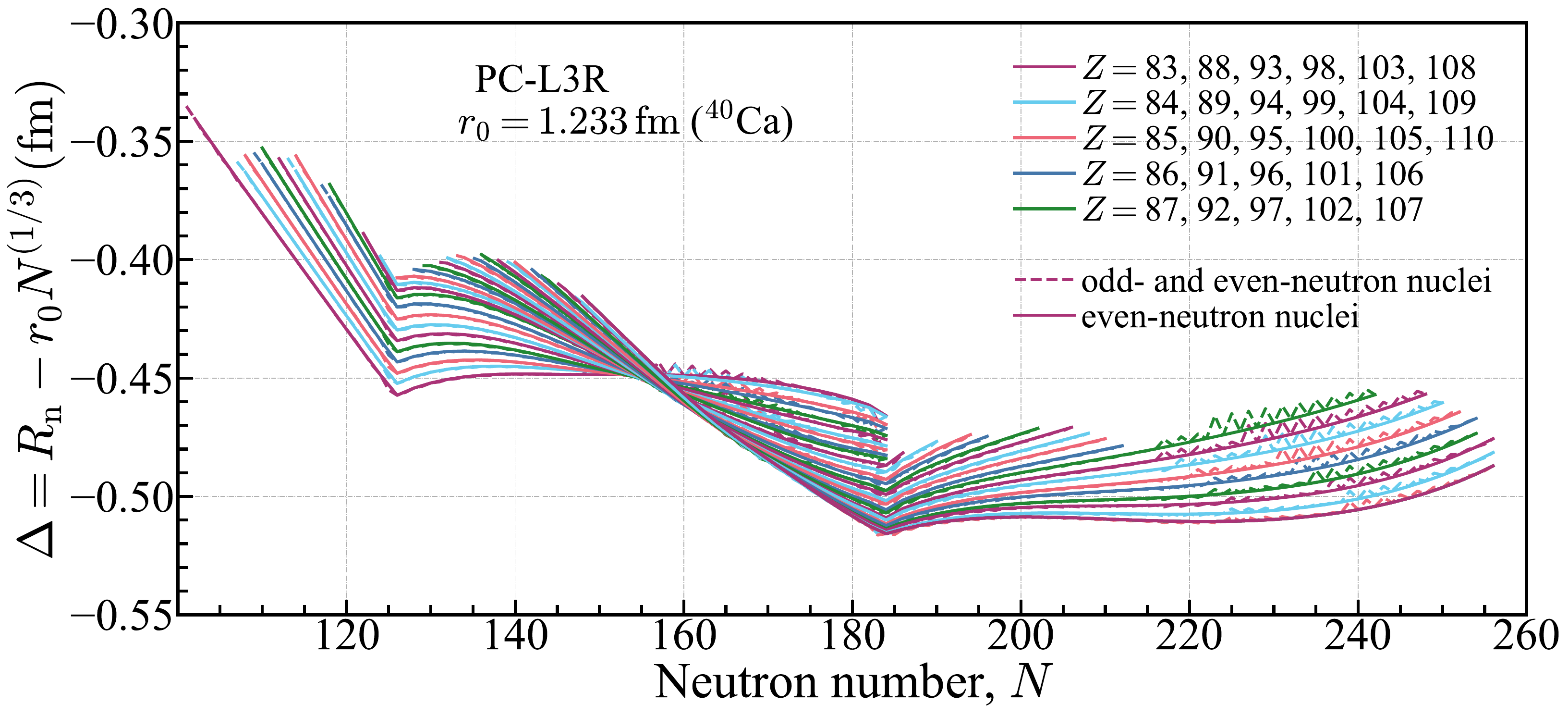}
\label{fig:RN-PCL3R-20-82-110}
\end{minipage}
\begin{minipage}{0.33\textwidth}
\centering
\includegraphics[width=\linewidth, angle=0]{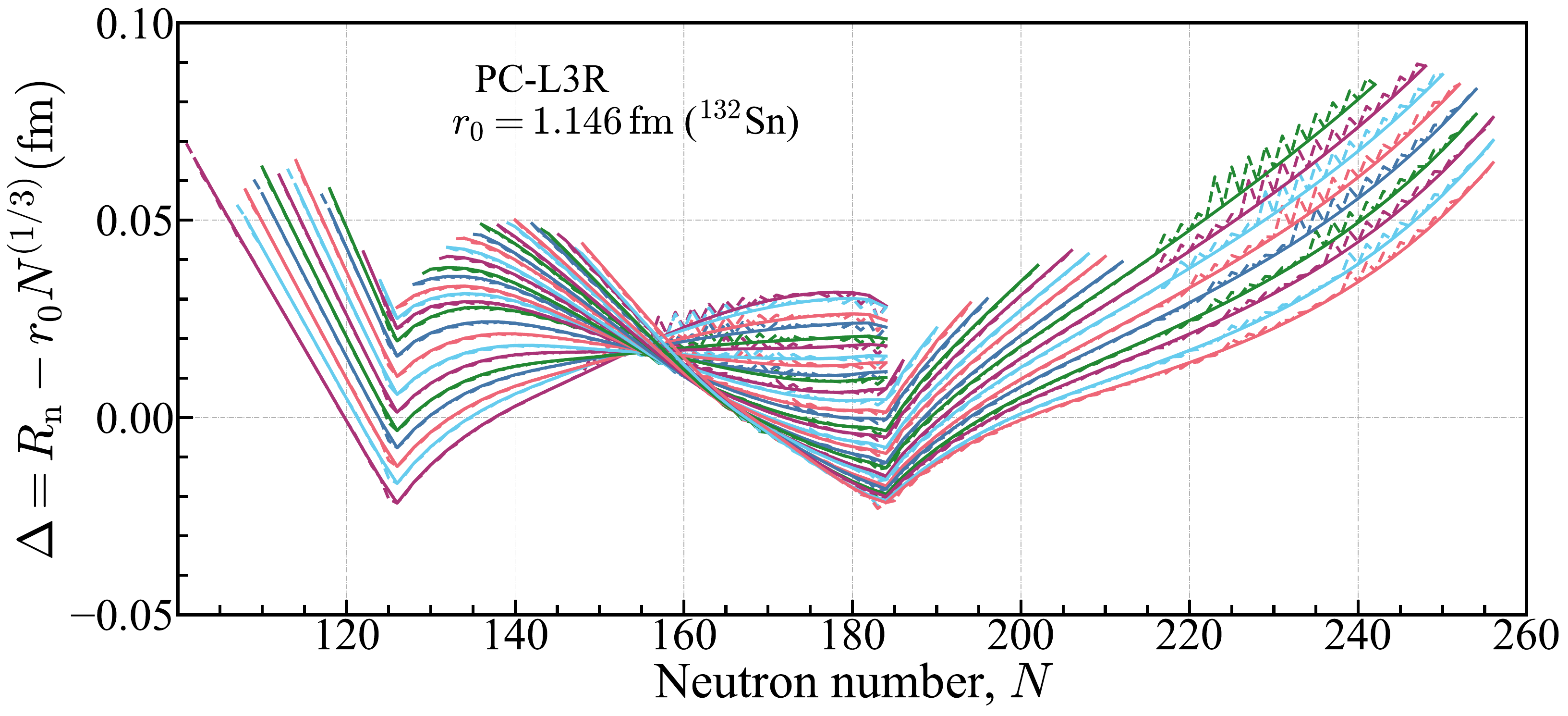}
\label{fig:RN-PCL3R-82-82-110}
\end{minipage}
\begin{minipage}{0.33\textwidth}
\centering
\includegraphics[width=\linewidth, angle=0]{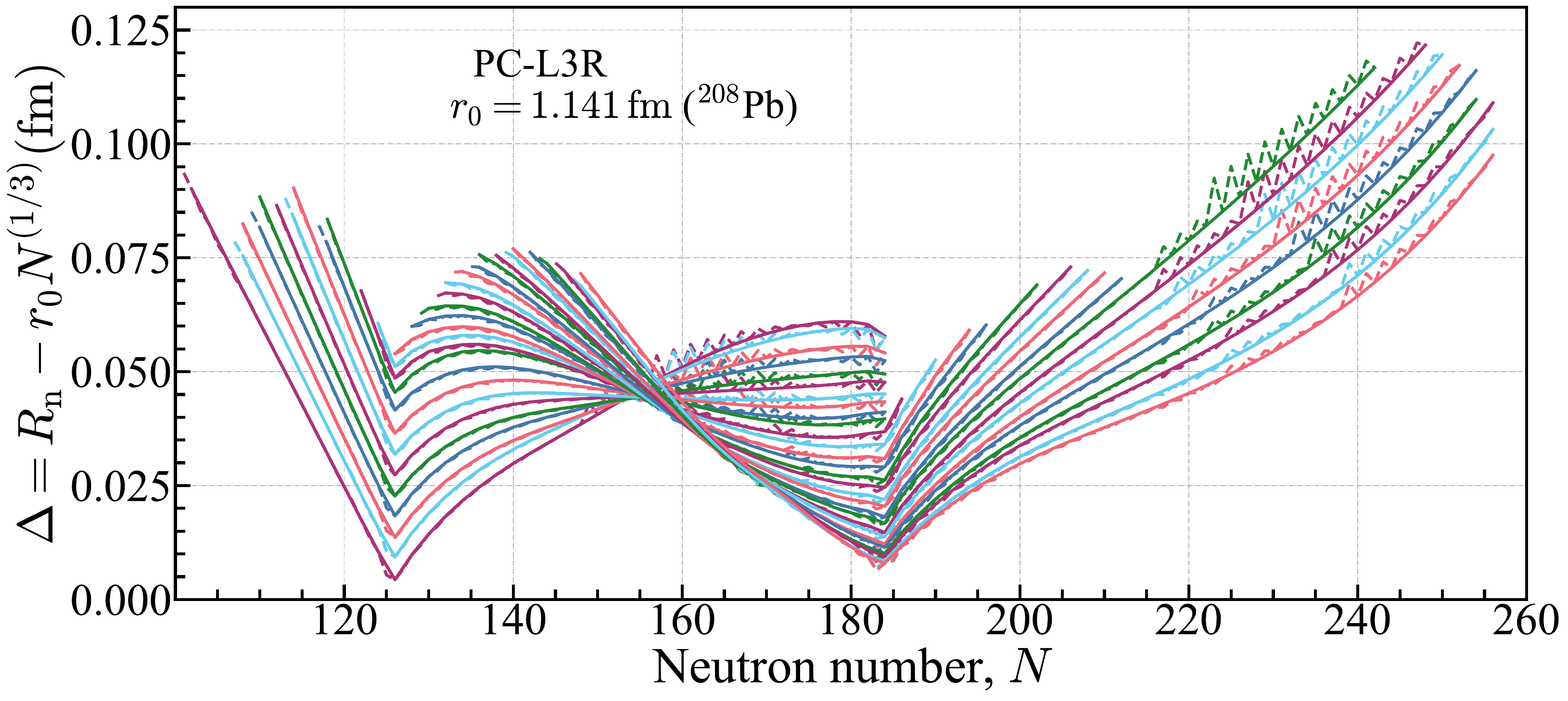}
\label{fig:RN-PCL3R-126-82-110}
\end{minipage}
\vspace{-2mm}
\centering
\begin{minipage}{\textwidth}
\caption{The deviations ($\Delta$) of theoretical root-mean-square radius of neutron distributions ($R_\mathrm{n}$) and the empirical $R_\mathrm{n}\!=\!r_{0}N^{1/3}$. The calculations are based on the RHB approach with the PC-L3R interaction for all isotopic chains in the range of $8\!\leq\! Z\!\leq\!110$. The $\Delta$ values of five ranges of $8\!\leq\! Z\!\leq\!20$, $21\!\leq\! Z\!\leq\!28$, $29\!\leq\! Z\!\leq\!50$, $51\!\leq\! Z\!\leq\!82$, and $83\!\leq\! Z\!\leq\!110$ are plotted accordingly from the top to bottom panels. Three empirical $R_\mathrm{n}$ curves are plotted using the $r_{0}$ of $^{40}$Ca, of $^{132}$Sn, and of $^{208}$Pb.}
\label{fig:RN}
\end{minipage}
\end{figure*}

\begin{figure*}
\centering
\begin{minipage}{0.33\textwidth}
\centering
\includegraphics[width=\linewidth, angle=0]{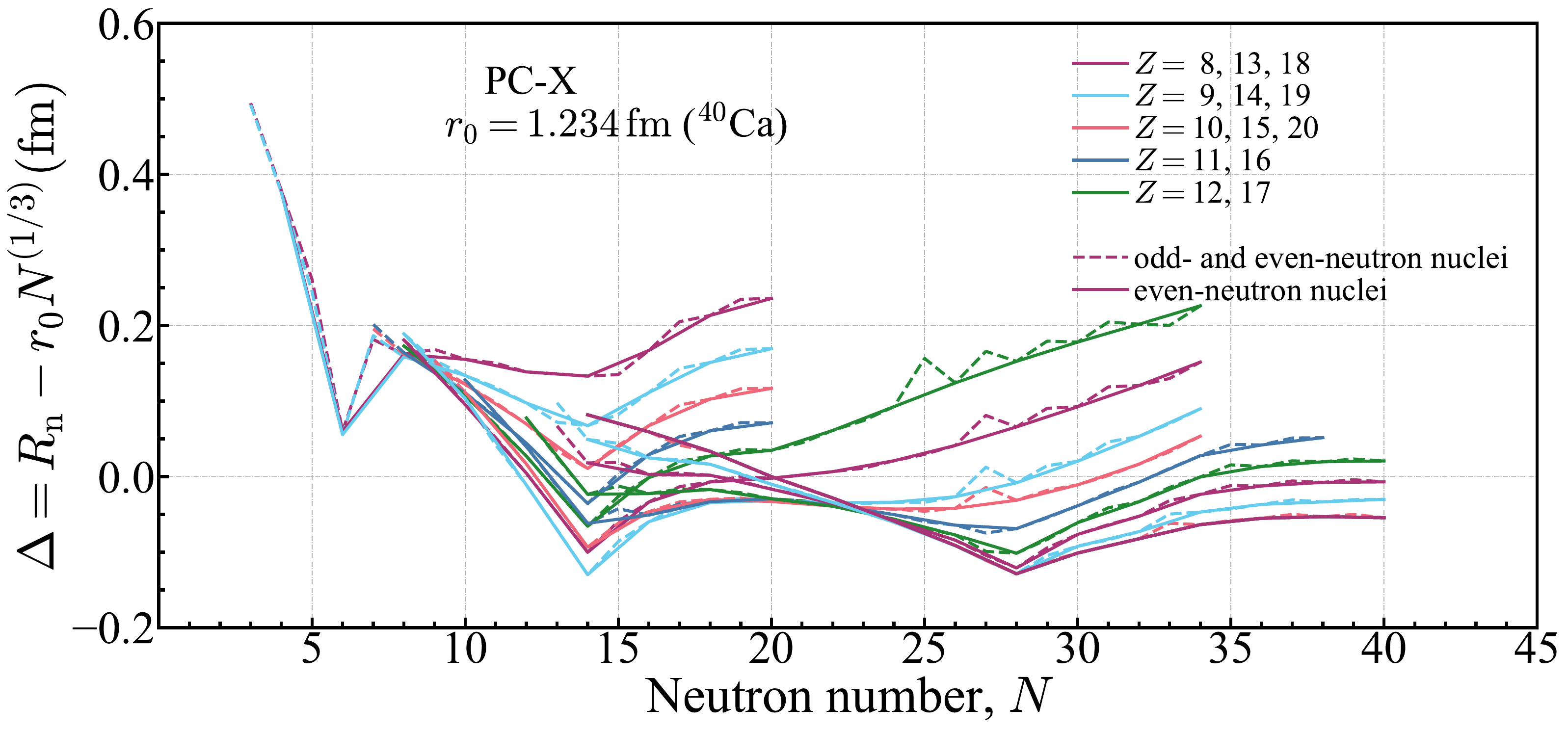}
\label{fig:RN-PCX-20-8-20}
\end{minipage}
\begin{minipage}{0.33\textwidth}
\centering
\includegraphics[width=\linewidth, angle=0]{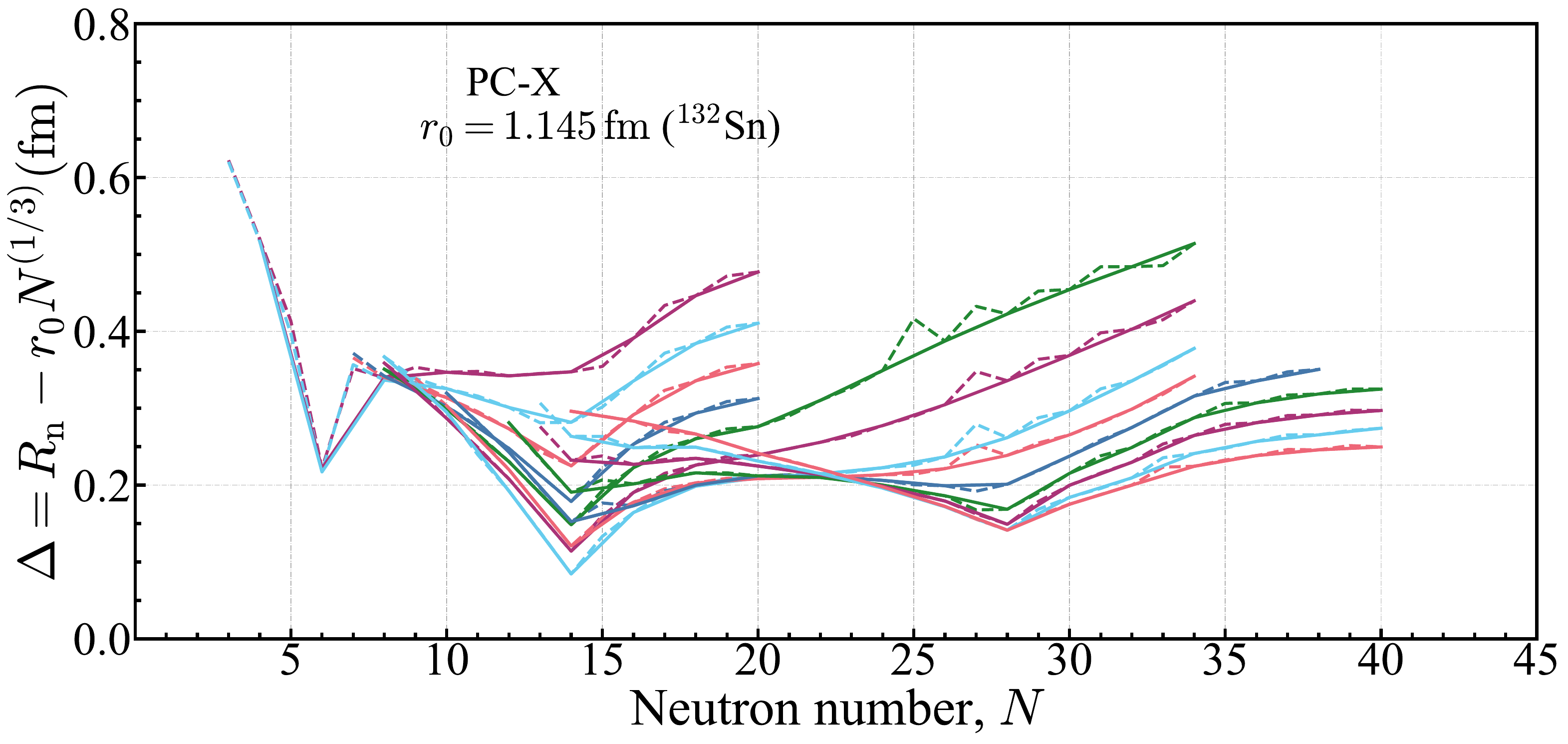}
\label{fig:RN-PCX-82-8-20}
\end{minipage}
\begin{minipage}{0.33\textwidth}
\centering
\includegraphics[width=\linewidth, angle=0]{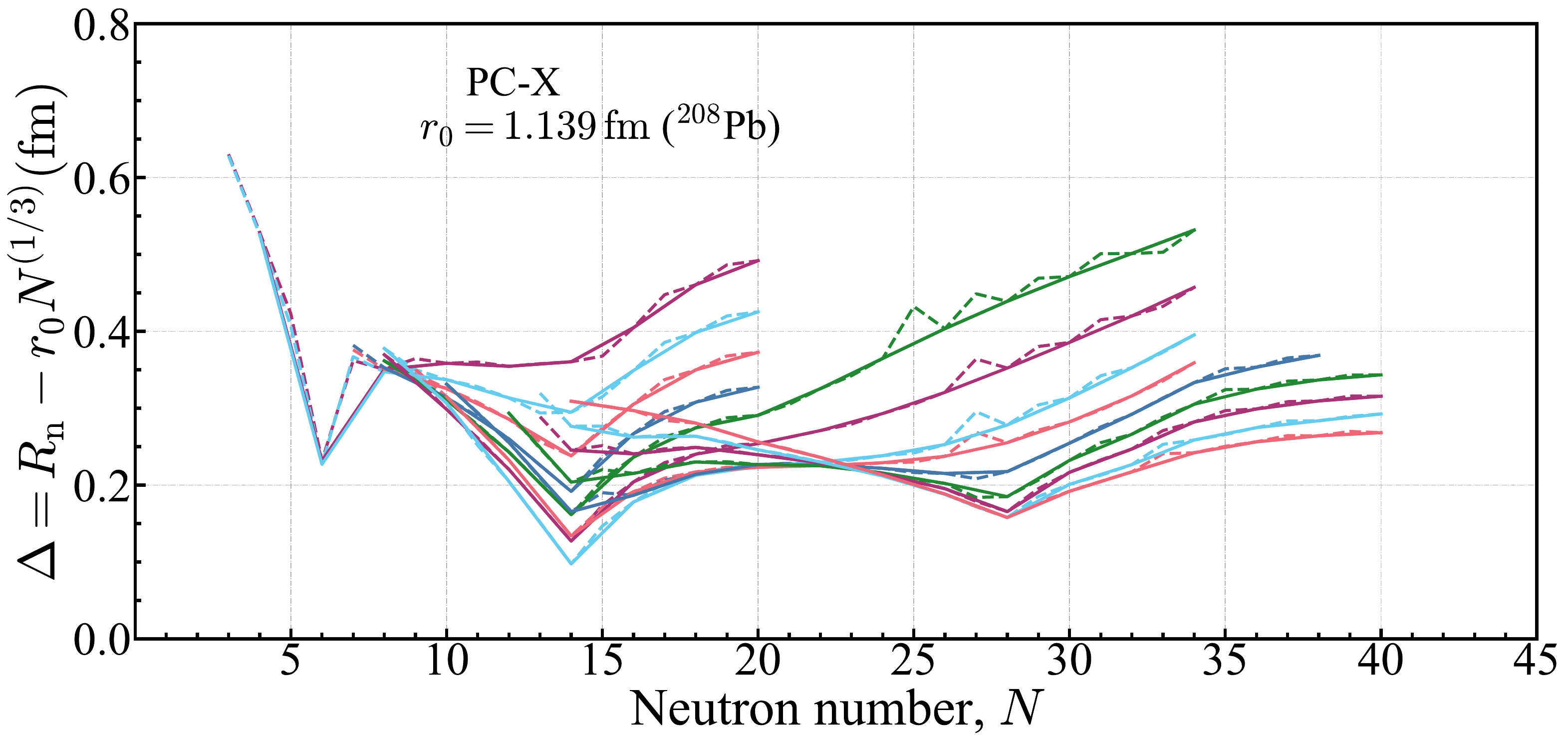}
\label{fig:RN-PCX-126-8-20}
\end{minipage}
\vspace{-2mm}
\begin{minipage}{0.33\textwidth}
\centering
\includegraphics[width=\linewidth, angle=0]{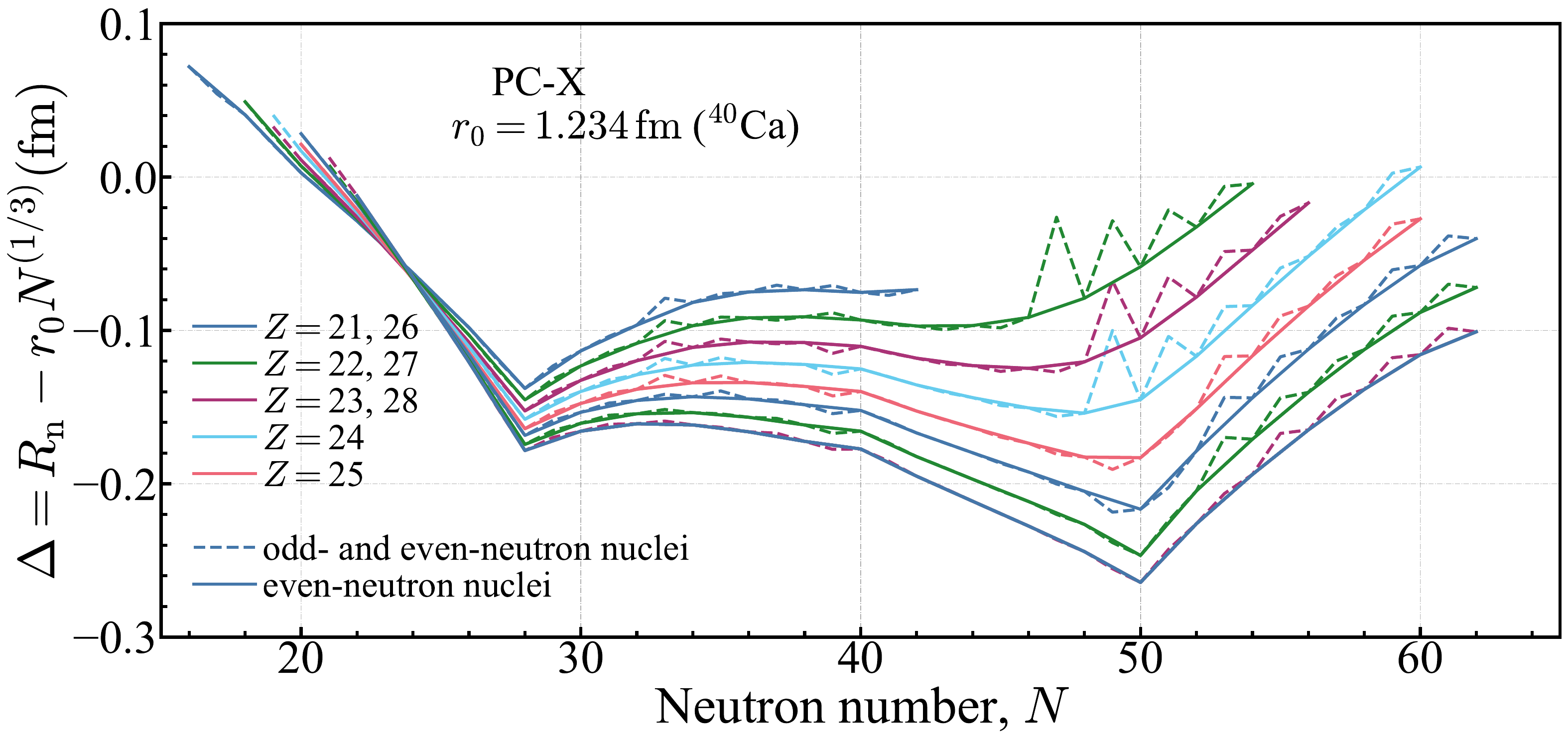}
\label{fig:RN-PCX-20-20-28}
\end{minipage}
\begin{minipage}{0.33\textwidth}
\centering
\includegraphics[width=\linewidth, angle=0]{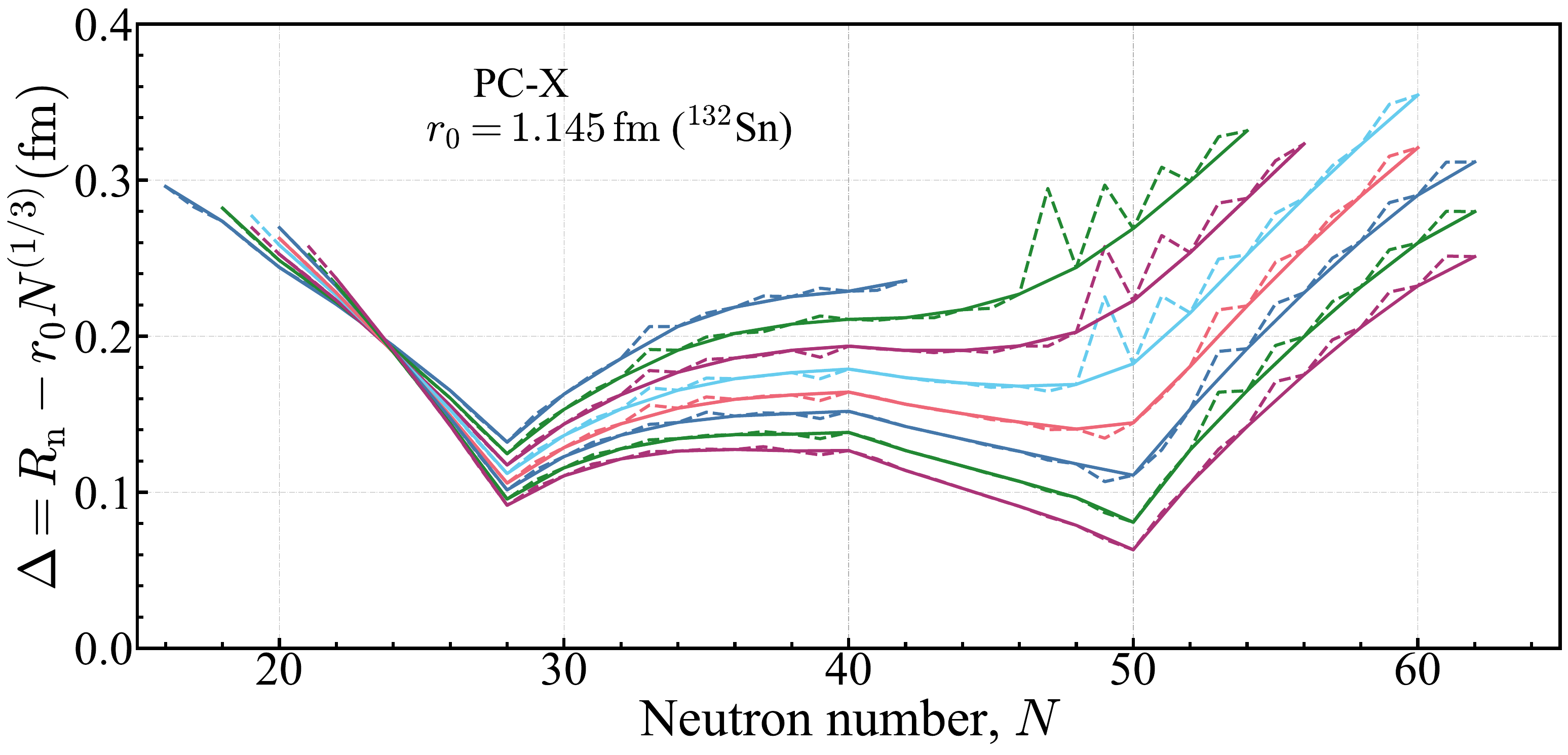}
\label{fig:RN-PCX-82-20-28}
\end{minipage}
\begin{minipage}{0.33\textwidth}
\centering
\includegraphics[width=\linewidth, angle=0]{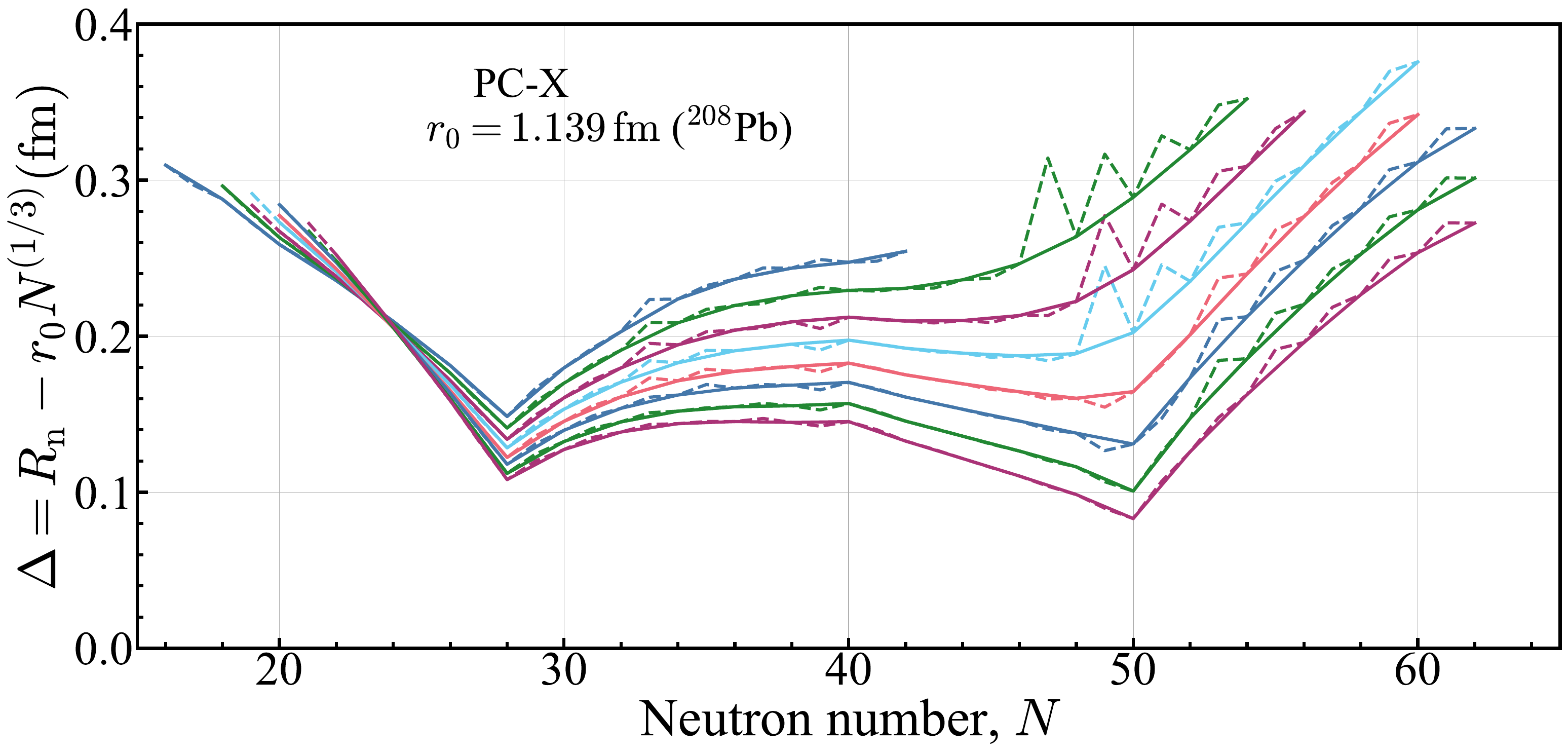}
\label{fig:RN-PCX-126-20-28}
\end{minipage}
\vspace{-2mm}
\begin{minipage}{0.33\textwidth}
\centering
\includegraphics[width=\linewidth, angle=0]{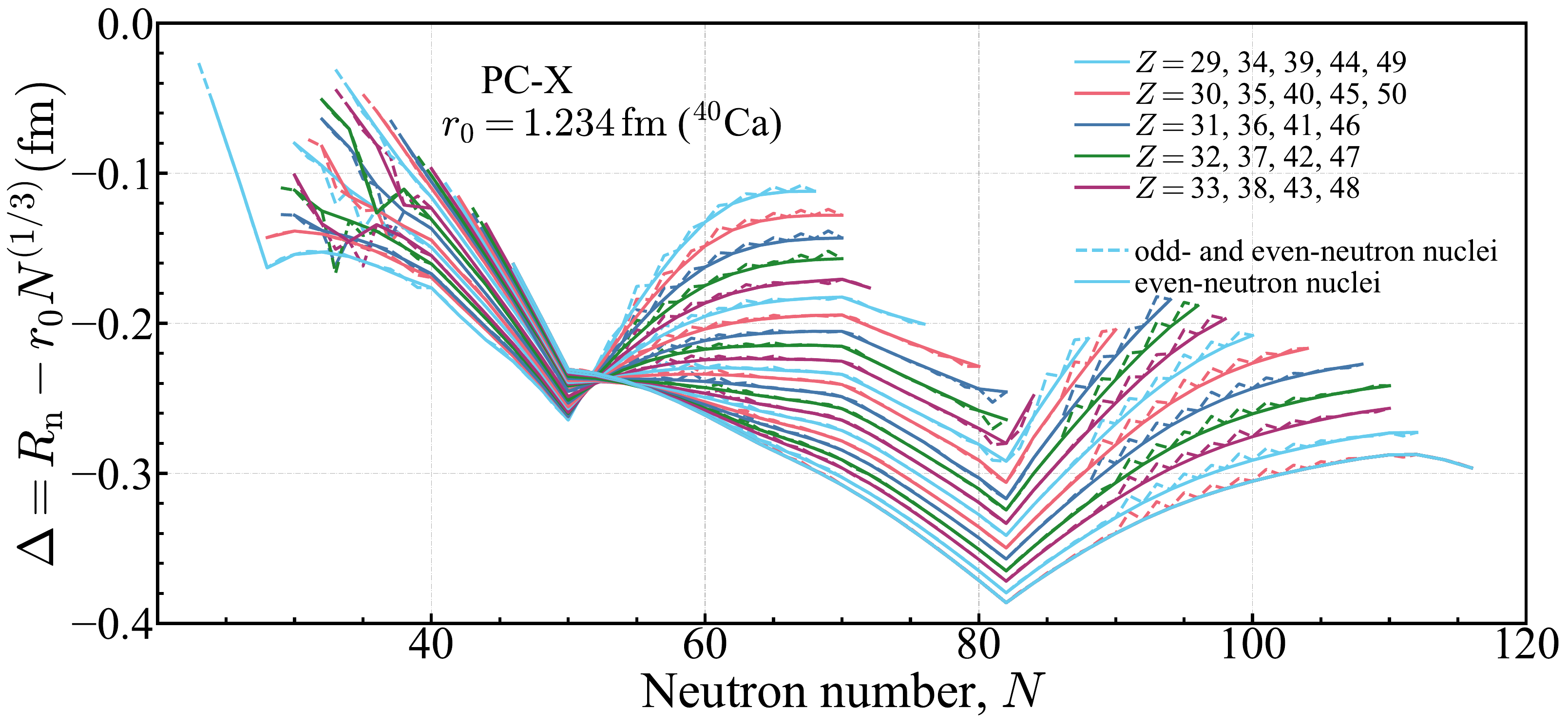}
\label{fig:RN-PCX-20-28-50}
\end{minipage}
\begin{minipage}{0.33\textwidth}
\centering
\includegraphics[width=\linewidth, angle=0]{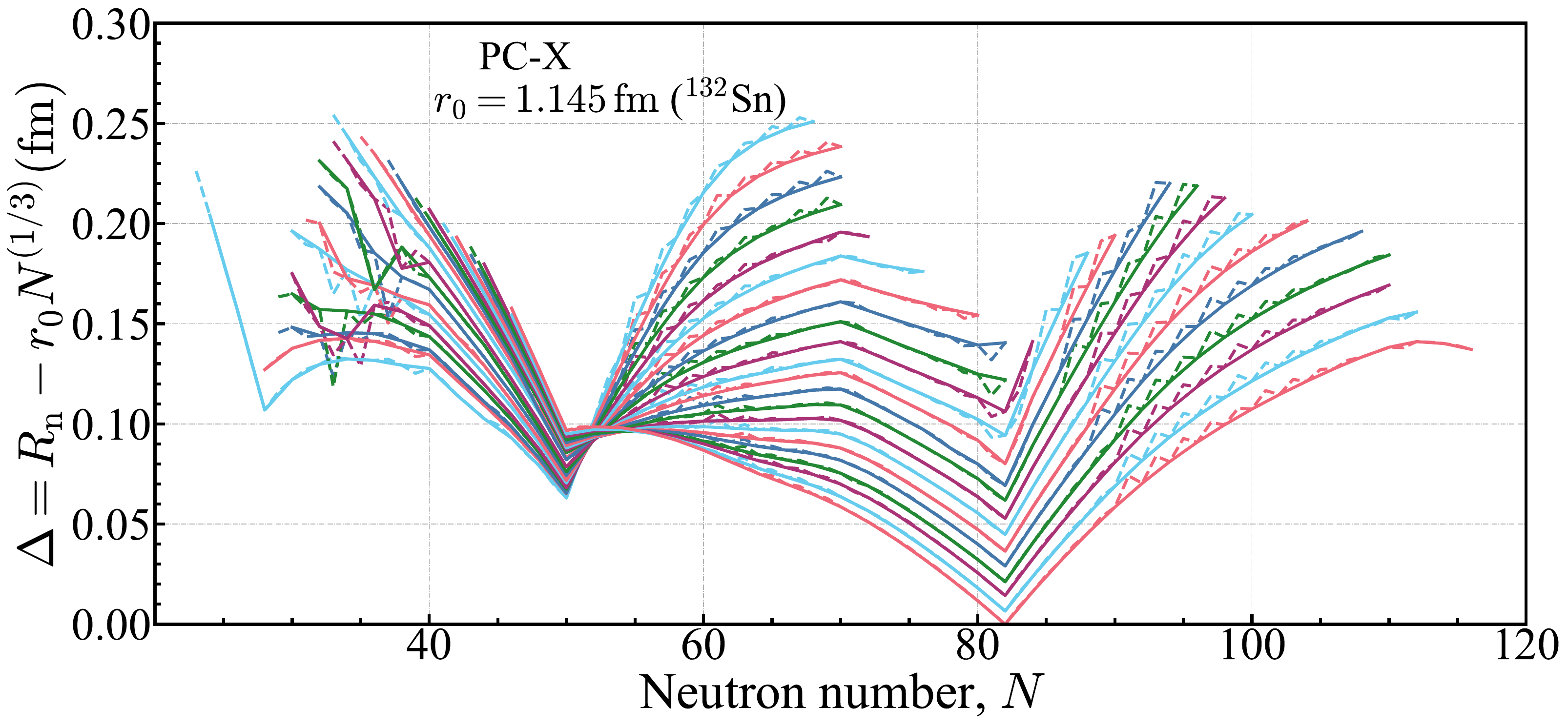}
\label{fig:RN-PCX-82-28-50}
\end{minipage}
\begin{minipage}{0.33\textwidth}
\centering
\includegraphics[width=\linewidth, angle=0]{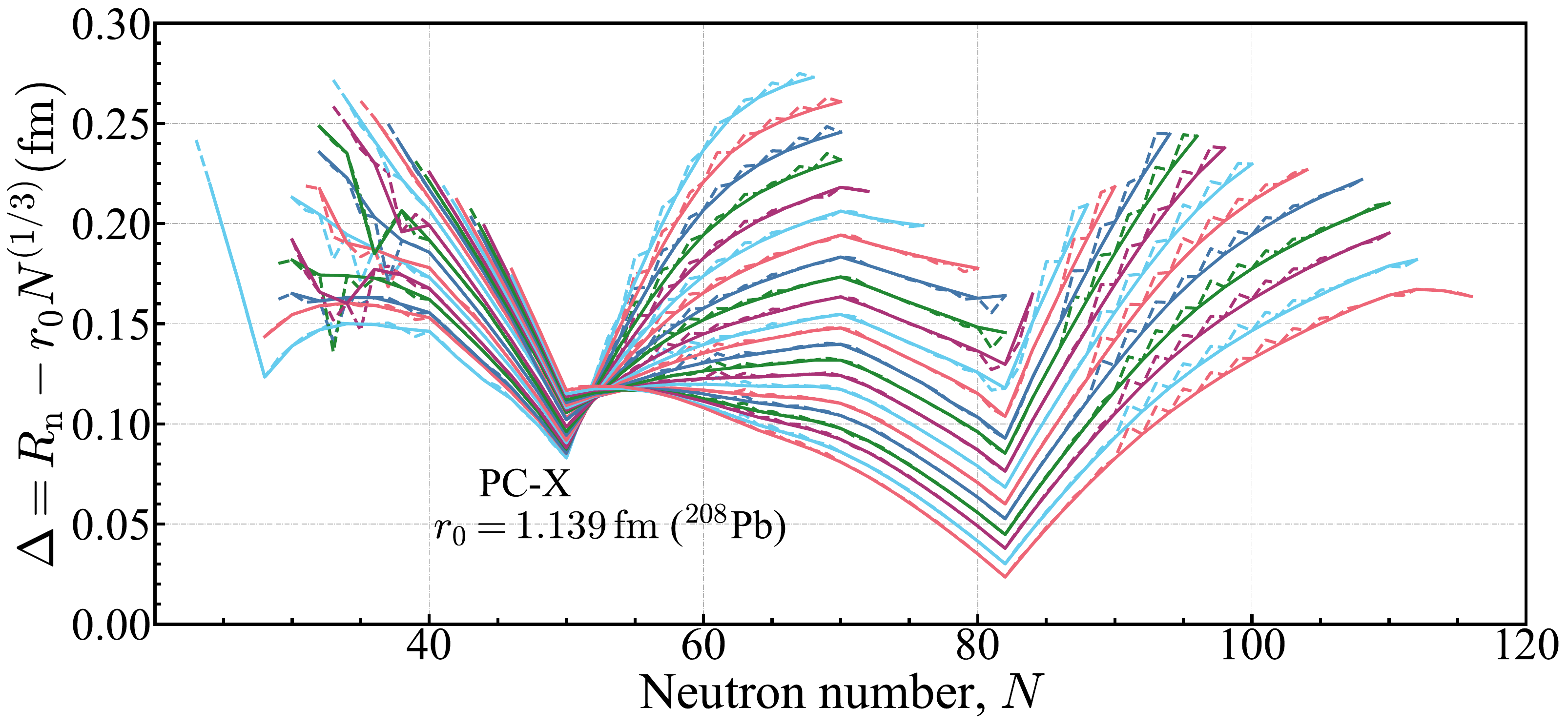}
\label{fig:RN-PCX-126-28-50}
\end{minipage}
\vspace{-2mm}
\begin{minipage}{0.33\textwidth}
\centering
\includegraphics[width=\linewidth, angle=0]{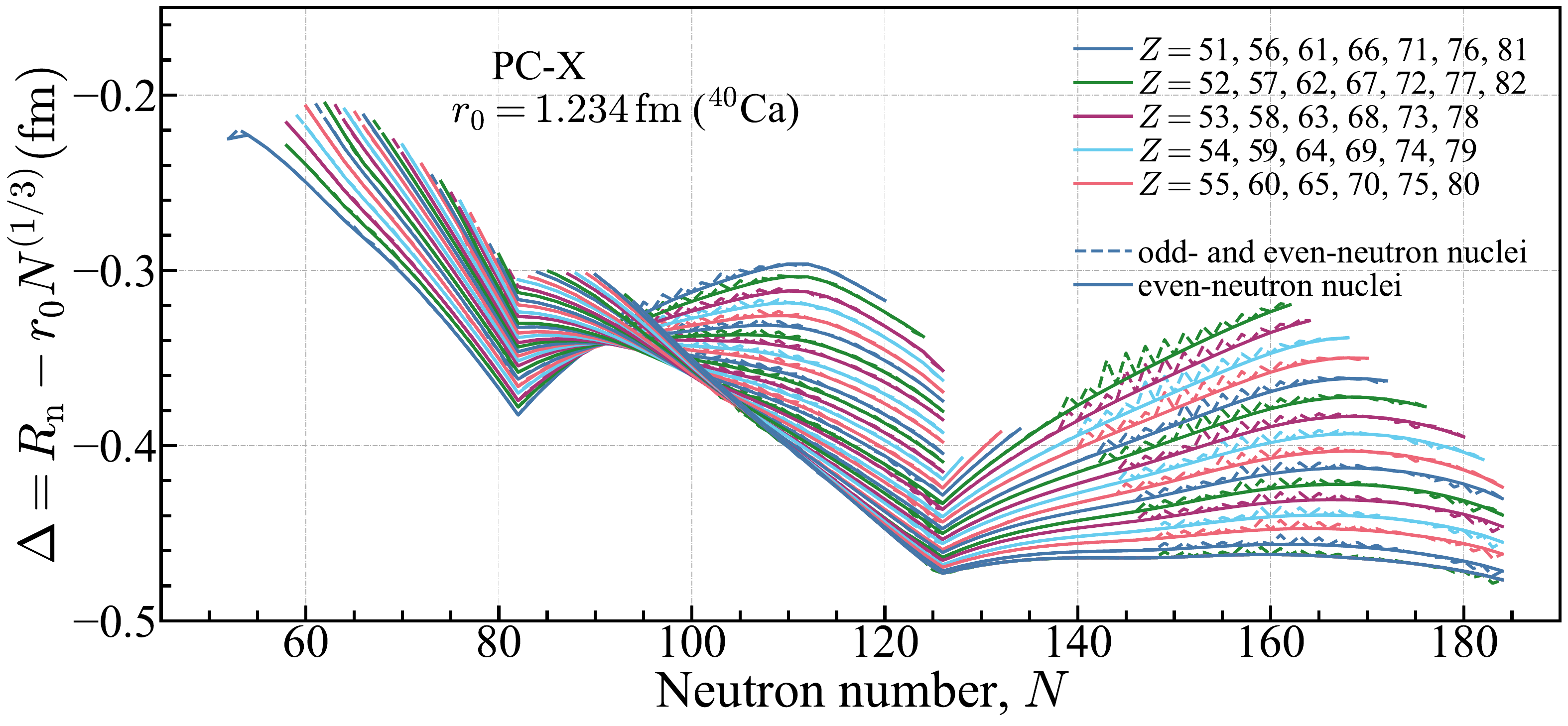}
\label{fig:RN-PCX-20-50-82}
\end{minipage}
\begin{minipage}{0.33\textwidth}
\centering
\includegraphics[width=\linewidth, angle=0]{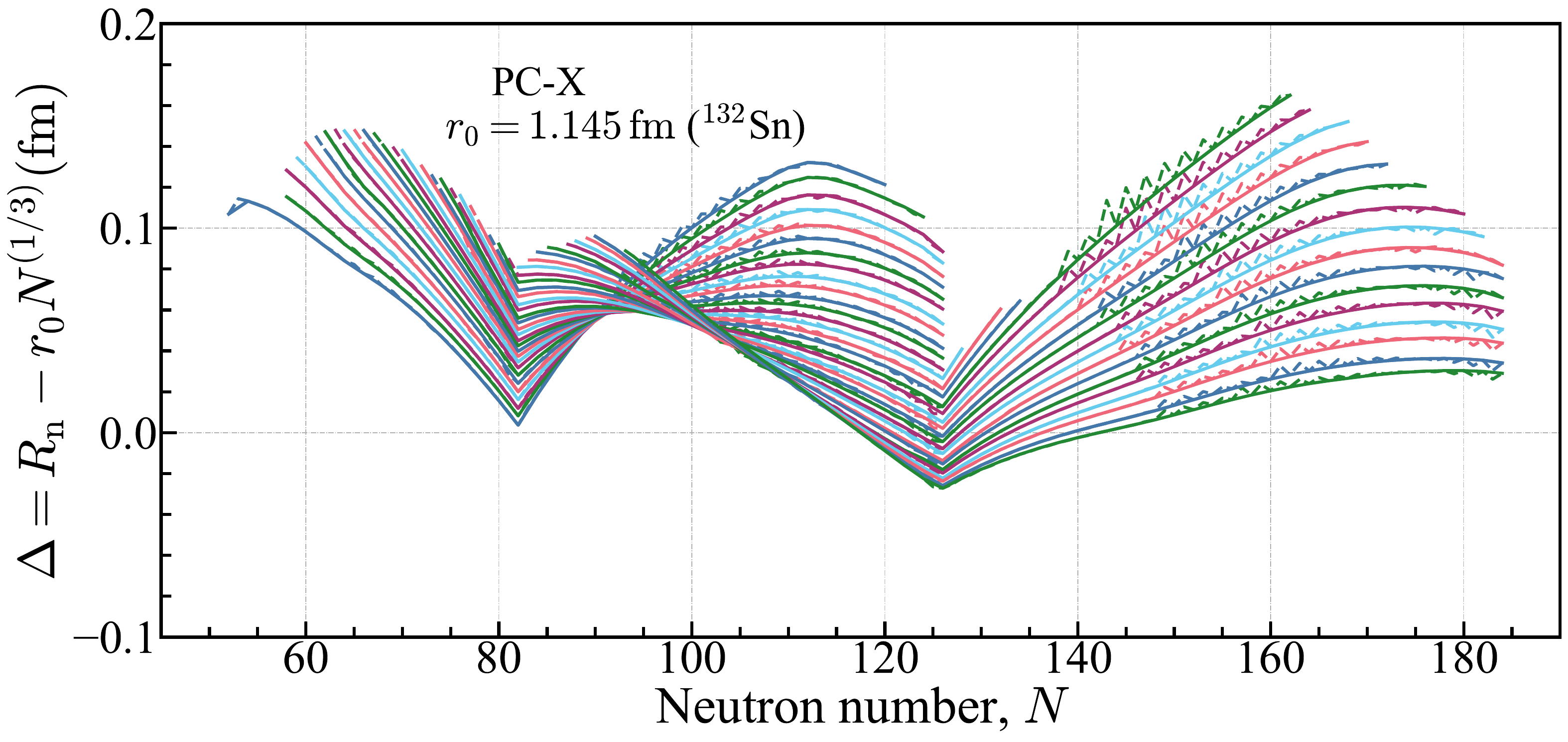}
\label{fig:RN-PCX-82-50-82}
\end{minipage}
\begin{minipage}{0.33\textwidth}
\centering
\includegraphics[width=\linewidth, angle=0]{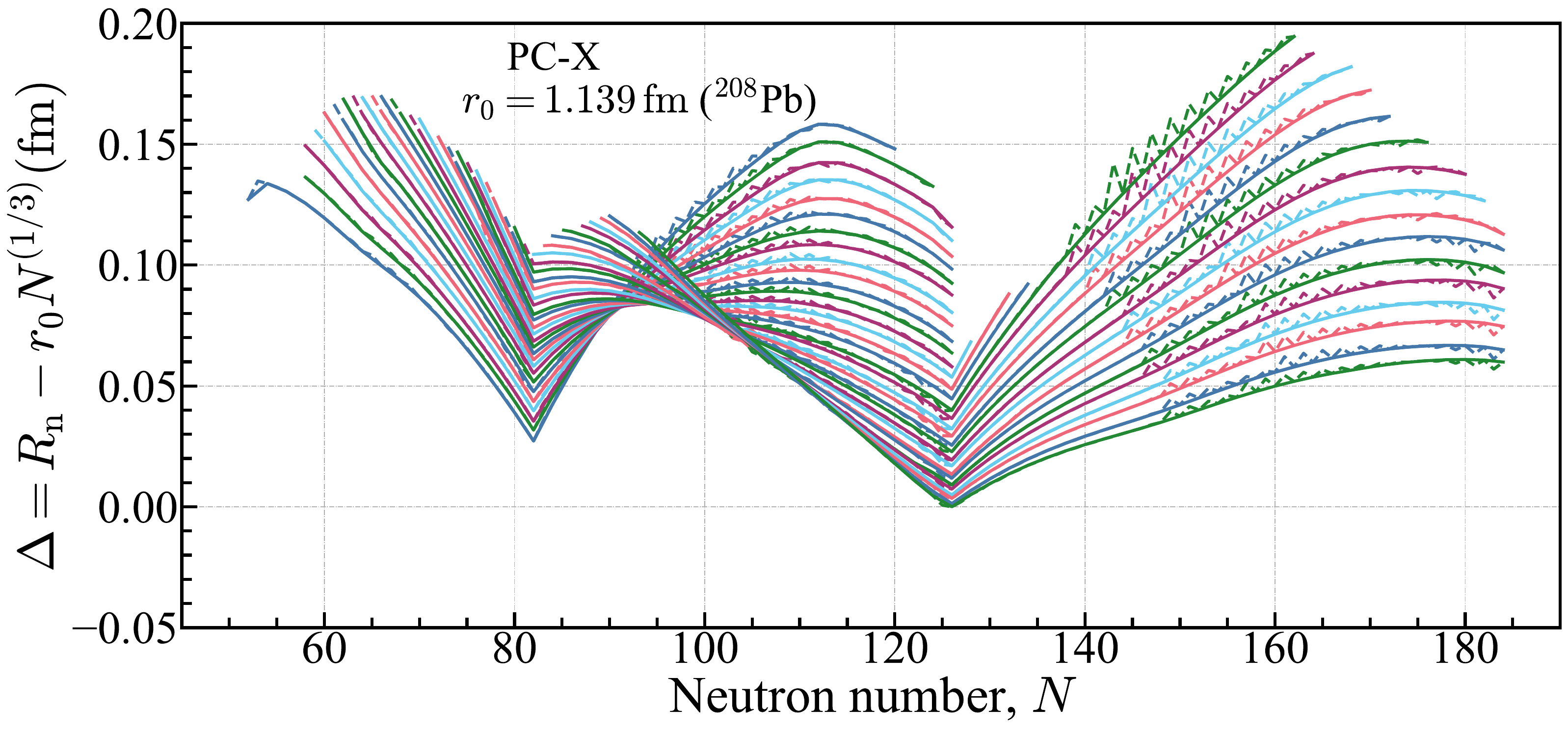}
\label{fig:RN-PCX-126-50-82}
\end{minipage}
\vspace{-2mm}
\begin{minipage}{0.33\textwidth}
\centering
\includegraphics[width=\linewidth, angle=0]{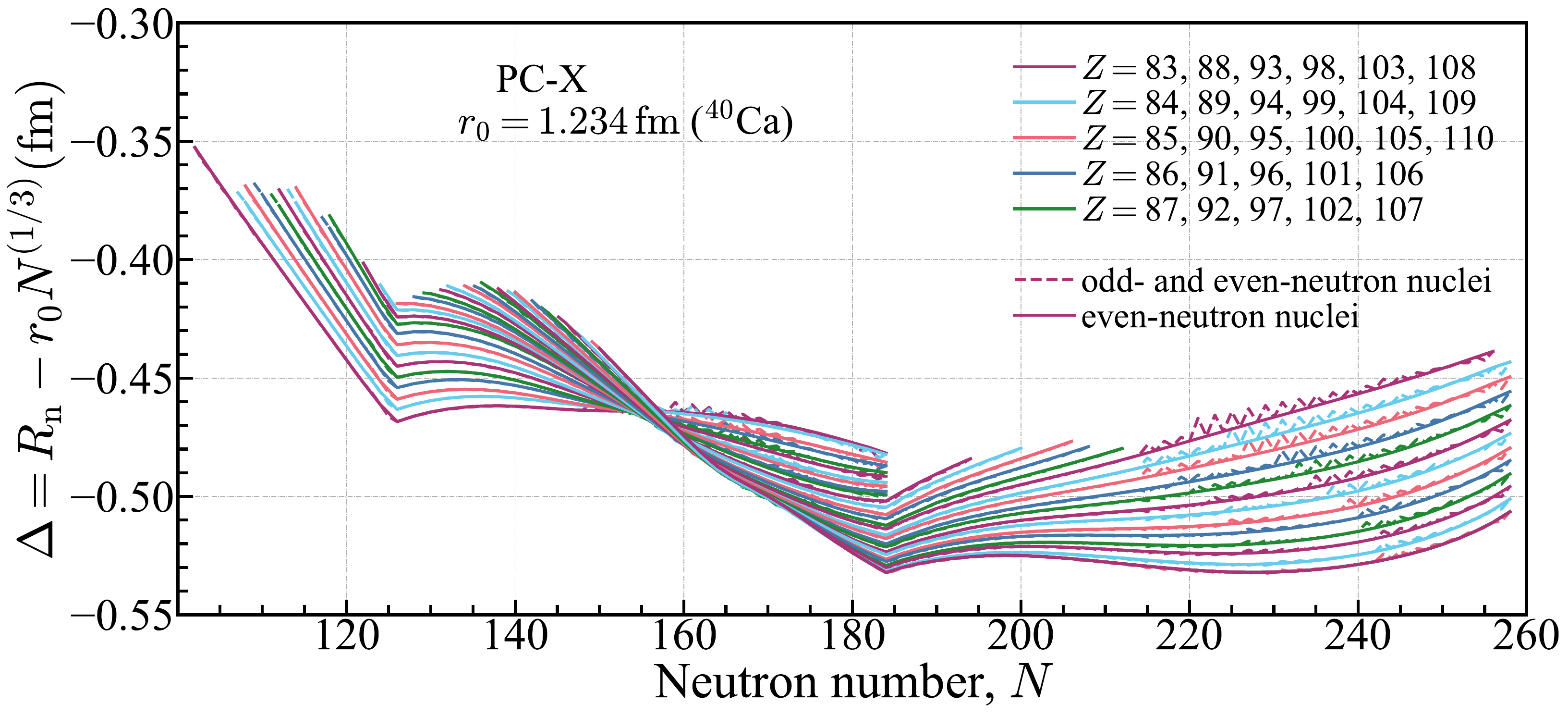}
\label{fig:RN-PCX-20-82-110}
\end{minipage}
\begin{minipage}{0.33\textwidth}
\centering
\includegraphics[width=\linewidth, angle=0]{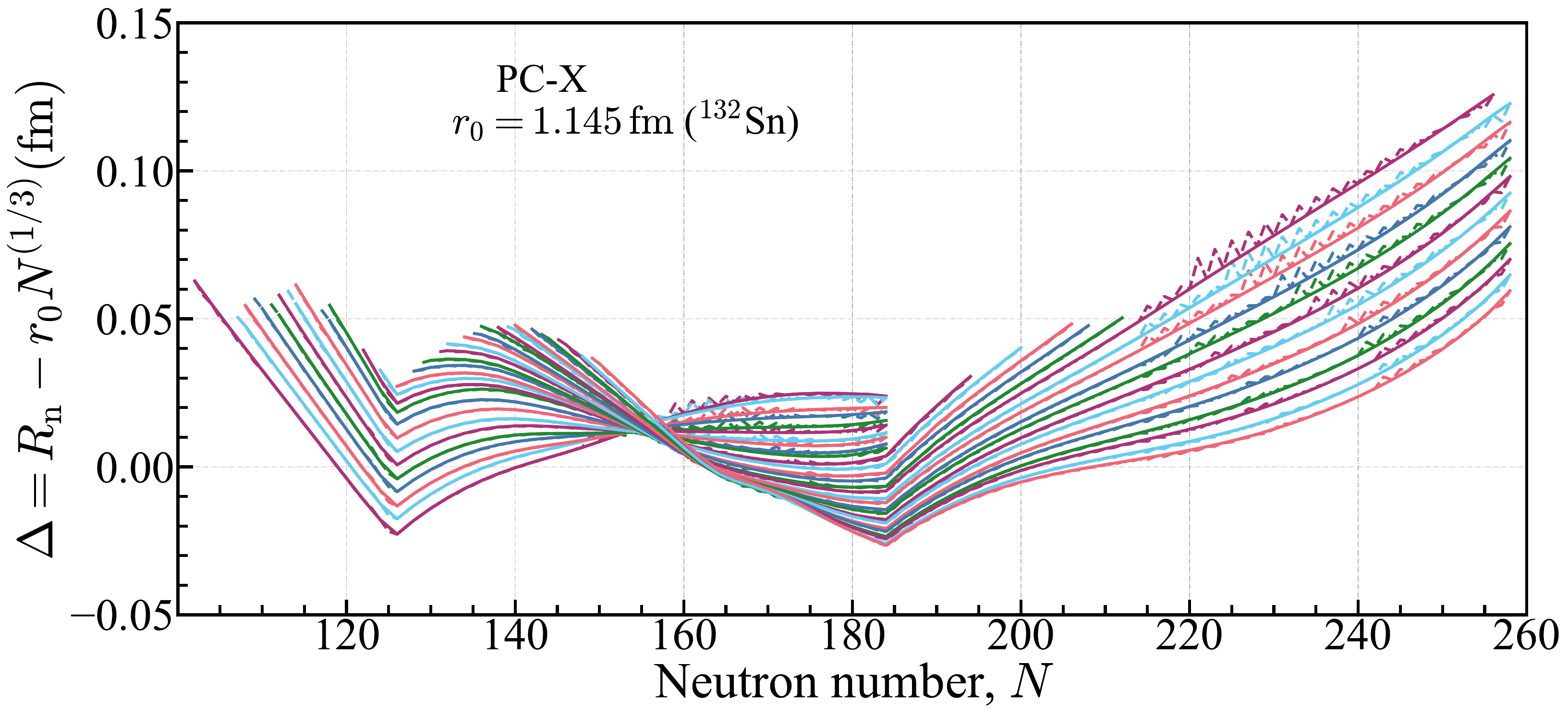}
\label{fig:RN-PCX-82-82-110}
\end{minipage}
\begin{minipage}{0.33\textwidth}
\centering
\includegraphics[width=\linewidth, angle=0]{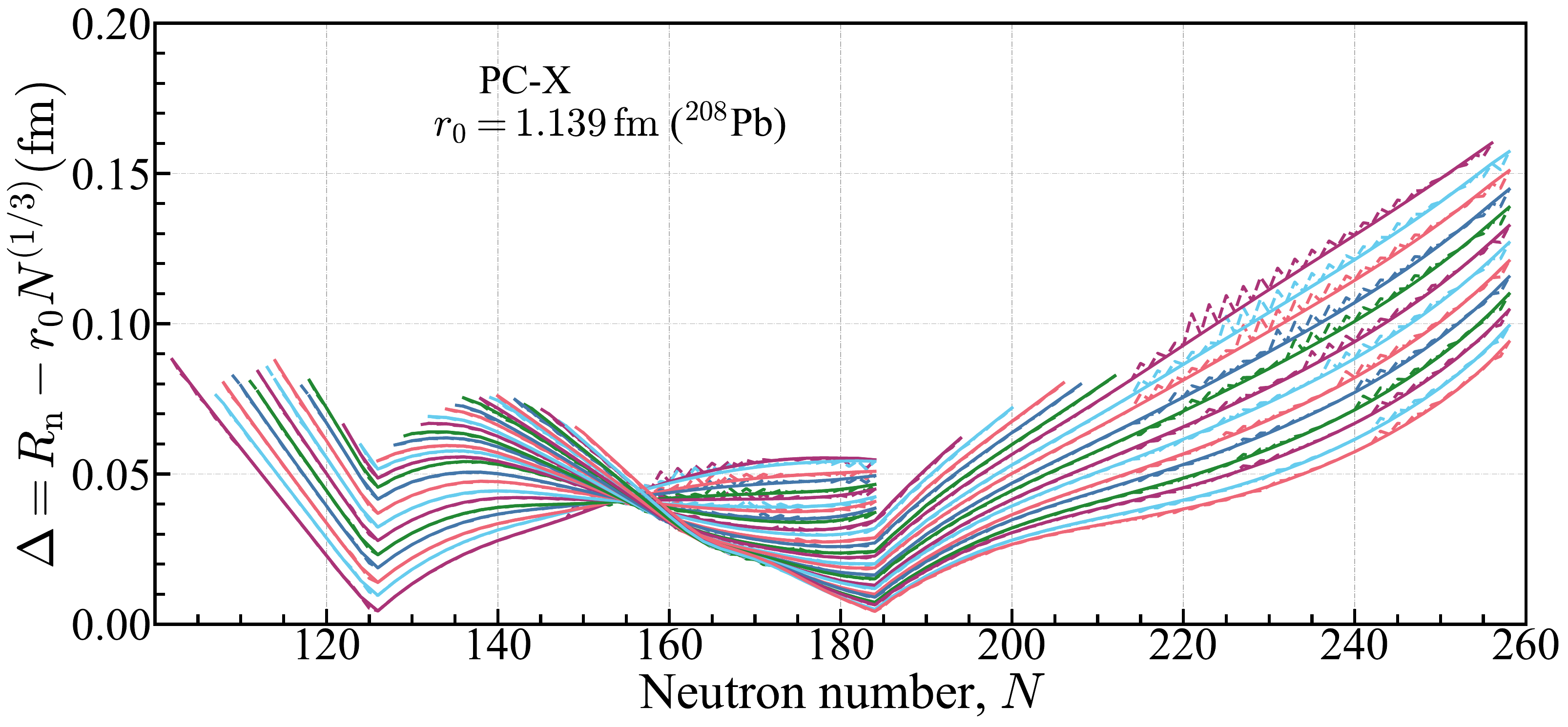}
\label{fig:RN-PCX-126-82-110}
\end{minipage}
\vspace{-2mm}
\centering
\begin{minipage}{\textwidth}
\caption{
The deviations ($\Delta$) of theoretical root-mean-square radius of neutron distributions ($R_\mathrm{n}$) and the empirical $R_\mathrm{n}\!=\!r_{0}N^{1/3}$. The calculations are based on the RHB approach with the PC-X interaction for all isotopic chains in the range of $8\!\leq\! Z\!\leq\!110$.
See Fig.~\ref{fig:RN} for further descriptions.}
\label{fig:RN-PCX}
\end{minipage}
\end{figure*}

\begin{figure*}
\centering
\begin{minipage}{0.33\textwidth}
\centering
\includegraphics[width=\linewidth, angle=0]{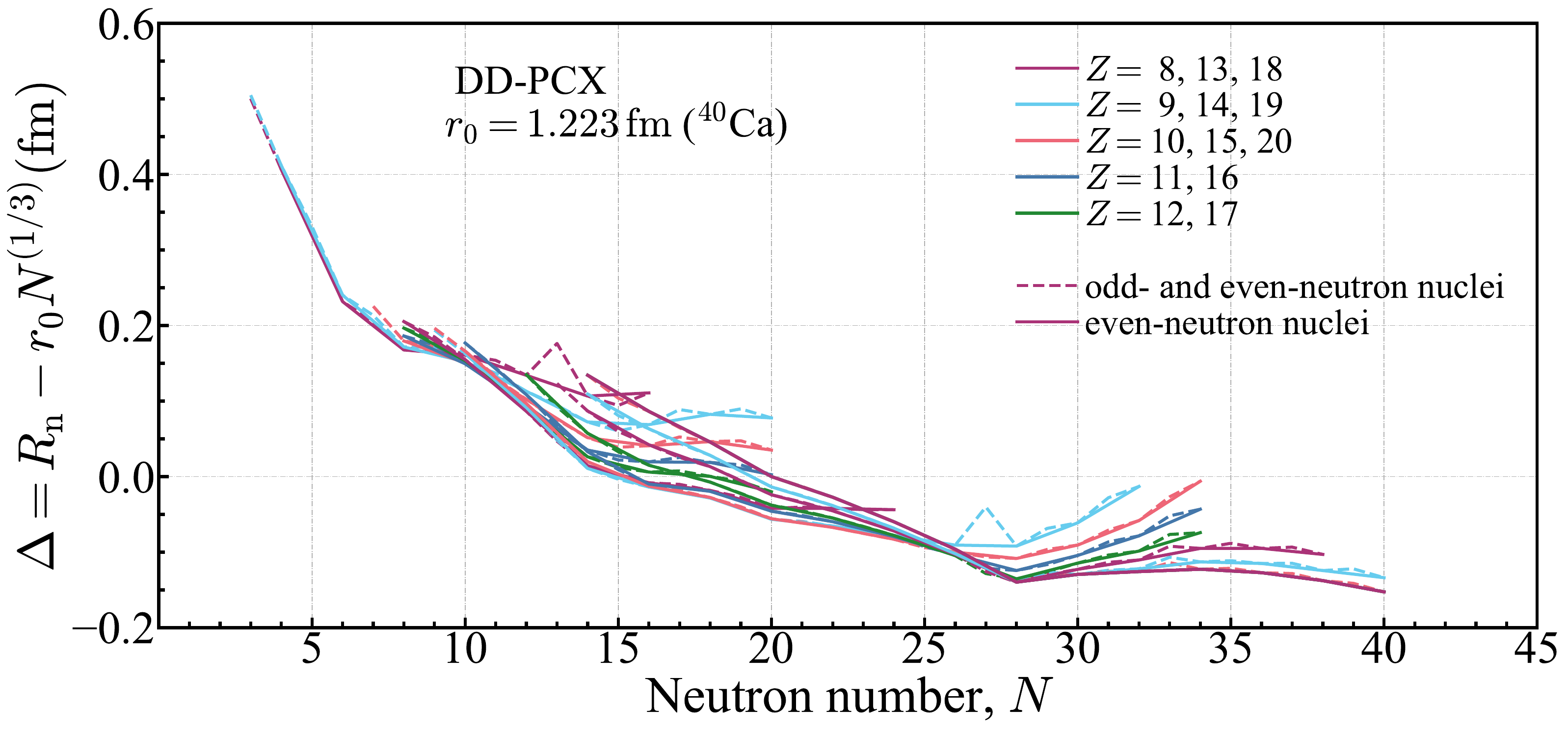}
\label{fig:RN-DDPCX-20-8-20}
\end{minipage}
\begin{minipage}{0.33\textwidth}
\centering
\includegraphics[width=\linewidth, angle=0]{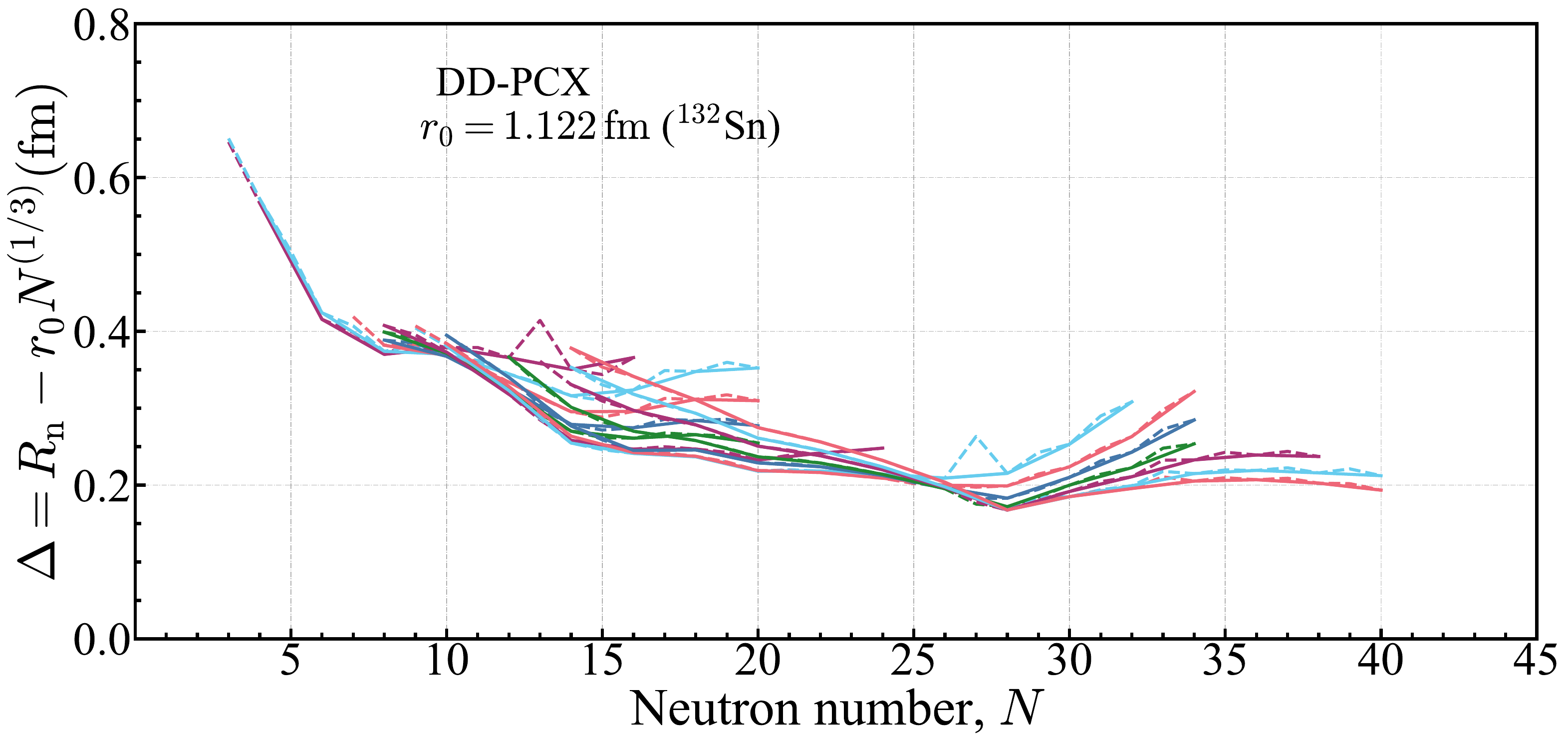}
\label{fig:RN-DDPCX-82-8-20}
\end{minipage}
\begin{minipage}{0.33\textwidth}
\centering
\includegraphics[width=\linewidth, angle=0]{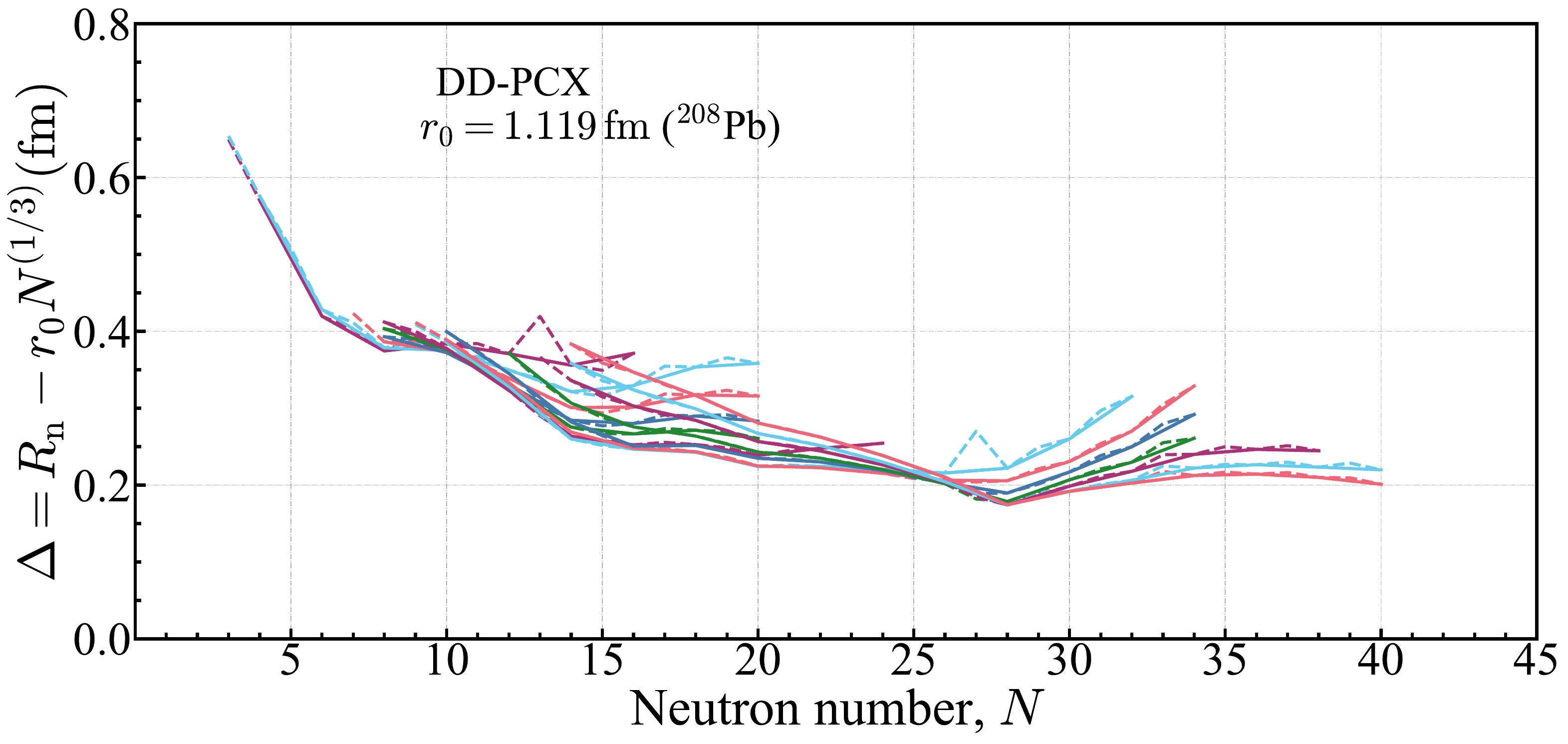}
\label{fig:RN-DDPCX-126-8-20}
\end{minipage}
\vspace{-2mm}
\begin{minipage}{0.33\textwidth}
\centering
\includegraphics[width=\linewidth, angle=0]{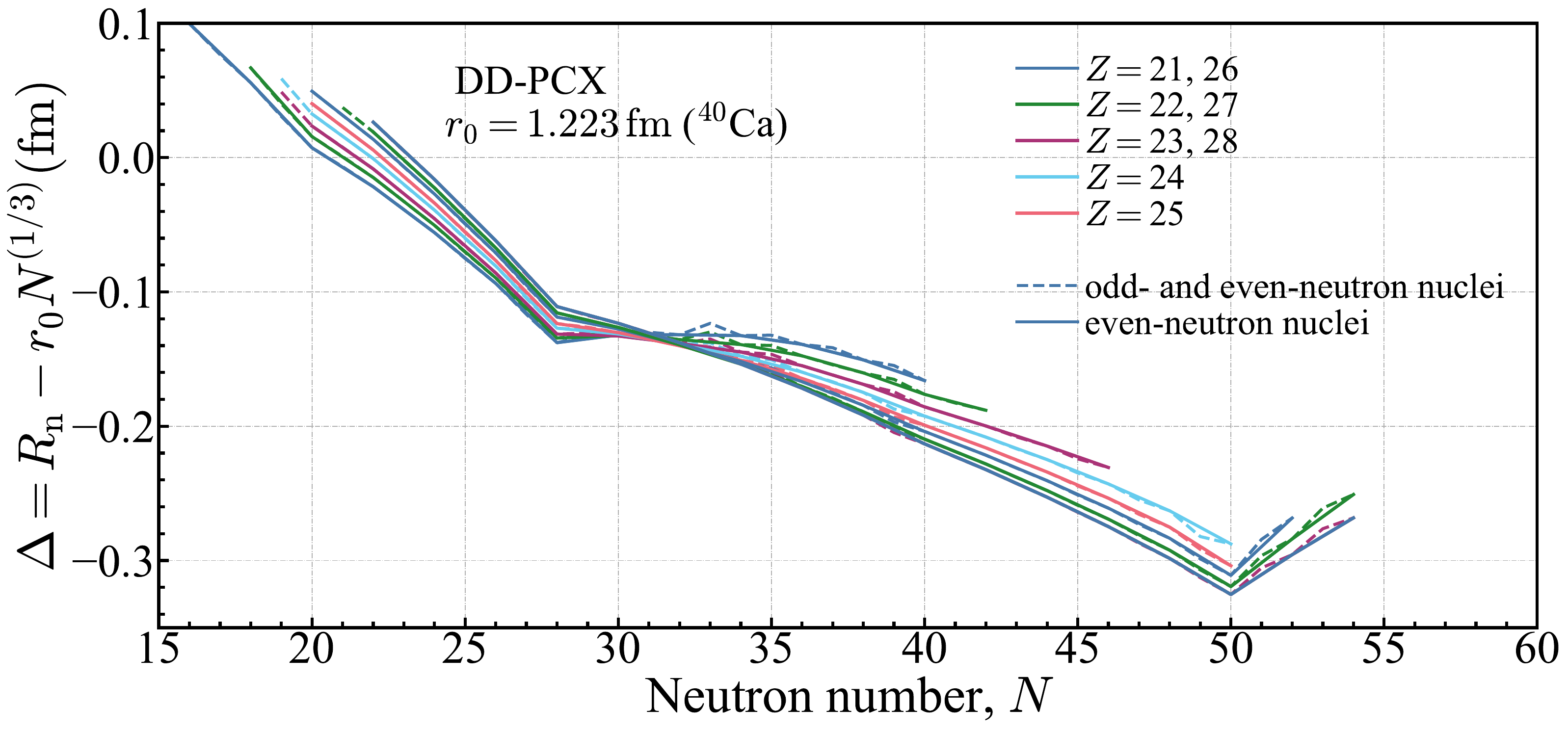}
\label{fig:RN-DDPCX-20-20-28}
\end{minipage}
\begin{minipage}{0.33\textwidth}
\centering
\includegraphics[width=\linewidth, angle=0]{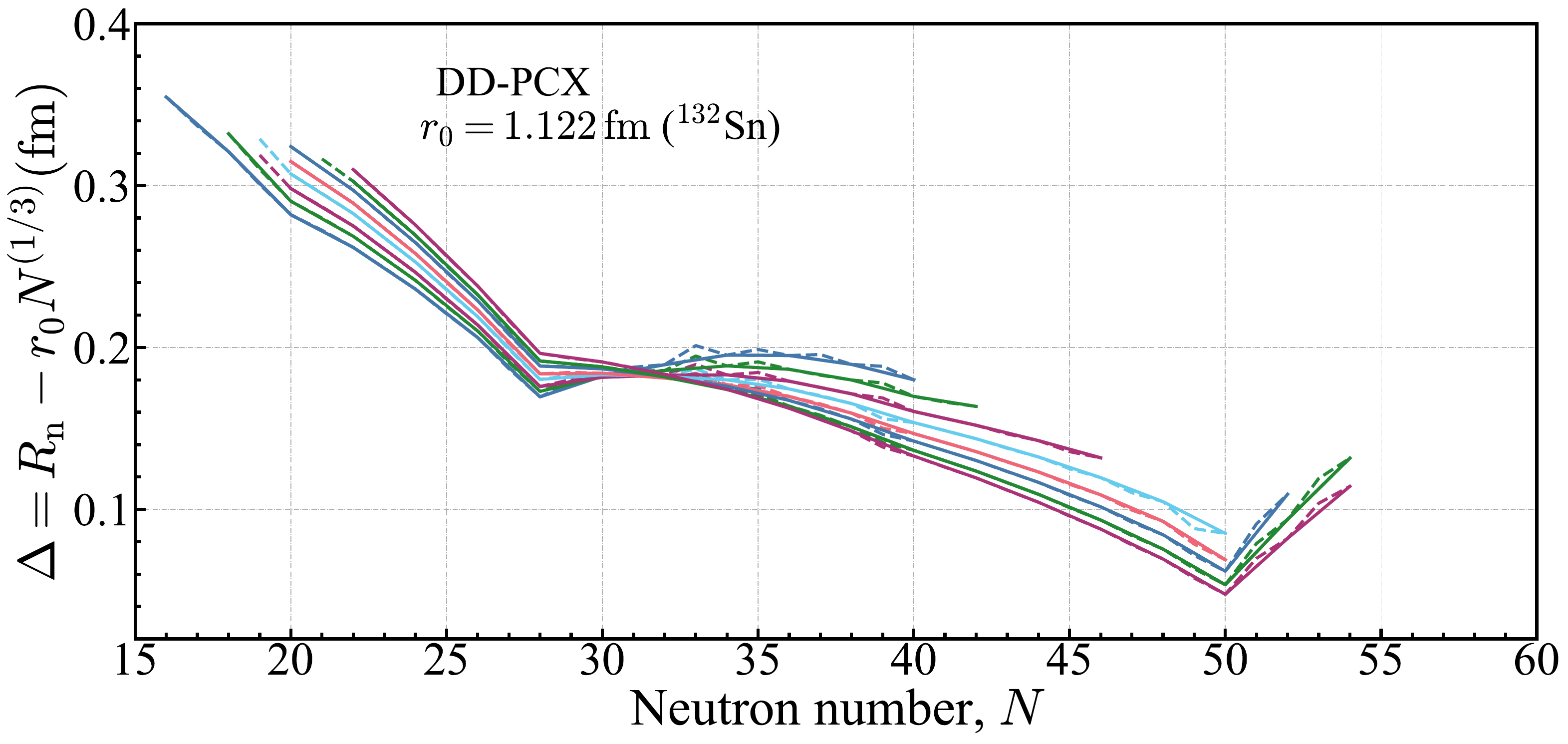}
\label{fig:RN-DDPCX-82-20-28}
\end{minipage}
\begin{minipage}{0.33\textwidth}
\centering
\includegraphics[width=\linewidth, angle=0]{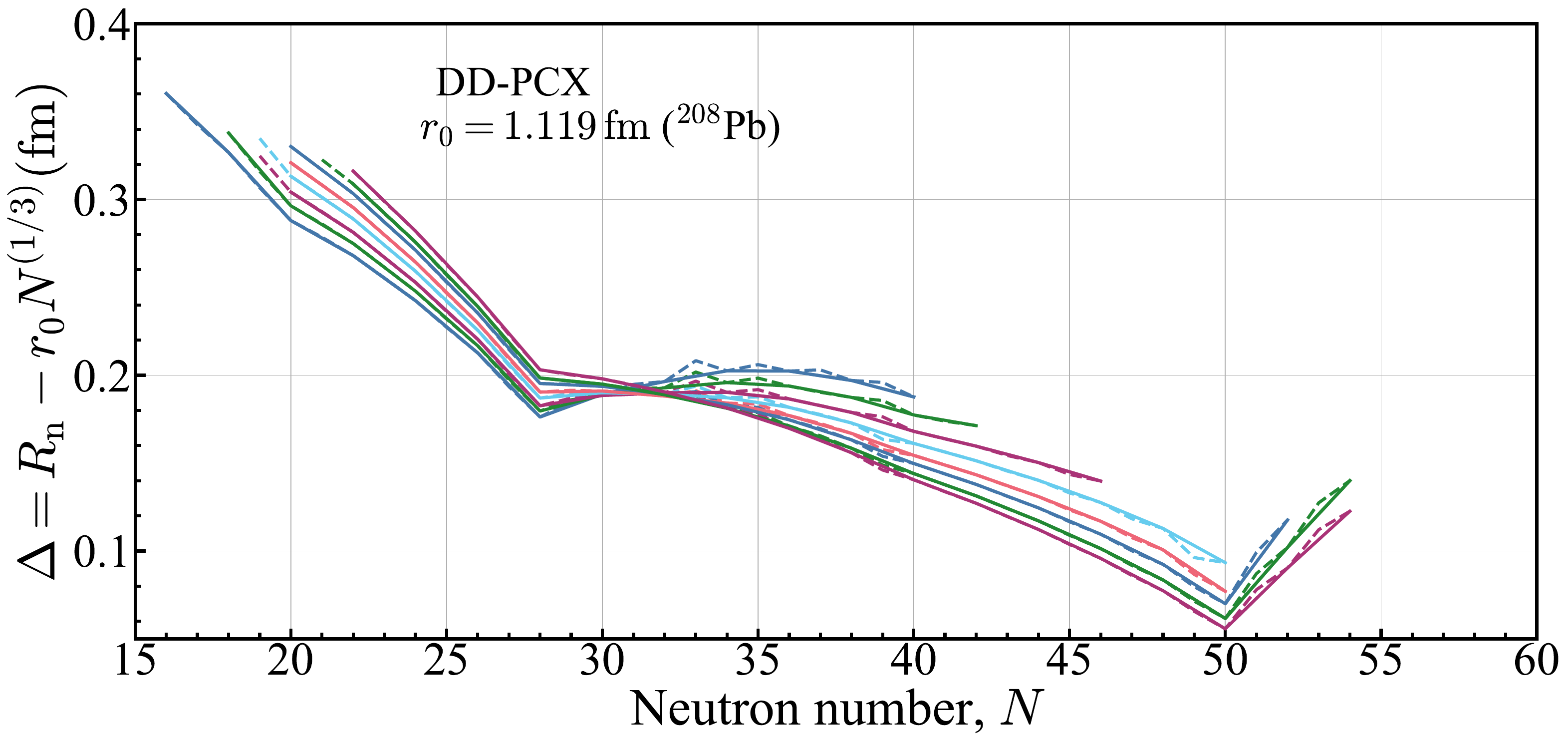}
\label{fig:RN-DDPCX-126-20-28}
\end{minipage}
\vspace{-2mm}
\begin{minipage}{0.33\textwidth}
\centering
\includegraphics[width=\linewidth, angle=0]{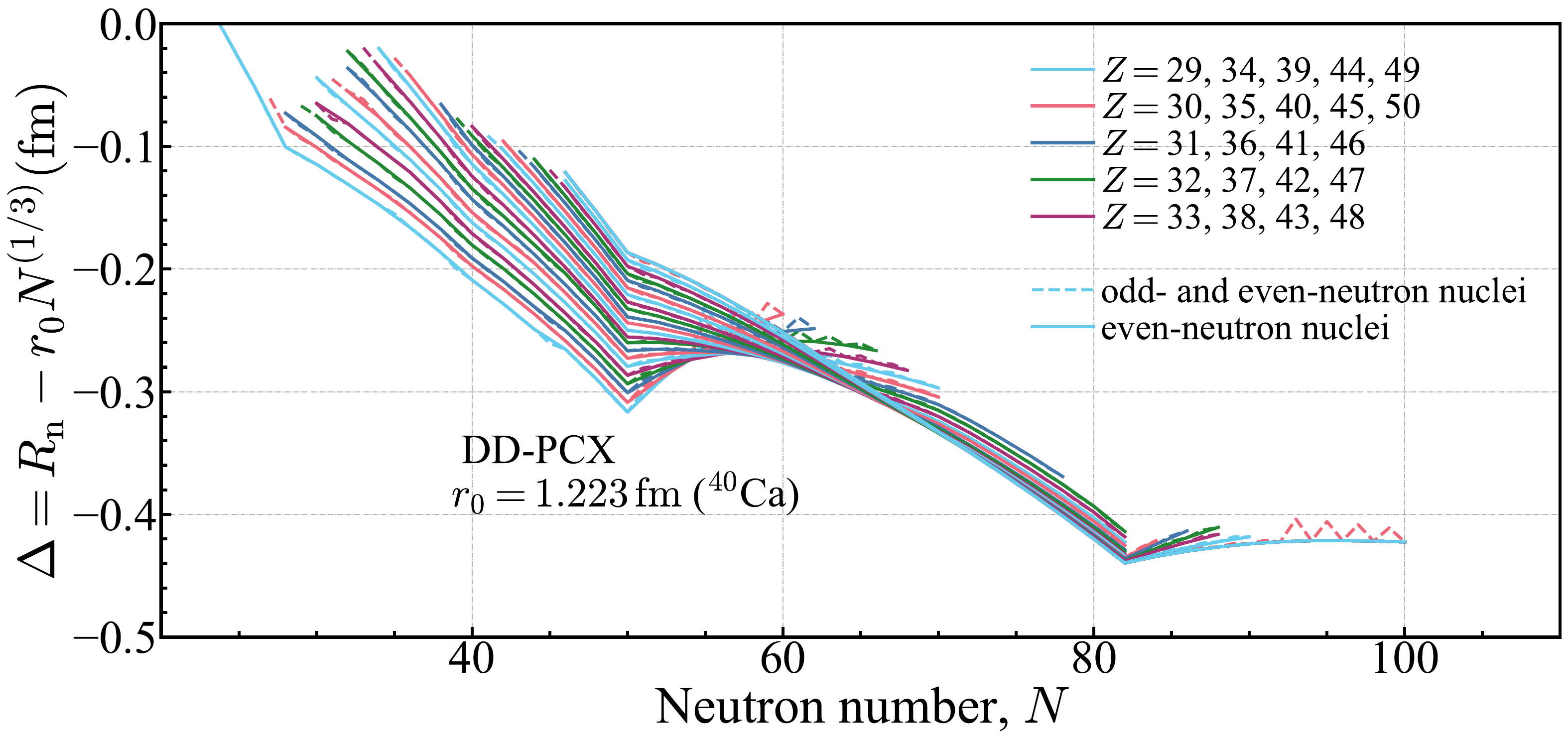}
\label{fig:RN-DDPCX-20-28-50}
\end{minipage}
\begin{minipage}{0.33\textwidth}
\centering
\includegraphics[width=\linewidth, angle=0]{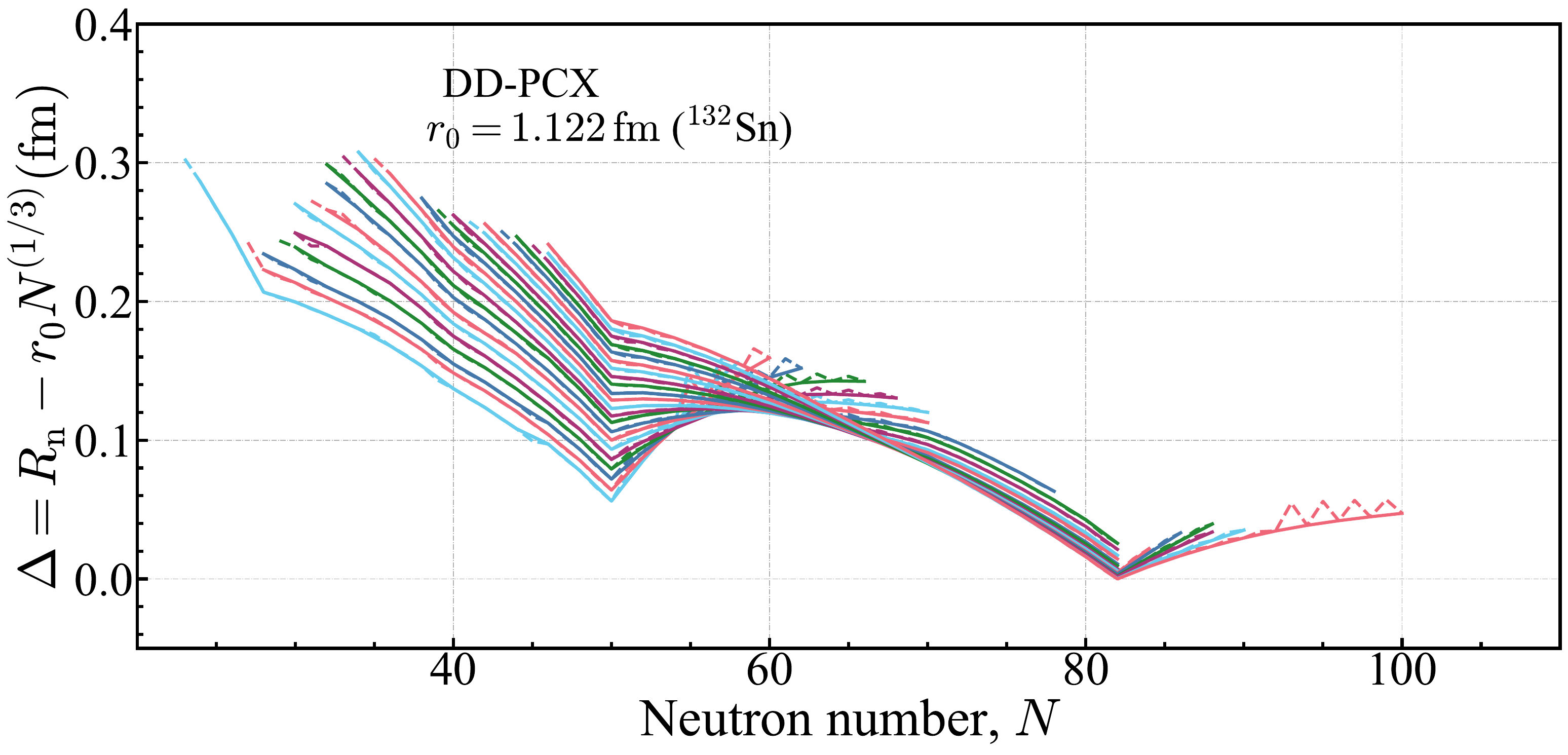}
\label{fig:RN-DDPCX-82-28-50}
\end{minipage}
\begin{minipage}{0.33\textwidth}
\centering
\includegraphics[width=\linewidth, angle=0]{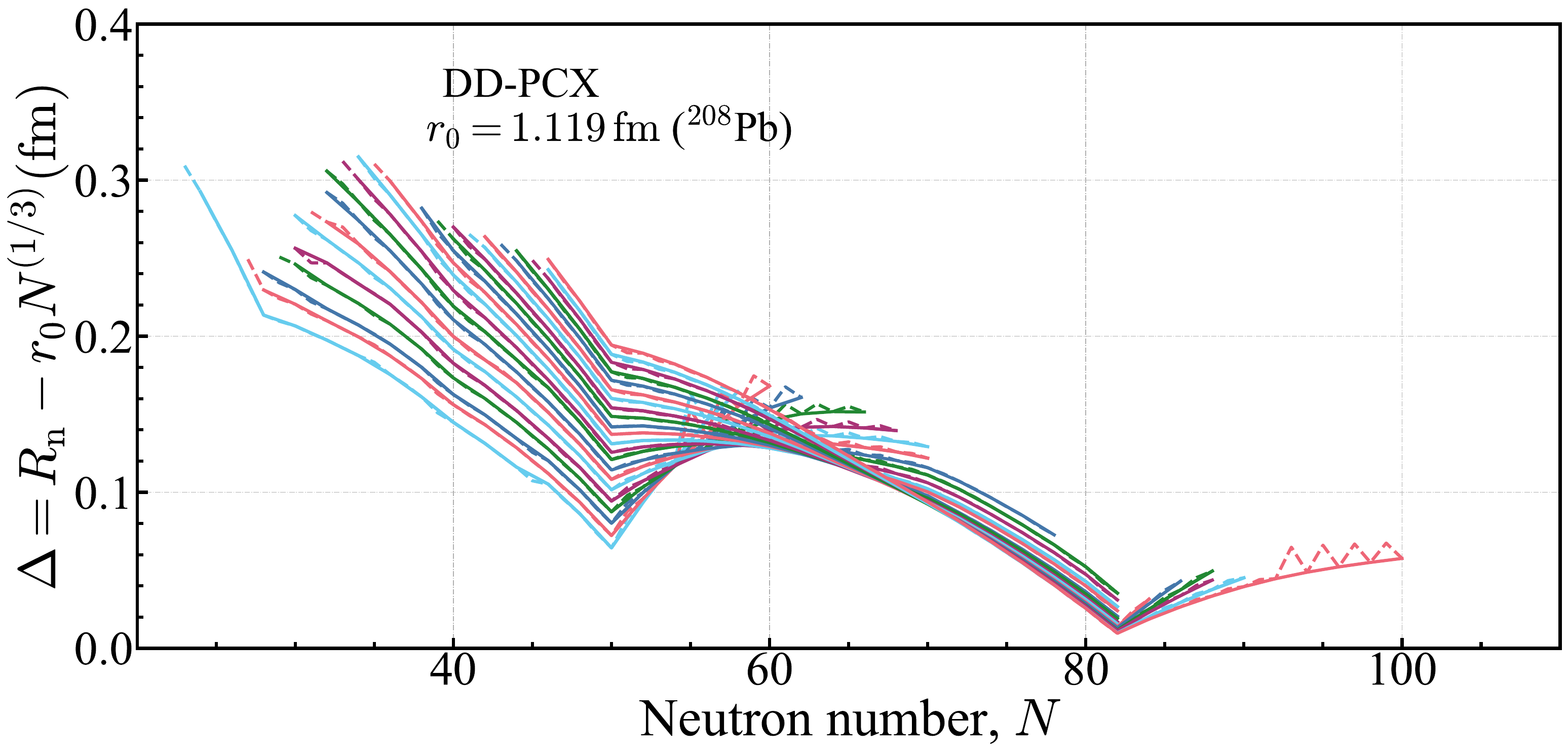}
\label{fig:RN-DDPCX-126-28-50}
\end{minipage}
\vspace{-2mm}
\begin{minipage}{0.33\textwidth}
\centering
\includegraphics[width=\linewidth, angle=0]{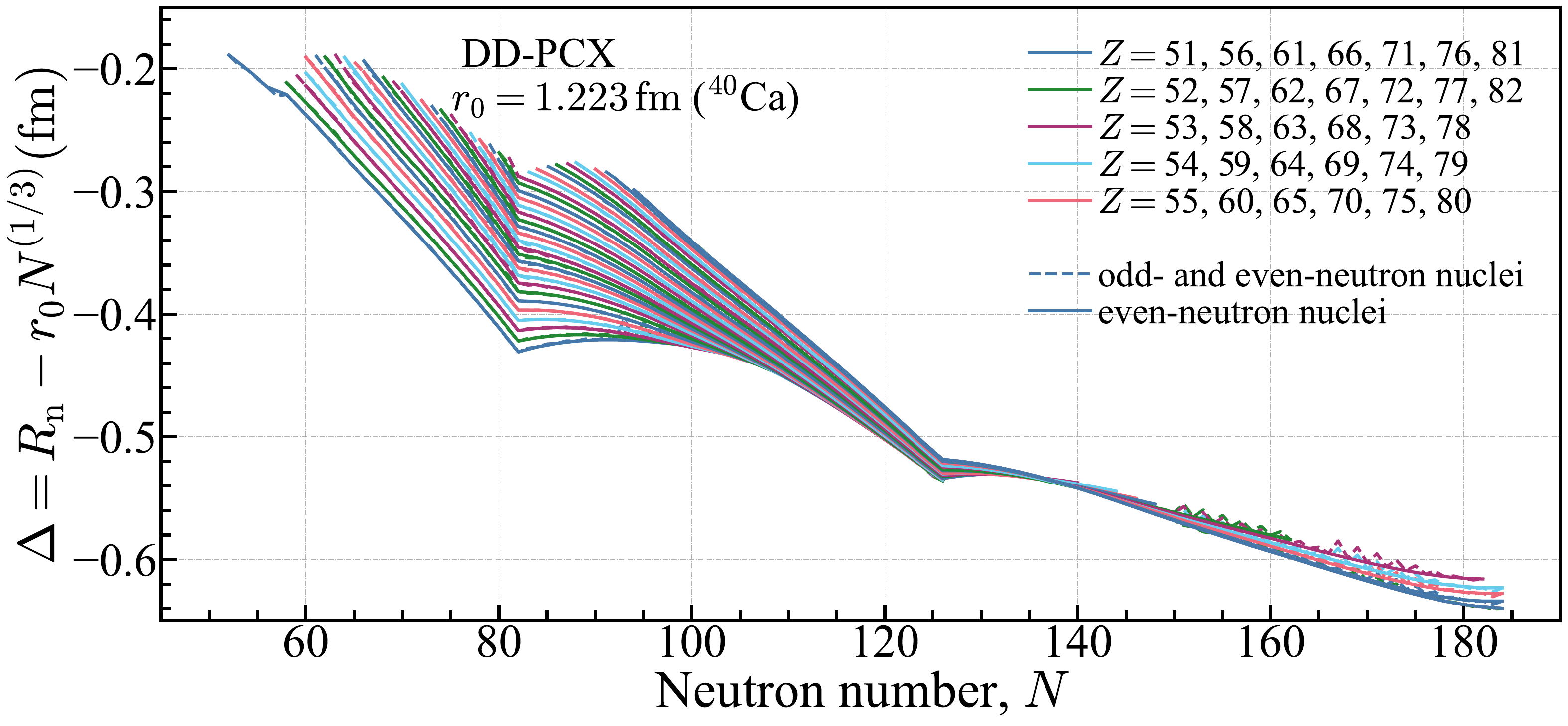}
\label{fig:RN-DDPCX-20-50-82}
\end{minipage}
\begin{minipage}{0.33\textwidth}
\centering
\includegraphics[width=\linewidth, angle=0]{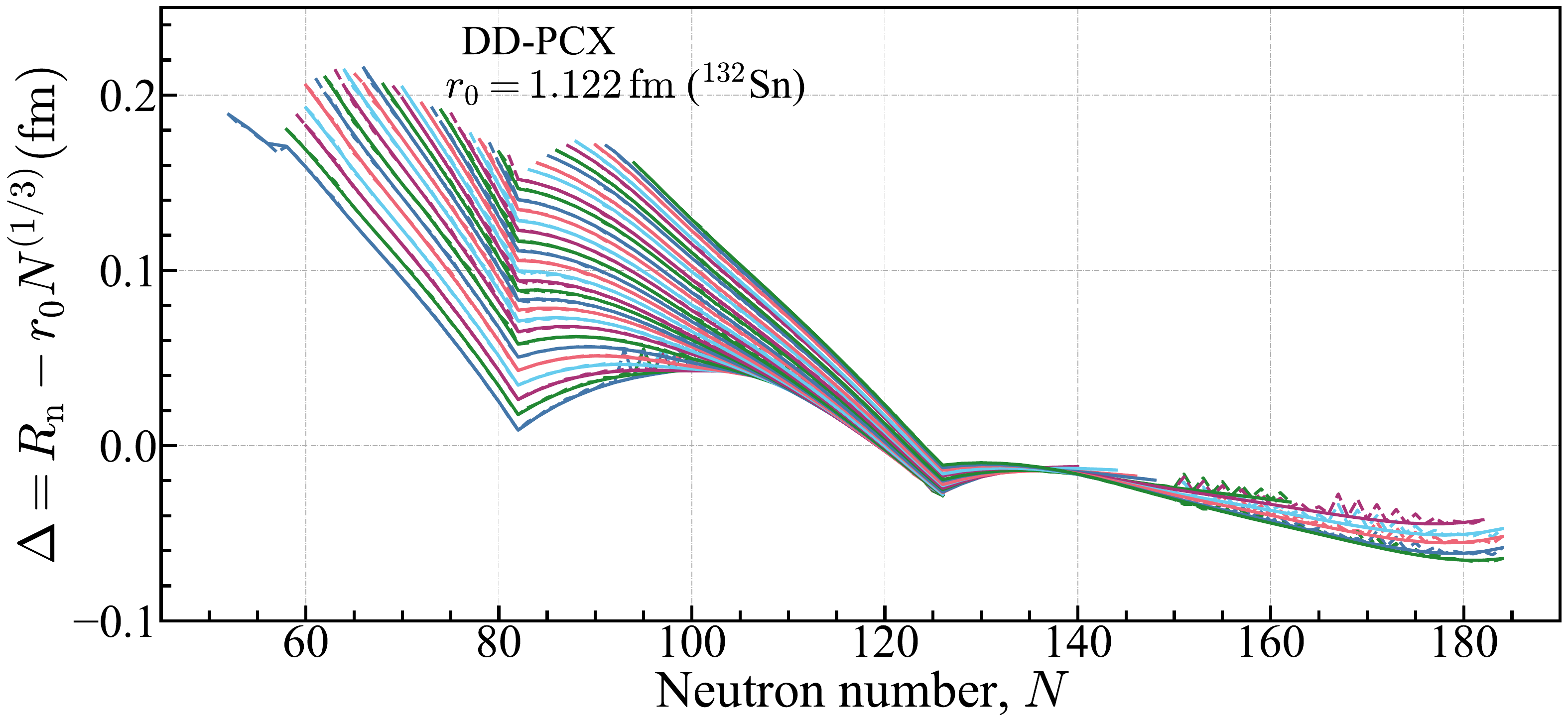}
\label{fig:RN-DDPCX-82-50-82}
\end{minipage}
\begin{minipage}{0.33\textwidth}
\centering
\includegraphics[width=\linewidth, angle=0]{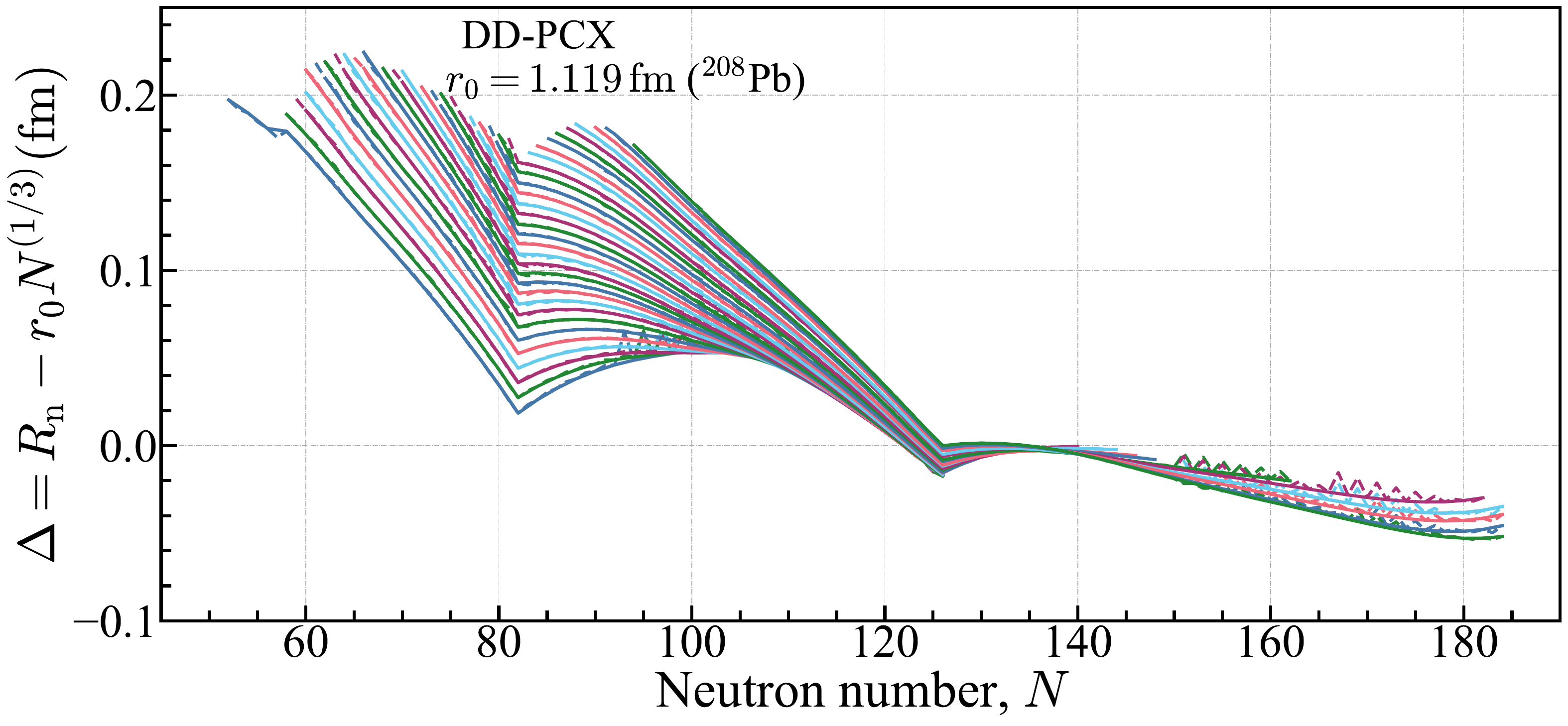}
\label{fig:RN-DDPCX-126-50-82}
\end{minipage}
\vspace{-2mm}
\begin{minipage}{0.33\textwidth}
\centering
\includegraphics[width=\linewidth, angle=0]{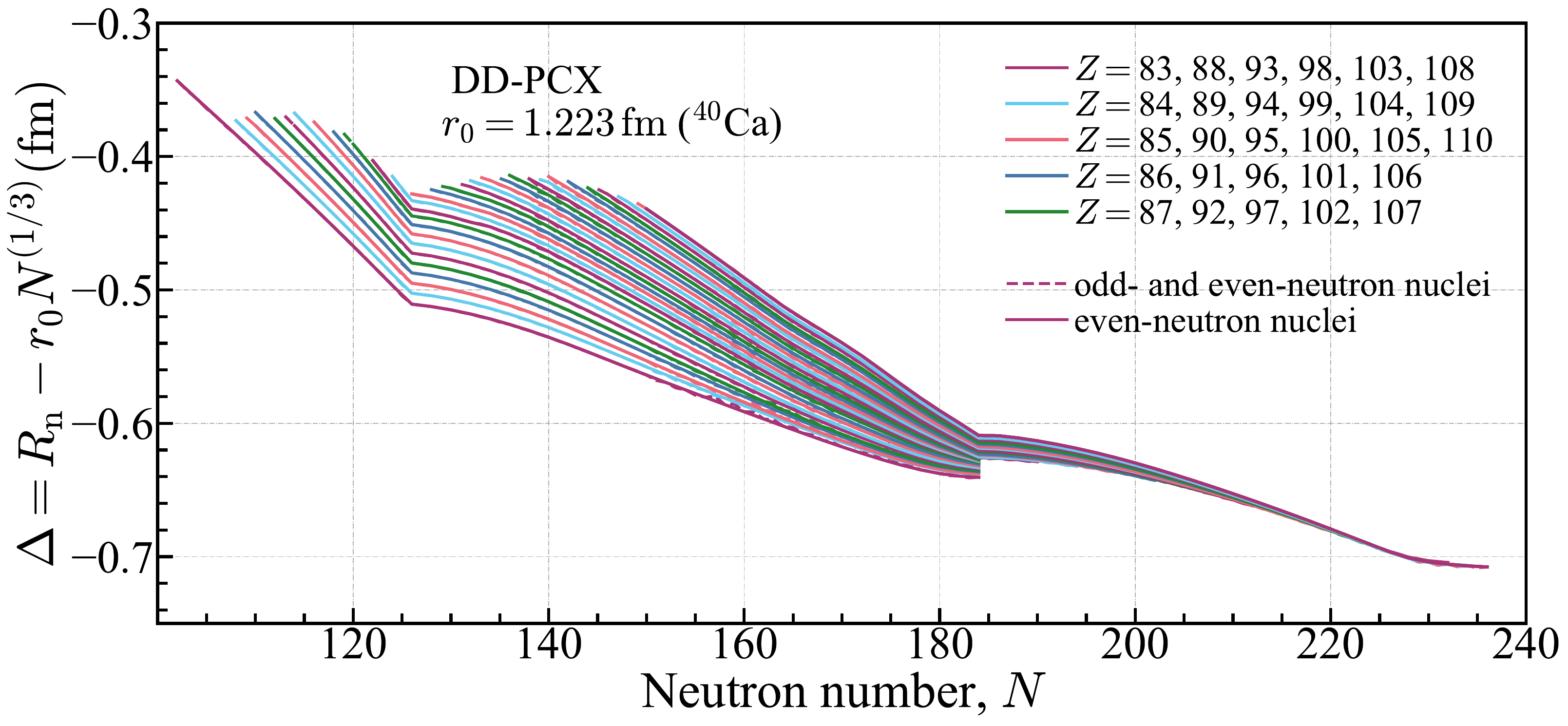}
\label{fig:RN-DDPCX-20-82-110}
\end{minipage}
\begin{minipage}{0.33\textwidth}
\centering
\includegraphics[width=\linewidth, angle=0]{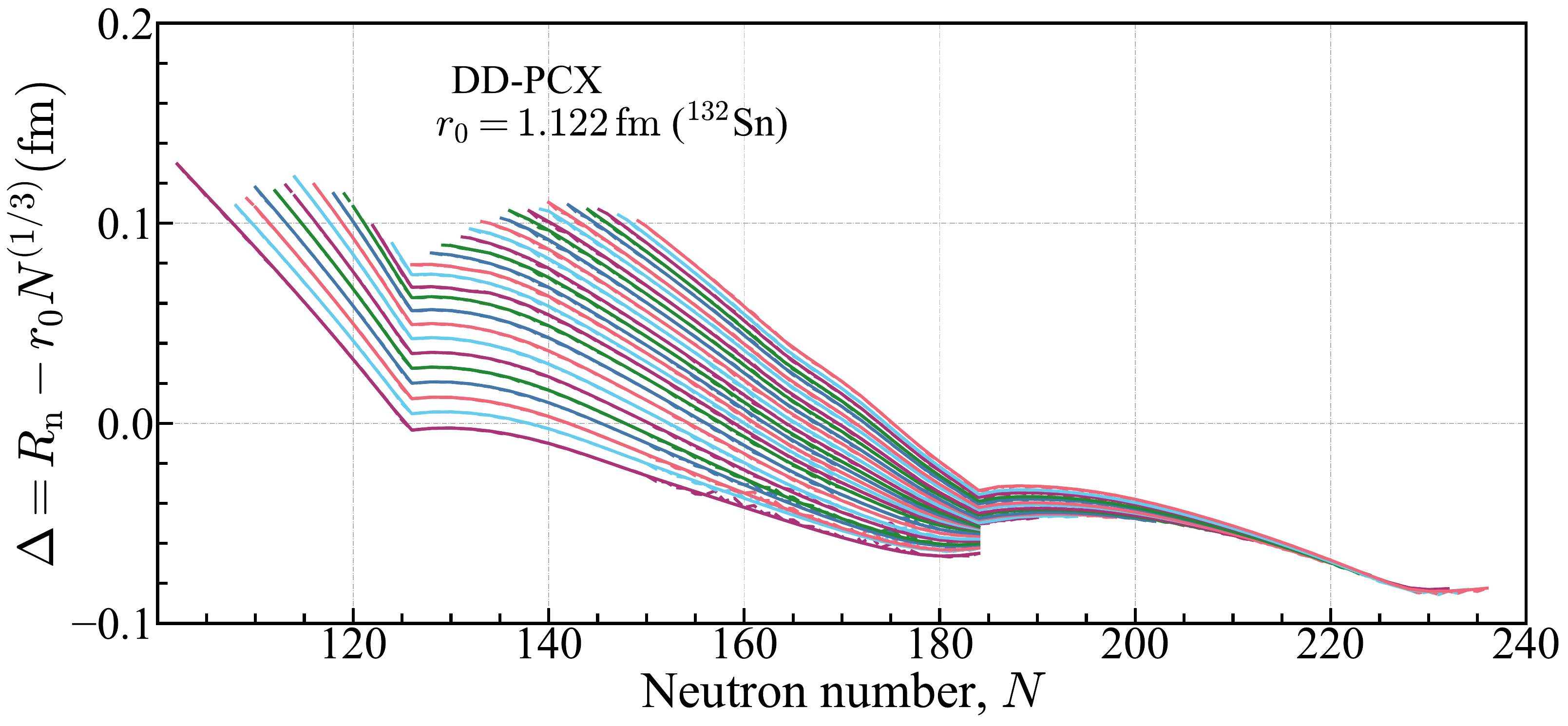}
\label{fig:RN-DDPCX-82-82-110}
\end{minipage}
\begin{minipage}{0.33\textwidth}
\centering
\includegraphics[width=\linewidth, angle=0]{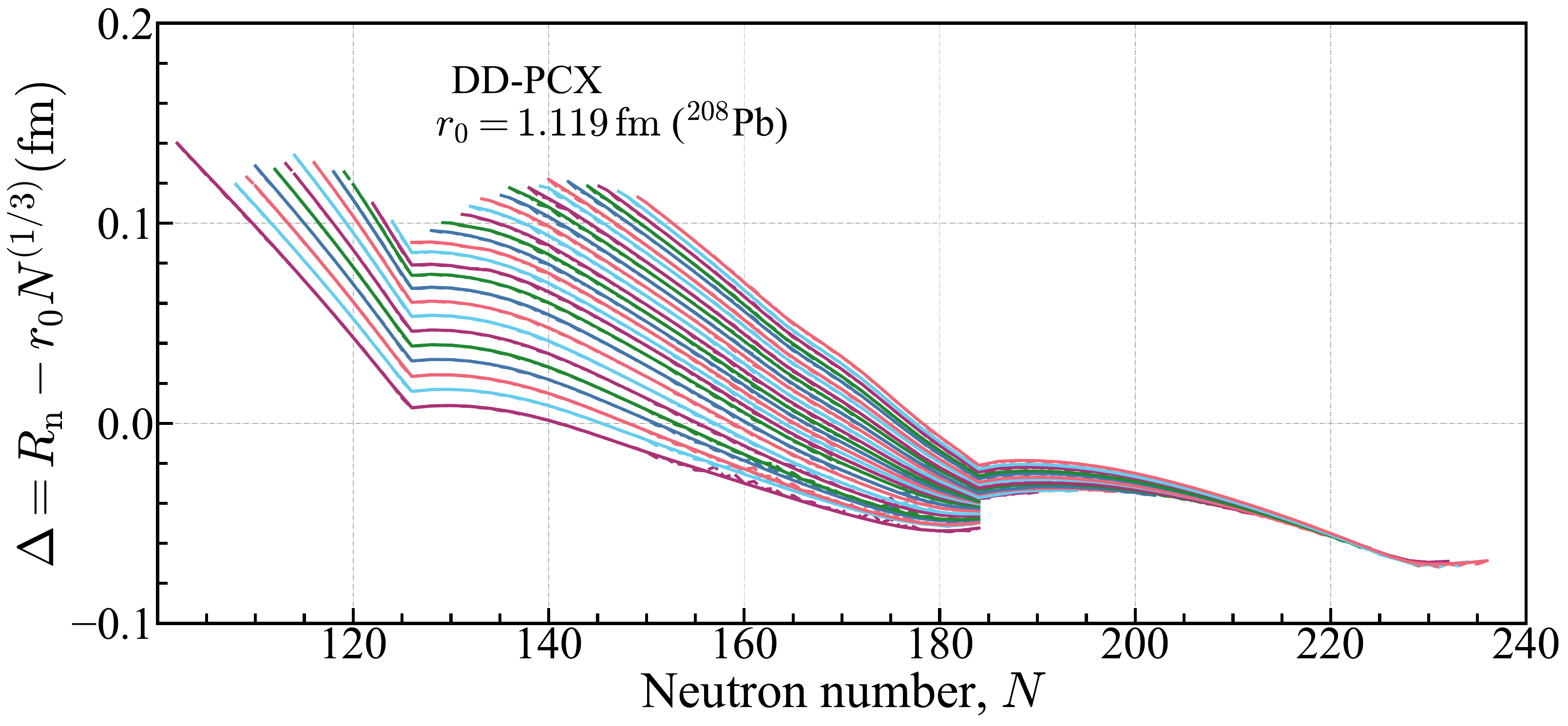}
\label{fig:RN-DDPCX-126-82-110}
\end{minipage}
\vspace{-2mm}
\centering
\begin{minipage}{\textwidth}
\caption{The deviations ($\Delta$) of theoretical root-mean-square radius of neutron distributions ($R_\mathrm{n}$) and the empirical $R_\mathrm{n}\!=\!r_{0}N^{1/3}$. The calculations are based on the RHB approach with the DD-PC interaction for all isotopic chains in the range of $8\!\leq\! Z\!\leq\!110$.
See Fig.~\ref{fig:RN} for further descriptions.}
\label{fig:RN-DDPCX}
\end{minipage}
\end{figure*}

\begin{figure*}
\centering
\begin{minipage}{0.33\textwidth}
\centering
\includegraphics[width=\linewidth, angle=0]{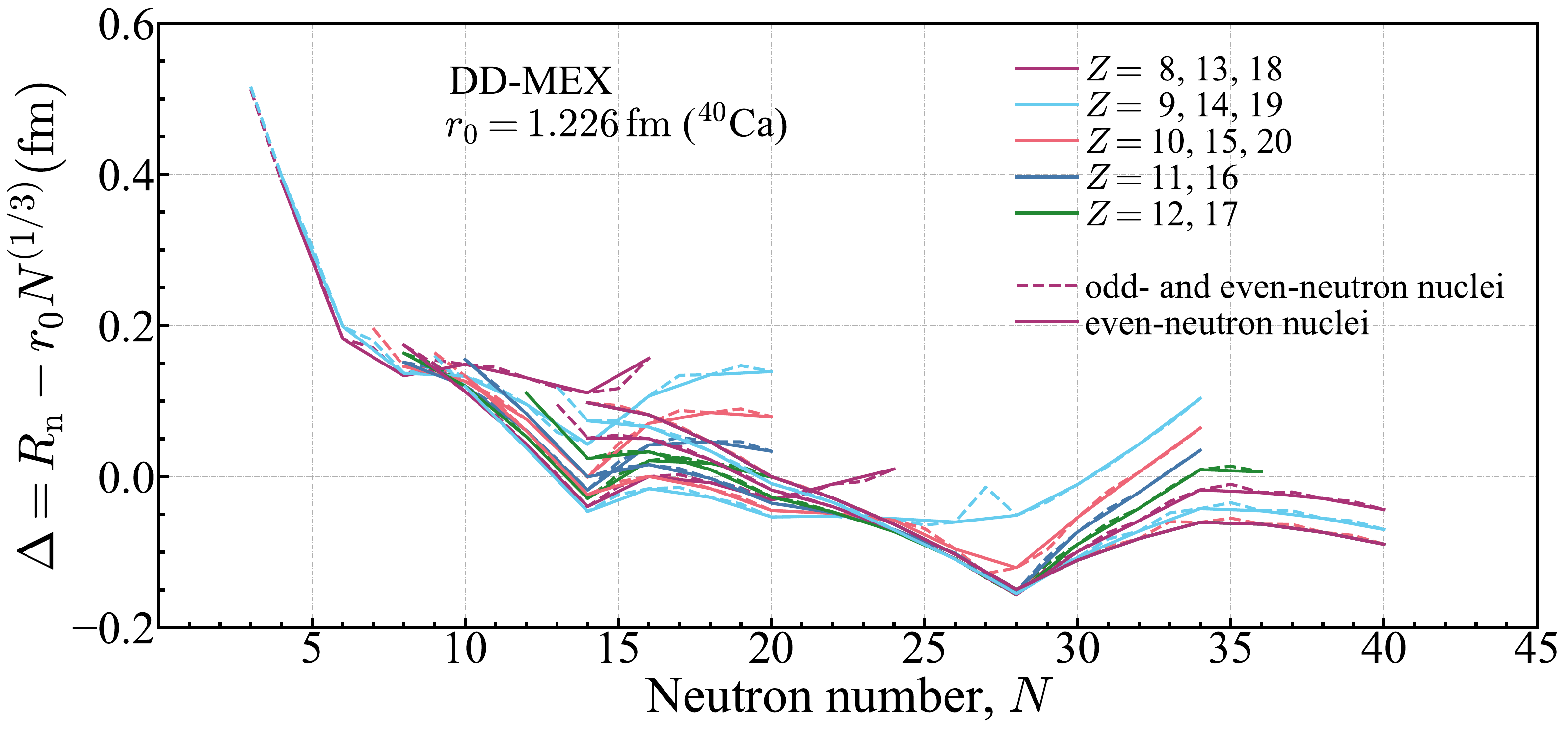}
\label{fig:RN-DDMEX-20-8-20}
\end{minipage}
\begin{minipage}{0.33\textwidth}
\centering
\includegraphics[width=\linewidth, angle=0]{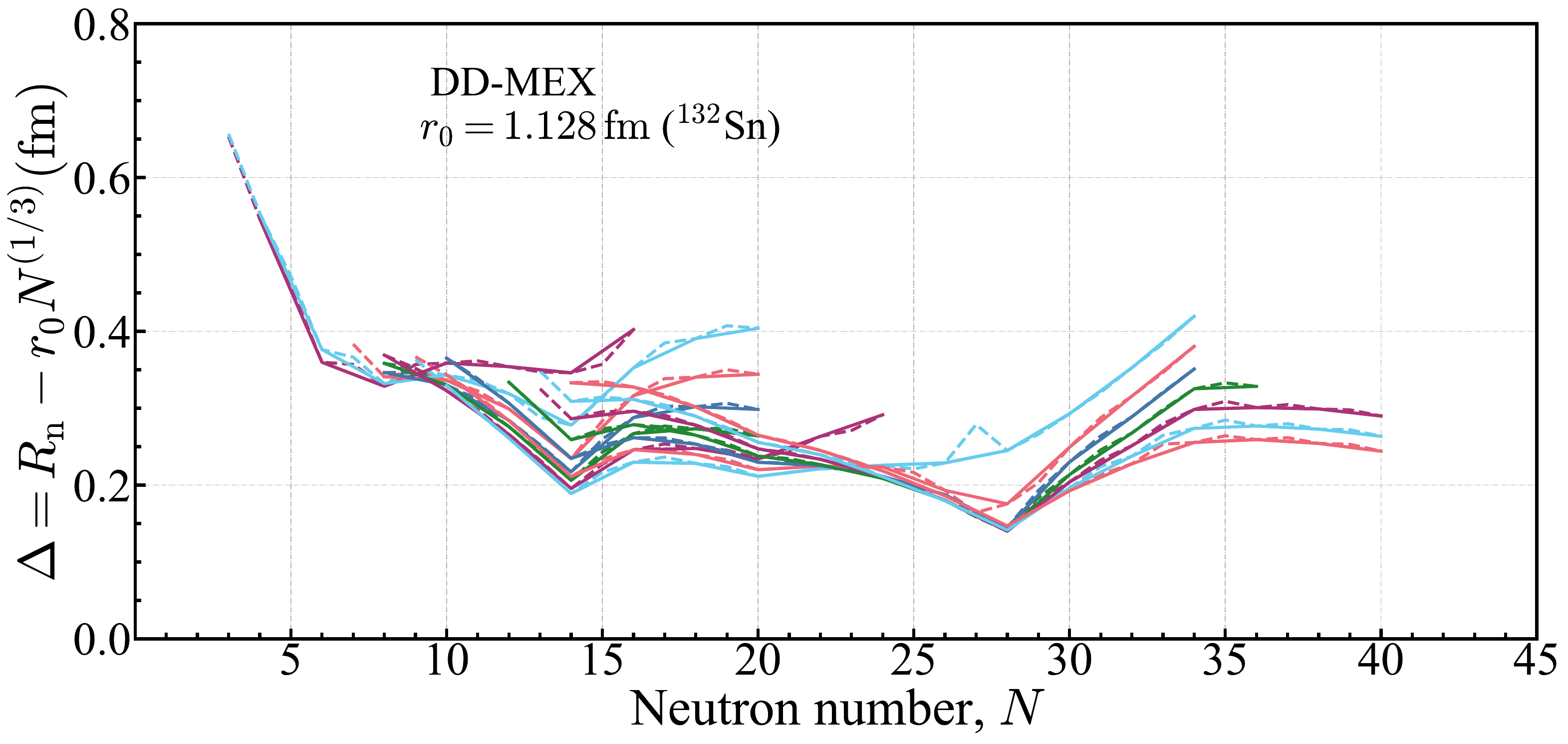}
\label{fig:RN-DDMEX-82-8-20}
\end{minipage}
\begin{minipage}{0.33\textwidth}
\centering
\includegraphics[width=\linewidth, angle=0]{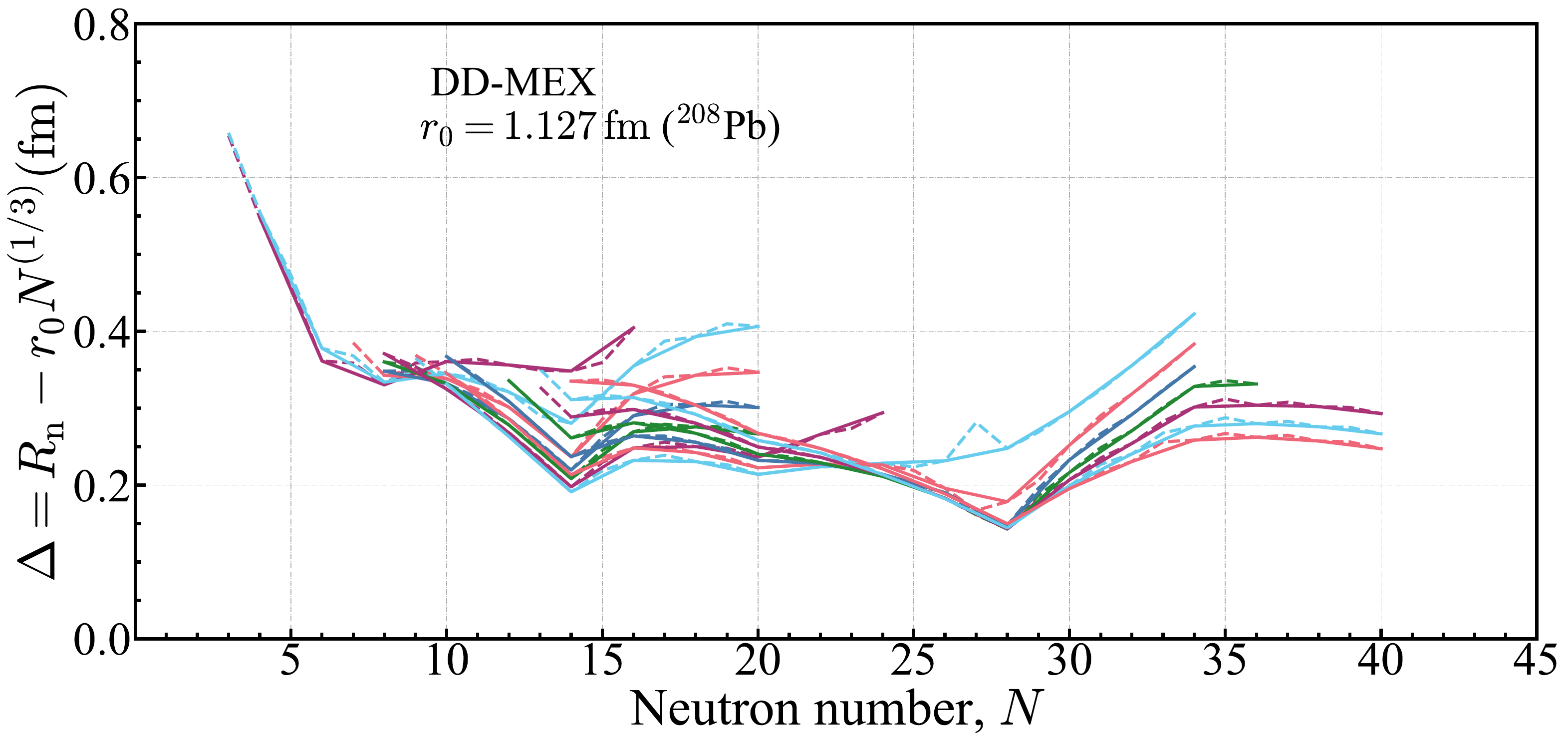}
\label{fig:RN-DDMEX-126-8-20}
\end{minipage}
\vspace{-2mm}
\begin{minipage}{0.33\textwidth}
\centering
\includegraphics[width=\linewidth, angle=0]{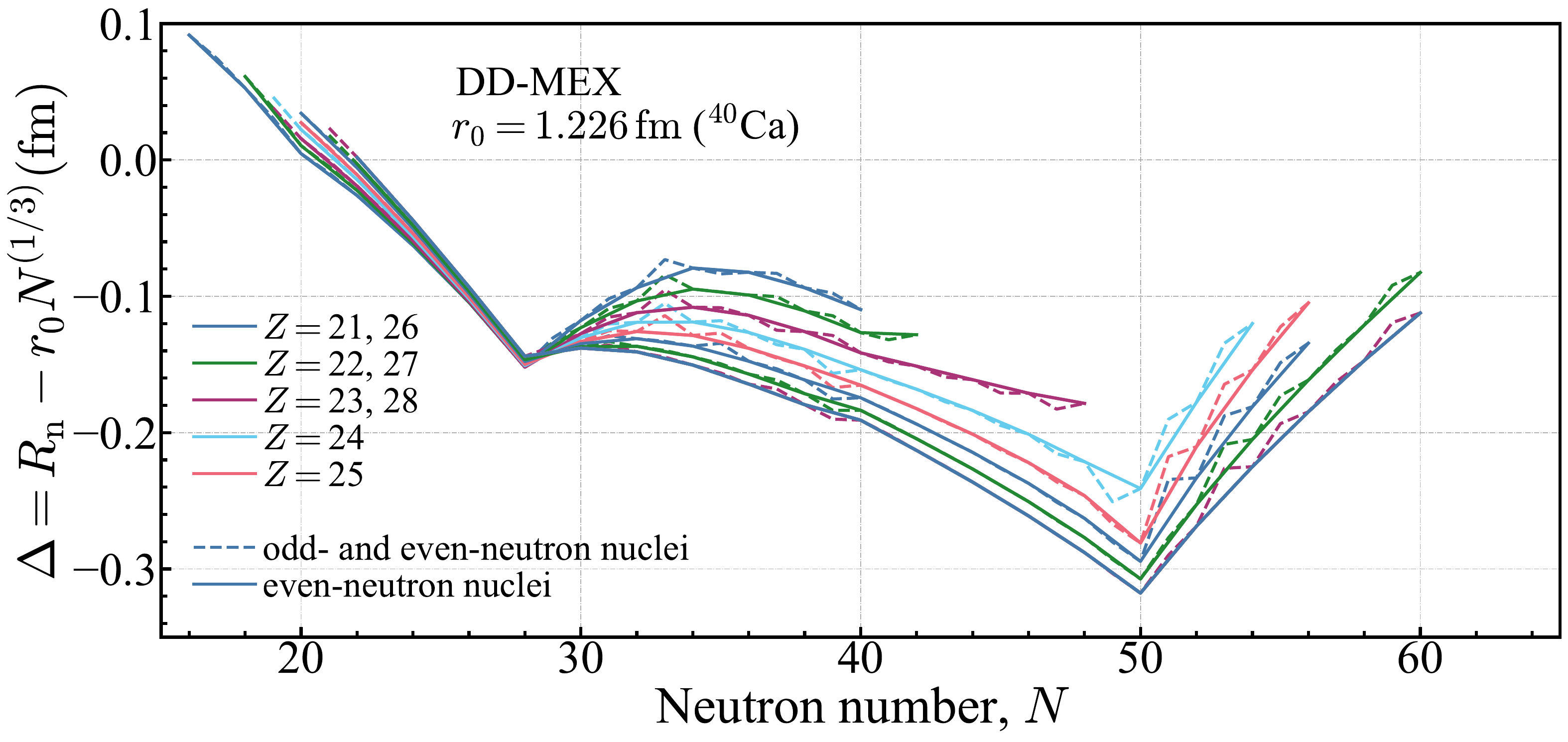}
\label{fig:RN-DDMEX-20-20-28}
\end{minipage}
\begin{minipage}{0.33\textwidth}
\centering
\includegraphics[width=\linewidth, angle=0]{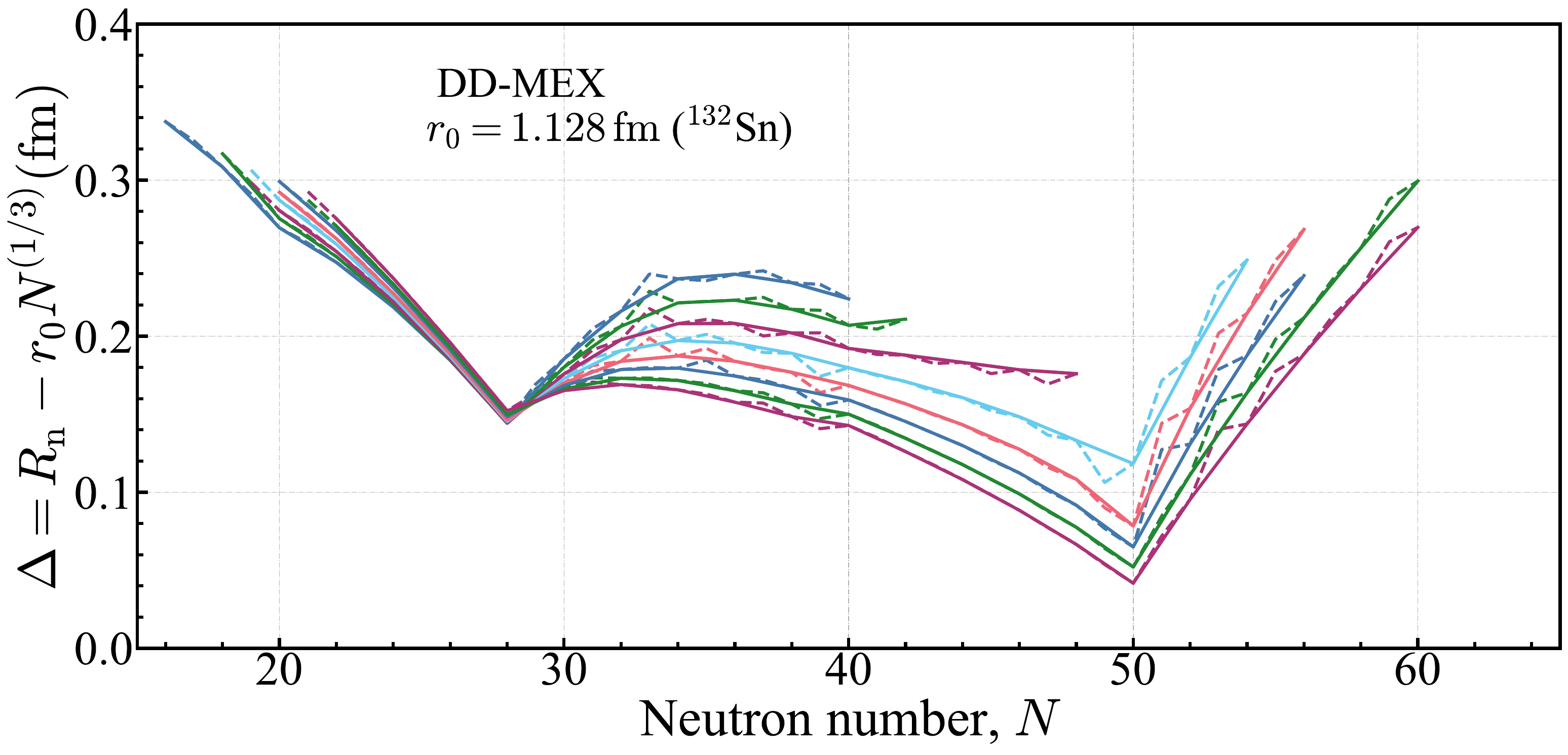}
\label{fig:RN-DDMEX-82-20-28}
\end{minipage}
\begin{minipage}{0.33\textwidth}
\centering
\includegraphics[width=\linewidth, angle=0]{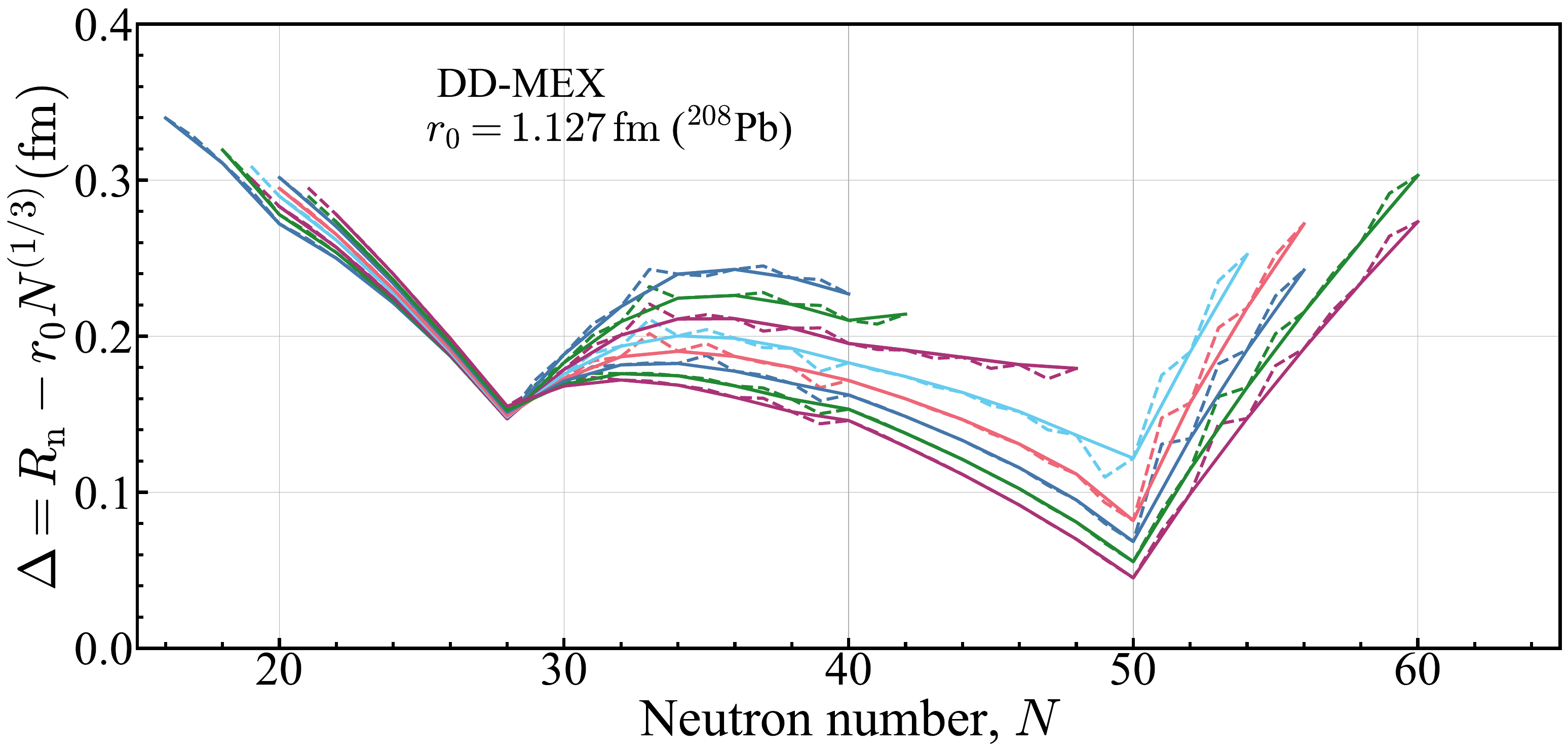}
\label{fig:RN-DDMEX-126-20-28}
\end{minipage}
\vspace{-2mm}
\begin{minipage}{0.33\textwidth}
\centering
\includegraphics[width=\linewidth, angle=0]{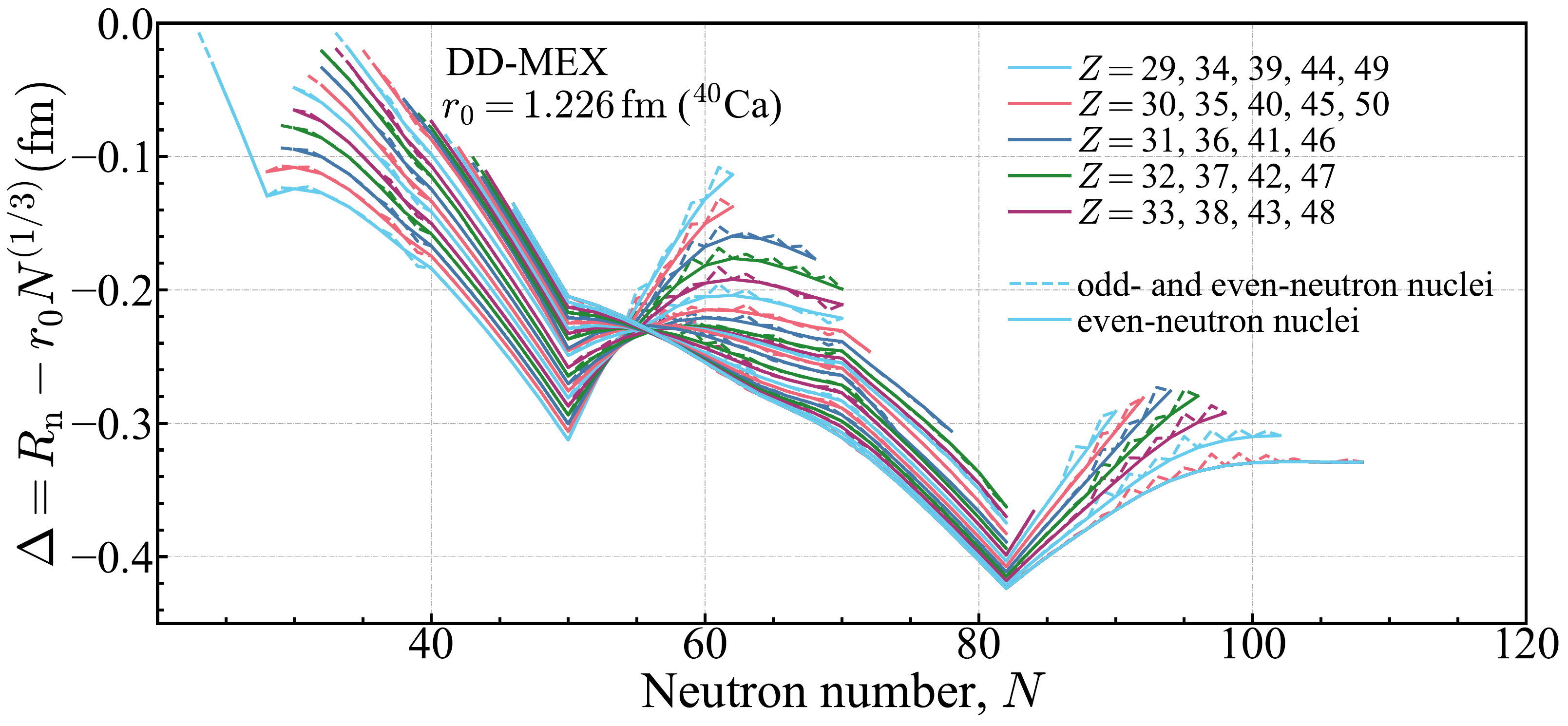}
\label{fig:RN-DDMEX-20-28-50}
\end{minipage}
\begin{minipage}{0.33\textwidth}
\centering
\includegraphics[width=\linewidth, angle=0]{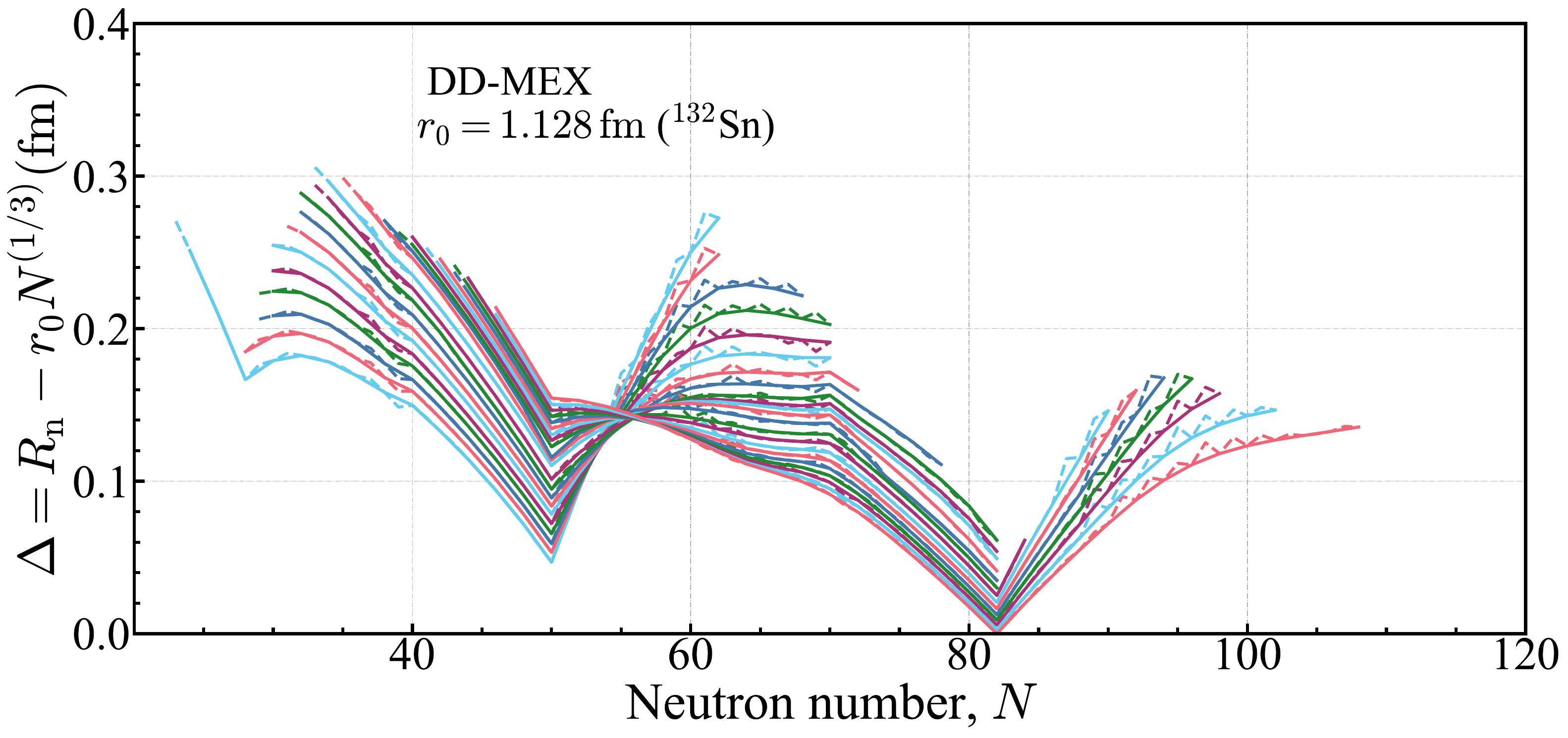}
\label{fig:RN-DDMEX-82-28-50}
\end{minipage}
\begin{minipage}{0.33\textwidth}
\centering
\includegraphics[width=\linewidth, angle=0]{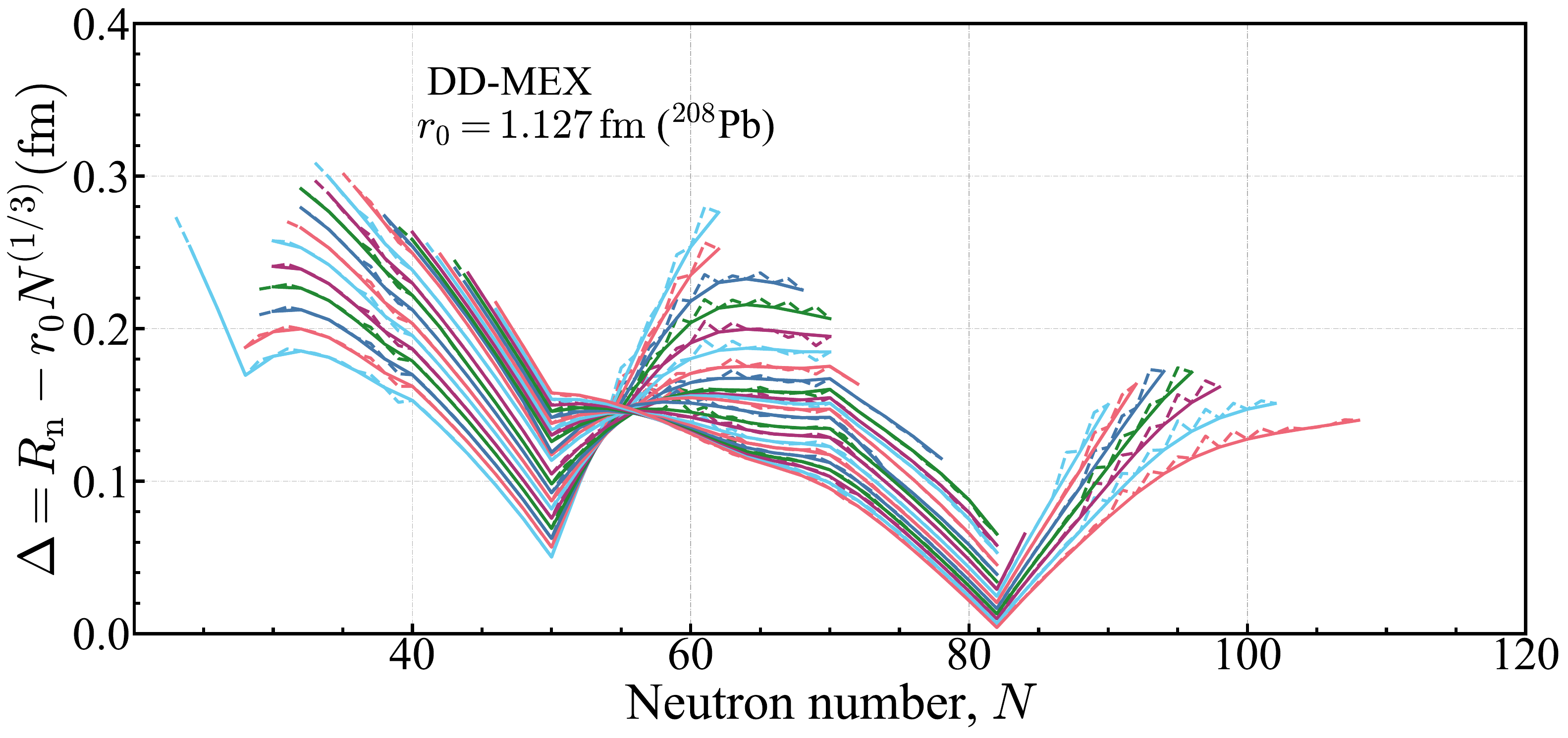}
\label{fig:RN-DDMEX-126-28-50}
\end{minipage}
\vspace{-2mm}
\begin{minipage}{0.33\textwidth}
\centering
\includegraphics[width=\linewidth, angle=0]{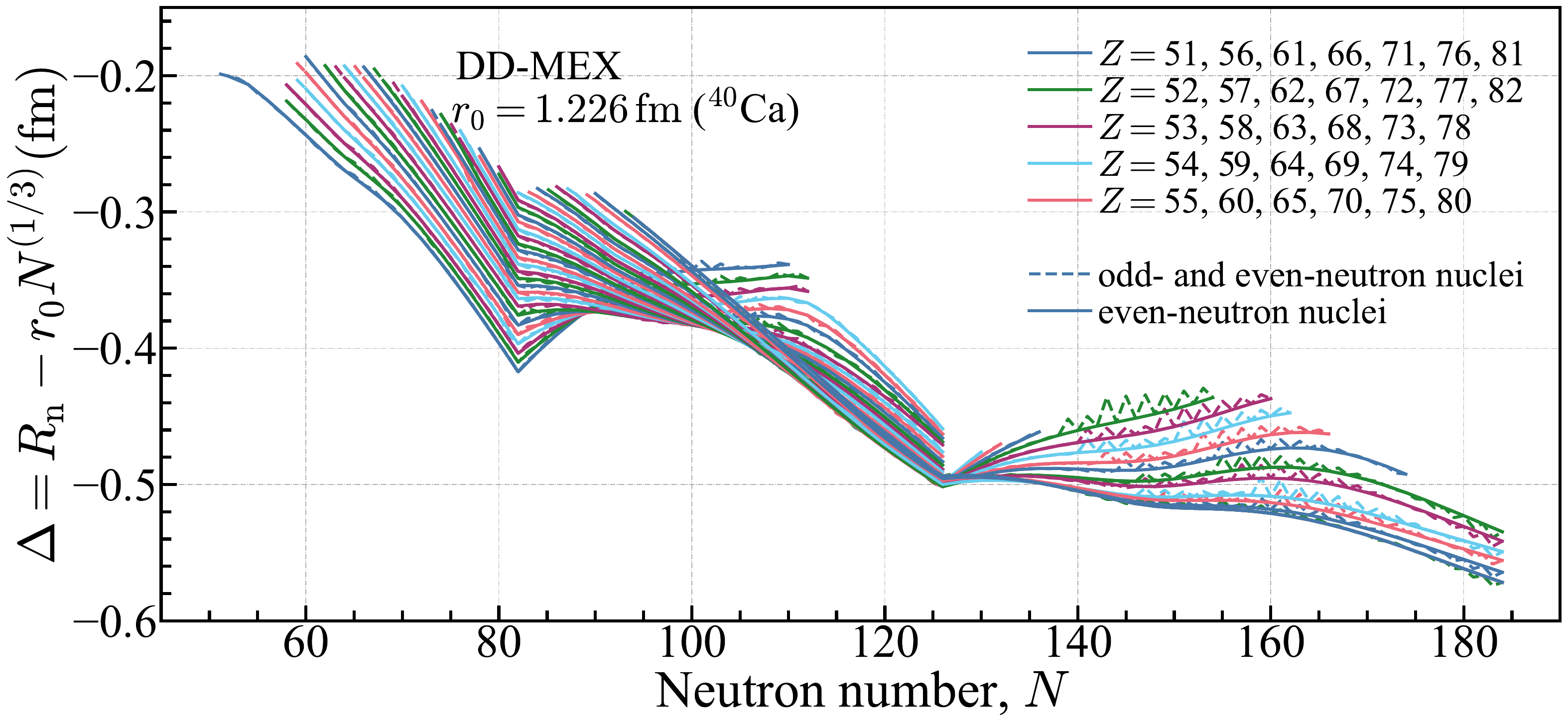}
\label{fig:RN-DDMEX-20-50-82}
\end{minipage}
\begin{minipage}{0.33\textwidth}
\centering
\includegraphics[width=\linewidth, angle=0]{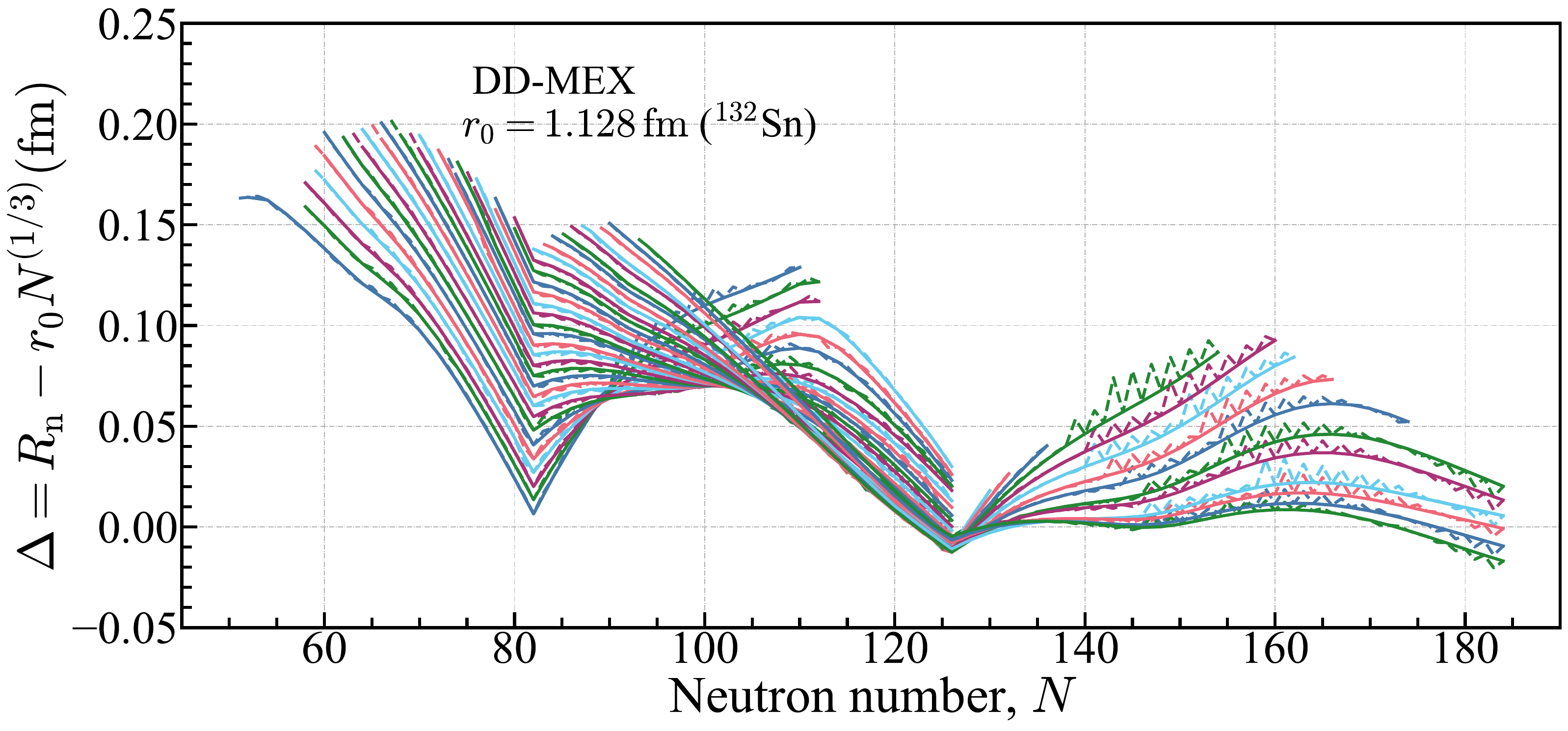}
\label{fig:RN-DDMEX-82-50-82}
\end{minipage}
\begin{minipage}{0.33\textwidth}
\centering
\includegraphics[width=\linewidth, angle=0]{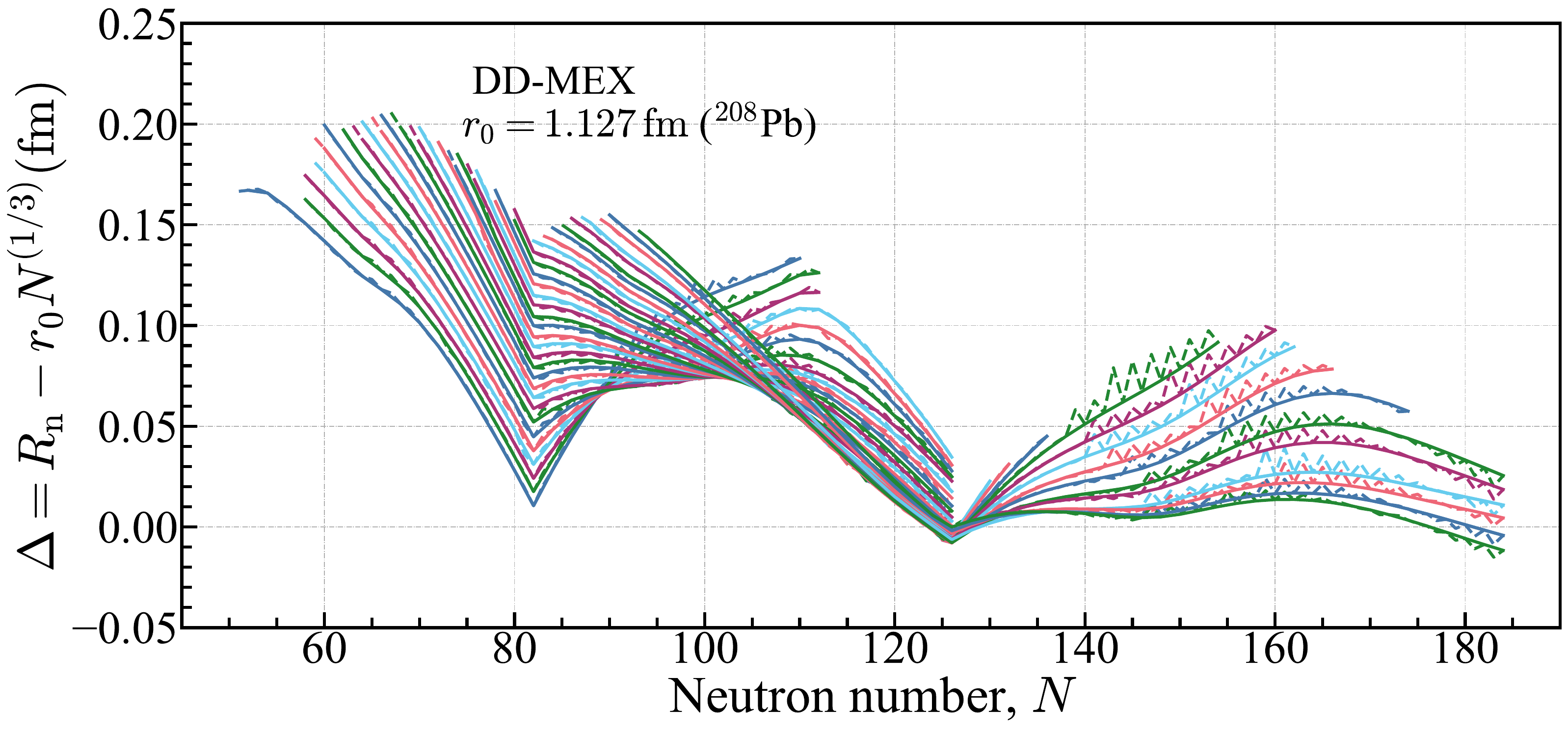}
\label{fig:RN-DDMEX-126-50-82}
\end{minipage}
\vspace{-2mm}
\begin{minipage}{0.33\textwidth}
\centering
\includegraphics[width=\linewidth, angle=0]{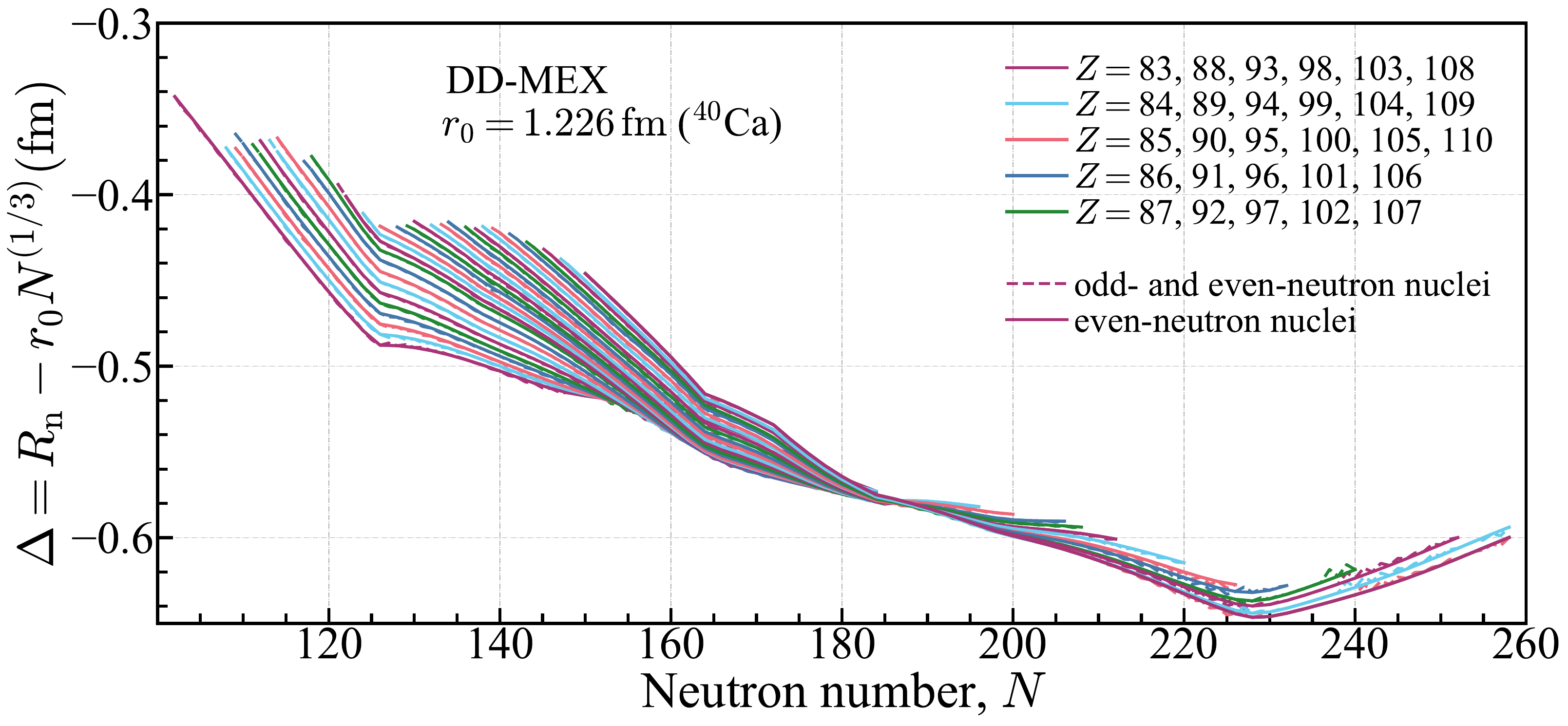}
\label{fig:RN-DDMEX-20-82-110}
\end{minipage}
\begin{minipage}{0.33\textwidth}
\centering
\includegraphics[width=\linewidth, angle=0]{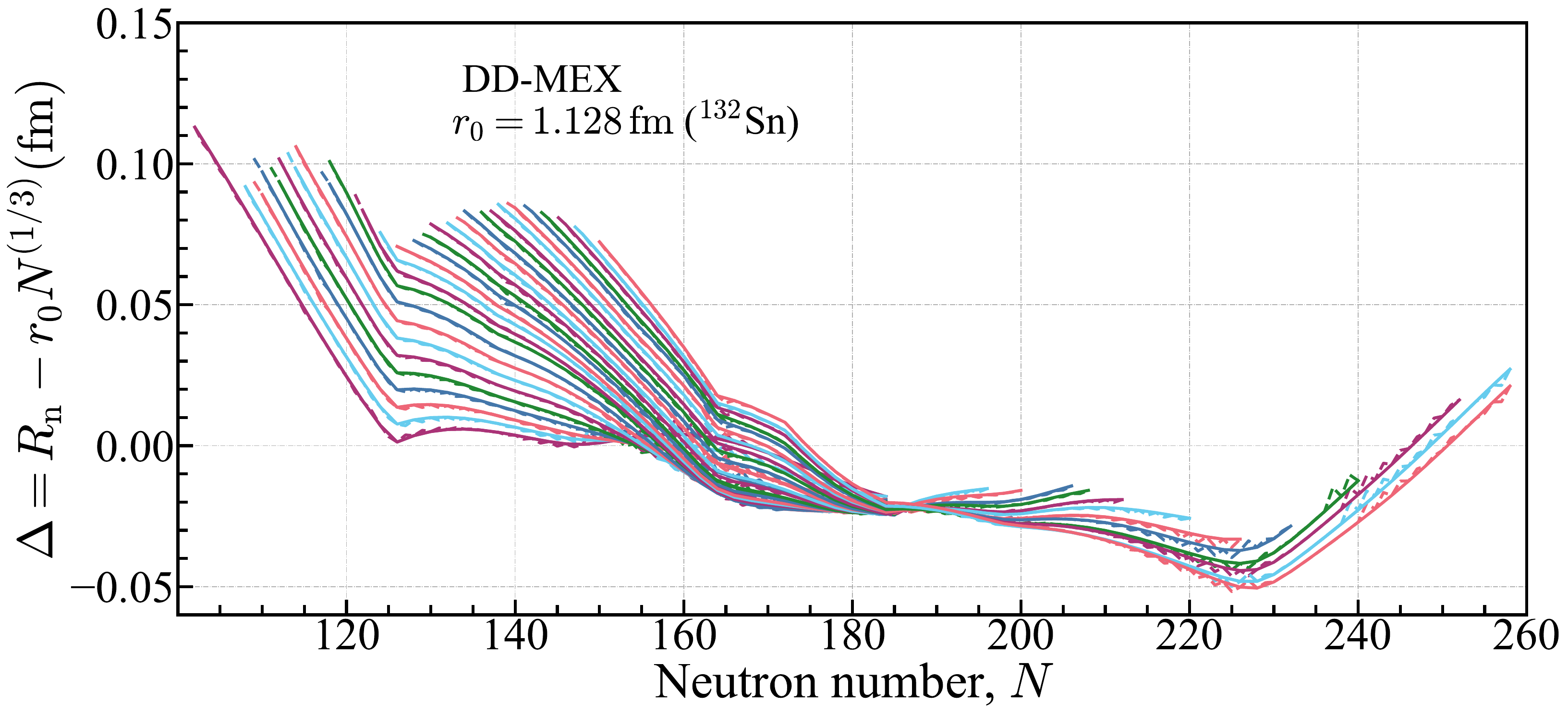}
\label{fig:RN-DDMEX-82-82-110}
\end{minipage}
\begin{minipage}{0.33\textwidth}
\centering
\includegraphics[width=\linewidth, angle=0]{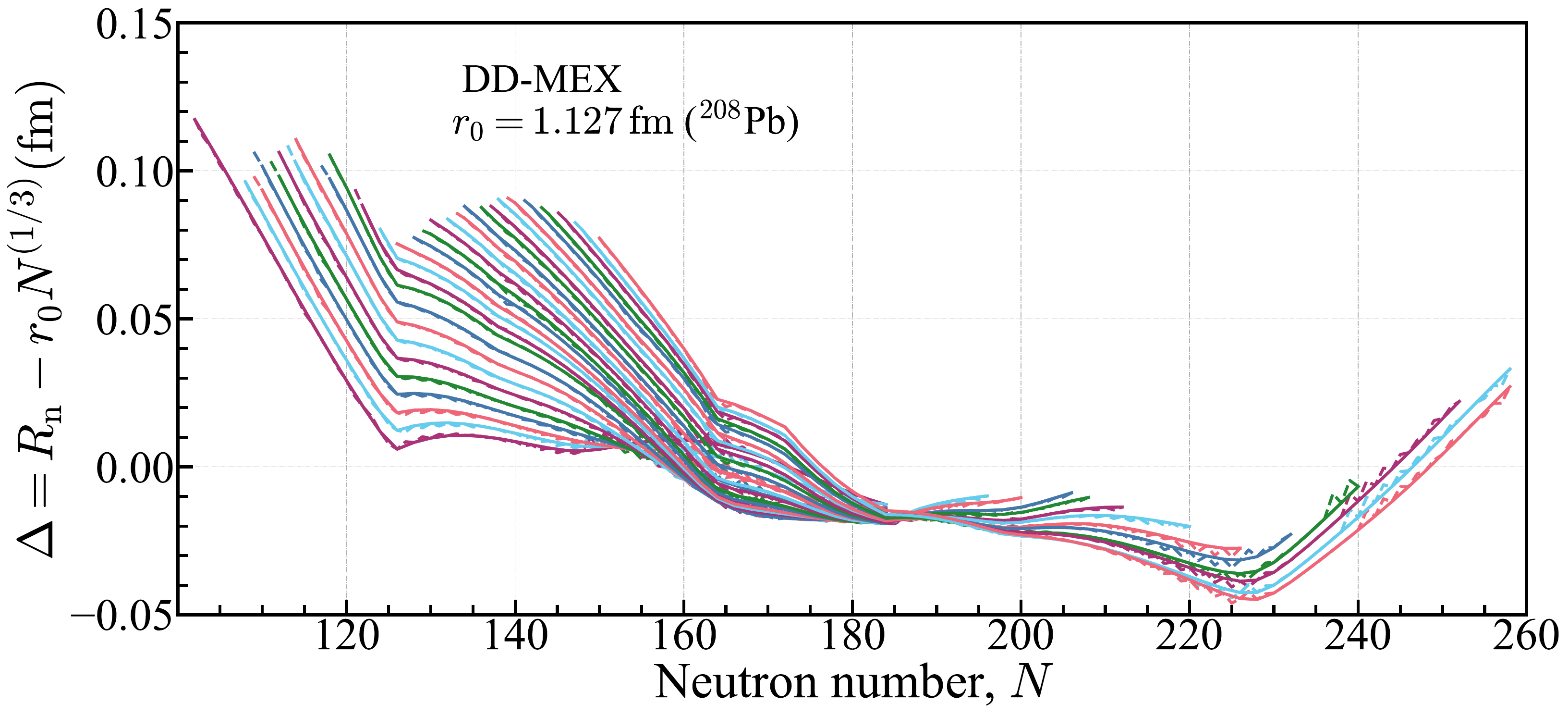}
\label{fig:RN-DDMEX-126-82-110}
\end{minipage}
\vspace{-2mm}
\centering
\begin{minipage}{\textwidth}
\caption{The deviations ($\Delta$) of theoretical root-mean-square radius of neutron distributions ($R_\mathrm{n}$) and the empirical $R_\mathrm{n}\!=\!r_{0}N^{1/3}$. The calculations are based on the RHB approach with the DD-MEX interaction for all isotopic chains in the range of $8\!\leq\! Z\!\leq\!110$.
See Fig.~\ref{fig:RN} for further descriptions.}
\label{fig:RN-DDMEX}
\end{minipage}
\end{figure*}

We present the differences between the empirical and theoretical $R_{\rm n}$, $\Delta\!=\! R_\mathrm{n}-r_{0}N^{1/3}$, resulted from PC-L3R, PC-X, DD-MEX, and DD-PCX in Figs.~\ref{fig:RN}, \ref{fig:RN-PCX}, \ref{fig:RN-DDPCX} and \ref{fig:RN-DDMEX}. Overall, for each isotopic chain, the trend of $\Delta$ decreases from the isotope at the proton edge, then it reverses at the neutron magic numbers $28$, $50$, $82$, $126$, and $184$. 
For the proton-rich nuclei of light isotopic chains, we find that the reversing points are $N\!=\!6$ and $14$, of which the neutron single-particle states 1p$_{3/2}$ and 1d$_{5/2}$ are filled, suggesting the possible sub-shell closure at $N\!=\!6$. 


We notice some pronounced deviations at nuclei near the neutron drip line of some isotopic chains, i.e., Mg, Ca, and Kr, indicating the possible existence of the halo or giant halo phenomena. The pronounced deviations imply that some possible halo structures could be found in the extreme neutron excess of these isotopic chains, of which the respective neutron separation energies are close to zero or the Fermi surface near the continuum threshold. The future study of exploring the possible halo structures in these nuclei requires the consideration of deformation.

\begin{table}[pos=h!,width=\columnwidth]
\scriptsize
\caption{\label{tab:r0}\footnotesize The coefficient $r_{0}$ of the empirical $R_{\rm n}$ determined by the theoretical $R_{\rm n}$ of $^{40}$Ca, of $^{132}$Sn, and of $^{208}$Pb, which are calculated in the RHB approach with the PC-L3R~\cite{PLB2023Liu}, or PC-X~\cite{PLB2020Taninah}, or DD-MEX~\cite{PLB2020Taninah}, or DD-PCX~\cite{PRC2019Yuksel} interactions.} 
\begin{tabular}{@{}l@{\hspace{4mm}}c@{\hspace{2mm}}c@{\hspace{2mm}}c@{\hspace{2mm}}c@{\hspace{2mm}}c@{\hspace{2mm}}}
\toprule
\midrule
   & \multicolumn{4}{c}{$r_{0}$ (fm)} \\
   & PC-L3R &  PC-X & DD-MEX & DD-PCX \\
\midrule
  $^{40}$Ca   &   $1.233$ &     $1.234$     &  $1.226  $ & $ 1.223 $ \\
  $^{132}$Sn  &   $1.146$ &     $1.145 $    &  $ 1.128 $ & $ 1.122 $ \\
  $^{208}$Pb  &   $1.141$ &     $1.139 $    &  $ 1.127 $ & $ 1.119 $ \\
\bottomrule
\end{tabular}
\end{table}

\section{Summary}
\label{sec:summary}

In conclusion, we have calculated the properties of all bound nuclei of the isotopic chains from $Z\!=\!8$-$110$ based on the RHB approach with the PC-L3R~\cite{PLB2023Liu}, PC-X~\cite{PLB2020Taninah}, DD-MEX~\cite{PLB2020Taninah}, and DD-PCX~\cite{PRC2019Yuksel} interactions. In this work, the RHB equation for nucleons is solved by using the spherical harmonic oscillator bases. The nuclear ground state properties are tabulated accordingly. These properties consists of binding energies ($E_\mathrm{b}$), one- and two-neutron separation energies ($S_\mathrm{n}$ and $S_\mathrm{2n}$), root-mean-square radii of matter ($R_\mathrm{m}$), of neutron ($R_\mathrm{n}$), of proton ($R_\mathrm{p}$), and of charge ($R_\mathrm{c}$) distributions, Fermi surfaces ($\lambda$), ground-state spins ($J$) and parities ($\pi$). 
The number of bound nuclei from oxygen ($Z\!=\!8$) to darmstadtium ($Z\!=\!110$) predicted by the covariant density functionals, PC-L3R, PC-X, DD-MEX, and DD-PCX are 9004, 9162, 7112, and 6799, respectively. 
Compared with the density-dependent point-coupling interactions, the nonlinear point-coupling interactions, PC-L3R and PC-X, give a more extended border of the neutron-rich region (Table.~\ref{tab:Neu-dp} and Fig.~\ref{fig:nuclear_chart}). Meanwhile, we also compare the presently predicted neutron drip lines with the ones from the mass tables, FRDM~\cite{MOLLER1995}, WS4~\cite{PLB2014Wang}, HFB-21~\cite{PRC2010Goriely}, UNEDF~\cite{Nature2012Erler}, TMA~\cite{PTP2005Geng}, NL3*~\cite{PLB2013Afanasjev}, and PC-PK1~\cite{ADNDT2018Xia} in Fig.~\ref{fig:nuclear_chart}.

\begin{landscape}
\begin{figure}
\centering
\centering
\includegraphics[height=0.93\textheight, angle=0]{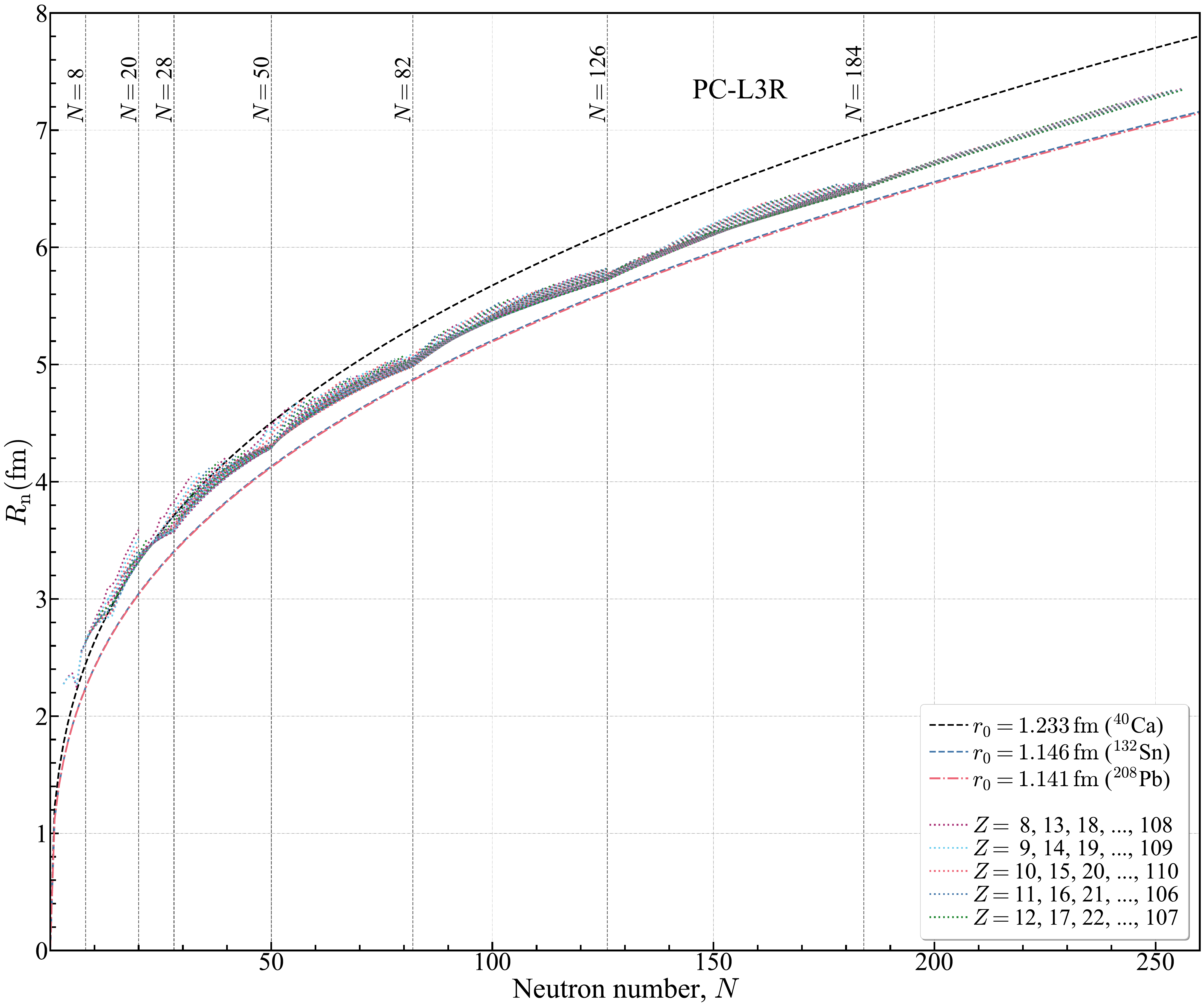}
\centering
\caption{The root-mean-square radius of neutron as a function of the neutron number for all isotopic chains of $8\!\leq Z\!\leq 110$ calculated from the RHB approach with the PC-L3R interaction. The empirical $R_{\rm n}\!=\!r_{0} N^{1/3}$ are plotted using the $r_{0}(^{40}\mathrm{Ca})$, $r_{0}(^{132}\mathrm{Sn})$ and $r_{0}(^{208}\mathrm{Pb})$ calculated from PC-L3R (Table~\ref{tab:r0}).}
\label{fig:Rn_PCL3R_all}
\end{figure}
\end{landscape}

\begin{landscape}
\begin{figure}
\centering
\centering
\includegraphics[height=0.93\textheight, angle=0]{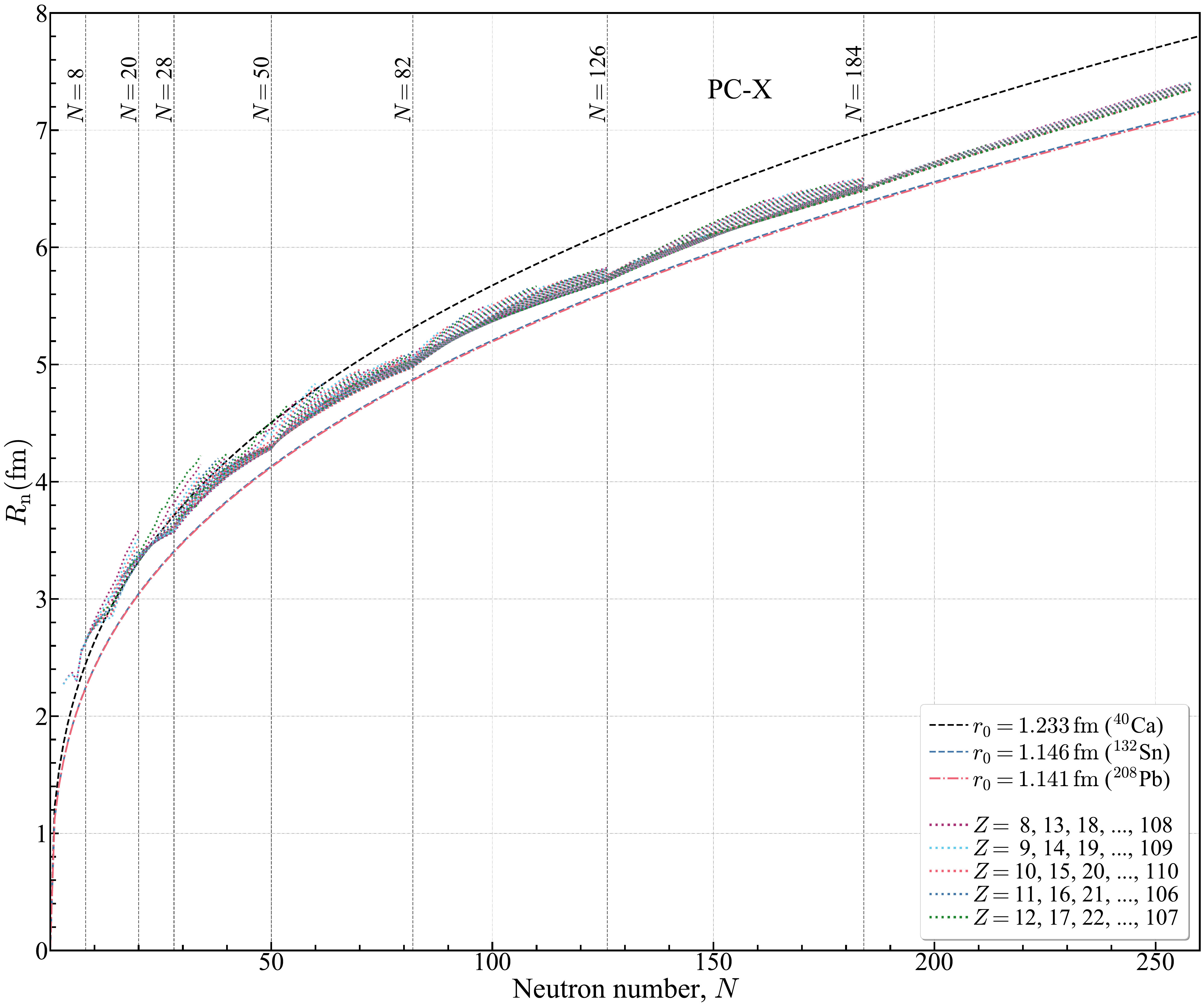}
\caption{
The root-mean-square radius of neutron as a function of the neutron number for all isotopic chains of $8\!\leq Z\!\leq 110$ calculated from the RHB approach with the PC-X interaction. The empirical $R_{\rm n}\!=\!r_{0} N^{1/3}$ are plotted using the $r_{0}(^{40}\mathrm{Ca})$, $r_{0}(^{132}\mathrm{Sn})$ and $r_{0}(^{208}\mathrm{Pb})$ calculated from PC-X (Table~\ref{tab:r0}).}
\label{fig:Rn_PCX_all}
\end{figure}
\end{landscape}

\begin{landscape}
\begin{figure}
\centering
\centering
\includegraphics[height=0.93\textheight, angle=0]{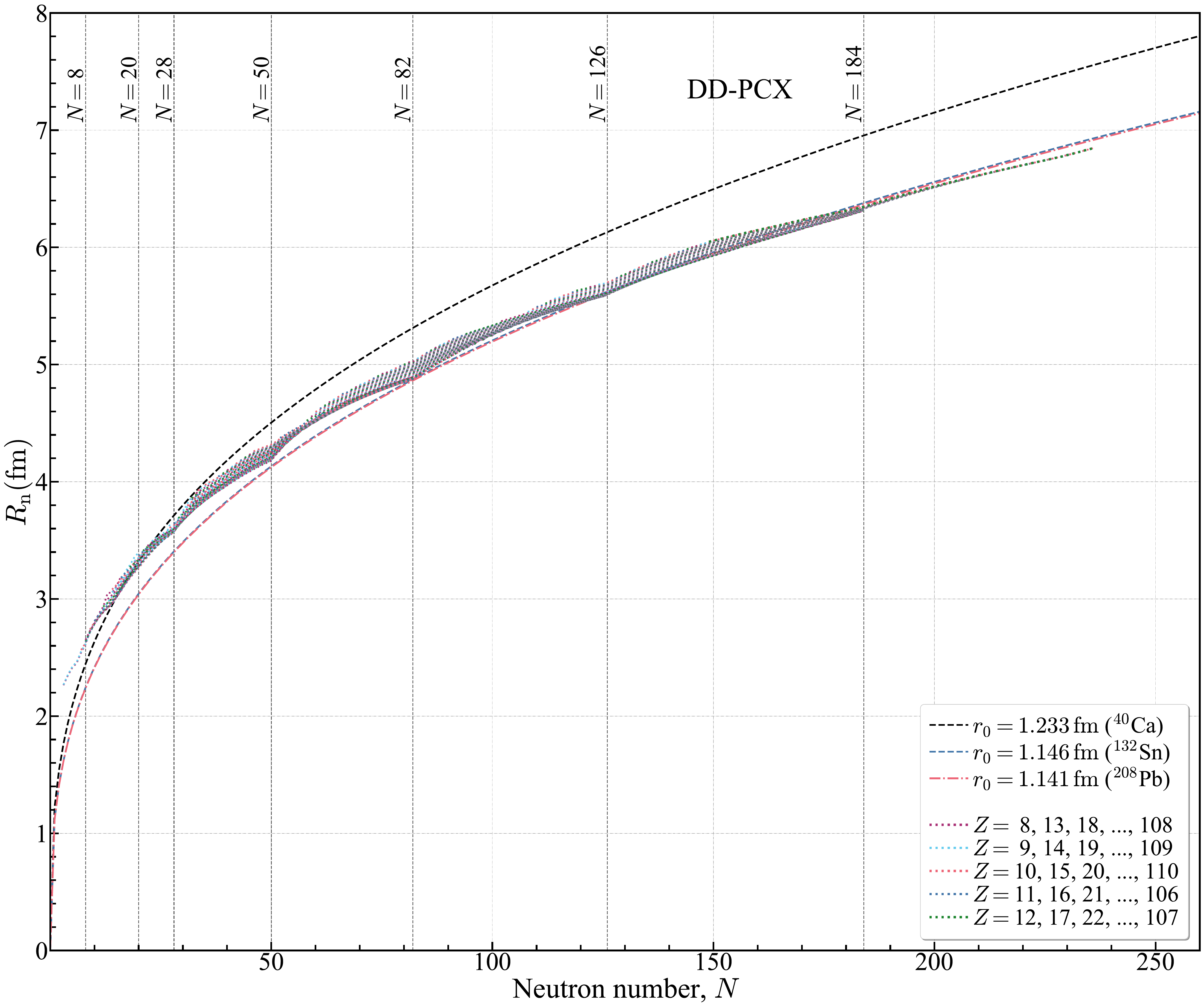}
\caption{
The root-mean-square radius of neutron as a function of the neutron number for all isotopic chains of $8\!\leq Z\!\leq 110$ calculated from the RHB approach with the DD-PCX interaction. The empirical $R_{\rm n}\!=\!r_{0} N^{1/3}$ are plotted using the $r_{0}(^{40}\mathrm{Ca})$, $r_{0}(^{132}\mathrm{Sn})$ and $r_{0}(^{208}\mathrm{Pb})$ calculated from DD-PCX (Table~\ref{tab:r0}).}
\label{fig:Rn_DDPCX_all}
\end{figure}
\end{landscape}

\begin{landscape}
\begin{figure}
\centering
\centering
\includegraphics[height=0.93\textheight, angle=0]{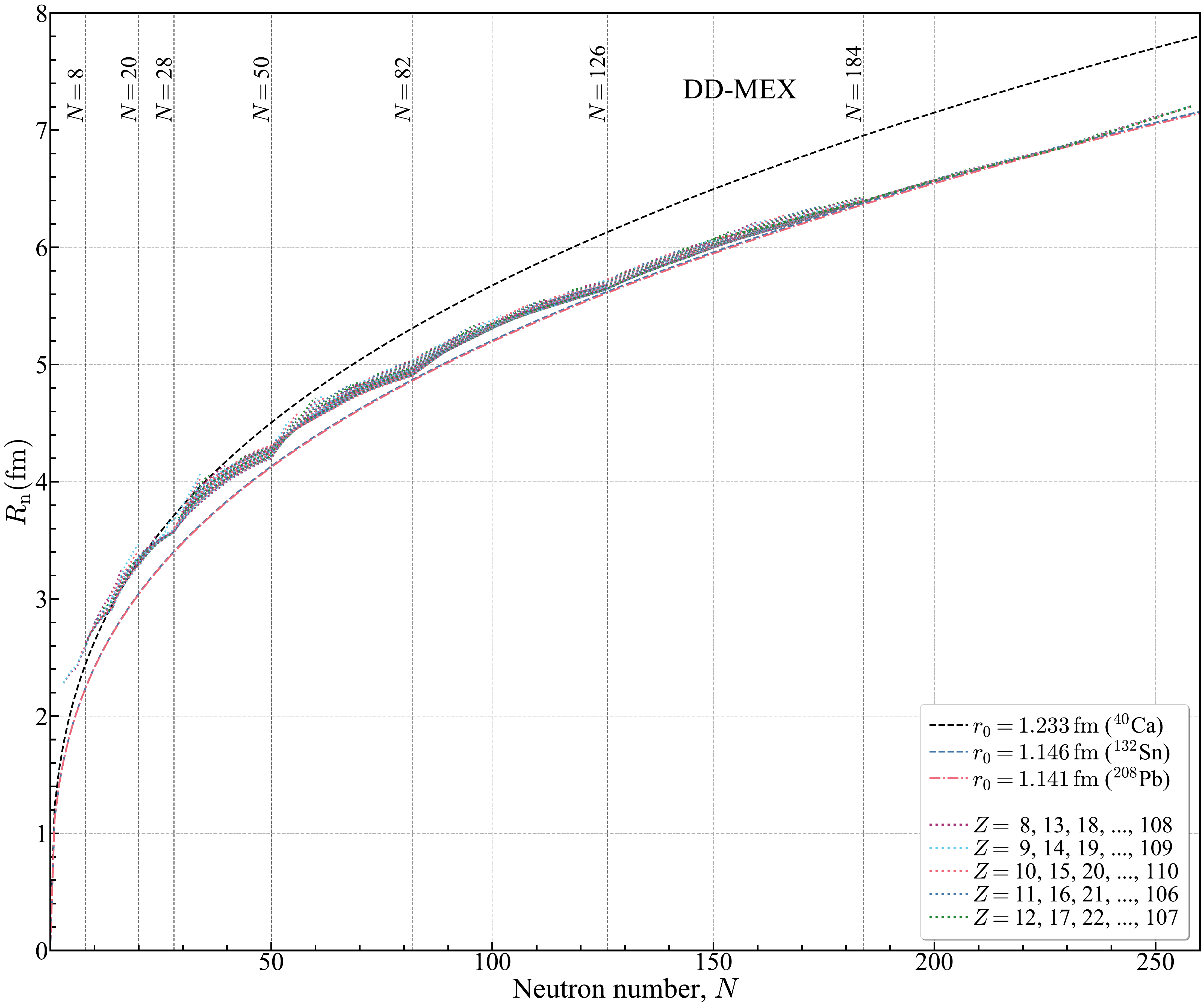}
\caption{
The root-mean-square radius of neutron as a function of the neutron number for all isotopic chains of $8\!\leq Z\!\leq 110$ calculated from the RHB approach with the DD-MEX interaction. The empirical $R_{\rm n}\!=\!r_{0} N^{1/3}$ are plotted using the $r_{0}(^{40}\mathrm{Ca})$, $r_{0}(^{132}\mathrm{Sn})$ and $r_{0}(^{208}\mathrm{Pb})$ calculated from DD-MEX (Table~\ref{tab:r0}).}
\label{fig:Rn_DDMEX_all}
\end{figure}
\end{landscape}


\noindent 
The root-mean-square deviations of one- (two-) neutron separation energies are $0.962$~MeV ($1.300$~MeV), $0.920$~MeV ($1.483$~MeV), $1.010$~MeV ($1.544$~MeV), and $0.993$~MeV ($1.753$~MeV) for PC-L3R, PCX, DD-MEX, and DD-PCX, respectively. The deviations of theoretical $S_\mathrm{n}$ and $S_\mathrm{2n}$ from the available experimental counterparts increase at the regions further away from the proton magic numbers, 8, 20, 28, 50, and 82, indicating the important role of deformation in these regions. 

We also calculate the root-mean-square radii of matter ($R_\mathrm{m}$), of neutron ($R_\mathrm{n}$), of proton ($R_\mathrm{p}$), and of charge ($R_\mathrm{c}$) distributions using the RHB approach with the PC-L3R, or PC-X, or DD-MEX, or DD-PCX interactions. Then, theoretical $R_\mathrm{c}$ are compared with the available experimental charge radii (932 nuclei) compiled by Angeli \emph{et al}.~\cite{ADNDT2013Angeli}. The rms of comparing the experimental and theoretical charge radii produced from PC-L3R, PC-X, DD-MEX, and DD-PCX are $0.035$~fm, $0.037$~fm, $0.034$~fm, and $0.035$~fm, respectively. Meanwhile, the rrs of comparing the experimental and theoretical charge radii generated from PC-L3R, PC-X, DD-MEX, or DD-PCX are in the range of $0.765$-$0.864$~\%. 
The root-mean-square radii of neutron distributions $R_\mathrm{n}$ for all nuclei of $8\!\leq\! Z\!\leq\! 110$ are shown in Fig.~\ref{fig:Rn_PCL3R_all}, \ref{fig:Rn_PCX_all}, \ref{fig:Rn_DDPCX_all}, and \ref{fig:Rn_DDMEX_all}. Theoretical $R_\mathrm{n}$ are plotted against the empirical $R_{\rm n}\!=\!r_{0} N^{1/3}$. The systematic trends of the theoretical root-mean-square radii of neutron distributions produced from PC-L3R and PC-X closely follow the empirical $R_\mathrm{n}$, except for the region of extreme neutron excess, whereas DD-PCX yields the trend lower than the empirical $R_\mathrm{n}$ at the $N\!<\!150$ region due to the soft symmetry energy property of DD-PCX.


Overall, we explore the nuclear ground state properties using the spherical RHB approach with the latest covariant density functionals, PC-L3R, PC-X, DD-MEX, and DD-PCX, and the respective separable pairing force. The borders of proton and neutron excesses are deduced from the theoretical nucleon separation energies and Fermi surfaces. At the spherical nuclear regions, the theoretical ground state properties are somehow encouraging as these predicted results are close to the experimental data. Meanwhile, the present work also implies the importance of considering deformation, continuum effects, and beyond mean-field in the future works. 

\bigskip
\noindent
{\bf Declaration of competing interest}\\

The authors declare that they have no known competing financial interests or personal relationships that could have appeared to influence the work reported in this paper.\\

\bigskip
\noindent
{\bf Data availability}\\

Electronic data (ASCII format) will be made available on request or be provided as downloadable data files at \href{https://github.com/lamyihua/nuclear-ground-state-properties}{https://github.com/lamyihua/nuclear-ground-state-properties}.\\

\newpage
\noindent
{\bf Acknowledgments}\\

We are deeply grateful to the anonymous reviewer for providing thoughtful and constructive suggestions on improving the manuscript. 
This work was financially supported by 
the Strategic Priority Research Program of Chinese Academy of Sciences (CAS, Grant Nos. XDB34020100) 
and National Natural Science Foundation of China (No. 11775277). 
We deeply appreciate the computing resources, i.e., PHYS\_T3 cluster and Distributed Cloud resources (FDR5 cluster), provided by the Institute of Physics and Academia Sinica Grid-computing Center of Academia Sinica (ASGC, Grant No. AS-CFII-112-103), Taiwan.
YHL gratefully acknowledges the financial supports from the Chinese Academy of Sciences President's International Fellowship Initiative (No. 2019FYM0002) and appreciates the laptop (Dell M4800) partially sponsored by Pin-Kok Lam and Fong-Har Tang during the pandemic of COVID-19.
PR gratefully acknowledges financial support from the Deutsche Forschungsgemeinschaft (DFG, German Research Foundation) under Germany's Excellence Strategy EXC-2094-390783311. 

\printcredits


\bibliographystyle{elsarticle-num-names}

\bibliography{bibliography}

\newpage
\clearpage


\bigskip
\renewcommand{\arraystretch}{1.0}

\section*{Explanation of Tables}
\label{Table_Explaination}

\begin{tabular*}{0.95\textwidth}{@{}@{\extracolsep{\fill}}lp{5.5in}@{}}
\multicolumn{2}{p{0.95\textwidth}}{{\bf Table~\ref{tab:BindingEnergy}}: Binding energies, one- and two-neutron separation energies of nuclei yielded from the RHB approach with the PC-L3R, PC-X, DD-MEX, or DD-PCX interaction, compared with available experimental (AME2020) binding energies. (Throughout this table, experimental uncertainties are omitted.)}\\

$A$	& Nuclear mass number\\

$N$	& Neutron number\\

Exp.   & The experimental data quoted from AME2020~\cite{AME2020}\\

PC-L3R & The RHB approach with the PC-L3R interaction~\cite{PLB2023Liu}\\

PC-X & The RHB approach with the PC-X interaction~\cite{PLB2020Taninah}\\

DD-MEX & The RHB approach with the DD-MEX interaction~\cite{PLB2020Taninah}\\

DD-PCX & The RHB approach with the DD-PCX interaction~\cite{PRC2019Yuksel}\\

$E_\mathrm{b}$  & Binding energy (MeV) \\

$S_\mathrm{n}$  & Single-neutron separation energy (MeV)\\

$S_\mathrm{2n}$ & Two-neutron separation energy (MeV)\\

$^*$            & The last bound nuclei of an isotopic chain based on the two-neutron and proton separation energies \\

$^\dagger$      & The last bound nuclei of an isotopic chain based on the nucleon Fermi surface, $\lambda_{i}\!>\!0, (i\!=\!\mathrm{n},\mathrm{p})$ \\

\end{tabular*}
\label{tableI}



\renewcommand{\arraystretch}{1.0}

\bigskip

\begin{tabular}{@{}p{1in}p{6in}@{}}
\multicolumn{2}{p{0.95\textwidth}}{{\bf Table~\ref{tab:radii}}: Radii of nuclei yielded from the RHB approach with the PC-L3R, PC-X, DD-MEX, or DD-PCX interaction, compared with available experimental charge radii. (Throughout this table, experimental uncertainties are omitted.)}\\
$A$	& Nuclear mass number\\

$N$	& Neutron number\\

Exp. & The experimental data quoted from the compilation of Angeli \& Marinova~\cite{ADNDT2013Angeli}\\

PC-L3R & The RHB approach with the PC-L3R interaction~\cite{PLB2023Liu}\\

PC-X   & The RHB approach with the PC-X interaction~\cite{PLB2020Taninah}\\

DD-MEX & The RHB approach with the DD-MEX interaction~\cite{PLB2020Taninah}\\

DD-PCX & The RHB approach with the DD-PCX interaction~\cite{PRC2019Yuksel}\\

$r_\mathrm{c}$ & Charge radius (fm) \\

$r_\mathrm{m}$ & Root-mean-square radius of matter (fm) \\

$r_\mathrm{n}$ & Root-mean-square radius of neutron (fm) \\

$r_\mathrm{p}$ & Root-mean-square radius of proton (fm) \\
\end{tabular}
\label{tableII}

\renewcommand{\arraystretch}{1.0}

\bigskip

\begin{tabular}{@{}p{1in}p{6in}@{}}
\multicolumn{2}{p{0.95\textwidth}}{{\bf Table~\ref{tab:Fermi_surface}}: Ground state properties of nuclei yielded from the RHB approach with the PC-L3R, PC-X, DD-MEX, or DD-PCX interaction.}\\
$A$	& Nuclear mass number\\

$N$	& Neutron number\\

PC-L3R & The RHB approach with the PC-L3R interaction~\cite{PLB2023Liu}\\

PC-X   & The RHB approach with the PC-X interaction~\cite{PLB2020Taninah}\\

DD-MEX & The RHB approach with the DD-MEX interaction~\cite{PLB2020Taninah}\\

DD-PCX & The RHB approach with the DD-PCX interaction~\cite{PRC2019Yuksel}\\

$\lambda^i_\mathrm{n}$ & Neutron Fermi surface energies (MeV) \\

$\lambda^i_\mathrm{p}$ & Proton Fermi surface energies (MeV) \\

$2j^\pi_\mathrm{n}$ & A factor of 2 times the total angular momentum of neutrons \\

$2j^\pi_\mathrm{p}$ & A factor of 2 times the total angular momentum of protons  \\
\end{tabular}
\label{tableIII}
\setlength{\LTleft}{0pt}
\setlength{\LTright}{0pt} 

\setlength{\tabcolsep}{0.5\tabcolsep}
\renewcommand{\arraystretch}{1.0}
\footnotesize 

\onecolumn
\newpage
\setlength{\LTleft}{0pt}
\setlength{\LTright}{0pt}
\setlength{\tabcolsep}{0.5\tabcolsep}
\renewcommand{\arraystretch}{1.0}
\footnotesize
\begin{longtable}{@{\extracolsep\fill}rrrrrrrrrrrrrrrrr@{}}
\caption{\label{tab:BindingEnergy}Binding energies, one- and two-neutron separation energies of nuclei yielded from the RHB approach with the PC-L3R, PC-X, DD-MEX, or DD-PCX interaction, compared with available experimental (AME2020) binding energies. See page \pageref{Table_Explaination} for Explanation of Tables. The electronic data (ASCII~format) is provided as a downloadable data file at\\ \url{https://github.com/lamyihua/nuclear-ground-state-properties}.} \\
\hline 
$A$&$N$&\multicolumn{3}{c}{Exp.}&\multicolumn{3}{c}{PC-L3R} & \multicolumn{3}{c}{PC-X} & \multicolumn{3}{c}{DD-MEX} & \multicolumn{3}{c}{DD-PCX} \\
 &  & $E_b$ & $S_\mathrm{n}$ & $S_\mathrm{2n}$ & $E_b$ & $S_\mathrm{n}$ & $S_\mathrm{2n}$ & $E_b$ & $S_\mathrm{n}$ & $S_\mathrm{2n}$ & $E_b$ & $S_\mathrm{n}$ & $S_\mathrm{2n}$ & $E_b$ & $S_\mathrm{n}$ & $S_\mathrm{2n}$ \\
\hline
\endfirsthead
\caption[]{(continued)}\\
\hline 
&&\multicolumn{3}{c}{Exp.}&\multicolumn{3}{c}{PC-L3R} & \multicolumn{3}{c}{PC-X} & \multicolumn{3}{c}{DD-MEX} & \multicolumn{3}{c}{DD-PCX} \\
$A$ & $N$ & $E_b$ & $S_\mathrm{n}$ & $S_\mathrm{2n}$ & $E_b$ & $S_\mathrm{n}$ & $S_\mathrm{2n}$ & $E_b$ & $S_\mathrm{n}$ & $S_\mathrm{2n}$ & $E_b$ & $S_\mathrm{n}$ & $S_\mathrm{2n}$ & $E_b$ & $S_\mathrm{n}$ & $S_\mathrm{2n}$ \\
\hline
\endhead
\multicolumn{15}{l}{ Oxygen $(Z= 8 )$} \\
 11  &    3  &   34.786  &          &          &   37.210$^{\dag }$  &          &          &   37.181$^{\dag }$  &          &          &   34.267$^{\dag }$  &          &          &   33.544$^{\dag }$  &          &          \\
 12  &    4  &   58.584  &   23.798  &          &   59.825$^{\dag }$  &   22.615  &          &   59.385$^{\dag }$  &   22.204  &          &   57.355$^{\dag }$ &   23.088  &          &   57.892$^{\dag }$  &   24.348  &          \\
 13  &    5  &   75.553  &   16.969  &   40.768  &   77.724  &   17.899  &   40.514  &   77.509  &   18.124  &   40.328  &   76.216  &   18.861  &   41.949  &   76.365  &   18.473  &   42.821  \\
 14  &    6  &   98.732  &   23.179  &   40.148  &  101.780  &   24.056  &   41.955  &  100.259  &   22.750  &   40.874  &   98.479  &   22.263  &   41.124  &   98.235  &   21.870  &   40.343  \\
 15  &    7  &  111.955  &   13.223  &   36.402  & 112.735  &  10.955  &  35.011  & 112.359  &  12.100  &  34.850  &  112.664  &   14.185  &   36.448  &  112.584  &   14.349  &   36.219  \\
 16  &    8  &  127.619  &   15.664  &   28.887  &  127.139  &  14.404  &   25.359  &  126.827  &  14.468  &   26.568  &  127.599  &   14.935  &   29.120  &  127.300  &   14.716  &   29.065  \\
 17  &    9  &  131.762  &    4.143  &   19.807  &  132.345  &    5.206  & 19.610  &  132.097  &    5.270  &  19.738  &  132.562  &    4.963  &   19.898  &  132.328  &    5.028  &   19.744  \\
\end{longtable}
\newpage

\newpage
\setlength{\LTleft}{0pt}
\setlength{\LTright}{0pt}
\setlength{\tabcolsep}{0.5\tabcolsep}
\renewcommand{\arraystretch}{1.0}
\footnotesize
\begin{longtable}{@{\extracolsep\fill}rrrrrrrrrrrrrrrrrrr@{}}
\caption{\label{tab:radii}Radii of nuclei yielded from the RHB approach with the PC-L3R, PC-X, DD-MEX, or DD-PCX interaction, compared with available experimental charge radii. See page \pageref{Table_Explaination} for Explanation of Tables. The electronic data (ASCII~format) is provided as a downloadable data file at\\ \url{https://github.com/lamyihua/nuclear-ground-state-properties}.}\\
\hline\\
&&\multicolumn{1}{c}{Exp.}&\multicolumn{4}{c}{PC-L3R} & \multicolumn{4}{c}{PC-X} & \multicolumn{4}{c}{DD-MEX} & \multicolumn{4}{c}{DD-PCX} \\
$A$ & $N$ & \multicolumn{1}{c}{$R_\mathrm{c}$} & \multicolumn{1}{c}{$R_\mathrm{c}$} & \multicolumn{1}{c}{$R_\mathrm{m}$} & \multicolumn{1}{c}{$R_\mathrm{n}$} & \multicolumn{1}{c}{$R_\mathrm{p}$} & \multicolumn{1}{c}{$R_\mathrm{c}$} & \multicolumn{1}{c}{$R_\mathrm{m}$} & \multicolumn{1}{c}{$R_\mathrm{n}$} & \multicolumn{1}{c}{$R_\mathrm{p}$} & \multicolumn{1}{c}{$R_\mathrm{c}$} & \multicolumn{1}{c}{$R_\mathrm{m}$} & \multicolumn{1}{c}{$R_\mathrm{n}$} & \multicolumn{1}{c}{$R_\mathrm{p}$} & \multicolumn{1}{c}{$R_\mathrm{c}$} & \multicolumn{1}{c}{$R_\mathrm{m}$} & \multicolumn{1}{c}{$R_\mathrm{n}$} & \multicolumn{1}{c}{$R_\mathrm{p}$} \\
\hline\\
\endfirsthead
\caption[]{(continued)}\\
\hline\\
&&\multicolumn{1}{c}{Exp.}&\multicolumn{4}{c}{PC-L3R} & \multicolumn{4}{c}{PC-X} & \multicolumn{4}{c}{DD-MEX} & \multicolumn{4}{c}{DD-PCX} \\
$A$ & $N$ & \multicolumn{1}{c}{$R_\mathrm{c}$} & \multicolumn{1}{c}{$R_\mathrm{c}$} & \multicolumn{1}{c}{$R_\mathrm{m}$} & \multicolumn{1}{c}{$R_\mathrm{n}$} & \multicolumn{1}{c}{$R_\mathrm{p}$} & \multicolumn{1}{c}{$R_\mathrm{c}$} & \multicolumn{1}{c}{$R_\mathrm{m}$} & \multicolumn{1}{c}{$R_\mathrm{n}$} & \multicolumn{1}{c}{$R_\mathrm{p}$} & \multicolumn{1}{c}{$R_\mathrm{c}$} & \multicolumn{1}{c}{$R_\mathrm{m}$} & \multicolumn{1}{c}{$R_\mathrm{n}$} & \multicolumn{1}{c}{$R_\mathrm{p}$} & \multicolumn{1}{c}{$R_\mathrm{c}$} & \multicolumn{1}{c}{$R_\mathrm{m}$} & \multicolumn{1}{c}{$R_\mathrm{n}$} & \multicolumn{1}{c}{$R_\mathrm{p}$} \\
\hline\\
\endhead
\multicolumn{15}{l}{ Oxygen $(Z= 8 )$} \\
 11  &    3  &           &  3.2340 &  2.9242 &  2.2737 &  3.1335 &  3.2416 &  2.9302 &  2.2734 &  3.1413 &  3.3461 &  3.0159 &  2.2804 &  3.2490 &  3.2319 &  2.9204 &  2.2639 &  3.1313 \\
 12  &    4  &           &  3.0218 &  2.7351 &  2.3364 &  2.9140 &  3.0270 &  2.7391 &  2.3371 &  2.9194 &  3.0276 &  2.7396 &  2.3374 &  2.9200 &  2.9670 &  2.6978 &  2.3472 &  2.8571 \\
 13  &    5  &           &  2.8761 &  2.6170 &  2.3655 &  2.7626 &  2.8837 &  2.6240 &  2.3708 &  2.7705 &  2.8614 &  2.6149 &  2.3879 &  2.7472 &  2.8415 &  2.6108 &  2.4141 &  2.7266 \\
 14  &    6  &           &  2.7860 &  2.4964 &  2.2461 &  2.6687 &  2.7787 &  2.5133 &  2.3016 &  2.6611 &  2.7584 &  2.5437 &  2.4096 &  2.6398 &  2.7702 &  2.5689 &  2.4535 &  2.6522 \\
 15  &    7  &           &  2.7696 &  2.5978 &  2.5350 &  2.6516 &  2.7744 &  2.6034 &  2.5413 &  2.6566  &  2.7425 &  2.5733 &  2.5150 &  2.6233 &  2.7594 &  2.5961 &  2.5440 &  2.6409 \\
 16  &    8  &   2.6991  &   2.7676  &   2.6379  &   2.6263  &   2.6494  &   2.7698  &   2.6404  &   2.6291  &   2.6518  &   2.7350  &   2.6001  &   2.5847  &   2.6153  &   2.7594  &   2.6270  &   2.6131  &   2.6409  \\
 17  &    9  &   2.6932  &   2.7597  &   2.6903  &   2.7332  &   2.6412  &   2.7615  &   2.6918  &   2.7343  &   2.6431  &   2.7322  &   2.6610  &   2.7035  &   2.6125  &   2.7584  &   2.6823  &   2.7196  &   2.6398  \\
\end{longtable}
\newpage

\newpage
\setlength{\LTleft}{0pt}
\setlength{\LTright}{0pt}
\setlength{\tabcolsep}{0.5\tabcolsep}
\renewcommand{\arraystretch}{1.0}
\footnotesize
\begin{longtable}{@{\extracolsep\fill}rrrrrrrrrrrrrrrrrr@{}}
\caption{\label{tab:Fermi_surface}Ground state properties of nuclei yielded from the RHB approach with the PC-L3R, PC-X, DD-MEX, or DD-PCX interaction. See page \pageref{Table_Explaination} for Explanation of Tables. The electronic data (ASCII~format) is provided as a downloadable data file at\\ \url{https://github.com/lamyihua/nuclear-ground-state-properties}.}\\
\hline\\
&&\multicolumn{4}{c}{PC-L3R} & \multicolumn{4}{c}{PC-X} & \multicolumn{4}{c}{DD-MEX} & \multicolumn{4}{c}{DD-PCX} \\
$A$ & $N$ & $\lambda^i_\mathrm{n}$ & $\lambda^i_\mathrm{p}$   & $2j^\pi_\mathrm{n}$ & $2j^\pi_\mathrm{p}$ & $\lambda^j_\mathrm{n}$ & $\lambda^j_\mathrm{p}$ & $2j^\pi_\mathrm{n}$ & $2j^\pi_\mathrm{p}$ & $\lambda^k_\mathrm{n}$ & $\lambda^k_\mathrm{p}$ & $2j^\pi_\mathrm{n}$ & $2j^\pi_\mathrm{p}$ & $\lambda^l_\mathrm{n}$ & $\lambda^l_\mathrm{p}$ & $2j^\pi_\mathrm{n}$ & $2j^\pi_\mathrm{p}$ \\
\hline\\
\endfirsthead
\caption[]{(continued)}\\
\hline\\
&&\multicolumn{4}{c}{PC-L3R} & \multicolumn{4}{c}{PC-X} & \multicolumn{4}{c}{DD-MEX} & \multicolumn{4}{c}{DD-PCX} \\
$A$ & $N$ & $\lambda^i_\mathrm{n}$ & $\lambda^i_\mathrm{p}$   & $2j^\pi_\mathrm{n}$ & $2j^\pi_\mathrm{p}$ & $\lambda^j_\mathrm{n}$ & $\lambda^j_\mathrm{p}$ & $2j^\pi_\mathrm{n}$ & $2j^\pi_\mathrm{p}$ & $\lambda^k_\mathrm{n}$ & $\lambda^k_\mathrm{p}$ & $2j^\pi_\mathrm{n}$ & $2j^\pi_\mathrm{p}$ & $\lambda^l_\mathrm{n}$ & $\lambda^l_\mathrm{p}$ & $2j^\pi_\mathrm{n}$ & $2j^\pi_\mathrm{p}$ \\
\hline\\
\endhead
\multicolumn{15}{l}{ Oxygen $(Z= 8 )$} \\
 11  &    3  &  -24.806 &   1.810 &  3$^-$ &  0$^+$ & -24.857 &   1.852 &  3$^-$ &  0$^+$ & -25.889 &   2.165 &  3$^-$ &  0$^+$ & -27.207 &   2.748 &  3$^-$ &  0$^+$ \\
 12  &    4  &  -19.467 &   0.845 &  0$^+$ &  0$^+$ & -19.380 &   0.777 &  0$^+$ &  0$^+$ & -20.199 &   0.838 &  0$^+$ &  0$^+$ & -20.848 &   0.679 &  0$^+$ &  0$^+$ \\
 13  &    5  &  -16.994 &  -0.390 &  3$^-$ &  0$^+$ & -16.851 &  -0.248 &  3$^-$ &  0$^+$ & -17.637 &  -0.309 &  3$^-$ &  0$^+$ & -17.866 &  -0.142 &  3$^-$ &  0$^+$ \\
 14  &    6  &  -15.430 &  -0.843 &  0$^+$ &  0$^+$ & -15.509 &  -1.023 &  0$^+$ &  0$^+$ & -17.352 &  -2.551 &  0$^+$ &  0$^+$ & -17.329 &  -3.130 &  0$^+$ &  0$^+$ \\
 15  &    7  &  -12.991 &  -4.459 &  1$^-$ &  0$^+$ & -12.782 &  -4.511 &  1$^-$ &  0$^+$ & -13.426 &  -4.449 &  1$^-$ &  0$^+$ & -12.819 &  -4.348 &  1$^-$ &  0$^+$ \\
 16  &    8  &  -10.420 &  -6.815 &  0$^+$ &  0$^+$ & -10.392 &  -6.787 &  0$^+$ &  0$^+$ & -11.314 &  -6.856 &  0$^+$ &  0$^+$ & -10.429 &  -6.796 &  0$^+$ &  0$^+$ \\
 17  &    9  &  -10.273 &  -8.751 &  5$^+$ &  0$^+$ & -10.710 &  -8.779 &  5$^+$ &  0$^+$ & -10.323 &  -8.859 &  5$^+$ &  0$^+$ & -10.129 &  -8.807 &  5$^+$ &  0$^+$ \\
\end{longtable}
\newpage




\end{document}